\documentclass[twocolumn,superscriptaddress,amsmath,amssymb]{revtex4}
\pdfoutput=1
\usepackage{graphicx}
\usepackage{dcolumn,xcolor}
\usepackage{amsmath} 
\usepackage{amssymb}
\usepackage{amsfonts}
\usepackage{bm}
\usepackage{overpic}
\usepackage{array}
\usepackage{hyperref}
\usepackage[latin1]{inputenc}
\usepackage{graphicx}
\usepackage{csvsimple}
\usepackage{ifsym,stmaryrd}
\usepackage{wasysym}
\usepackage{pifont}
\DeclareFontEncoding{LS1}{}{}
\DeclareFontSubstitution{LS1}{stix}{m}{n}
\DeclareSymbolFont{symbols4}{LS1}{stixbb}{m}{it}
\DeclareMathSymbol{\varhexagonblack}{\mathord}{symbols4}{"DD}

\begin{document}

\title{Spreading, pinching, and coalescence: the Ohnesorge units}

\author{Marc A.~Fardin}
\altaffiliation[Corresponding author ]{}
\email{marc-antoine.fardin@ijm.fr}
\affiliation{Universit\'{e} de Paris, CNRS, Institut Jacques Monod, F-75013 Paris, France}
\affiliation{The Academy of Bradylogists}
\author{Mathieu~Hautefeuille}
\affiliation{Universit\'{e} de Paris, CNRS, Institut Jacques Monod, F-75013 Paris, France}
\affiliation{Facultad de Ciencias, Departamento de Fisica, Universidad Nacional Aut\'{o}noma de M\'{e}xico,Ciudad Universitaria, DF 04510, Mexico}
\affiliation{Institut de Biologie Paris Seine, UMR 7622, Sorbonne Universit\'{e}, 7 quai Saint Bernard, 75005 Paris, France}
\author{Vivek~Sharma}
\affiliation{Department  of Chemical  Engineering, University of Illinois at Chicago, Chicago, Illinois 60608, UnitedStates}
\affiliation{The Academy of Bradylogists}

\date{\today}

\begin{abstract}
Understanding the kinematics and dynamics of spreading, pinching, and coalescence of drops is critically important for a diverse range of applications involving spraying, printing, coating, dispensing, emulsification, and atomization. Hence experimental studies visualize and characterize the increase in size over time for drops spreading over substrates, or liquid bridges between coalescing drops, or the decrease in the radius of pinching necks during drop formation. Even for Newtonian fluids, the interplay of inertial, viscous, and capillary stresses can lead to a number of scaling laws, with three limiting self-similar cases: visco-inertial (VI), visco-capillary (VC) and inertio-capillary (IC). Though experiments are presented as examples of the methods of dimensional analysis, the lack of precise values or estimates for pre-factors, transitions, and scaling exponents presents difficulties for quantitative analysis and material characterization. In this tutorial review, we reanalyze and summarize an elaborate set of landmark published experimental studies on a wide range of Newtonian fluids. We show that moving beyond VI, VC, and IC units in favor of intrinsic timescale and lengthscale determined by all three material properties (viscosity, surface tension and density), creates a complementary system that we call the Ohnesorge units. We find that in spite of large differences in topological features, timescales, and material properties, the analysis of spreading, pinching and coalescing drops in the Ohnesorge units results in a remarkable collapse of the experimental datasets, highlighting the shared and universal features displayed in such flows.
\end{abstract}

\maketitle

In a 1936 study on ``the formation of drops at nozzles'', Ohnesorge showed that jetting of Newtonian fluids can be classified into three cases on a plot with two dimensionless groups~\cite{Ohnesorge1936}: the Reynolds number ($\text{Re}_*$) as the x-axis and $Z=Z(\text{Re}_*)$, referred in the modern literature as Ohnesorge number (Oh) as the y-axis~\cite{Eggers1997,Basaran2002,McKinley2011}. The succinct plot incorporated: (i) axisymmetric breakup, investigated theoretically and experimentally by Rayleigh for inviscid fluids~\cite{Rayleigh1879}, (ii) wavy breakup studied experimentally by Haenlein~\cite{Haenlein1931}, and theoretically by Weber~\cite{Weber1931}, and (iii) atomization~\cite{Haenlein1931,Weber1931}. The transition from a laminar to a turbulent jet with increasing imposed velocity ($U$) occurs with enhancement in $\text{Re}_*\equiv \rho D U/\eta$ (the star subscript is here to distinguish this Reynolds number from the one associated with boundary layers, which we will discuss in the review). However, the Ohnesorge plot Re$_*$ vs Oh shows that the transitions between jetting regimes depend on a dimensionless group, $\text{Oh}=\eta/(\rho \Gamma D)^\frac{1}{2}$ that is independent of $U$. The Ohnesorge number Oh incorporates three \textit{intrinsic} fluid properties: viscosity ($\eta$, with dimensions ${M}.{L}^{-1}.{T}^{-1}$), surface tension ($\Gamma$, ${M}.{T}^{-2}$), and density ($\rho$, ${M}.{L}^{-3}$), and includes one \textit{extrinsic} lengthscale ($D$, which can be the nozzle diameter or drop size).  The Ohnesorge number can alternatively be represented as the square-root of the ratio of an intrinsic length, $\ell_o\equiv \eta^2/\rho\Gamma$ and the extrinsic length $D$. Fascinatingly, Haenlein and Weber discussed the breakup time of viscous fluids in a dimensionless form, scaled by an intrinsic time, $\tau_o\equiv \eta^3/\rho\Gamma^2$. 

In this review, we revisit and reanalyze the experimental universe of spreading, coalescence and pinching drops to show that the two intrinsic measures of length and time, $\ell_o$ and $\tau_o$, referred to as the Ohnesorge units, provide a cohesive, universal, and succinct representation of the kinematics and interpretation of governing dynamics of Newtonian fluids. 

Spreading, pinching and coalescence of drops are examples of interfacial or free surface flows, primarily governed by three stresses: inertial, viscous, and capillary. The rich and complex dynamics that arise from the interplay of these three stresses are described in many excellent books and reviews~\cite{Leger1992,Stone1994,Middleman1995,Eggers1997,Oron1997,Basaran2002,McKinley2005,Yarin2006,Leal2007,Villermaux2007,Eggers2008,Bonn2009,McKinley2011,PGG2013,Snoeijer2013,Eggers2015,Kavehpour2015}. The motivations for investigating the formation of drops and their coalescence and spreading behavior ranges from an innate curiosity about the physical world around us to necessity, especially as countless applications involve jetting, spraying, atomization, coating, printing, or dispensing of liquids. The underlying challenges and opportunities lie in the intricate mathematics of nonlinear differential equations, similarity solutions, perturbation analysis, singularities, and topology~\cite{Eggers1997,Eggers2008,Barenblatt2003,Eggers2015,McKinley2011}. Even for trained fluid mechanicians, the physicochemical hydrodynamics underlying spreading, coalescence, and pinching drops requires advanced methods to describe and track interfaces and interfacial effects. The multiverse of length and time scales that must be tackled to obtain a realistic description of the phenomena make most numerical and computations efforts into arduous exercises that often require validation from experiments. High-speed imaging as well as visualization tricks (involving lighting, sparks, or strobes) provide picturesque experimental data and insights into their mathematics, hydrodynamics, and applications.

In this review, we restrict our attention to spreading, pinching, and coalescence phenomena where the influence of $\rho$, $\eta$ and $\Gamma$ can be captured using parameters of one liquid. Thus, we systematically exclude studies that investigate the influence of non-Newtonian viscosity, viscoelasticity, adsorption kinetics, Marangoni flows, disjoining pressure, and in the case of emulsions, density or viscosity ratios of inner and outer fluids. Likewise, we exclude from discussion any phenomena with externally imposed velocity, including drop impact on substrates, and pinching and coalescence of drops moving under the influence of an external flow. The choices restrict us to the three intrinsic parameters ($\rho$, $\Gamma$, $\eta$) and one extrinsic lengthscale ($D$), or to the world of Ohnesorge units, such that each experimental observation included in the study corresponds to a fixed value of Oh, which will be color-coded in all figures, with blue shades corresponding to $\text{Oh}>1$ and red shades corresponding to $\text{Oh}<1$ (the full color scale is given in the final figure of the manuscript).

Experiments that visualize and analyze the change in contact area of a drop undergoing spreading, a liquid neck experiencing pinching or break-up, or liquid bridge expanding due to coalescence, often reveal that the size variation follows power laws of the form: $d\sim t^\alpha$. Even though most experimental analysis show that values of $\alpha$ equal to $\frac{2}{3}$ (inertio-capillary), 1 (visco-capillary), and $\frac{1}{2}$ (visco-inertial) are typically manifested if the variable size $d$ is much smaller than the drop size $D$, significant disagreements remain about the precise value of pre-factors. Dynamics such that $d$ is comparable with $D$ usually reveal a greater diversity of scalings depending explicitly on $D$. Often more than one power law is manifested in the same experiment~\cite{Castrejon2015,Dinic2019,Dinic2020,Eggers1997,McKinley2005}. In such cases, many questions arise about the dominant or contributing stresses, the role or measurement of material properties, and choosing the lengthscales and timescales to observe universalities that help to elucidate underlying physical mechanisms. 

Motivated by these longstanding challenges, we took up the arduous but rewarding task of collecting, replotting, reanalyzing, and collating experimental data sets that explore spreading, coalescence, and pinching. Here we provide an extensive and unprecedented stock of data and analysis, with a detailed investigation of pre-factors, transition lengths and times, and power law exponents (see ESI online$^\dag$, with plots as well as the numerical value of data sets and material properties included). We proceed to show that by analyzing spreading, coalescing, and pinching drops in Ohnesorge units, the experimental data sets measured using a wide range of liquids collapse onto universal scaling laws. The final figure here illustrates the beauty of the science of scaling, or dimensional analysis, and a homage to Ohnesorge. We acknowledge an immense debt to many exquisite experiments we chose to employ here and to extensive numerical and computation work that have helped advance our understanding. We anticipate that this review will facilitate a deeper understanding of the pragmatic, pedagogical, and physically intuitive description of the power laws underlying spreading, pinching, and coalescence encountered in our daily life, in nature and industry.

\section{Visco-inertial scaling}
\begin{figure*}
\centering
\begin{overpic}[abs,unit=1mm,width=15cm]{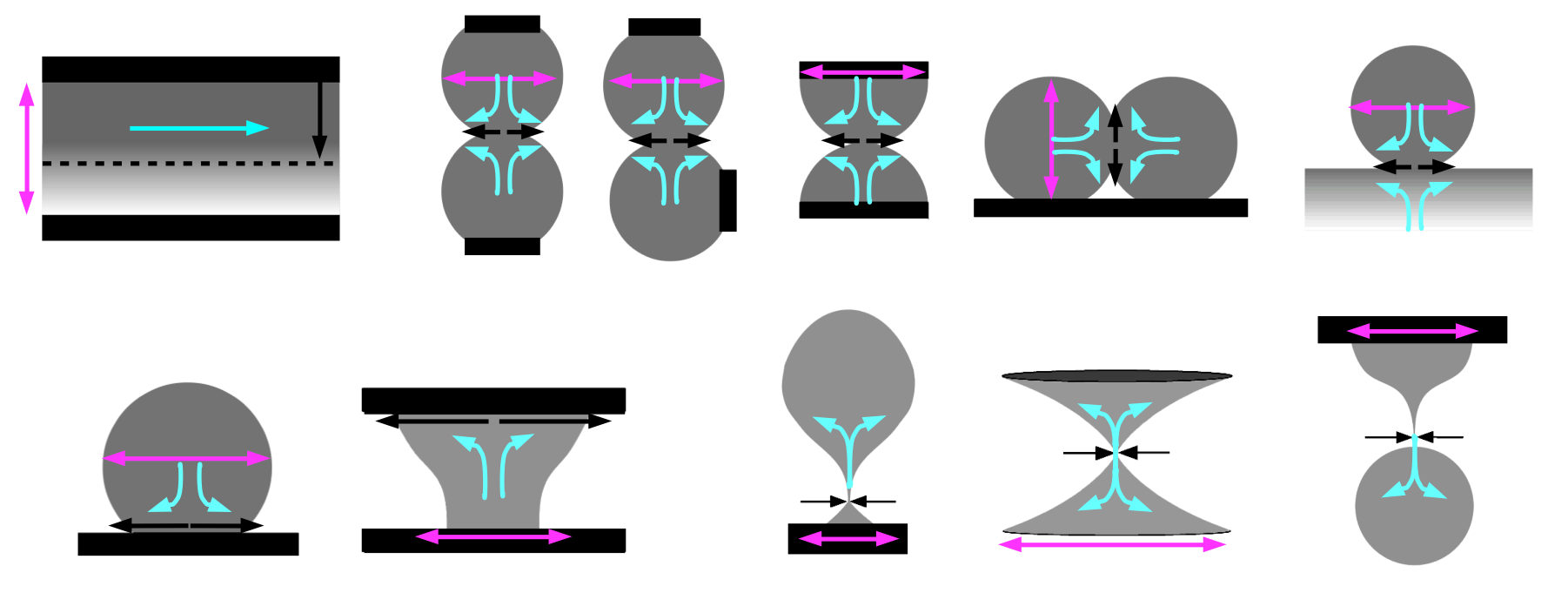}
\put(0,55){(a)}
\put(35,55){(b)}
\put(0,25){(c)}
\put(70,25){(d)}
\end{overpic}
\caption{Schematics of the experimental setups discussed in the review: (a) Confined boundary layer, (b) coalescence, (c) spreading, (d) pinching. The magenta arrows highlight the extent of the extrinsic length $D$. For all schematics the arrow has length $2D$, except for the boundary layer where the arrow has length $D$. For spreading set-ups, the precise extrinsic length is $D=(3\Omega/4\pi)^\frac{1}{3}$, where $\Omega$ is the volume of the drop. The black arrows show the direction of extension of the length $d$. For pinching the arrows are reversed. The blue arrows schematize the main flow in each configuration. 
\label{figsetups}}
\end{figure*}
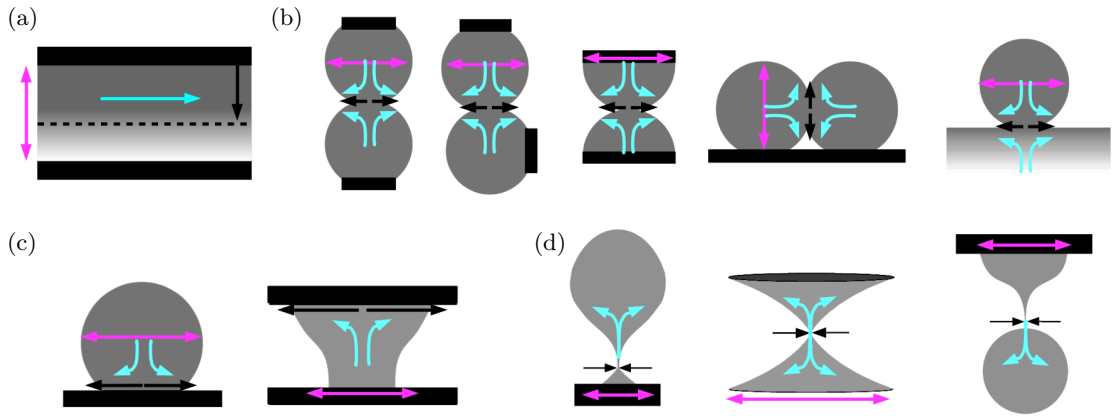 
The ratio between viscosity and density is the kinematic viscosity $\nu\equiv\eta/\rho$. For instance, for water $\nu\simeq 10^{-6}$~m$^2$/s. The dimensions of this quantity are $[\nu]={L}^{2}.{T}^{-1}$, and so the kinematic viscosity can also be understood as the momentum diffusivity. If space is associated with a single size $d$ and time $t$, one can express the kinematic viscosity as $\nu\propto d^2.t^{-1}$, where we have used $d$ and $t$ to `measure' the dimensions ${L}$ and ${T}$. This dimensional relationship can be recast as a `simple spreading-like law'~\cite{Landau1959}: 
\begin{equation}
d = \delta_{vi} \Big( \frac{\eta}{\rho}\Big)^\frac{1}{2} t^\frac{1}{2}
\label{BL}
\end{equation}
\noindent where $\delta_{vi}$ is a dimensionless `constant of order 1', which more rigorously means that the variations of $\delta_{vi}$ with $d$, $t$, $\eta$ or $\rho$ can be at most logarithmic. The subscript `vi' stands for `visco-inertial'. Note that in fluid dynamics, the adjective `inertial' is often used to describe dynamics depending explicitly on the density, and we shall use this convention too. The scaling law in Eq.~\ref{BL} describes the spreading of a boundary layer, \textit{i.e.} the size of the sheared part of a fluid, at time $t$ after it started being sheared. The exact value of the pre-factor $\delta_{vi}$ will depend on the type of boundary conditions. For instance, if the fluid is sheared by a flat plate $\delta_{vi}\simeq 5$~\cite{Blasius1907}. 

Because $d$ and $t$ are connected by a power law $d\sim t^\alpha$, the speed of the leading edge of the boundary layer is $v=\partial d/\partial t= \alpha d/t$, here with $\alpha=\frac{1}{2}$. With this definition, the spreading law of the boundary layer can also be expressed by a famous dimensionless number:
\begin{align}
&d=\delta_{vi} (\eta/\rho)^{\frac{1}{2}}  t^{\frac{1}{2}}\\
\Leftrightarrow\quad &d= \delta_{vi} (\eta/\rho)^{\frac{1}{2}}  (d/2v)^{\frac{1}{2}}\\
\Leftrightarrow\quad &(\rho/\eta)^{\frac{1}{2}} d^{\frac{1}{2}} v^{\frac{1}{2}} =   \delta_{vi}/2^\frac{1}{2}\\
\Leftrightarrow\quad &\frac{\rho d v}{\eta}\equiv  \text{Re} =\frac{\delta_{vi}^2}{2}
\label{Rey}
\end{align}
The visco-inertial scaling describes dynamics keeping the Reynolds number constant. Note that the Reynolds number defined here depends on $d$ and $v$ rather than on the extrinsic length $D$ and imposed velocity $U$ as in Re$_*$. Even though Re depends on two variables, it combines them in such a way that the product is constant. If the dynamics only depend on $\eta$ and $\rho$ the number $\delta_{vi}^2/2$ is a constant `of order 1'. Note that in practice, the values of these dimensionless constants can be substantially different from 1. For instance, if $\delta_{vi}\simeq 5$, then $\delta_{vi}^2/2 \simeq 12.5$. Nevertheless, we will most often forget these factors when using the sign `$\propto$' instead of `='. For instance, we will say that the boundary layer spreading is such that Re$\propto 1$. Note that in the last line of the previous equations, we squared the left-hand side because dimensionless numbers are defined modulo an overall power. By convention, the power chosen is usually such that the factors of the dimensionless number have integer rather than fractional powers. 
  
If a flow is confined one can only expect the scaling $d\sim t^\frac{1}{2}$ to be valid up to a distance $D$ set by the confinement, as sketched in Fig.~\ref{figsetups}a. This asymptotic size of the boundary layer would be reached after a time ${\tau}_{vi} \propto \rho D^2/\eta$. Actually, the quantities $D$, $\eta$ and $\rho$ can be combined to produce a full set of units $\lbrace \rho,\eta,D\rbrace$, which may be used to derive expressions for a mass, length and time, hereafter called the visco-inertial units: 
\begin{align}
{m}_{vi}&\equiv \rho D^3\\
{\ell}_{vi} &\equiv D\\
{\tau}_{vi} &\equiv \frac{\rho D^2}{\eta}
\end{align}
With these units one can express any quantity. For instance, one can construct a stress ${\Sigma}_{vi}={m}_{vi}.{\ell}_{vi}^{-1}.{\tau}_{vi}^{-2}=\eta^2/\rho D^2 = \eta/\tau_{vi}$. 

In visco-inertial units, Eq.~\ref{BL} can be written as $d/\ell_{vi} \propto (t/\tau_{vi})^\frac{1}{2}$. Obviously, this scaling can only be valid if $d/D\lesssim 1$. This dimensionless geometric ratio will reappear throughout this review, where we call it the `size ratio' $\Lambda \equiv d/D$. This dimensionless number describes the ratio between the time-dependent size $d$ and the fixed extrinsic size $D$, which are sketched in Fig.~\ref{figsetups} for all setups discussed in the review. For $\Lambda\gtrsim 1$, $d\propto D t^0$. 

For water, assuming $D=1$~mm gives ${\tau}_{vi}\simeq 1~$s. For air, assuming $D=1$~m gives ${\tau}_{vi}\simeq 1~$day. Note that the actual crossover time depends on the pre-factor $\delta_{vi}$, since the equation $\delta_{vi} (\eta/\rho)^\frac{1}{2} t^\frac{1}{2}=D$ leads to $t=\tau_{vi}/\delta_{vi}^2$. Moreover, the transition from the diffusive regime to the asymptotic state typically includes logarithmic corrections due to `finite size effects'~\cite{Schlichting2016}, which soften the transition form $d\sim t^\frac{1}{2}$ to $d\sim t^0$, a point we shall discuss in more detail for the visco-capillary scaling.

\section{Visco-capillary scaling}
\begin{figure*}
\centering
\begin{overpic}[abs,unit=1mm,width=17cm]{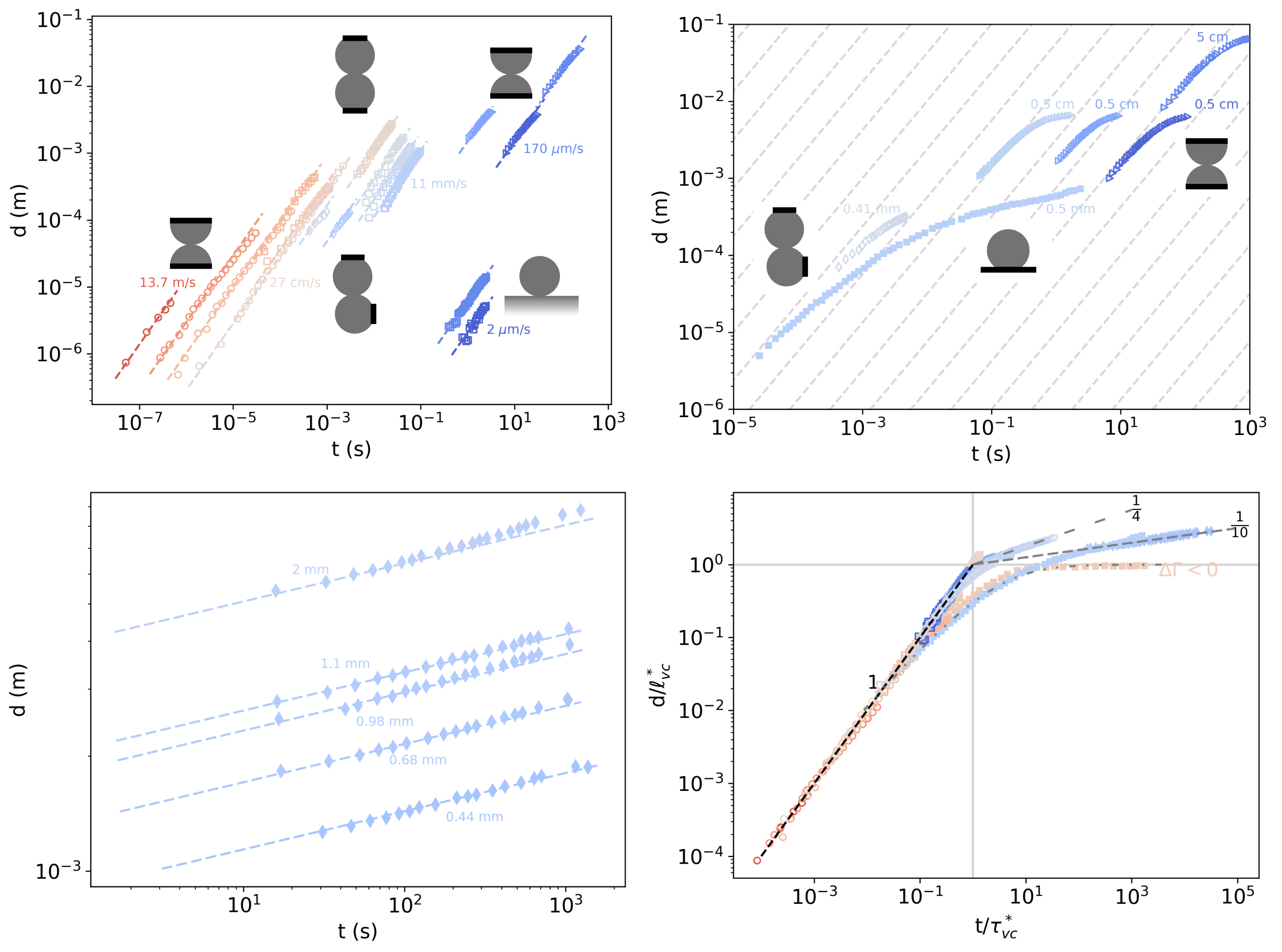}
\put(15,118){(a)}
\put(100,118){(b)}
\put(15,56){(c)}
\put(100,56){(d)}
\end{overpic}
\caption{Visco-capillary spreading (filled symbols) and coalescence (open symbols). (a) Purely visco-capillary regime of coalescence following Eq.~\ref{ViscoCap} for fluids of different viscosities and surface tensions, plotted in standard units. A few values of visco-capillary speeds are highlighted. All fitted values of $\delta_{vc}$ are given in ESI-Fig. 5$^\dag$. Note that only the portions of the data exhibiting the visco-capillary regime are shown here. This is particularly the case for the data by Paulsen \textit{et al.}~\cite{Paulsen2011}, which will be shown in their entirety in Fig.~\ref{fig3}-\ref{fig5}. Data reproduced from Aarts \textit{et al.}~\cite{Aarts2005,Aarts2008} ($\square$), Paulsen \textit{et al.} ($\circ$)~\cite{Paulsen2011}, Rahman \textit{et al.} ($\Diamond$)~\cite{Rahman2019} and Yao \textit{et al.} ($\triangleright$)~\cite{Yao2005}. (b) Approaches to the asymptotic regime for a few examples of coalescence from panel-a (open symbols) and for an example of spreading from Eddi \textit{et al.} ($\blacksquare$)~\cite{Eddi2013}. The labels give the values of the drop size $D$ for each experiment. The parallel dashed lines follow $d\sim t$. (c) Tanner regime for the late spreading of viscous drops of different sizes $D$, reproduced from Cazabat \textit{et al.} ($\blacklozenge$)~\cite{Cazabat1986}. The values of ${\delta}_{Tan}$ are given in ESI-Fig. 6a$^\dag$. Eventual departure from Tanner's regime is due to gravity~\cite{Cazabat1986}. (d) Data from panels a to c are replotted together with visco-capillary units based on the system $\lbrace \Gamma,\eta,D\rbrace$. The units are $\tau_{vc}^* \equiv \gamma_1 \tau_{vc}$, with $\tau_{vc}\equiv  \eta D/\Gamma$, and $\ell_{vc}^* \equiv \gamma_2 \ell_{vc}$, with $\ell_{vc}\equiv D$. The values of $\gamma_1$ and $\gamma_2$ for all curves are obtained from the pre-factors $\delta_{vc}$ and $\delta_{Tan}$ (see ESI section 2E for details$^\dag$). With these units the two spreading laws become $d/\ell_{vc}^* = t/\tau_{vc}^*$ (black dashed line) and $d/\ell_{vc}^*= (t/\tau_{vc}^*)^\frac{1}{10}$ (gray dashed line). Also included are coalescence data in between two plates~\cite{Yokota2011} ($\varhexagon$), in which case the late spreading follows $d/\ell_{vc}^*= (t/\tau_{vc}^*)^\frac{1}{4}$ (gray spaced dashed line). In this case, the extrinsic size $D$ is the geometric mean between the in-plane drop size and the spacing between the two plates. The dotted dashed line includes the logarithmic correction  $d/\ell_{vc}^* = (t/\tau_{vc}^*)(-0.25 \log(d/\ell_{vc}^*))$~\cite{Eddi2013}. The data set corresponding to a spreading in partial wetting conditions, is labeled by `$\Delta\Gamma<0$'~\cite{Eddi2013}, for which the asymptotic regime is $d\propto D$. Note that by definition $d/\ell_{vc}^* =\Lambda/\gamma_2$. Details of the fluid properties for all data sets in the panels of this figure are given in ESI section 1$^\dag$. The schematics on the panels illustrate the different experimental set-ups, which are discussed in more detail in ESI section 6$^\dag$. The color used for each data set gives to the value of the Ohnesorge number $\text{Oh}=\eta/(D\Gamma\rho)^\frac{1}{2}$. The full color scale is given in Fig.~\ref{fig5}.
\label{fig1}}
\end{figure*}
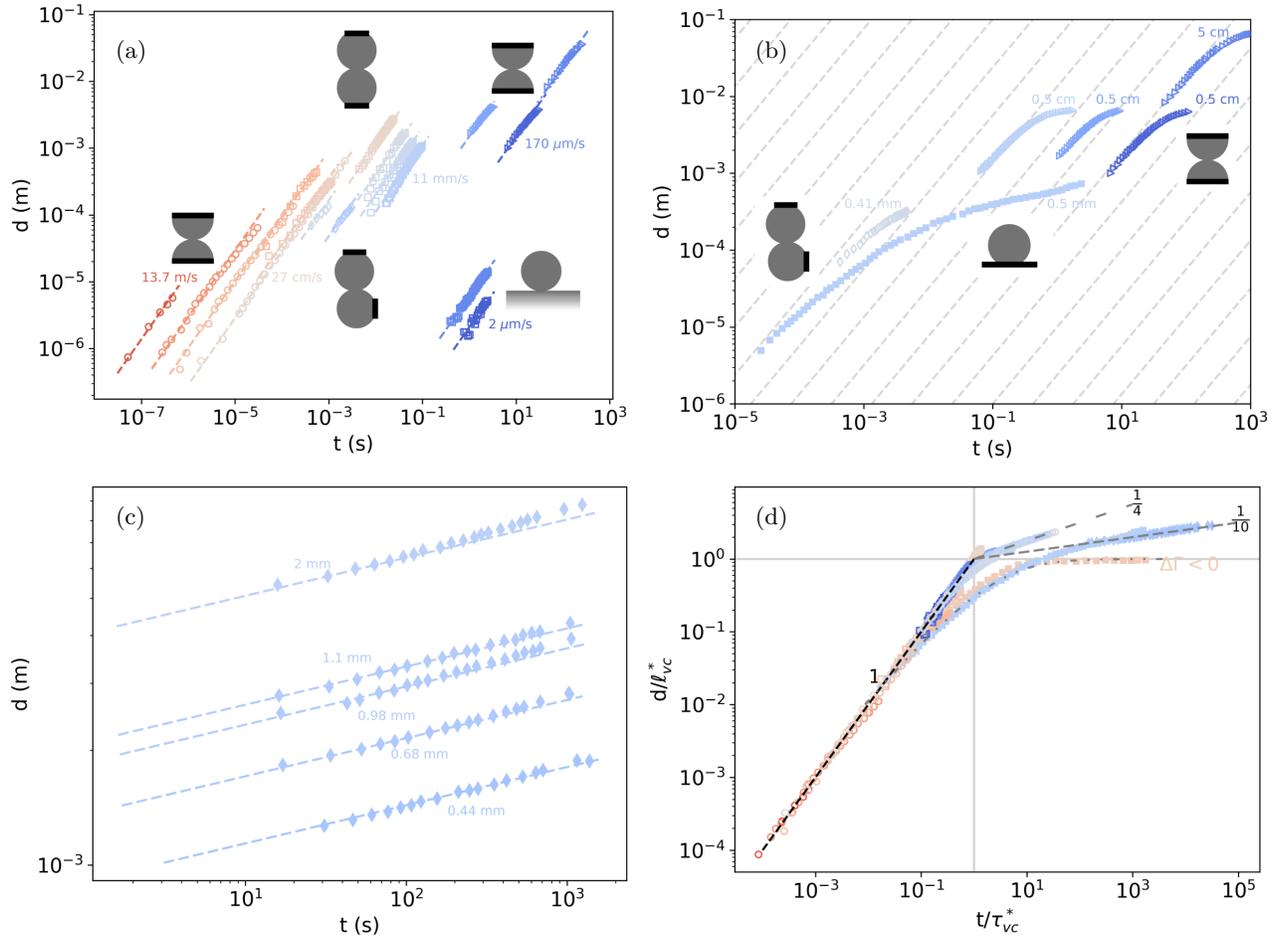 
In the example of the boundary layer, the relevant material parameters were the viscosity $\eta$ and density $\rho$. Here, we wish to discuss dynamics where density has a negligible effect, but where surface tension $\Gamma$ plays a role. Whereas the ratio of viscosity and density produces a diffusion coefficient, the ratio of surface tension and viscosity produces a speed $c\equiv\Gamma/\eta$, often called the visco-capillary speed. For instance, in water $c\simeq 10^2$~m/s (over 200 miles/hour!). The interplay of dimensions between viscosity and surface tension can be expressed as a simple spreading-like law: 
\begin{equation}
d = \delta_{vc} \frac{\Gamma}{\eta}  t
\label{ViscoCap}
\end{equation}
\noindent where the subscript `vc' stands for `visco-capillary'. This regime corresponds to a constant capillary number:
\begin{equation}
\text{Ca}\equiv \frac{\eta v}{\Gamma}=\delta_{vc}
\end{equation}
 \noindent where $v$ is the leading edge speed. 

Whereas the spreading of a boundary layer described the motion of the edge between sheared and unsheared portions of the same fluid, the visco-capillary spreading describes the motion of the front between an advancing fluid and its surrounding medium.

Striking experimental illustrations of the simple visco-capillary regime of Eq.~\ref{ViscoCap} are found in studies of drop coalescence. In this context, the growing size $d$ is that of the neck between two drops, or between a drop and a pool of the same fluid. Instances of the visco-capillary regime for fluids of different viscosities and surface tensions are shown in Fig.~\ref{fig1}a, corresponding to visco-capillary speeds between a dozen meters per second and a few microns per second~\cite{Aarts2005,Aarts2008,Paulsen2011,Rahman2019,Yao2005}. All fitted values of the pre-factor $\delta_{vc}$ are given in ESI section 2D$^\dag$. Usually $\delta_{vc}$ is slightly lower than 1. 

As in the boundary layer case, the visco-capillary regime of Eq.~\ref{ViscoCap} is only expected to last if $\Lambda\lesssim 1 $, where $D$ is now the radius of the drop before contact, as sketched in Fig.~\ref{figsetups}. When $\Lambda\gtrsim 1$, the size and shape (curvature) of the drop starts to have a significant influence on the coalescence or spreading. In Fig.~\ref{fig1}b we show later data points belonging to some experiments from Fig.~\ref{fig1}a, showing how the spreading slows down as $d$ approaches $D$. Also shown is a curve for the spreading of a viscous drop on a flat substrate (filled symbols)~\cite{Eddi2013}. In that case, $d$ is the contact radius. 

In the confined boundary layer case, the asymptotic regime was $d\propto D$. Here, the situation is more complex since the radius $d$ can actually grow beyond the initial drop radius $D$. This is most clearly illustrated for spreading droplets. Whereas the initial spreading depends only on the value of the surface tension $\Gamma$ related to the interface between the drop and the surrounding fluid, the final spreading can depend on the surface energies between the drop and the substrate ($\Gamma'$), and between the substrate and the surrounding fluid ($\Gamma''$). One usually defines a `spreading parameter' $\Delta\Gamma=\Gamma''-\Gamma-\Gamma'$~\cite{Leger1992,PGG2013}. For partial wetting, \textit{i.e.} if $\Delta\Gamma<0$, the spreading usually stops for $\Lambda\propto 1$ (as labeled in Fig.~\ref{fig1}d). The contact line can even recede in some cases~\cite{Eddi2013}. For total wetting, \textit{i.e.} if $\Delta\Gamma>0$, the surface tension $\Gamma$ between the drop and the surrounding medium dominates. In this case, the long-time behavior of the spreading often follows what is usually referred to as `Tanner's law'~\cite{Tanner1979,Leger1992}: 
\begin{equation}
d = {\delta}_{Tan} \Big(\frac{\Gamma}{\eta}\Big)^\frac{1}{10} D^\frac{9}{10} t^\frac{1}{10}
\label{Tanner}
\end{equation}
This trend keeps the following dimensionless product constant: $\text{Ca}\Lambda^9=\delta_{Tan}^{10}/10$. 

In Fig.~\ref{fig1}c Tanner's trend is shown for drops of the same fluid for various values of $D$~\cite{Cazabat1986}. Note that this scaling cannot be obtained by simple dimensional analysis, since it depends on three rather than two parameters. The value of the exponent actually depends on a crossover between the bulk of the drop and a very thin precursor film~\cite{Leger1992,Oron1997,PGG2013}. Note that gravity can also alter the late spreading~\cite{Cazabat1986}, as is apparent in Fig.~\ref{fig1}c, and as we shall discuss in the last section. 

To systematically describe the spreading and coalescence of viscous fluids, one can introduce visco-capillary units $\lbrace \Gamma,\eta,D\rbrace$: 
\begin{align}
{m}_{vc} &\equiv \frac{\eta^2 D^2}{\Gamma}\\
{\ell}_{vc} &\equiv D\\
{\tau}_{vc} &\equiv \frac{\eta D}{\Gamma}
\end{align}
Note that the stress in this system is Laplace's pressure, ${\Sigma}_{vc}={m}_{vc}.{\ell}_{vc}^{-1}.{\tau}_{vc}^{-2}= \Gamma/D$. In this system of units, the mass is not that of the volume $D^3$, instead it can be understood from Newton's law as $m_{vc}=F_{vc}/a_{vc}$, where the visco-capillary force and acceleration are respectively $F_{vc}=\Gamma D$ and $a_{vc}=\ell_{vc}.\tau_{vc}^{-2}=c^2/D$. The effective visco-capillary mass can also be understood from $E_{vc}= m_{vc} c^2$, where $c=\Gamma/\eta$ and where $E_{vc}=\Gamma D^2$ is the capillary energy.  

In visco-capillary units the timescale gives the crossover between the early and late spreading. For instance, if the viscosity and surface tensions are $\eta\simeq 60$~mPa.s and $\Gamma\simeq 60$~mN/m, and $D\simeq 0.5$~mm, one has $\tau_{vc}\simeq 0.5$~ms. As in the boundary layer example, the actual crossover radius and time must include dimensionless constants. Matching Eq.~\ref{ViscoCap} and~\ref{Tanner} would yield $\tau_{vc}^* \equiv \gamma_1 {\tau}_{vc}$ and $\ell_{vc}^* \equiv \gamma_2 {\ell}_{vc}$, with $\gamma_1\equiv ({\delta}_{Tan}/\delta_{vc})^\frac{10}{9}$ and $\gamma_2\equiv ({\delta}_{Tan}^{10}/\delta_{vc})^\frac{1}{9}$. For instance, if $\delta_{vc}\simeq 0.5$ and ${\delta}_{Tan}\simeq 0.8$, then $\tau_{vc}^* \simeq 1.7  {\tau}_{vc}$. All values of $\gamma_1$ and $\gamma_2$ are given in ESI section 2E$^\dag$, together with a detailed procedure on how they are derived for all plots of this article. 

In Fig.~\ref{fig1}d, data sets from Fig.~\ref{fig1}a-c are replotted using visco-capillary units. Additional data sets of viscous coalescence or spreading are also included. In visco-capillary units, Eq.~\ref{ViscoCap} and~\ref{Tanner} are written as $d/\ell_{vc}^* = (t/\tau_{vc}^*)^\alpha$, respectively with $\alpha=1$ and $\alpha=\frac{1}{10}$. More broadly, any spreading law of the form $d/\ell_{vc}^* = (t/\tau_{vc}^*)^\alpha$ is dimensionally sound and so \textit{a priori} possible. In particular, the geometry of the late spreading can significantly alter the value of the exponent $\alpha$. For instance, we included in Fig.~\ref{fig1}d data on the coalescence between two parallel plates, in which case $\alpha=\frac{1}{4}$~\cite{Yokota2011}. 

In between the early and late regimes, some data show an intermediate trend similar to what we discussed for the boundary layer when finite size effects come into play. The equations of motion themselves suggest that the pre-factor $\delta_{vc}$ in Eq.~\ref{ViscoCap} actually includes a logarithmic correction, \textit{i.e.} $\delta_{vc}\propto \log(d/D)$~\cite{Eggers1999,Hopper1990,Eddi2013}. This correction is shown by the dotted-dashed line in Fig.~\ref{fig1}d. This trend agrees quite well with some spreading drops. We will see later that the transition from early to late spreading/coalescence can also be altered by inertia. 

\section{Inertio-capillary scaling} 
\begin{figure*}
\centering
\begin{overpic}[abs,unit=1mm,width=17cm]{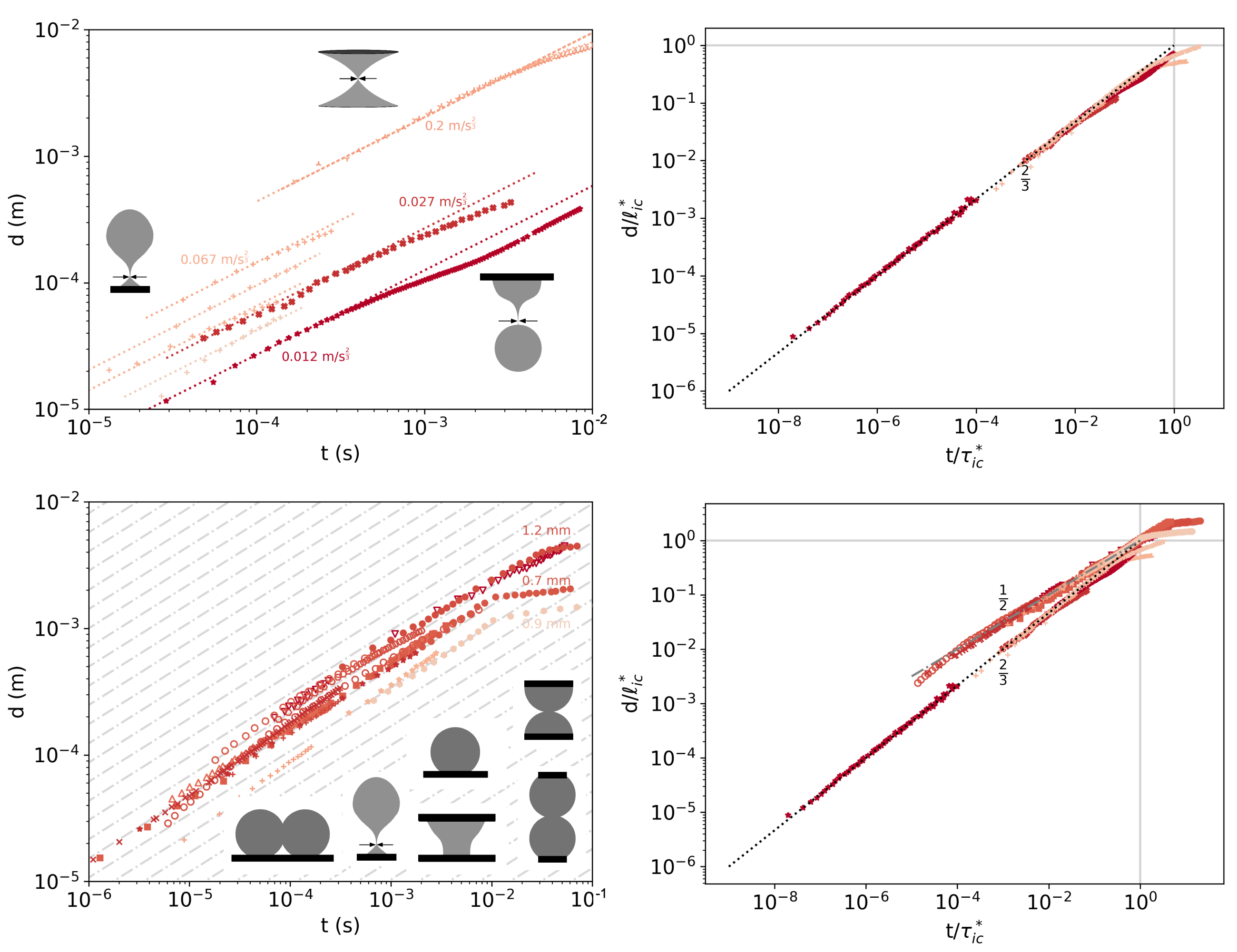}
\put(15,123){(a)}
\put(100,123){(b)}
\put(15,58){(c)}
\put(100,58){(d)}
\end{overpic}
\caption{Inertio-capillary pinching (crosses and stars), coalescence (open symbols) and spreading (filled symbols).   (a) Purely inertio-capillary regime of pinching following Eq.~\ref{IC} for fluids of different densities and surface tensions, plotted in standard units. A few values of $\delta_{ic}(\Gamma/\rho)^\frac{1}{3}$ are highlighted. All fitted values of $\delta_{ic}$ are given in ESI section 2E$^\dag$. Data reproduced from Bolanos \textit{et al.} ($+$)~\cite{Bolanos2009}, Burton \textit{et al.} ($\star$)~\cite{Burton2007}, Chen \textit{et al.} ($\Yup$) \cite{Chen1997}, Chen \textit{et al.} (\ding{54})~\cite{Chen2002} and Goldstein \textit{et al.} ($\Ydown$)~\cite{Goldstein2010}. (b) Data from panel-a reproduced in inertio-capillary units. Additional data sets are included, in particular for the pinching of mercury from Burton \textit{et al.} ($\star$)~\cite{Burton2004}. (c) Illustrations of the Rayleigh regime of Eq.~\ref{IC2} for spreading from Biance \textit{et al.} ($\bullet$)~\cite{Biance2004}, Eddi \textit{et al.} ($\blacksquare$)~\cite{Eddi2013}, Chen \textit{et al.} ($\varhexagonblack$)~\cite{Chen2014}, for coalescence from Menchaca-Rocha \textit{et al.} ($\triangledown$)~\cite{Menchaca2001}, Thoroddsen \textit{et al.} ($\hexagon$)~\cite{Thoroddsen2005,Thoroddsen2005b}, Paulsen \textit{et al.} ($\circ$)~\cite{Paulsen2011,Paulsen2014}, Soto \textit{et al.} ($\triangle$)~\cite{Soto2018}, and for pinching of bubbles from Burton \textit{et al.} ($\star$)~\cite{Burton2005}, Bolanos \textit{et al.} ($+$)~\cite{Bolanos2009} and  Keim \textit{et al.} ($\times$)~\cite{Keim2006}. The parallel dashed lines follow $d\sim t^\frac{1}{2}$. (d) All data from panels a to c are replotted together with inertio-capillary units. The units are $\tau_{ic}^* \equiv \gamma_1 \tau_{ic}$, with $\tau_{ic}\equiv  (\rho D^3/\Gamma)^\frac{1}{2}$, and $\ell_{ic}^* \equiv \gamma_2 \ell_{ic}$, with $\ell_{ic}\equiv D$. The values of $\gamma_1$ and $\gamma_2$ for all curves are obtained from the pre-factors $\delta_{ic}$ and $\delta_{Ray}$ (see ESI section 2E for details$^\dag$). In these units the dotted and dotted-dashed lines correspond to $d/\ell_{ic}^*= (t/\tau_{ic}^*)^\alpha$, respectively with $\alpha=\frac{2}{3}$ and $\frac{1}{2}$. Note that by definition $d/\ell_{ic}^* =\Lambda/\gamma_2$. Details of the fluid properties for all data sets in the panels of this figure are given in ESI section 1$^\dag$. The schematics on the panels illustrate the different experimental set-ups, which are discussed in more detail in ESI section 6$^\dag$. The color used for each data set gives to the value of the Ohnesorge number $\text{Oh}=\eta/(D\Gamma\rho)^\frac{1}{2}$. The full color scale is given in Fig.~\ref{fig5}.
\label{fig2}}
\end{figure*}
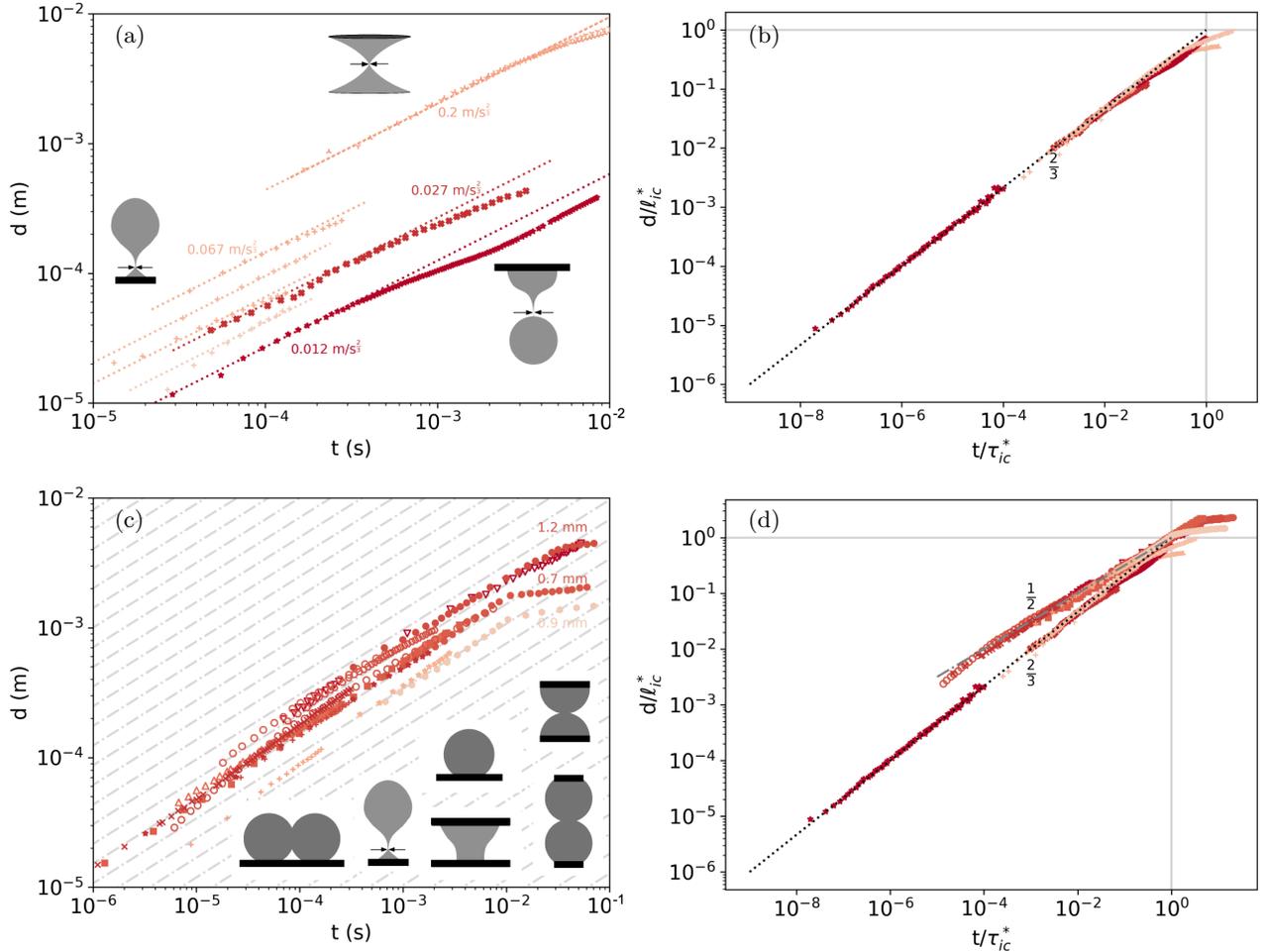
In the example of the boundary layer, the relevant material parameters were the viscosity $\eta$ and density $\rho$. For visco-capillary spreading and coalescence, the parameters were $\eta$ and $\Gamma$. Now we wish to discuss capillary dynamics where inertia assumes prominence over viscous effects, considering $\rho$ instead of $\eta$. The ratio between a surface tension and a density produces a kinematic quantity $\kappa\equiv \Gamma/\rho$, with dimensions $[\kappa]={L}^3.{T}^{-2}$. Such ratio is not as well known as a diffusion coefficient like $\nu$, or a speed like $c$, but it is no less fundamental. This interplay of dimensions between density and surface tension can be expressed as a simple spreading-like law~\cite{Keller1983}: 
\begin{equation}
d = \delta_{ic} \Big(\frac{\Gamma}{\rho}\Big)^\frac{1}{3}  t^\frac{2}{3}
\label{IC}
\end{equation}
The subscript now stand for `inertio-capillary'. This regime corresponds to a constant Weber number:
\begin{equation}
\text{We}\equiv \frac{\rho d v^2}{\Gamma} = (2/3)^2 \delta_{ic}^3
\end{equation}
The reader is referred to the ESI section 3$^\dag$ for details on this derivation. 

The most vivid examples of the inertio-capillary regime described by Eq.~\ref{IC} are actually found in the pinching dynamics of drops, bubbles and soap films~\cite{Day1998,Chen1997}. If $\tilde{t}$ is the standard time `running forward', and $\tilde{t}_c$ is the instant of pinch-off, then one can define a time $t\equiv \tilde{t}_c  - \tilde{t}$, `running backward' from the pinch-off instant. In this frame of reference, the pinch-off instant becomes analogous to the first contact in regular spreading or coalescence. 

In Fig.~\ref{fig2}a, a set of inertio-capillary pinching dynamics is shown to exhibit the $\frac{2}{3}$ scaling of Eq.~\ref{IC}~\cite{Bolanos2009,Burton2007,Chen1997,Chen2002,Goldstein2010}. All pre-factors $\delta_{ic}$ are of order 1. Note again that the actual value of the pre-factor $\delta_{ic}$ can include logarithmic dependencies~\cite{Deblais2018,Dinic2019b} (see ESI section 2D for details$^\dag$). For pinching dynamics, the nozzle size or the initial bridge radius provide an extrinsic length $D$, as sketched in Fig.~\ref{figsetups}d. In exactly the same way we followed for the boundary layer and for the visco-capillary spreading, we can build a system of units based on this additional length $\lbrace \Gamma,\rho, D\rbrace$: 
\begin{align}
{m}_{ic} &\equiv \rho D^3\\
{\ell}_{ic} &\equiv D\\
{\tau}_{ic} &\equiv \Big(\frac{\rho D^3}{\Gamma}\Big)^\frac{1}{2}
\end{align}
In this system, the characteristic stress is still the Laplace pressure, ${\Sigma}_{ic}={m}_{ic}.{\ell}_{ic}^{-1}.{\tau}_{ic}^{-2}= \Gamma/D$. The data from Fig.~\ref{fig2}a are replotted in these inertio-capillary units in Fig.~\ref{fig2}b. Most of the data collapse on the trend $d/{\ell}_{ic} \propto (t/{\tau}_{ic})^\frac{2}{3}$, except when $\Lambda \propto 1$, where the extrinsic length starts to have an influence. As shown in Fig.~\ref{fig2}a and b, multiple behaviors have been observed depending on the type of system. For instance, the data set at the bottom of Fig.~\ref{fig2}a corresponds to the pinching of a drop of superfluid Helium~\cite{Burton2005}. In this case, as the size of the neck gets closer to the extrinsic length $D$, the geometry of the drop can alter the value of the pre-factor $\delta_{ic}$ in the $\frac{2}{3}$ scaling of Eq.~\ref{IC}. This transition from the self-similar regime to a so-called `roll-off regime' was first described in the context of soap films by Chen and Steen~\cite{Chen1997}. Some of their data are reproduced in Fig.~\ref{fig2}a and b. Also shown are data from the collapse of a soap film on a M{\"o}bius strip~\cite{Goldstein2010}. At first, for $\Lambda\ll 1$, the geometry of the soap film has no influence and the data overlap with the regular soap bridge. Only in the end do the data diverge. 

The $\frac{2}{3}$ scaling of Eq.~\ref{IC} has been observed in a vast array of systems, but some pinching dynamics deviate from it. For instance, a bubble surrounded by water will be pinching according to a power law with an exponent close to $\frac{1}{2}$, as can be seen in Fig.~\ref{fig2}c (star symbols). Theory predicts an exponent slightly varying with time, around a value close to 0.56~\cite{Eggers2007}. In practice, experiments may have difficulties differentiating such non-trivial exponent from $\frac{1}{2}$ if logarithmic corrections are allowed (see ESI section 3A for details$^\dag$). For inertio-capillary dynamics with exponents close to $\frac{1}{2}$, an important historical guide and good approximation has been `Rayleigh's law'~\cite{Rayleigh1879}: 
\begin{equation}
d ={\delta}_{Ray} \Big(\frac{\Gamma D}{\rho}\Big)^\frac{1}{4}  t^\frac{1}{2}
\label{IC2}
\end{equation}
This trend keeps the following dimensionless product constant: $\text{We} \Lambda\delta_{Ray}^{4}/4$. 

This non-trivial scaling is not universal in the sense that it depends on $D$, but it is most definitely widespread, having been observed for pinching (crosses and stars), coalescence (open symbols) and spreading (filled symbols), as shown in Fig.~\ref{fig2}c. In these different contexts, the pre-factor of the $\frac{1}{2}$ scaling depends on density, surface tension and size, in a compounded way. For instance, in Fig.~\ref{fig2}c the data with $D=0.9$~mm lie below the data with $D=0.7$~mm, because of differences in surface tension and density. Note that wettability of the substrate can also influence this regime~\cite{Courbin2009,Eddi2013}. All pre-factors $\delta_{Ray}$ are of order 1 (see ESI section 2D for details$^\dag$). 

In inertio-capillary units the $\frac{1}{2}$ regime is written as $d/{\ell}_{ic} \propto (t/{\tau}_{ic})^\frac{1}{2}$. When plotted in dimensionless form in Fig.~\ref{fig2}d, the data from Fig.~\ref{fig2}c partially collapse for their portions abiding to Eq.~\ref{IC2}. For $\Lambda\gtrsim 1$, the spreading data transition to Tanner's regime for total wetting conditions. The behavior at long time is similar to that of more viscous droplets described in Fig.~\ref{fig1}~\cite{Biance2004,Carlson2012}. The pinching dynamics following the $\frac{2}{3}$ regime are also plotted in Fig.~\ref{fig2}d. The purely inertio-capillary regime and the size-dependent inertio-capillary regime of Rayleigh's law run concurrently rather than consecutively, in contrast to the purely visco-capillary regime and the size-dependent visco-capillary regime of Tanner's law. 

\section{The Ohnesorge number}
\begin{figure*}
\centering
\begin{overpic}[abs,unit=1mm,width=8.5cm]{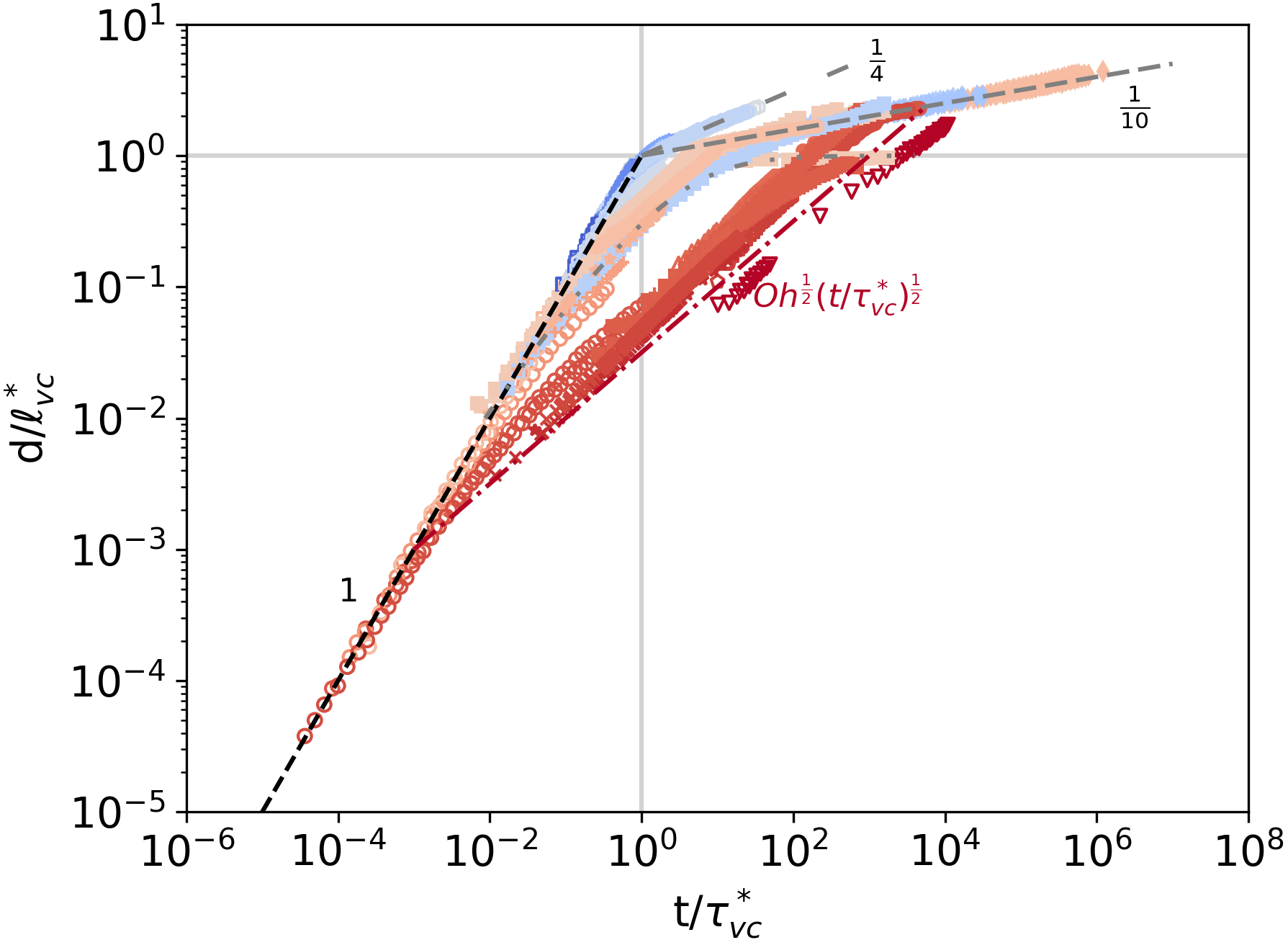}
\put(17,55){(a)}
\end{overpic}
\begin{overpic}[abs,unit=1mm,width=8.5cm]{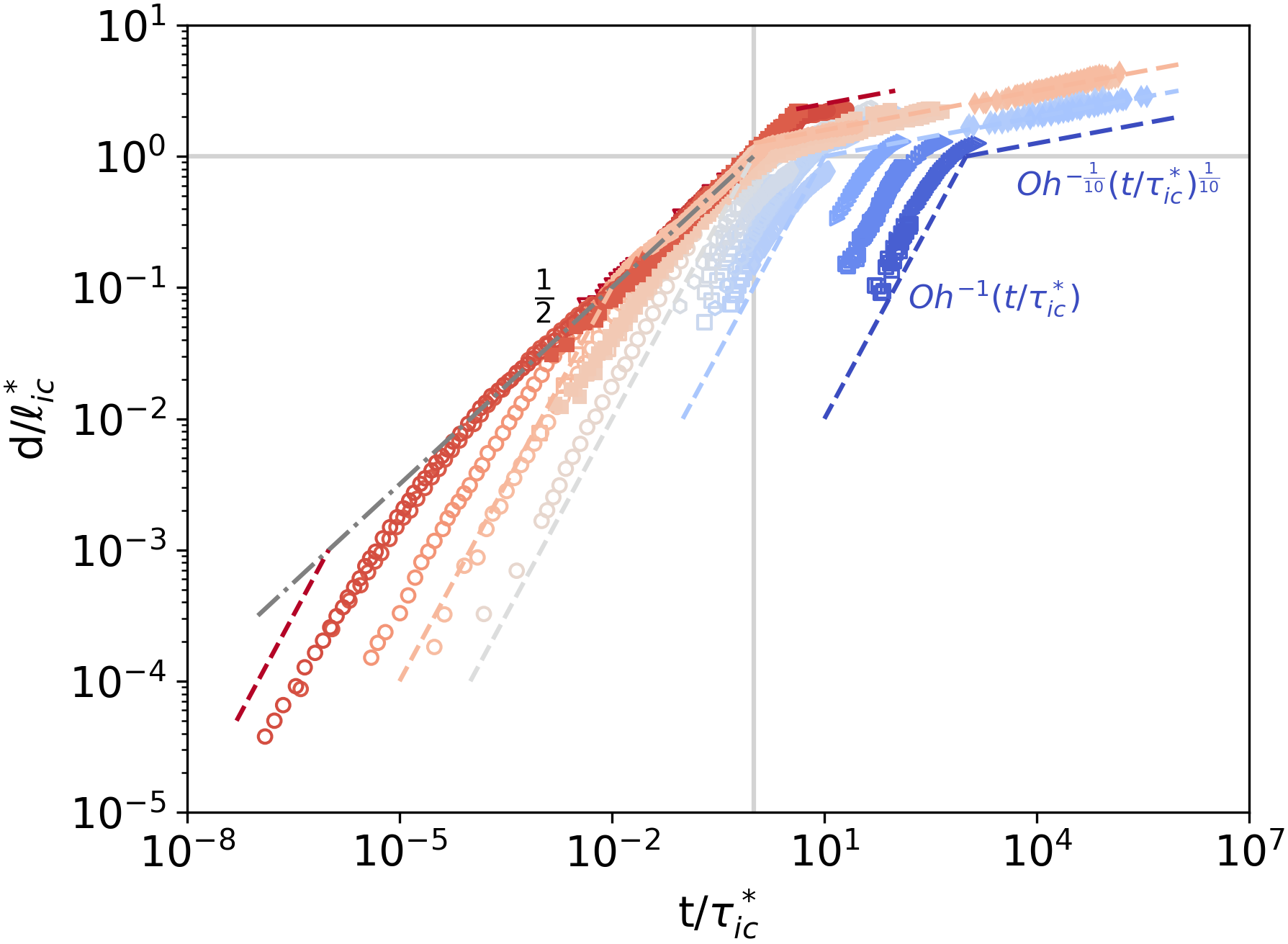}
\put(17,55){(b)}
\end{overpic}
\caption{Spreading (filled symbols), coalescence (open symbols) and bubble pinching (crosses and stars) experiments for fluids with different degrees of inertia, viscosity and surface tension. All data from Fig.~\ref{fig1} and~\ref{fig2} are shown, except for dynamics following the $\frac{2}{3}$ scaling, which are shown in ESI-Fig. 2 and 3$^\dag$. The color used for each data set gives to the value of the Ohnesorge number, with Oh$>1$ corresponding to blue shades and Oh$<1$ corresponding to red shades. The full color scale is given in Fig.~\ref{fig5}. (a) In visco-capillary units Rayleigh's regime of Eq.~\ref{IC2} corresponds to an intermediate regime between early and late spreading, following $d/\ell_{vc}^*=\text{Oh}^\frac{1}{2} (t/\tau_{vc}^*)^\frac{1}{2}$. The red dotted-dashed line corresponds to $\text{Oh}=10^{-3}$. As the value of Oh gets closer to 1, the extent of the intermediate regime shrinks. (b) In inertio-capillary units the different values of viscosity associated with each experiment are revealed by different points of departure with the $\frac{1}{2}$ regime. In the early dynamics, the purely visco-capillary regime of Eq.~\ref{ViscoCap} corresponds to parallel dashed lines following $d/\ell_{ic}^*=\text{Oh}^{-1} (t/\tau_{ic}^*)$. Different values of Ohnesorge number are also manifested in the late dynamics of spreading abiding to Tanner's law, which can be expressed as $d/\ell_{ic}^*=\text{Oh}^{-\frac{1}{10}} (t/\tau_{ic}^*)^\frac{1}{10}$. Details of the fluid properties for all data sets in the panels of this figure are given in ESI section 1$^\dag$. Schematic versions of these plots are available in ESI-Fig. 1a and 1b$^\dag$. Animated versions of these figures are provided in ESI$^\dag$. 
\label{fig3}}
\end{figure*} 
So far, we have described capillary dynamics by distinguishing fluids dominated by inertia or by viscosity. In practice, fluids can display both inertia and viscosity, in conjunction with surface tension. In general, the interplay between $\eta$, $\Gamma$, $\rho$ and an extrinsic size $D$ can be described by the Ohnesorge number~\cite{McKinley2011}:
\begin{equation}
{\text{Oh}} \equiv \frac{\eta}{(\rho \Gamma D)^\frac{1}{2}}
\end{equation}
For each experiment a value of Ohnesorge number can be computed. The colors used for all data sets shown in this contribution actually give the values of the Ohnesorge number, with Oh$>1$ corresponding to blue shades and Oh$<1$ corresponding to red shades. The full color scale is given in Fig.~\ref{fig5}. 

The Ohnesorge number can be understood as a Reynolds number or a Weber number for which the characteristic length is $D$, and the characteristic speed is the visco-capillary speed $\Gamma/\eta$. Alternatively, the Ohnesorge number can be understood as a capillary number for which the characteristic speed is the inertio-capillary speed $\ell_{ic}/\tau_{ic}=(\Gamma/\rho D)^\frac{1}{2}$ (also called `Taylor-Culick speed'). 

So far, we have used four kinds of dimensionless numbers: the Reynolds, Capillary and Weber numbers, built from the variables $d$ and $t$, and the size ratio $\Lambda$, which gives the ratio between the variable $d$ and the constant $D$. These numbers can be used as factors of the Ohnesorge number~\cite{Ohnesorge1936,McKinley2011}: 
\begin{equation}
\text{Oh}^2=\frac{\text{Ca}}{\text{Re}}\Lambda = \frac{\text{We}}{\text{Re}^2}\Lambda = \frac{\text{Ca}^2}{\text{We}}\Lambda
\label{ohcare}
\end{equation} 
\noindent where we have used the following identity: 
\begin{equation}
\text{We}=\text{CaRe}
\end{equation}
This identity comes with a convenient mnemonic device, since it can be read as `we care', complementing the German `ohne sorge', which can be translated as `without worries'. The ESI section 3$^\dag$ provides a more in-depth discussion of the connections between the different dimensionless quantities used in this article.  
 
Another way to understand the Ohnesorge number is as a translation factor between the three timescales we have introduced so far: $\tau_{vc}=\text{Oh}  \tau_{ic}=\text{Oh}^2  \tau_{vi}$. Roughly, Oh$>$1 corresponds to more viscous dynamics and Oh$<1$ corresponds to more inertial dynamics. The Ohnesorge number actually states that higher surface tension, higher density or extrinsic size $D$ have the same effect as smaller viscosity. 

The translations formulas between timescales can be used to express visco-capillary spreadings in inertio-capillary units or inertio-capillary spreadings in visco-capillary units. For instance, Rayleigh's regime of Eq.~\ref{IC2} can be written in visco-capillary units as $d/D \propto  \text{Oh}^\frac{1}{2} (t/\tau_{vc})^\frac{1}{2}$. This new expression for Rayleigh's regime can be understood by plotting the data from Fig.~\ref{fig2}c in the visco-capillary units of Fig.~\ref{fig1}d. This combination is done in Fig.~\ref{fig3}a. The $\frac{1}{2}$ trend shown in the figure corresponds to Oh=$10^{-3}$. As the value of the Ohnesorge number increases, \textit{i.e.} as the shade of red gets paler, the curves moves toward the origin where $d=\ell_{vc}^*$ and $t=\tau_{vc}^*$. The origin is reached for Oh=1. An animated detailed version of this figure is provided in ESI$^\dag$. In addition, the figure is drawn in more detail in ESI-Fig. 2a$^\dag$ for all data sets such that Oh$<1$. A similar combination can be done for the purely inertio-capillary $\frac{2}{3}$ regime, as shown in ESI-Fig. 2b$^\dag$. Formulas for the points where the $\frac{1}{2}$ or $\frac{2}{3}$ regimes intersect with the visco-capillary regimes are given in ESI-Fig. 1a$^\dag$. All intersections can be expressed as powers of Oh. 

A similar translation scheme can be followed to express viscous dynamics (Oh$>1$) in inertio-capillary units, as displayed in Fig.~\ref{fig3}b. For instance, the purely visco-capillary regime of Eq.~\ref{ViscoCap} can be written as $d/D \propto \text{Oh}^{-1} t/\tau_{ic}$. These trends are shown for a few values of Oh in Fig.~\ref{fig3}b. Here again, all intersections between trends can be expressed as powers of Oh, as shown in ESI-Fig. 1b$^\dag$. All data sets following the purely inertio-capillary scaling of Eq.~\ref{IC} were excluded from this figure for clarity, as they overlap with the rest of the data. A plot restricted to these data sets is given in ESI-Fig. 3a$^\dag$.   

\begin{figure}
\centering
\includegraphics[width=8.5cm,clip]{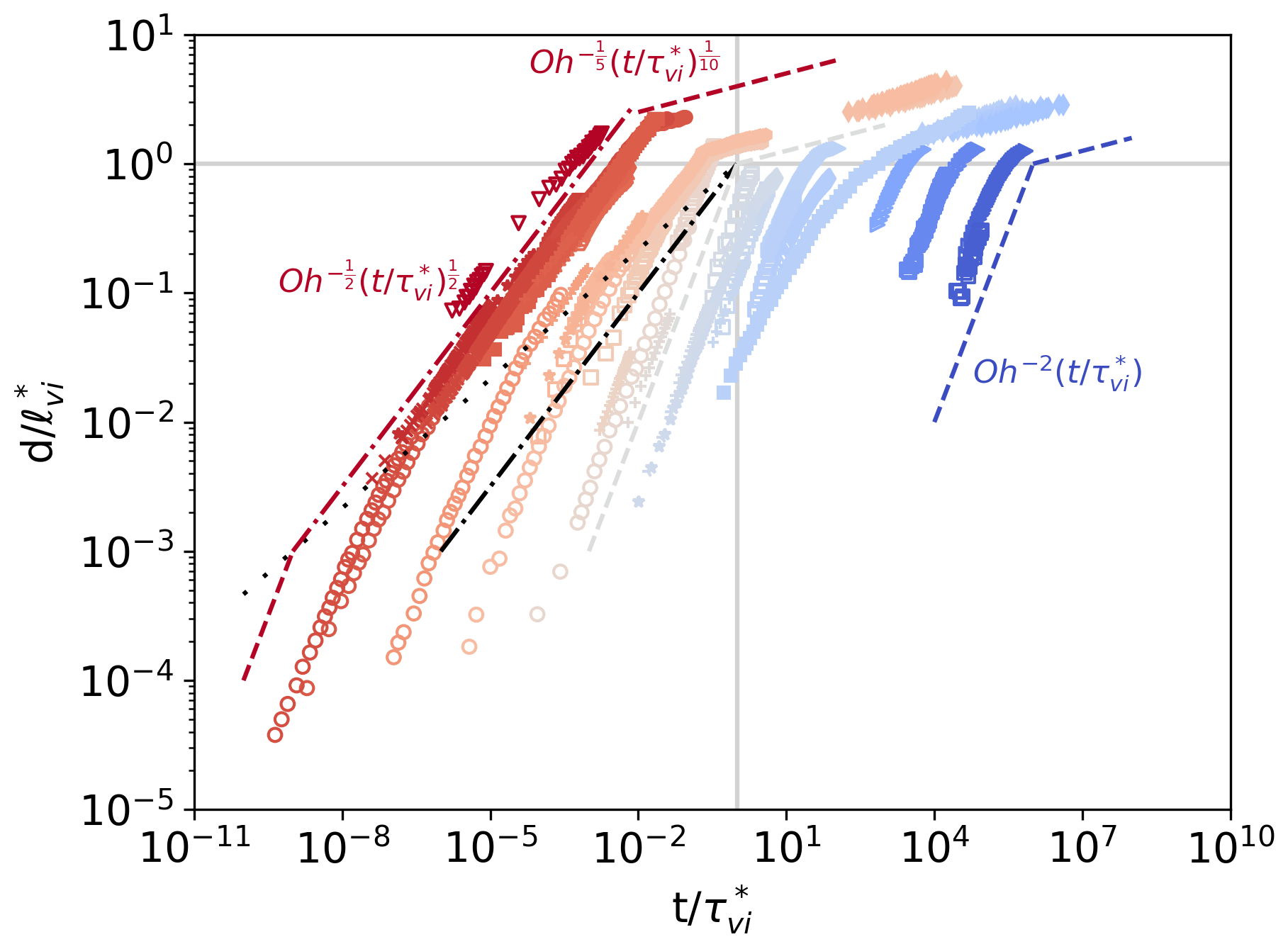}
\caption{Spreading (filled symbols), coalescence (open symbols) and bubble pinching  (crosses and stars) experiments for fluids with different degrees of inertia, viscosity and surface tension, represented in visco-inertial units. The dotted-dashed line gives the boundary layer scaling of Eq.~\ref{BL}. The color used for each data set gives to the value of the Ohnesorge number (full color scale in Fig.~\ref{fig5}). The dashed lines give three examples of the purely visco-capillary scaling of Eq.~\ref{ViscoCap} and Tanner's regime of Eq.~\ref{Tanner}, for $\text{Oh}=10^{-3}$, 1 and $10^3$. The red dotted-dashed line gives Rayleigh's regime of Eq.~\ref{IC2} for $\text{Oh}=10^{-3}$. The black loosely dotted line follows $d/\ell_{vi}^*= (t/\tau_{vi}^*)^\frac{1}{3}$, which gives the loci of the intersections between the purely visco-capillary regime and Rayleigh's regime. The black dotted dashed line is the boundary layer scaling $d/\ell_{vi}^*= (t/\tau_{vi}^*)^\frac{1}{2}$. Details of the fluid properties for all data sets in the panels of this figure are given in ESI section 1$^\dag$. A schematic version of this plot is available in ESI-Fig. 1c$^\dag$. An animated version of this figures is provided in ESI$^\dag$. 
\label{fig4}}
\end{figure}

In Fig.~\ref{fig3}a or b, it is quite clear that Rayleigh's $\frac{1}{2}$ scaling only has a non-vanishing extent if Oh$<1$, because of the existence of the visco-capillary regime at earlier times. However, if such linear early regime is not present, one may notice that in the limit $\text{Oh}=1$, Rayleigh's regime becomes identical to the boundary layer scaling. This is seen clearly by expressing Eq.~\ref{IC2} in visco-inertial units as $d/D \propto \text{Oh}^{-\frac{1}{2}} (t/\tau_{vi})^\frac{1}{2}$, where $d/D \propto (t/\tau_{vi})^\frac{1}{2}$ is just Eq.~\ref{BL}, \textit{i.e.} $d\propto (\nu t)^\frac{1}{2}$. To illustrate this point, we can replot the data from Fig.~\ref{fig3} in visco-inertial units, where we know that the boundary layer dynamics are most naturally expressed. Fig.~\ref{fig4} shows such combination. Again, the purely inertio-capillary dynamics have been excluded for clarity, they are shown in ESI-Fig. 4c$^\dag$. If the boundary layer scaling is understood as corresponding to $\text{Oh}=1$, this implies that the associated effective surface tension of a boundary layer can be defined as $\Gamma_{vi}=\eta^2/\rho D={m}_{vi}.{\tau}_{vi}^{-2}$. For instance, for water, this would give $\Gamma_{vi}\simeq 10^{-6}$~N/m if $D=1$~mm. This value is a few orders of magnitude lower than a typical surface tension between a liquid and air, but it is close to liquid-liquid interfacial tension~\cite{Israelachvili2015}. 

\section{The Ohnesorge units}
\begin{figure*}
\centering
\includegraphics[width=17cm,clip]{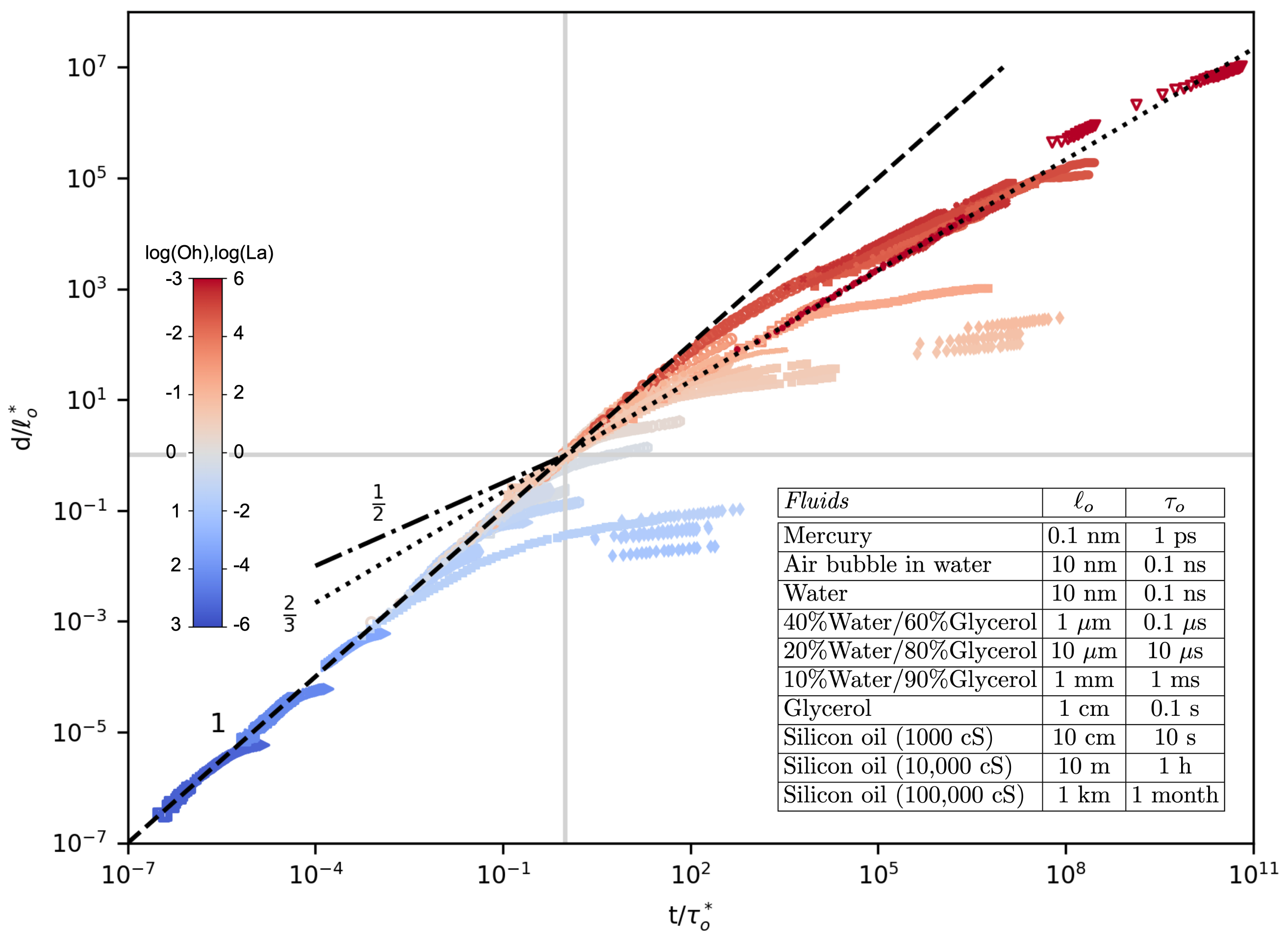}
\caption{Spreading (filled symbols), coalescence (open symbols) and pinching (crosses and stars) experiments for fluids with different degrees of inertia, viscosity and surface tension, represented in Ohnesorge units. The color used for each data set gives to the value of the Ohnesorge or Laplace numbers. The few data sets with $\text{Oh}<10^{-3}$ are saturated and displayed with the color corresponding to $\text{Oh}=10^{-3}$. The inset table gives orders of magnitudes for $\ell_o$ and $\tau_o$ for a set of commonly used fluids (assuming normal temperature and pressure). Note that the dimensionless size and time include the dimensionless pre-factors, such that $\tau_{o}^* \equiv \gamma_1 \tau_{o}$  and $\ell_{o}^* \equiv \gamma_2 \ell_{o}$. The values of $\gamma_1$ and $\gamma_2$ for all curves are obtained from the pre-factors $\delta_{vc}$, $\delta_{Tan}$, $\delta_{ic}$ and $\delta_{Ray}$ (see ESI section 2E for details$^\dag$). Details of the fluid properties for all data sets in the panels of this figure are given in ESI section 1$^\dag$. Schematic and animated versions of this plot are also available in ESI$^\dag$. 
\label{fig5}}
\end{figure*} 
So far, we have mentioned three systems of units: the visco-inertial units $\lbrace \eta,\rho, D\rbrace$, the visco-capillary units $\lbrace \Gamma,\eta, D\rbrace$ and the inertio-capillary units $\lbrace \Gamma,\rho, D\rbrace$. Using examples from boundary layers, spreading drops, coalescence and pinching we have illustrated the use of these units in association with three simple spreading laws, where `spreading' is understood in a broad sense: $d\propto (\eta/\rho)^\frac{1}{2} t^\frac{1}{2}$, $d\propto (\Gamma/\eta) t$, and $d \propto (\Gamma/\rho)^\frac{1}{3} t^\frac{2}{3}$. These scaling laws are `simple' in the sense that they can be obtained directly from dimensional analysis. These laws are also called `universal' because they do not depend on any external parameter like the size of the drop $D$. One can verify that the visco-inertial regime corresponds to the only choice of exponent $\alpha$, such that $d/\ell_{vi} \propto (t/\tau_{vi})^\alpha$ does not depend on $D$. Similarly, the visco-capillary regime and the inertio-capillary regime correspond to the only exponents canceling the influence of $D$ in visco-capillary and inertio-capillary units respectively. 

The three simple scalings intersect at the same spatio-temporal location: 
\begin{equation}
\Big(\frac{\eta}{\rho}\Big)^\frac{1}{2} t^\frac{1}{2} = \frac{\Gamma}{\eta}  t =  \Big(\frac{\Gamma}{\rho}\Big)^\frac{1}{3} t^\frac{2}{3}   \rightarrow  t=\tau_o \quad \text{and} \quad d=\ell_o
\end{equation}
The coordinates of the intersection between these three regimes are the time and lengthscale of a fourth system of units $\lbrace \Gamma,\eta,\rho\rbrace$, which we call the Ohnesorge units:  
\begin{align}
m_o &\equiv \frac{\eta^6}{\Gamma^3 \rho^2}\\
\ell_o &\equiv \frac{\eta^2}{\Gamma \rho}\\
\tau_o &\equiv \frac{\eta^3}{\Gamma^2 \rho}
\end{align}  
The Ohnesorge units have been invoked directly or indirectly in a number of studies, and date back to Haenlein, Weber and Ohnesorge, as stated in the introduction~\cite{Haenlein1931,Weber1931,Ohnesorge1936}.    

The magnitude of these units can vary widely depending on the properties of the fluid, as shown on a few examples in the inset of Fig.~\ref{fig5}. A complete table with the values computed for all experiments used in this article is given in ESI$^\dag$. The Ohnesorge length and time vary from $\ell_o\simeq 3~\AA$ and $\tau_o\simeq 1$~ps for mercury~\cite{Burton2004}, and $\ell_o\simeq 1~$km and $\tau_o\simeq 4$~months for a viscous silicon oil~\cite{Yao2005}. 

In Ohnesorge units, the scalings of the form $d/\ell_o \propto (t/\tau_o)^\alpha$ admit three special values: for $\alpha=\frac{1}{2}$ the dynamics are independent of $\Gamma$ (Eq.~\ref{BL}), for $\alpha=1$ the dynamics are independent of $\rho$ (Eq.~\ref{ViscoCap}), and for $\alpha=\frac{2}{3}$ the dynamics are independent of $\eta$ (Eq.~\ref{IC}). 

In contrast to the three systems introduced so far, the Ohnesorge units are purely intrinsic and do not depend on any extrinsic length $D$. Being at the crossroad of viscosity, inertia and surface tension, these units can be understood in a few different ways. For instance, the Ohnesorge time $\tau_o$ can be connected to the three other timescales by the Ohnesorge number as: 
\begin{equation}
\tau_{vi} \xrightarrow{\times \text{Oh}} \tau_{ic} \xrightarrow{\times \text{Oh}} \tau_{vc} \xrightarrow{\times \text{Oh}^2} \tau_{o}
\label{Ohorder2}
\end{equation}
% units with tvc*Oh?
Thus, the two possible orderings for the four timescales are either $\tau_{vi}<\tau_{ic}<\tau_{vc}<\tau_{o}$ if Oh$>1$, or $\tau_{vi}>\tau_{ic}>\tau_{vc}>\tau_o$ if Oh$<1$. The Ohnesorge time can be understood as a visco-inertial time, or a visco-capillary time, or an inertio-capillary time when the distance $D$ is replaced by $\ell_o$. 

The Ohnesorge length $\ell_o$ can be expressed as $\ell_o= \text{Oh}^2 D$. Conversely, one can use the Ohnesorge length to define the Ohnesorge number as a ratio between intrinsic and extrinsic lengthscales, $\text{Oh}=(\ell_o/D)^\frac{1}{2}$. Alternatively, one can use the Laplace or Suratman number~\cite{McKinley2011}: 
\begin{equation}
\text{La}\equiv \frac{D}{\ell_o} = \frac{1}{\text{Oh}^2}
\end{equation}
Large values of the Laplace number correspond to more inertial dynamics, whereas small values correspond to more viscous ones. Note that the Ohnesorge number can also be expressed from the masses of the different systems of units, as $m_o=m_{vc} \text{Oh}^4=m_{ic} \text{Oh}^6$, where $m_{ic}=m_{vi}=\rho D^3$ is the actual mass (neglecting numerical factors).  

In the Ohnesorge units, the characteristic stress is ${\Sigma}_{o}={m}_{o}.{\ell}_{o}^{-1}.{\tau}_{o}^{-2}= \Gamma^2 \rho/\eta^2$. Again, this formula can be understood in a few ways.  From an inertial perspective, one can write the stress as a dynamic pressure $\Sigma_{o}= \rho u_o^2$, where $u_o = \ell_o/\tau_o \propto \Gamma/\eta$ is the visco-capillary speed. From a viscous perspective, one can write the stress in a Newtonian way as $\Sigma_{o} = \eta (u_o/\ell_o)$. From a capillary perspective, one can write the stress as a Laplace pressure, $\Sigma_{o}= \Gamma/\ell_o$. All these perspectives lead back to the same formula.

The Ohnesorge units are ideal to represent spreading, coalescence and pinching dynamics from a purely intrinsic perspective. All data reproduced in this article are shown in Ohnesorge units in Fig.~\ref{fig5}. These intrinsic units allow to circumvent the challenges faced with the overlap of data sets in Fig.~\ref{fig3} and~\ref{fig4}, where the purely inertio-capillary data had to be plotted separately. Here, all data can be shown simultaneously, including 71 separate curves $d(t)$ obtained from 25 different studies~\cite{Cazabat1986,Biance2004,Eddi2013,Chen2014,Menchaca2001,Wu2004,Yao2005,Thoroddsen2005,Thoroddsen2005b,Aarts2005,Aarts2008,Yokota2011,Paulsen2011,Paulsen2014,Soto2018,Rahman2019,Chen1997,McKinley2000,Chen2002,Burton2004,Burton2005,Burton2007,Keim2006,Bolanos2009,Goldstein2010}, spanning 7 orders of magnitude in the value of Oh, and respectively 14 and 18 orders of magnitude in the values of $d/\ell_o$ and $t/\tau_o$. In this representation, the different values of extrinsic length $D$ emerge as different points of departure from comparatively more universal trends. 

For $\text{Oh}>1$, \textit{i.e.} $\text{La}<1$, the data seem to converge on the linear scaling when $d/\ell_o \ll \text{La}$, \textit{i.e.} when $\Lambda\ll 1$. If $\Lambda \gtrsim 1$, different departures are possible depending on the boundary conditions. For instance, for total wetting, spreading data can follow Tanner's law $d/\ell_o \propto \text{Oh}^{-\frac{9}{5}} (t/\tau_o)^\frac{1}{10}$. 

For $\text{Oh}<1$, \textit{i.e.} $\text{La}>1$, the asymptotic departures similarly correspond to different values of the extrinsic size $D$. Whereas the linear scaling seems quite attractive to all dynamics for $d/\ell_o \ll \text{La} <1$, multiple trends are possible for $d/\ell_o \ll \text{La} >1$. For instance, spreading, coalescence and bubble pinching can remain on the linear scaling as long as $d/\ell_o<\text{Oh}^{-1}$, then follow the size-dependent inertio-capillary regime, now written as $d/\ell_o \propto \text{Oh}^{-\frac{1}{2}} (t/\tau_o)^\frac{1}{2}$. This intermediate regime reaches $d\propto D$ when crossing the $\frac{2}{3}$ trend. Not all data follow this path through the size-dependent inertio-capillary regime. In particular, we have seen that the pinching dynamics of liquids can transition from a purely visco-capillary regime to a purely inertio-capillary $\frac{2}{3}$ regime, as soon as $d>\ell_o$. A zoom on the quadrant $d>\ell_o$ and $t>\tau_o$ of Fig.~\ref{fig5} is provided in ESI-Fig. 11$^\dag$, to allow for easier distinction between the different possible paths. 

What are the conditions dynamics must meet to exhibit an intermediate size-dependent inertio-capillary $\frac{1}{2}$ regime instead of directly following the purely inertio-capillary $\frac{2}{3}$ regime? For instance, bubbles of air in water will exhibit the $\frac{1}{2}$ regime, whereas water drops will exhibit the $\frac{2}{3}$ regime~\cite{Eggers2008}. For spreading drops, results from Biance \textit{et al.} and Chen \textit{et al.} corresponding to $\text{Oh}\simeq 10^{-1}-10^{-3}$ exhibit the $\frac{1}{2}$  scaling~\cite{Biance2004,Chen2014}. However, we notice that data from Eddi \textit{et al.}~\cite{Eddi2013} for $\text{Oh}\simeq 10^{-1}-1$ can be seen to follow the $\frac{2}{3}$ regime (see ESI-Fig. 11 for details$^\dag$). The conditions bringing dynamics along the $\frac{1}{2}$ or $\frac{2}{3}$ regimes remain unclear, but may be related to the initial shape of the drop~\cite{Courbin2009,Eddi2013b}. Moreover, some dynamics even seem to follow the $\frac{1}{2}$ or $\frac{2}{3}$ scaling for $d<\ell_o$. These data are not shown in Fig.~\ref{fig5}, but are given in  ESI-Fig. 12$^\dag$. The coexistence of different parallel regimes was manifested in Fig.~\ref{fig2} for $\text{Oh}<1$, and dimensional analysis alone does not preclude the existence of multiples regimes for $\text{Oh}>1$. For instance, if the boundary layer dynamics are interpreted as having $\text{Oh}=1$, then $\ell_o=D$, and the data follow the $\frac{1}{2}$ scaling drawn with the dotted-dashed line in Fig.~\ref{fig5}. More surprising, the data on spreading drops from Eddi \textit{et al.}~\cite{Eddi2013}, with $\text{Oh}\gtrsim 1$ are quite close from the $\frac{2}{3}$ scaling drawn as a dotted line in Fig.~\ref{fig5}, depending on the pre-factors $\gamma_1$ and $\gamma_2$ that one may choose (see ESI-Fig. 12 for details$^\dag$). Note also that in the context of the coalescence of drops with $\text{Oh}\propto 1$, a regime combining inertial, viscosity and surface tension has been evidenced~\cite{Paulsen2012}. 

In reviewing the literature on spreading, pinching, and coalescence, we thought that spreading and coalescence would show a stronger dependency on geometry and display the size-dependent $\frac{1}{2}$ regime for $\text{Oh}<1$, whereas pinching would follow the more universal $\frac{2}{3}$ regime. Experiments seem to paint a more nuanced picture. 

\section{Departures from Ohnesorge's units}
All data discussed in this article abide quite well to the Ohnesorge units. Despite the fact that the data cover an unprecedentedly large spectrum, they only represent a small fraction of the possible dynamics influenced by viscosity, density and surface tension. For future studies to include more data, we see two complementary avenues set by answering the following questions: what kind of additional data would fit within the Ohnesorge units, and what kind would not? 

First, one may wonder about dynamics abiding to Ohnesorge units but in non-trivial ways. For instance, in the context of the spreading of liquid-on-liquid, one may encounter scalings of the form $d/\ell_o \propto (t/\tau_o)^\alpha$, with $\alpha$ different from 1, $\frac{1}{2}$ or $\frac{2}{3}$. In particular, $\alpha=\frac{3}{4}$ has been described in several instances~\cite{Wang2015,Berg2009}. Such non-trivial regime in Ohnesorge units is analogous to Tanner's law or to Rayleigh's law, in the sense that it cannot be derived directly from dimensional analysis. 

For the data represented in Fig.~\ref{fig5}, departures from the three simple scalings ($1$, $\frac{1}{2}$ and $\frac{2}{3}$) are all associated with the existence of an extrinsic or `integral' lengthscale $D$. In all experiments considered here, the spreading, coalescence or pinching dynamics are `free', in the sense that they happen spontaneously, without any imposed speed $U$ or acceleration $G$. If non-negligible extrinsic speed or acceleration are present, the `integral scale' is not solely characterized by a size $D$. Hence, in these `forced' cases of spreading, coalescence or pinching, one would expect the departures from Ohnesorge's units to be more varied. Already in Fig.~\ref{fig1}c we noticed that gravity, \textit{i.e.} an acceleration $G=g$ can alter Tanner's law at long time, leading to $d\sim t^\frac{1}{8}$ instead of $d\sim t^\frac{1}{10}$~\cite{Cazabat1986}. Similarly, an extrinsic speed can have a very significant impact, as was indeed demonstrated for jetting by Ohnesorge himself~\cite{Ohnesorge1936}.  

Departures from the Ohnesorge units can also be due to additional intrinsic mechanisms beyond viscosity, inertia and capillarity. For instance, in the case of viscoelastic fluids an elasticity $\Sigma$ becomes relevant and generate an intrinsic `relaxation time' $\tau_{ve}=\eta/\Sigma$, which can lead to exponential regimes, so far mostly described in the context of pinching~\cite{McKinley2005,Dinic2019,Dinic2020}.

In general if the departure from Ohnesorge units can be traced back to a quantity $Q$, then a dimensionless number $N=Q/Q_o$ can be built, where $Q_o$ has the same dimensions than $Q$ and is given in Ohnesorge units. For instance, if $Q=D$, then $N=D/\ell_o=\text{La}=\text{Oh}^{-2}$. If $Q=G$, then $N=G/G_o=G \eta^4 /\Gamma^3 \rho=\text{Bo} \text{Oh}^4$, where $\text{Bo}=\rho G D^2/\Gamma$ is the Bond number~\cite{PGG2013}. If $Q=\Sigma$, then $N=\Sigma/\Sigma_o=\Sigma \eta^2 /\Gamma^2 \rho=\text{Ec}^{-1} \text{Oh}^2$, where $\text{Ec}=\Gamma/\Sigma D$ is the elasto-capillary number~\cite{McKinley2005}. 

Another way in which dynamics can differ from those depicted in Fig.~\ref{fig5} is in the presence of competing choices of densities, viscosities or surface tensions. In this article, we focused on cases where a single choice of material parameters was possible, but it is not necessarily the case. In general, ratios of the material properties of the inner and outer fluids can have an impact on the dynamics. This issue has been investigated for spreading~\cite{Bazazi2018,Jose2017}, coalescence~\cite{Paulsen2014,Jose2017} and pinching~\cite{Verbeke2020} and we hope to be able to include these studies in the future. Some of these dynamics may fit well with the Ohnesorge units with minimal adjustments. For instance, the coalescence of bubbles in a viscous fluid can give rise to a scaling $d/\ell_{vc}\propto (t/\tau_{vc})^\frac{1}{2}$, when the outer viscosity is used to compute the visco-capillary time $\tau_{vc}$~\cite{Paulsen2014}.

\section{Conclusion}
Because lengths, durations and masses are so engraved in our understanding of physical reality, we tend to forget that their fundamental stature is somewhat of a convention. International standards encourage the use of units based on the dimensions of mass, length and time, reminding for example that a viscosity of 1~poise stands for 0.1~kg.m$^{-1}$.s$^{-1}$. This approach to physical quantities like viscosity, density or surface tension is rooted in metrology. For instance, if one needs to measure viscosity, one will have to rely on some associated measurements of size (e.g. dimensions of the rheometer), time (e.g. in measuring speeds) and mass (e.g. if stress is measured through a torque, itself measured by a mass and pulley system). In everyday life, distances, durations and to some extent masses are the most easily available measures of the physical world. However, the microcosm unfolding at the scale of drops of fluid, spreading, pinching and coalescing, is quite different from ours. There, space, time and mass are derived quantities, produced by the interplay of three basic dimensions called viscosity, surface tension and density. By elevating these three quantities as fundamental dimensions, recent studies have greatly advanced our understanding of capillary phenomena. Building on years of experimental studies on spreading, pinching and coalescence, they have shown how the different classes of behaviors are quantitatively related by the use of appropriate units based on $\eta$, $\Gamma$ and $\rho$. 

The three simple spreading laws discussed in this article (Eq.~\ref{BL}, \ref{ViscoCap} and \ref{IC}) provide the three essential ways in which space-time emerges in the universe of droplets, the three ways in which length and duration are coupled. One way to interpret the Ohnesorge units is as giving the dimensions of length, time and mass as derived from those of viscosity, surface tension and density. For instance, $[t]=[\eta]^3.[\Gamma]^{-2}.[\rho]^{-1}$. This equation on the dimensions is always true, even beyond drops of fluids. What is most striking is that if the equation is used without the brackets, \textit{i.e.} to yield numerical values, then the derived timescale turns out to be very significant to the dynamics in the world of drops. As shown in Fig.~\ref{fig5}, the Ohnesorge time $\tau_o$ is at the crossroads of an array of different phenomena and often marks the turning point between distinct dynamical regimes. 

The Ohnesorge units formalize the essential intrinsic properties of a world governed by viscosity, surface tension and density. However, this universe is not unbounded, and its limit overlaps with our more familiar realm through the influence of the `integral' or `extrinsic' length $D$. In this article, we have carefully shown how the intrinsic Ohnesorge units can be transformed by the addition of the length $D$, and how size-dependent dynamical regimes like those of Tanner or Rayleigh can become preponderant. Together, the four systems of units described in this article provide the four ways to choose three parameters from the set $\lbrace \Gamma,\eta,\rho,D\rbrace$. These four parameters are connected to each other by the Ohnesorge number, which provides a natural way to distinguish between `more viscous' ($\text{Oh}>1$) and `more inertial' ($\text{Oh}<1$) dynamics.  

\section*{Conflicts of interest}
There are no conflicts to declare.

\section*{Acknowledgements}
VS acknowledges funding support from NSF CBET 1806011. M.H. thanks DGAPA-UNAM for funding his sabbatical within the joyful atmosphere of the Ladoux-M\`{e}ge lab. The authors thank Carina Martinez and G.H. McKinley for critically reviewing the manuscript. Ernesto Horne is thanked for his endorsement.

\providecommand*{\mcitethebibliography}{\thebibliography}
\csname @ifundefined\endcsname{endmcitethebibliography}
{\let\endmcitethebibliography\endthebibliography}{}

\end{document}

%\widowpenalty=1000
%\clubpenalty=1000

%\preprint{APS/123-QED}

\title{Supplementary Information:\\ Spreading, pinching, and coalescence: the Ohnesorge units}

\author{M.A.~Fardin}
\altaffiliation[Corresponding author ]{}
\email{marc-antoine.fardin@ijm.fr}
\affiliation{The Academy of Bradylogists}
\affiliation{Universit\'{e} de Paris, CNRS, Institut Jacques Monod, F-75013 Paris, France}
\author{M.~Hautefeuille}
\affiliation{Universit\'{e} de Paris, CNRS, Institut Jacques Monod, F-75013 Paris, France}
\affiliation{Facultad de Ciencias, Departamento de Fisica, Universidad Nacional Aut\'{o}noma de M\'{e}xico,Ciudad Universitaria, DF 04510, Mexico}
\affiliation{Institut de Biologie Paris Seine, Sorbonne Universit\'{e}, 7 quai Saint Bernard, 75005 Paris, France}
\author{V.~Sharma}
\affiliation{The Academy of Bradylogists}
\affiliation{Department  of Chemical  Engineering, University of Illinois at Chicago, Chicago, Illinois 60608, UnitedStates}

\date{\today}

\maketitle
\tableofcontents
\newpage

\section{Experimental data summary}
Our study provides a meta-analysis of a number of experiments on spreading, coalescence and pinching of fluids of various properties. In this section we explain the protocol we followed to extract the data sets from the original articles and we give tables summarizing the properties of all experiments reproduced in the figures of the article. 

\subsection{Data extraction}
All data were extracted semi-manually from the figures of the original articles given in Table~\ref{table1}. For a given figure, the data points were identified manually on the free imagining software Fiji~\cite{Schindelin2012}. The coordinates of the data points were stored and converted to standard units (seconds and meters). The precision is expected to be on the order of the size of the symbols used in the original graphs. When the sampling was high and multiple data points overlapped we only selected a subsets of the original data points. We omitted data points with large error bars, which usually corresponded to the first few measurements at the limit of the resolution of the experiment. All extracted data sets ($t$ and $d$) are given as two-columns text-files in the supplementary archive `\textbf{DataSets.zip}'.   
 
\begin{table}
\begin{tabular}{|l|c|c|l|}
\hline
 \textbf{Label} & \textbf{Type} & \textbf{Symbol} & \textbf{Reference} \\ \hline\hline
Cazabat1986  & Spreading & $\blacklozenge$ & A. Cazabat and M. C. Stuart, \textit{J. Phys. Chem.} \textbf{90}, 5845 (1986) \\\hline
Biance2004  & Spreading & \large$\bullet$ & A.-L. Biance \textit{et al.}, \textit{Phys. Rev.
E} \textbf{69}, 016301 (2004) \\\hline
Eddi2013  & Spreading & $\blacksquare$ & A. Eddi \textit{et al.}, \textit{Phys. Fluids} \textbf{25}, 013102 (2013) \\\hline
Chen2014  & Spreading & $\varhexagonblack$ & L. Chen and E. Bonaccurso, \textit{Phys. Rev. E} \textbf{90},
022401 (2014)\\\hline
Menchaca2001  & Coalescence & $\triangledown$ & A. Menchaca-Rocha \textit{et al.}, \textit{Phys. Rev. E} \textbf{63}, 046309 (2001)\\\hline
Wu2004  & Coalescence & $\pentagon$ & M. Wu \textit{et al.}, \textit{Phys. Fluids} \textbf{16}, L51--L54 (2004)\\\hline
Yao2005  & Coalescence & $\triangleright$ & W. Yao \textit{et al.}, \textit{Phys. Rev. E} \textbf{71}, 016309 (2005)\\\hline
Thoroddsen2005  & Coalescence & $\hexagon$ & S.T. Thoroddsen \textit{et al.}, \textit{J. Fluid Mech.} \textbf{527}, 85--114 (2005) \\\hline
Thoroddsen2005b  & Coalescence & $\hexagon$ & S.T. Thoroddsen \textit{et al.}, \textit{Phys. Fluids} \textbf{17}, 071703 (2005) \\\hline
Aarts2005  & Coalescence & $\square$ & D. Aarts \textit{et al.}, \textit{Phys. Rev. Lett.} \textbf{95}, 164503 (2005) \\\hline
Aarts2008  & Coalescence & $\square$ & D. Aarts and H. Lekkerkerker, \textit{J. Fluid Mech.} \textbf{71}, 275--294 (2008)\\\hline
Yokota2011  & Coalescence & $\varhexagon$ & M. Yokota and K. Okumura, \textit{PNAS} \textbf{108}, 6395--6398 (2011)\\\hline
Paulsen2011  & Coalescence & \large$\circ$ & J. Paulsen  \textit{et al.}, \textit{Phys. Rev. Lett.} \textbf{106}, 114501 (2011)\\\hline
Paulsen2014  & Coalescence & \large$\circ$ & J. Paulsen  \textit{et al.}, \textit{Nat. Commun.} \textbf{5}, 1--7 (2014)\\\hline
Soto2018  & Coalescence & $\triangle$ & A.M. Soto  \textit{et al.}, \textit{J. Fluid Mech.} \textbf{846}, 143--165 (2018)\\\hline
Rahman2019  & Coalescence & $\Diamond$ & M. Rahman  \textit{et al.}, \textit{Phys. Fluids} \textbf{31}, 012104 (2019)\\\hline
Chen1997  & Pinching & $\Yup$ & Y.J. Chen and P.H. Steen, \textit{J. Fluid Mech.} \textbf{341}, 245--267 (1997)\\\hline
McKinley2000  & Pinching &  $\Yleft$ & G.H. McKinley and A. Tripathi, \textit{J. Rheol.} \textbf{44}, 653--670 (2000)\\\hline
Chen2002  & Pinching &  \ding{54} & A.U. Chen \textit{et al.}, \textit{Phys. Rev. Lett.} \textbf{88}, 174501 (2002)\\\hline
Burton2004  & Pinching &  $\star$ & J.C. Burton \textit{et al.}, \textit{Phys. Rev. Lett.} \textbf{92}, 244505 (2004)\\\hline
Burton2005  & Pinching &  $\star$ & J.C. Burton \textit{et al.}, \textit{Phys. Rev. Lett.} \textbf{94}, 184502 (2005)\\\hline
Burton2007  & Pinching &  $\star$ & J.C. Burton \textit{et al.}, \textit{Phys. Rev. E} \textbf{75}, 036311 (2007)\\\hline
Keim2006  & Pinching &  $\times$ & N.C. Keim \textit{et al.}, \textit{Phys. Rev. Lett.} \textbf{97}, 144503 (2006)\\\hline
Bolanos2009  & Pinching &  $+$ & R. Bolanos-Jim{\'e}nez \textit{et al.}, \textit{Phys. Fluids} \textbf{21}, 072103 (2009)\\\hline
Goldstein2010  & Pinching &  $\Ydown$ & R.E. Goldstein \textit{et al.}, \textit{PNAS} \textbf{107}, 21979--21984 (2010)\\\hline
\end{tabular}
\caption{Summary of the original studies reproduced in the article. 
\label{table1}}
\end{table}

\subsection{Data summary}
The values of the material parameters $\rho$, $\eta$ and $\Gamma$, of the extrinsic size $D$, and the associated value of the Ohnesorge number for spreading, coalescence and pinching experiments are given in Tables~\ref{table2s}, \ref{table2c} and \ref{table2p} respectively, which can be found in the supplementary file `\textbf{DataSummary.csv}'. More details on each data set are given in section~\ref{datacomments}. 

\begin{table}
\csvreader[tabular=|l|c|c|c|c|c|c|c|c|c|,respect all,
    table head=\hline \textbf{Label} & $\bm\rho$ (kg.m$^{-3}$) & $\bm\eta$ (kg.m$^{-1}$.s$^{-1}$) & $\bm\Gamma$ (kg.s$^{-2}$) & $\bm{D}$ (m) & \textbf{Oh} & $\bm{\tau_{vi}}$ (s) & $\bm{\tau_{ic}}$ (s) & $\bm{\tau_{vc}}$ (s) & $\bm{\tau_{o}}$ (s)  \\ \hline,
    late after line=\\\hline,
    filter=\equal{\Type}{Spreading}
    ]%
{./FiguresArxivFinal/FIGS/Summary.csv}{Label=\Label,Type=\Type,Density=\Density,Viscosity=\Viscosity, 5=\SufTens,7=\Size, 8=\Oh, 15=\tvi, 16=\tic, 17=\tvc, 18=\to}%
{\Label  &\Density &\Viscosity &\SufTens & \Size & \Oh & \tvi & \tic & \tvc & \to}
\caption{Summary of the properties of all spreading experiments reproduced in the article: density $\rho$, viscosity $\eta$, surface tension $\Gamma$, extrinsic size $D$, Ohnesorge number $\text{Oh}=\eta/(\rho\Gamma D)^\frac{1}{2}$, together with the values of the four time scales $\tau_{vi}$, $\tau_{ic}$, $\tau_{vc}$ and $\tau_o$. The content of this table is available in the supplementary file `\textbf{DataSummary.csv}'.  
\label{table2s}}
\end{table}

\begin{table}
\csvreader[tabular=|l|c|c|c|c|c|c|c|c|c|,respect all,
    table head=\hline \textbf{Label} & $\bm\rho$ (kg.m$^{-3}$) & $\bm\eta$ (kg.m$^{-1}$.s$^{-1}$) & $\bm\Gamma$ (kg.s$^{-2}$) & $\bm{D}$ (m) & \textbf{Oh} & $\bm{\tau_{vi}}$ (s) & $\bm{\tau_{ic}}$ (s) & $\bm{\tau_{vc}}$ (s) & $\bm{\tau_{o}}$ (s)  \\ \hline,
    late after line=\\\hline,
    filter=\equal{\Type}{Coalescence}
    ]%
{./FiguresArxivFinal/FIGS/Summary.csv}{Label=\Label,Type=\Type,Density=\Density,Viscosity=\Viscosity, 5=\SufTens,7=\Size, 8=\Oh, 15=\tvi, 16=\tic, 17=\tvc, 18=\to}%
{\Label  &\Density &\Viscosity &\SufTens & \Size & \Oh & \tvi & \tic & \tvc & \to }
\caption{Summary of the properties of all coalescence experiments reproduced in the article: density $\rho$, viscosity $\eta$, surface tension $\Gamma$, extrinsic size $D$, Ohnesorge number $\text{Oh}=\eta/(\rho\Gamma D)^\frac{1}{2}$, together with the values of the four time scales $\tau_{vi}$, $\tau_{ic}$, $\tau_{vc}$ and $\tau_o$.  The content of this table is available in the supplementary file `\textbf{DataSummary.csv}'.  
\label{table2c}}
\end{table}

\begin{table}
\csvreader[tabular=|l|c|c|c|c|c|c|c|c|c|,respect all,
    table head=\hline \textbf{Label} & $\bm\rho$ (kg.m$^{-3}$) & $\bm\eta$ (kg.m$^{-1}$.s$^{-1}$) & $\bm\Gamma$ (kg.s$^{-2}$) & $\bm{D}$ (m) & \textbf{Oh} & $\bm{\tau_{vi}}$ (s) & $\bm{\tau_{ic}}$ (s) & $\bm{\tau_{vc}}$ (s) & $\bm{\tau_{o}}$ (s)  \\ \hline,
    late after line=\\\hline,
    filter=\equal{\Type}{Pinching}
    ]%
{./FiguresArxivFinal/FIGS/Summary.csv}{Label=\Label,Type=\Type,Density=\Density,Viscosity=\Viscosity, 5=\SufTens,7=\Size, 8=\Oh, 15=\tvi, 16=\tic, 17=\tvc, 18=\to }%
{\Label  &\Density &\Viscosity &\SufTens & \Size & \Oh & \tvi & \tic & \tvc & \to}
\caption{Summary of the properties of all pinching experiments reproduced in the article: density $\rho$, viscosity $\eta$, surface tension $\Gamma$, extrinsic size $D$, Ohnesorge number $\text{Oh}=\eta/(\rho\Gamma D)^\frac{1}{2}$, together with the values of the four time scales $\tau_{vi}$, $\tau_{ic}$, $\tau_{vc}$ and $\tau_o$. The content of this table is available in the supplementary file `\textbf{DataSummary.csv}'.   
\label{table2p}}
\end{table}

\subsection{Extended figure legends}
The labels associated with each data set defined in Tables~\ref{table2s}, \ref{table2c} and \ref{table2p} are used in the legends of each figure present in the supplementary archive `\textbf{FigureLegends.zip}'. For each figure of the article and of the present supplementary material, the archive contain a txt file with the labels of all the data plotted in the given figure. The labels can then be used to recover the values of the material parameters, of $D$, and of any additional metric by using the various tables of this supplementary material.    

\clearpage

\section{Visco-inertio-capillary systems of units}
In this section, we provide additional details on each of the four systems of units introduced in the article: visco-inertial (vi), visco-capillary (vc), inertio-capillary (ic) and Ohnesorge units (Oh).  

\subsection{Physical quantities}
In the article we provided the characteristic length, time and mass scales associated with each system of units. From these quantities, the values of any quantity that can be expressed in units of mass, length and time can be derived. For instance, in the article we systematically gave the values of the stress $\Sigma=m.\ell^{-1}.\tau^{-2}$. In Table~\ref{table3} we give the expressions for a few other quantities of interest. 
 
\begin{table}[h]
\begin{tabular}{|l|c|c|c|c|}
\hline
 Quantity & \textbf{vi units} & \textbf{vc units} & \textbf{ic units} & \textbf{Oh units} \\ \hline\hline
Time  $\tau$ & $\frac{\rho D^2}{\eta}$ &$\frac{\eta D}{\Gamma}$  & $\Big(\frac{\rho D^3}{\Gamma}\Big)^{\frac{1}{2}}$  & $\frac{\eta^3}{\Gamma^2 \rho}$ \\\hline
Length  $\ell$ & $D$ &$D$  & $D$  & $\frac{\eta^2}{\Gamma \rho}$ \\\hline
Mass  $m$ & $\rho D^3$ &$\frac{\eta^2 D^2}{\Gamma}$  & $\rho D^3$  & $\frac{\eta^6}{\Gamma^3 \rho^2}$ \\\hline\hline
Viscosity  & $\eta$ &$\eta$  & $(\rho \Gamma D)^\frac{1}{2}$  & $\eta$ \\\hline
Surface tension  & $\frac{\eta^2}{\rho D}$ &$\Gamma$  & $\Gamma$  & $\Gamma$ \\\hline
Density  & $\rho$ &$\frac{\eta^2}{\Gamma D}$  & $\rho$  & $\rho$ \\\hline\hline

Speed  & $\frac{\eta}{\rho D}$ &$\frac{\Gamma}{\eta}$  & $\Big(\frac{\Gamma}{\rho D}\Big)^{\frac{1}{2}}$  & $\frac{\Gamma}{\eta}$ \\\hline
Acceleration  & $\frac{\eta^2}{\rho^2 D^3}$ &$\frac{\Gamma^2}{\eta^2 D}$  & $\frac{\Gamma}{\rho D^2}$  & $\frac{\Gamma^3\rho}{\eta^4}$ \\\hline

Energy  & $\frac{\eta^2 D}{\rho}$ &$\Gamma D^2$  & $\Gamma D^2$ & $\frac{\eta^4}{\rho^2 \Gamma}$ \\\hline
Force  & $\frac{\eta^2}{\rho}$ &$\Gamma D$  & $\Gamma D$  & $\frac{\eta^2}{\rho}$ \\\hline
Stress  & $\frac{\eta^2}{\rho D^2}$ &$\frac{\Gamma}{D}$  & $\frac{\Gamma}{D}$ &$\frac{\Gamma^2 \rho}{\eta^2}$ \\\hline
Power  & $\frac{\eta^3}{\rho^2 D}$ &$\frac{\Gamma^2 D}{\eta}$  & $\Big(\frac{\Gamma^3 D}{\rho}\Big)^{\frac{1}{2}}$  & $\frac{\eta \Gamma}{\rho}$ \\\hline
\end{tabular}
\caption{Summary of the main physical quantities in the four systems of units used in the article. 
\label{table3}}
\end{table}

\subsection{Scaling regimes}
In the article we mostly considered three simple scaling regimes (vi, vc and ic), and two non-trivial regimes, namely the size-dependent visco-capillary regime of Tanner's law and the size-dependent inertio-capillary regime of Rayleigh's law. These five regimes can be written in any of the four system of units as $d/\ell^*=Oh^\beta (t/\tau^*)^\alpha$, with $\tau^*=\gamma_1 \tau$ and $\ell^*=\gamma_2 \ell$, where $\tau$ and $\ell$ can be chosen from one of the four systems. Table~\ref{table4} give the values of Oh$^\beta$ for each regime and choice of units. Note that the prefactors $\gamma_1$ and $\gamma_2$ take into account the dimensionless coefficients of the scaling laws ($\delta_{vc}$, $\delta_{ic}$ etc) and will be discussed in section~\ref{prefact}.  

\begin{figure}
\centering
\begin{overpic}[abs,unit=1mm,width=8cm]{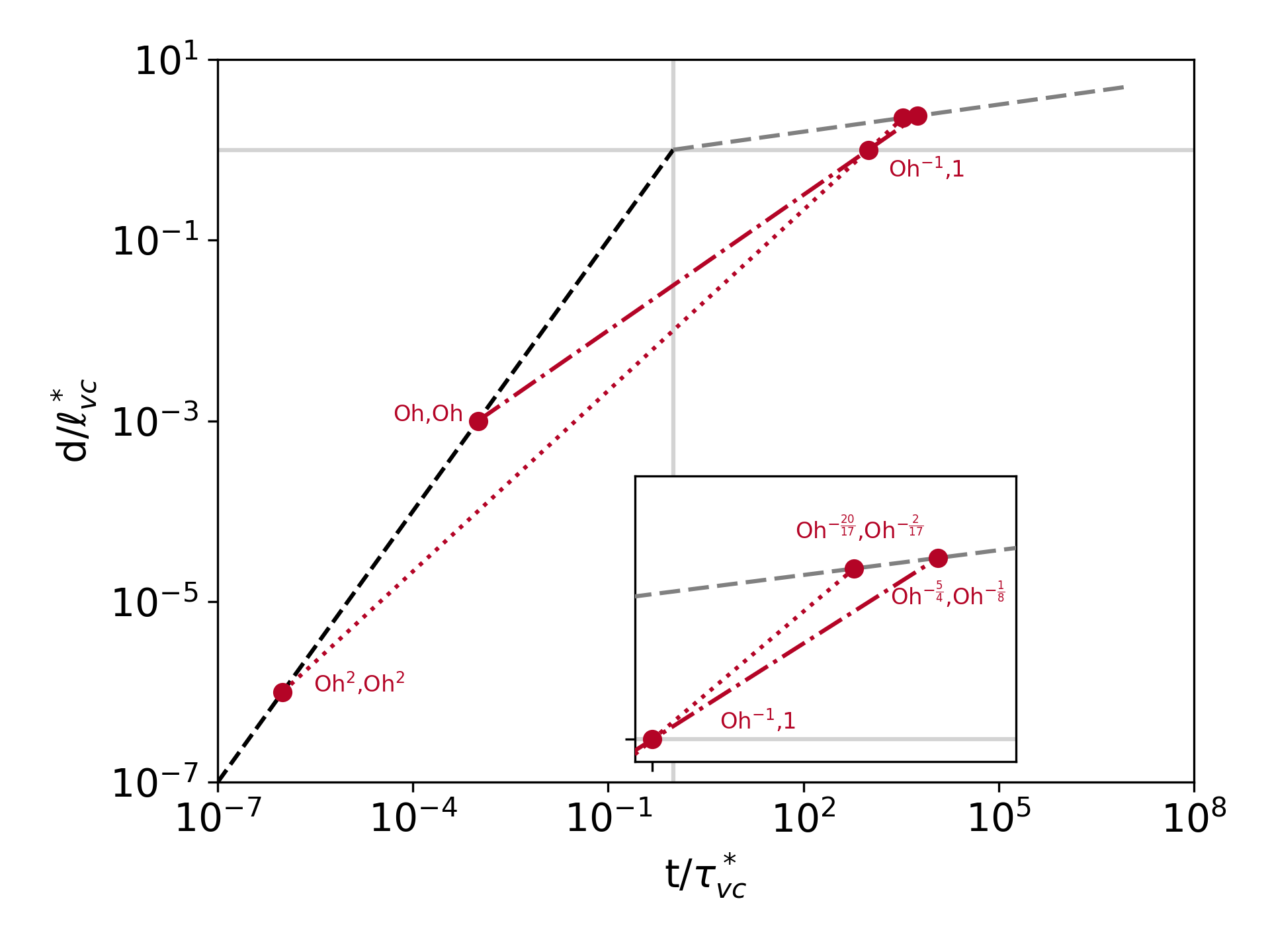}
\put(17,52){(a)}
\end{overpic}
\begin{overpic}[abs,unit=1mm,width=8cm]{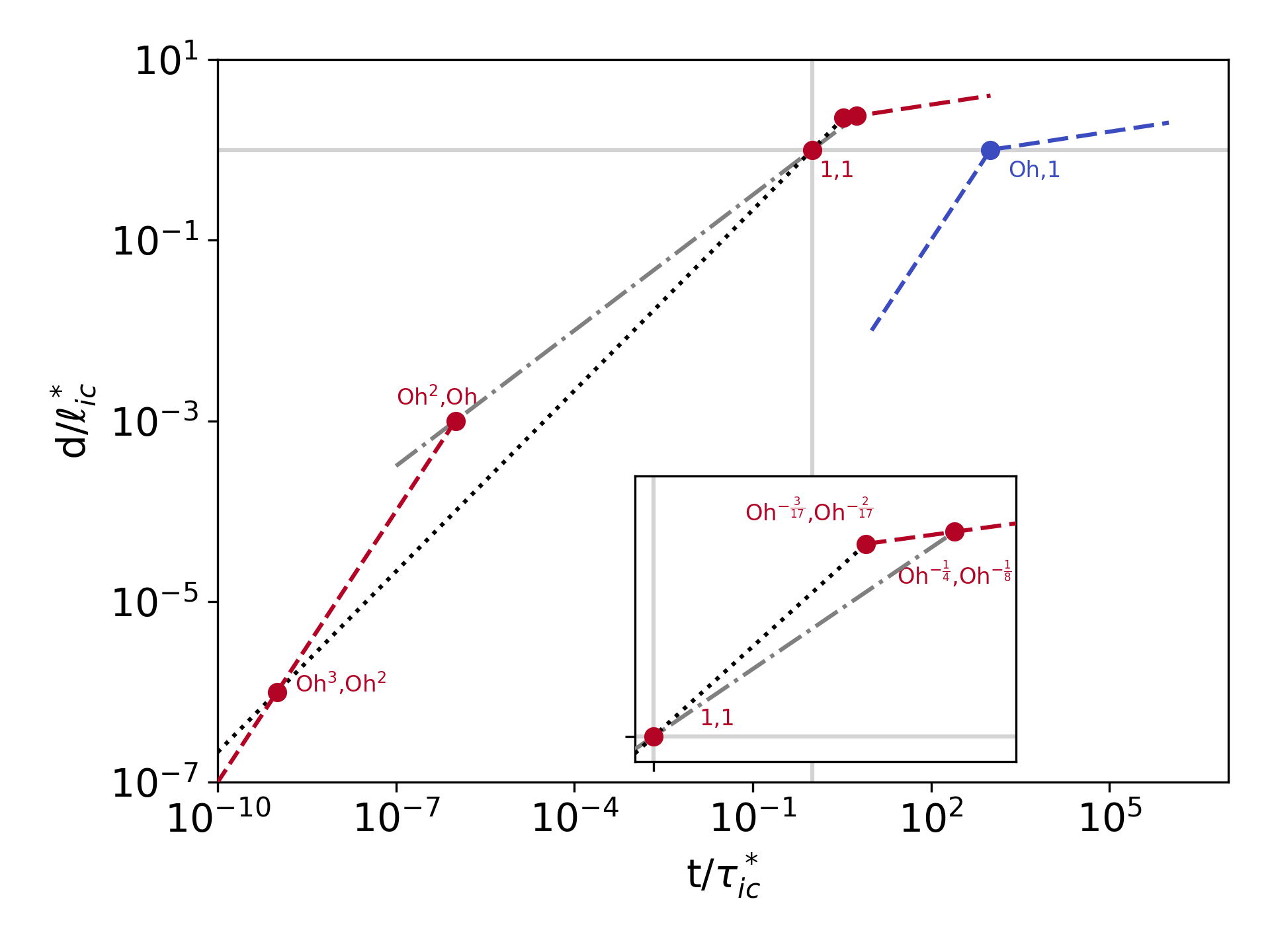}
\put(17,52){(b)}
\end{overpic}
\begin{overpic}[abs,unit=1mm,width=8cm]{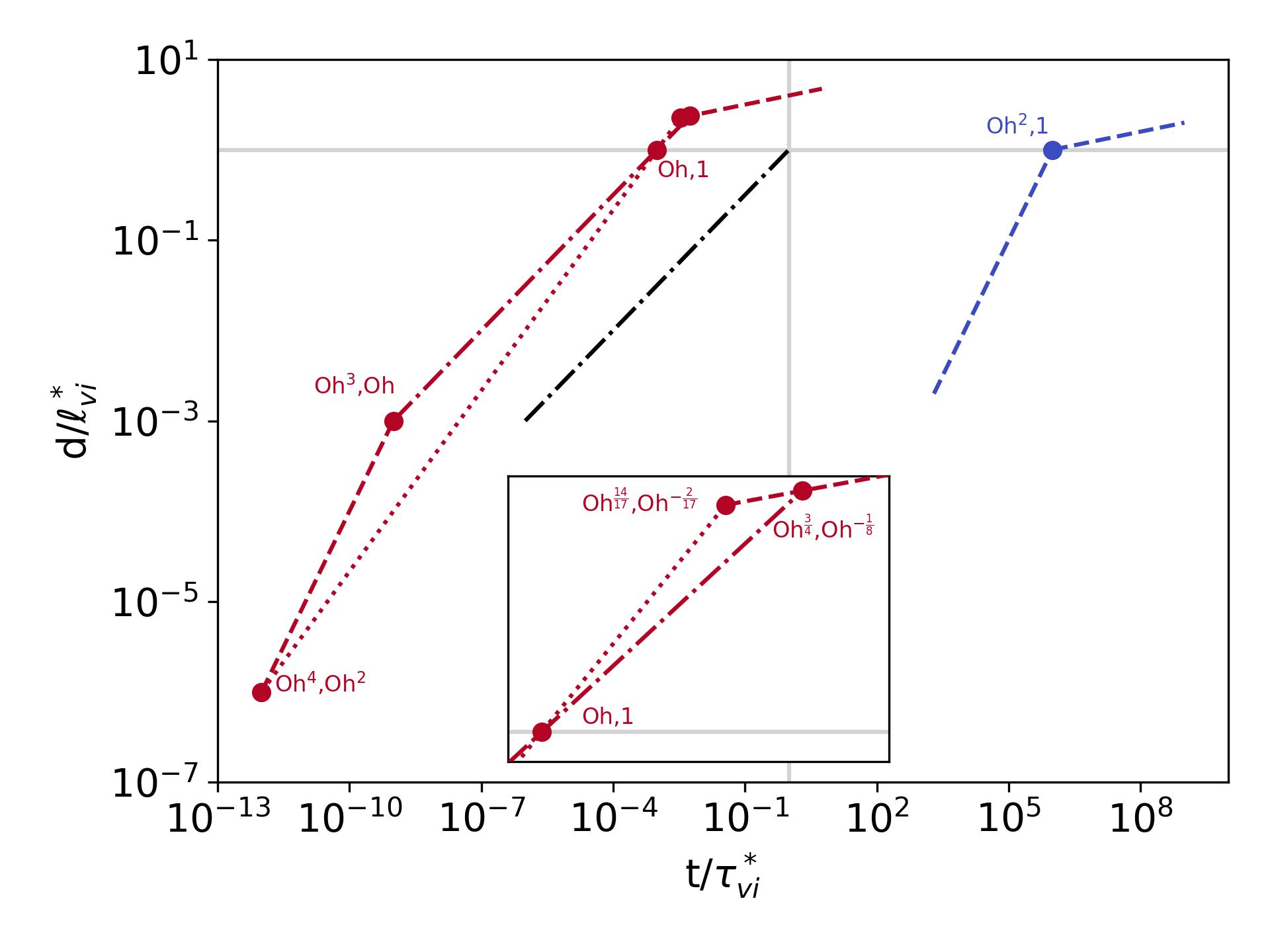}
\put(17,52){(c)}
\end{overpic}
\begin{overpic}[abs,unit=1mm,width=8cm]{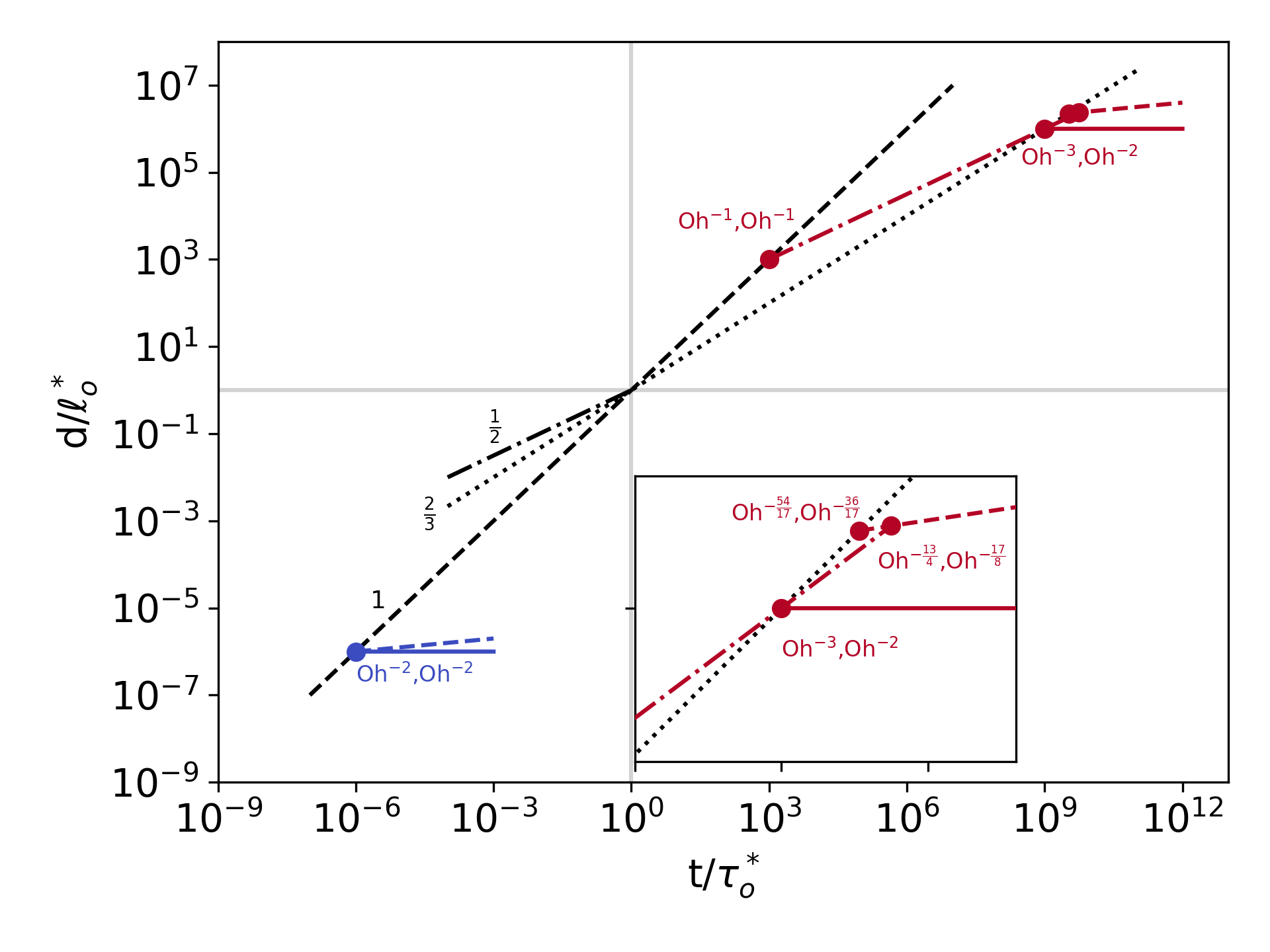}
\put(17,52){(d)}
\end{overpic}
\caption{Schematic plots of the main scaling laws discussed in the article, together with the coordinates of the their intersections, expressed in terms of the Ohnesorge number $\text{Oh}$. In each panel, the inset provides a close-up on the intersection between the $\frac{2}{3}$ regime, Rayleigh's $\frac{1}{2}$ regime and Tanner's $\frac{1}{10}$ regime. Note that the precision provided by current experiments does not offer ways to confirm or inform the slight differences between the intersections of the $\frac{1}{2}$ or $\frac{2}{3}$ regime with Tanner's law. For instance, for the visco-capillary units the intersection of the $\frac{2}{3}$ and $\frac{1}{10}$ regime is given by $t/\tau_{vc}^*=\text{Oh}^{-\frac{20}{17}}=\text{Oh}^{-\frac{5}{4}}\text{Oh}^{\frac{5}{68}}$ and $d/\ell_{vc}^*=\text{Oh}^{-\frac{2}{17}}=\text{Oh}^{-\frac{1}{8}}\text{Oh}^{\frac{1}{136}}$, which are here expressed from the coordinates of the intersection between the $\frac{1}{2}$ and $\frac{1}{10}$ regimes. Since these intersections are meaningful if $\text{Oh}<1$, the corrections factors $\text{Oh}^{\frac{5}{68}}$ and $\text{Oh}^{\frac{1}{136}}$ are always close to 1. The color of each curve gives the value of Oh (see Fig.~5 of main article).
\label{figSI_schema}}
\end{figure}

Schematic versions of the scaling regimes in vi, vc, ic and Oh units are given in SI-Fig.~\ref{figSI_schema} to illustrate how the coordinates of the intersection points can be expressed from the Ohnesorge number. 

\begin{table}[h!]
\begin{tabular}{|l|c|c|c|c|}
\hline
Law & \textbf{vi units} & \textbf{vc units} & \textbf{ic units} & \textbf{Oh units} \\ \hline\hline
$d\propto\Big(\frac{\eta}{\rho}\Big)^\frac{1}{2} t^\frac{1}{2}$&  1 & Oh & Oh$^\frac{1}{2}$  & 1 \\\hline
$d\propto\frac{\Gamma}{\eta} t$& Oh$^{-2}$ & 1 & Oh$^{-1}$  & 1 \\\hline
$d=\Big(\frac{\Gamma}{\rho}\Big)^\frac{1}{3} t^\frac{2}{3}$& Oh$^{-\frac{2}{3}}$ & Oh$^{\frac{2}{3}}$ & 1  & 1 \\\hline
$d\propto\Big(\frac{\Gamma D^9}{\eta}\Big)^\frac{1}{10} t^\frac{1}{10}$& Oh$^{-\frac{1}{5}}$ & 1 & Oh$^{-\frac{1}{10}}$  & Oh$^{-\frac{9}{5}}$ \\\hline
$d\propto\Big(\frac{\Gamma D}{\rho}\Big)^\frac{1}{4} t^\frac{1}{2}$& Oh$^{-\frac{1}{2}}$ &  Oh$^{\frac{1}{2}}$ & 1  & Oh$^{-\frac{1}{2}}$ \\\hline
\end{tabular}
\caption{Ohnesorge-based prefactors for the five main scaling laws discussed in the article. For instance, $d\propto (\Gamma/\eta)t$ can be written as $d/\ell_{vi} \propto \text{Oh}^{-2} (t/\tau_{vi})$, or $d/\ell_{vc} \propto  t/\tau_{vc}$, or $d/\ell_{ic} \propto \text{Oh}^{-1} (t/\tau_{ic})$, or $d/\ell_{o} \propto t/\tau_{o}$.
\label{table4}}
\end{table}

\clearpage

\subsection{Supplementary figures in vi, vc and ic units}
In this section we give different versions of the visco-capillary, inertio-capillary and visco-inertial plots of the article highlighting subsets of the data. 
\begin{figure}[h!]
\centering
\begin{overpic}[abs,unit=1mm,width=8cm]{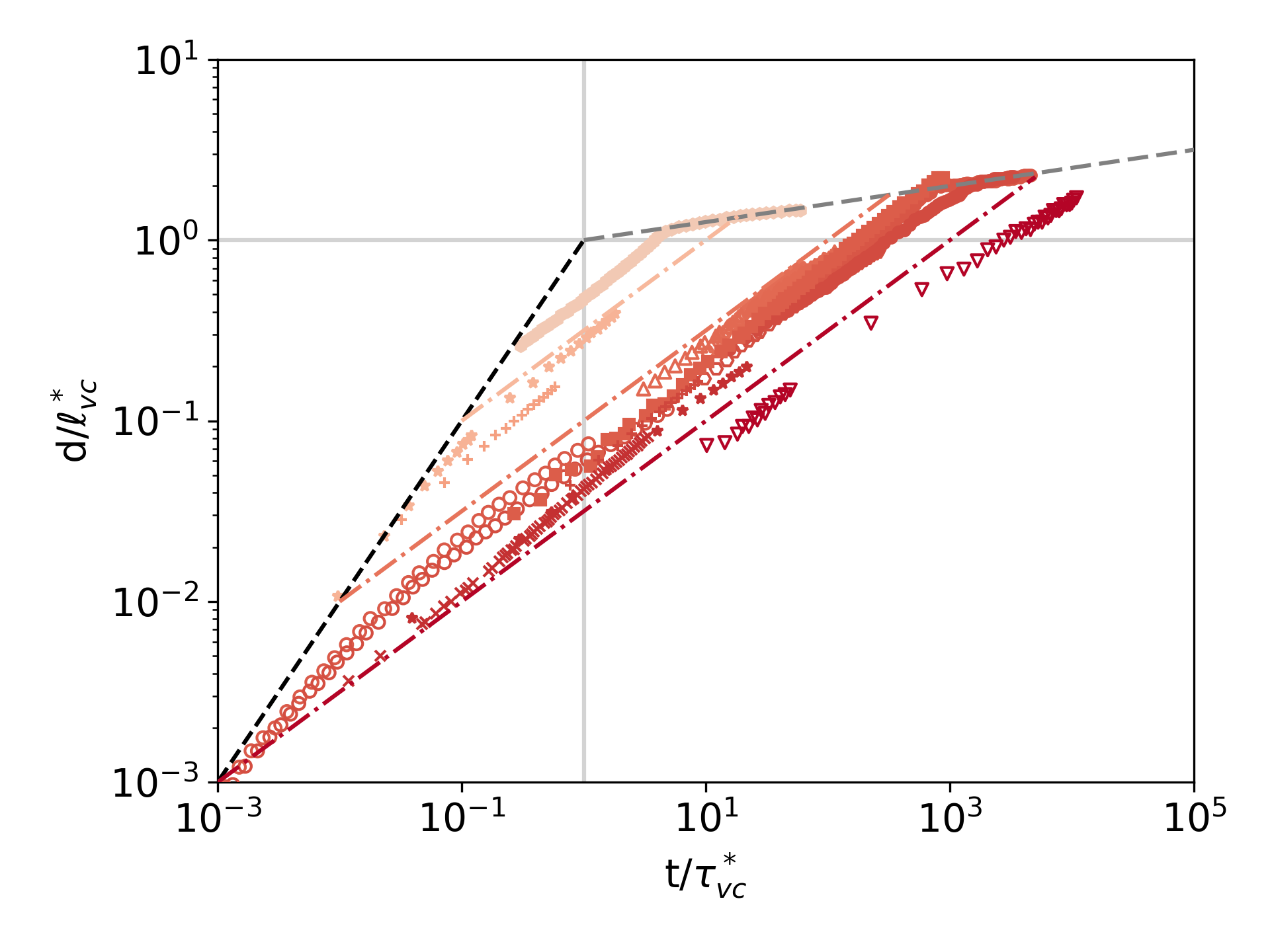}
\put(17,52){(a)}
\end{overpic}
\begin{overpic}[abs,unit=1mm,width=8cm]{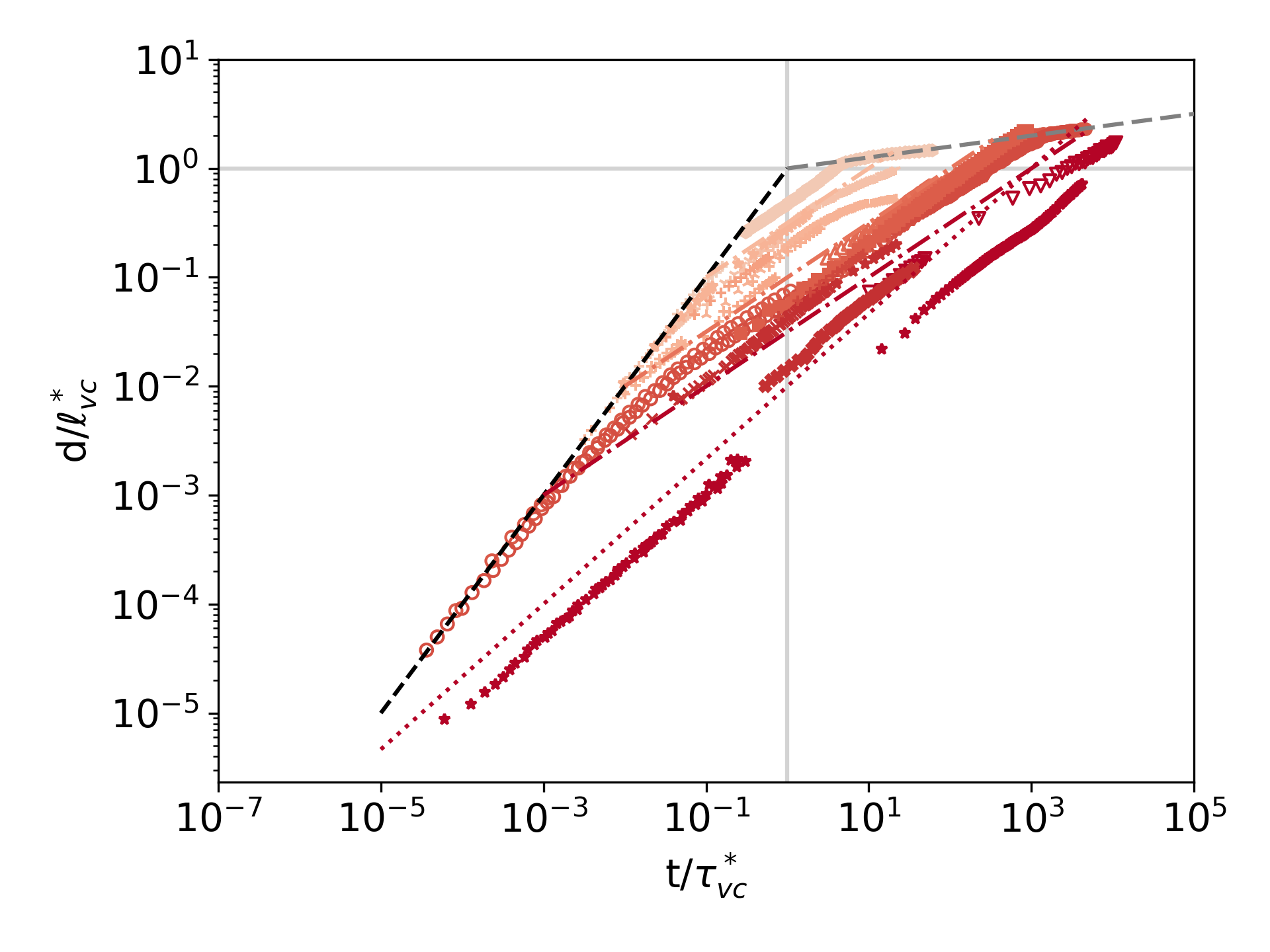}
\put(17,52){(b)}
\put(65,43){\small He}
\end{overpic}
\caption{Supplementary visco-capillary plots. (a) Data from Fig.~2c of the article, replotted in visco-capillary units. The three dashed lines follow Rayleigh's regime $d/\ell_{vc}^* = \text{Oh}^\frac{1}{2} (t/\tau_{vc}^*)^\frac{1}{2}$, for $\text{Oh}=10^{-1}$, $10^{-2}$ and $10^{-3}$. (b) Data from Fig.~2d of the article, replotted in visco-capillary units. In addition to the the three dashed lines following Rayleigh's regime, the dotted line follows the inertio-capillary regime $d/\ell_{vc}^* = \text{Oh}^\frac{2}{3} (t/\tau_{vc}^*)^\frac{1}{2}$, for $\text{Oh}=10^{-3}$. The data set labeled `He' corresponds to superfluid Helium, for which a viscosity is unavailable. We artificially set $\eta=10^{-6}$~Pa.s in order to represent the data on the visco-capillary plot. If $\eta\simeq 0$, the actual abscissa on this set should be infinite. It is probably more reasonable to assume that an effective viscosity would set in, due to a different dissipation mechanism~\cite{Donnelly1979}. The color of each curve gives the value of Oh (see Fig.~5 of the main article).
\label{Fig4SuppVC}}
\end{figure} 

\begin{figure}
\centering
\begin{overpic}[abs,unit=1mm,width=8cm]{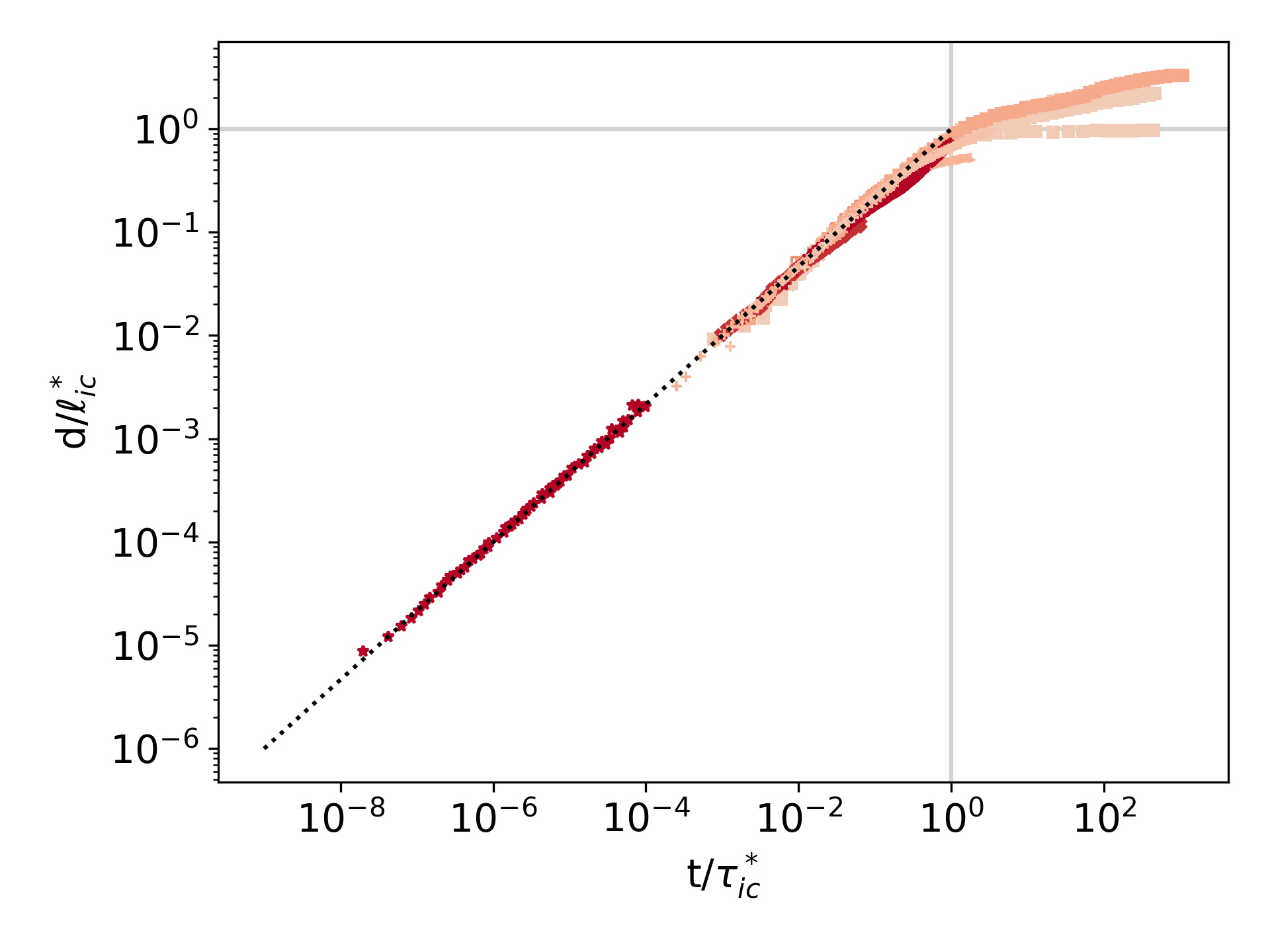}
\put(17,52){(a)}
\end{overpic}
\begin{overpic}[abs,unit=1mm,width=8cm]{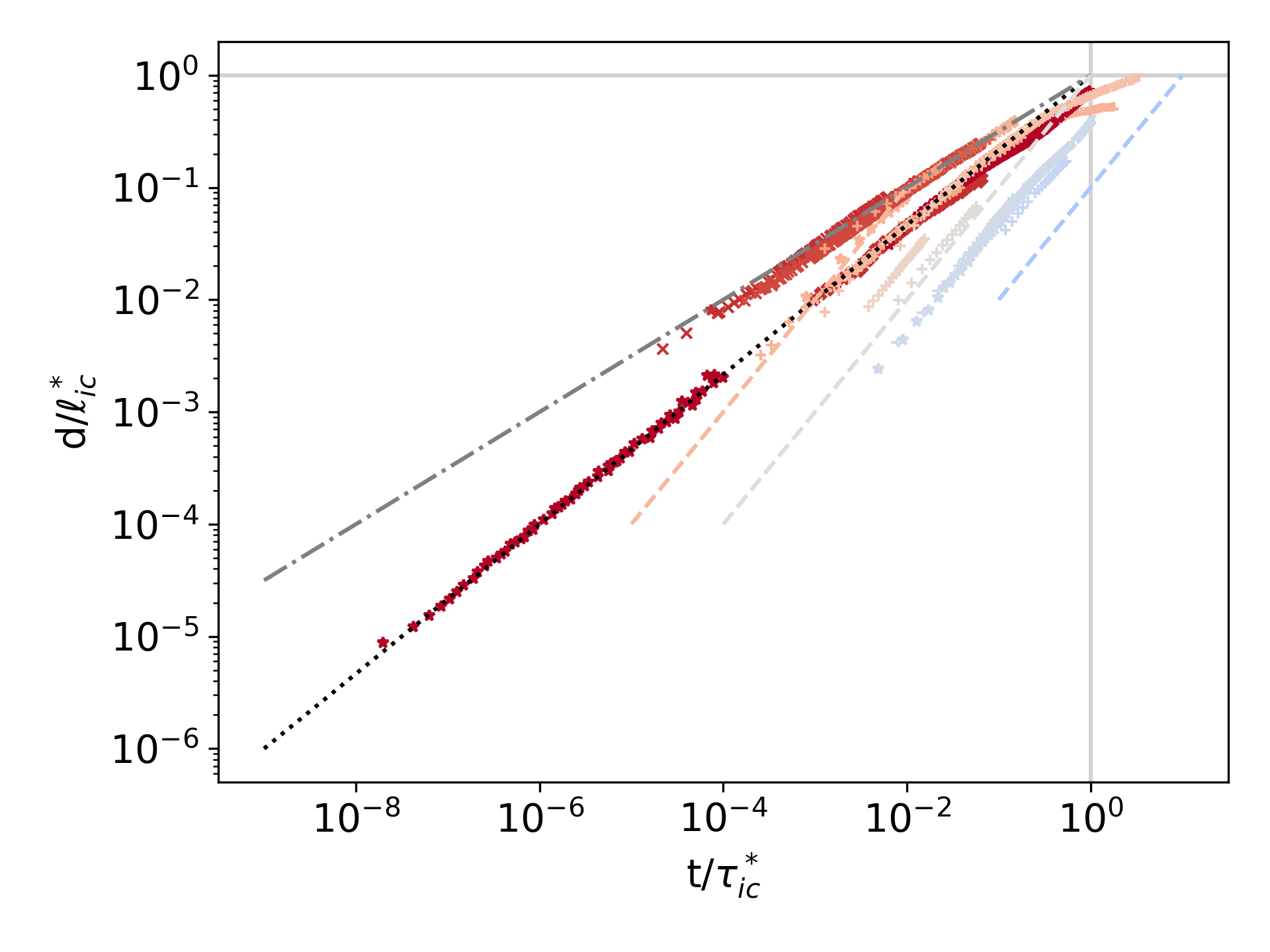}
\put(17,52){(b)}
\end{overpic}
\caption{Supplementary inertio-capillary plots. (a) All spreading, coalescence and pinching experiments reproduced in the article exhibiting a $\frac{2}{3}$ regime are replotted in inertio-capillary units (see Tables~\ref{table4s}-\ref{table4p} for values of $\delta_{ic}$). (b) All pinching experiments replotted in visco-capillary units. In addition to Rayleigh's regime (dotted-dashed line) and to the inertio-capillary regime (dotted line), the dashed lines follow the visco-capillary regime $d/\ell_{ic}^* = \text{Oh}^{-1} (t/\tau_{ic}^*)$, for $\text{Oh}=10^{-1}$, 1, $10^1$. The color of each curve gives the value of Oh (see Fig.~5 of the main article).
\label{Fig4SuppIC}}
\end{figure} 

\begin{figure}
\centering
\begin{overpic}[abs,unit=1mm,width=8cm]{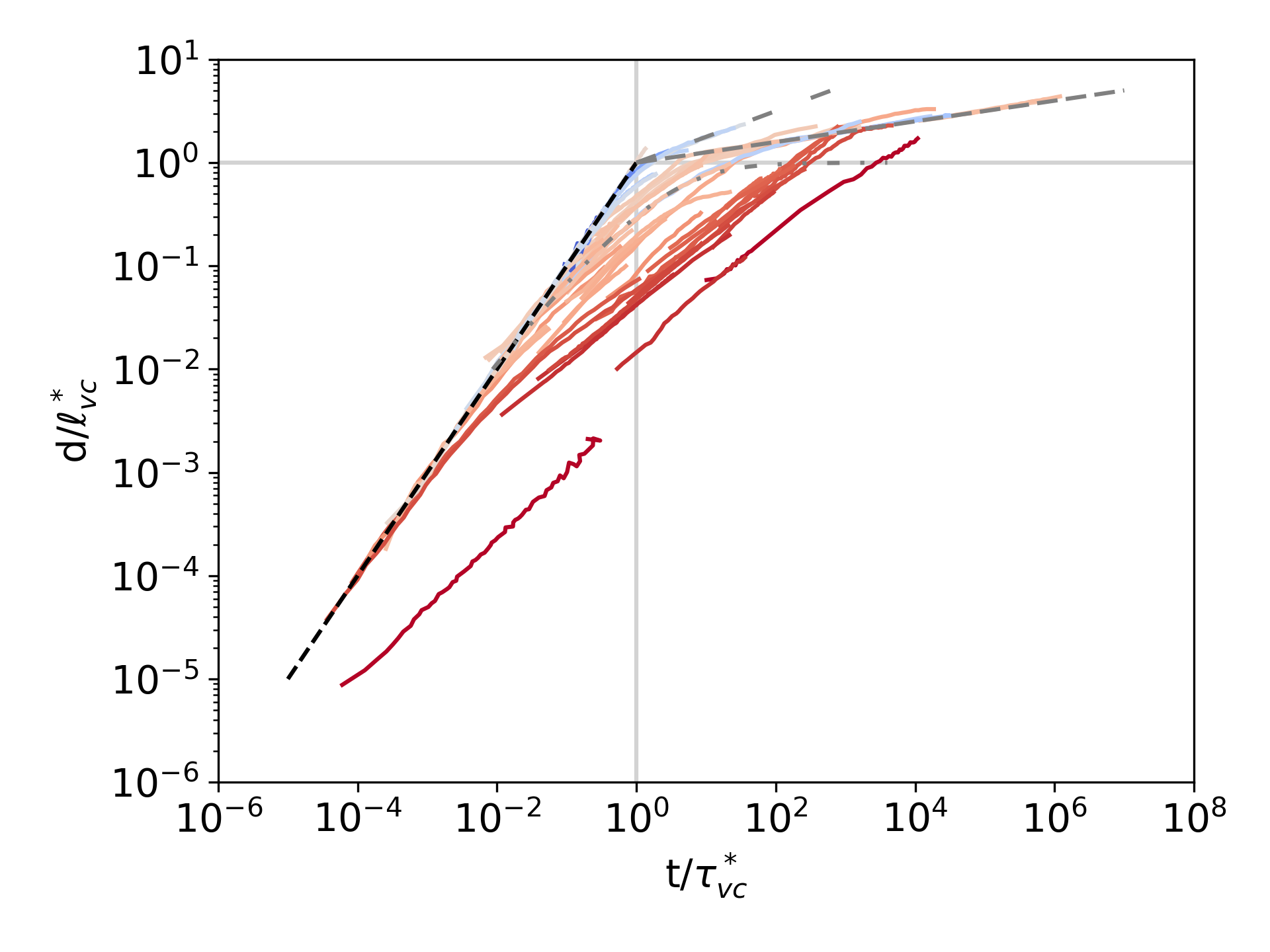}
\put(17,52){(a)}
\end{overpic}
\begin{overpic}[abs,unit=1mm,width=8cm]{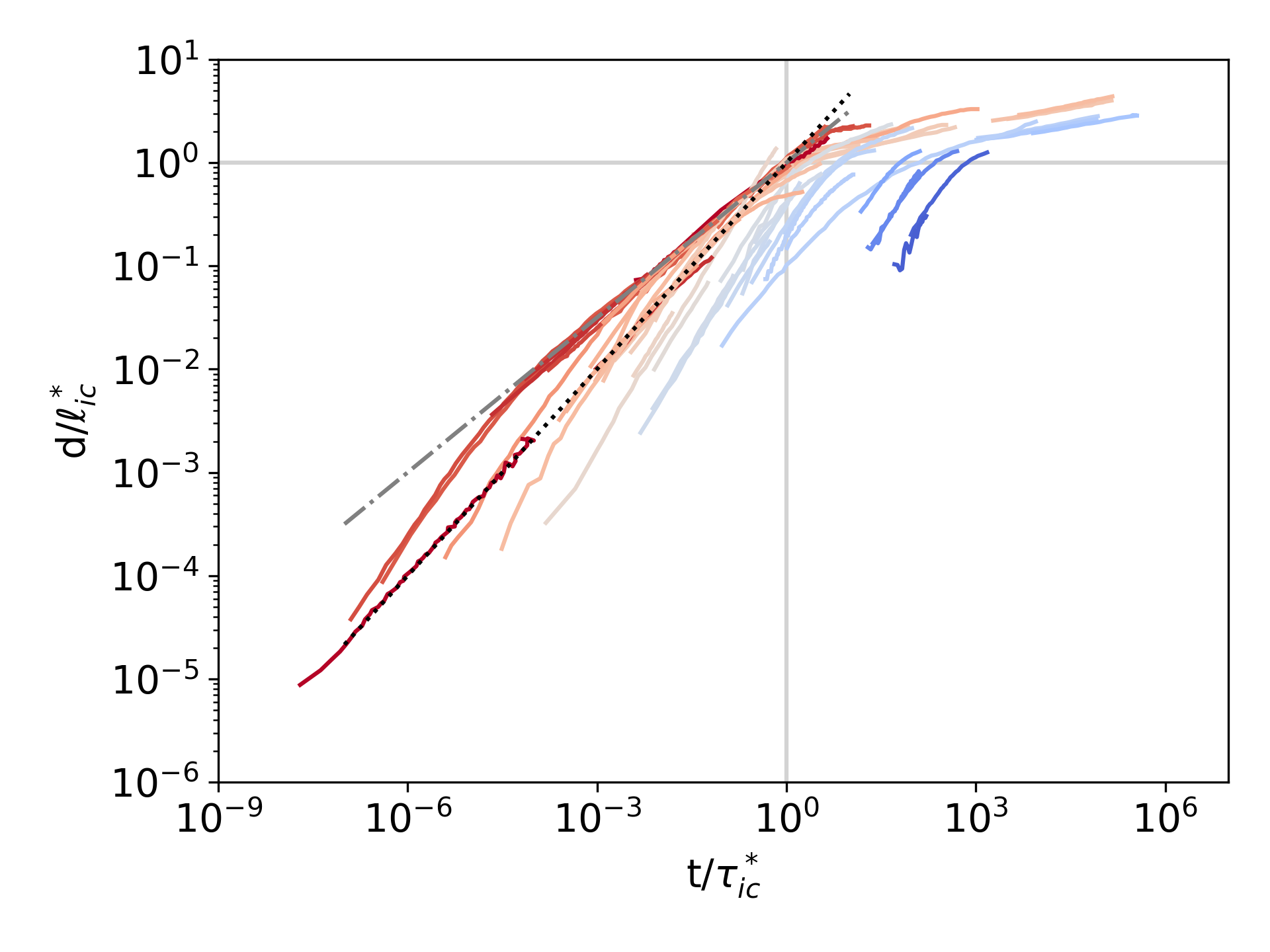}
\put(17,52){(b)}
\end{overpic}
\begin{overpic}[abs,unit=1mm,width=8cm]{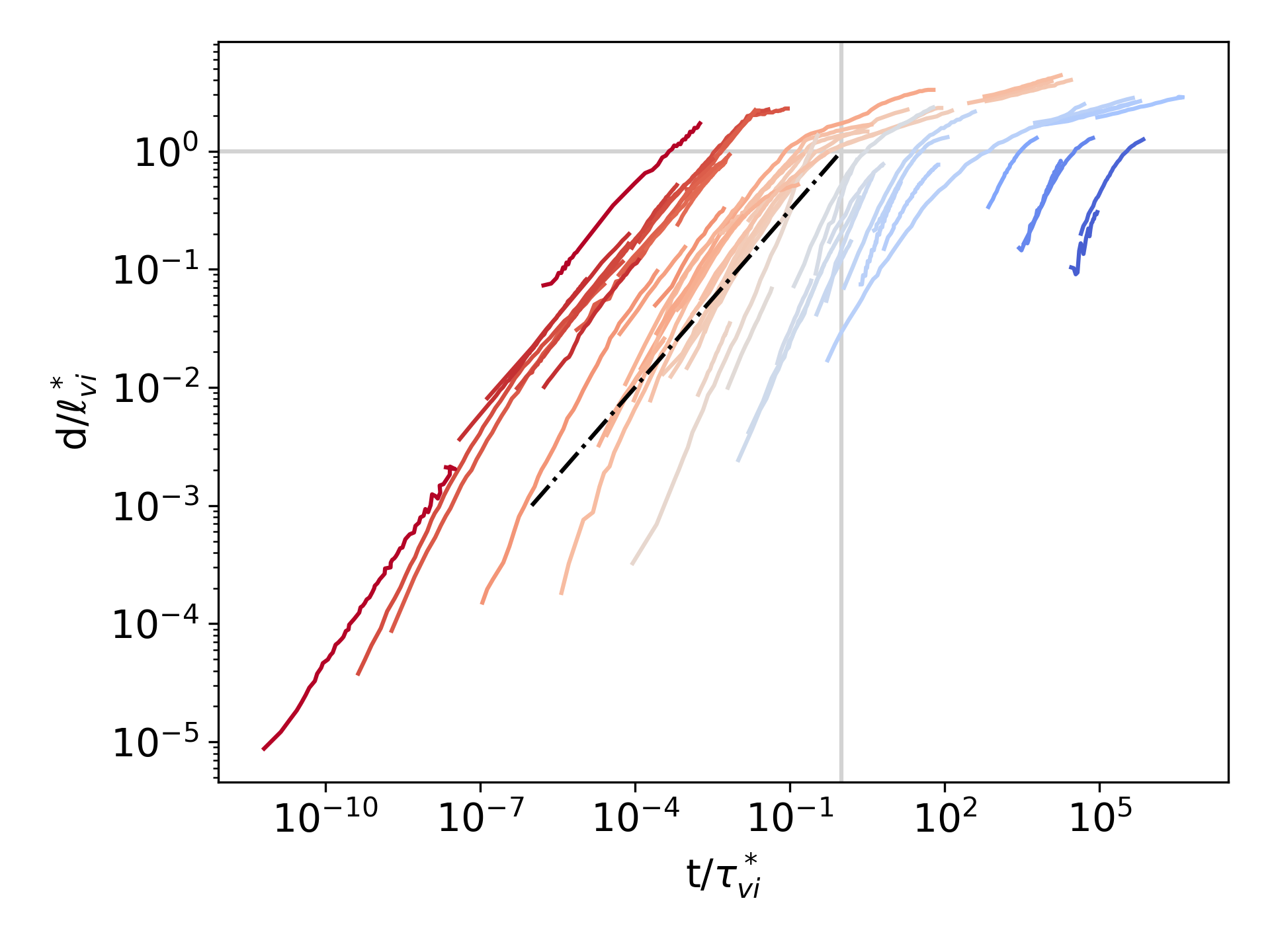}
\put(17,52){(c)}
\end{overpic}
\caption{All data sets from Fig.~5 of the main article reproduced in visco-capillary, inertio-capillary and visco-inertial units. See extended legend and animated figures for details. The color of each curve gives the value of Oh (see Fig.~5 of the main article). 
\label{FigallinOh}}
\end{figure}

\clearpage

\subsection{Dimensionless constants}
\begin{figure}
\centering
\includegraphics[width=17cm,clip]{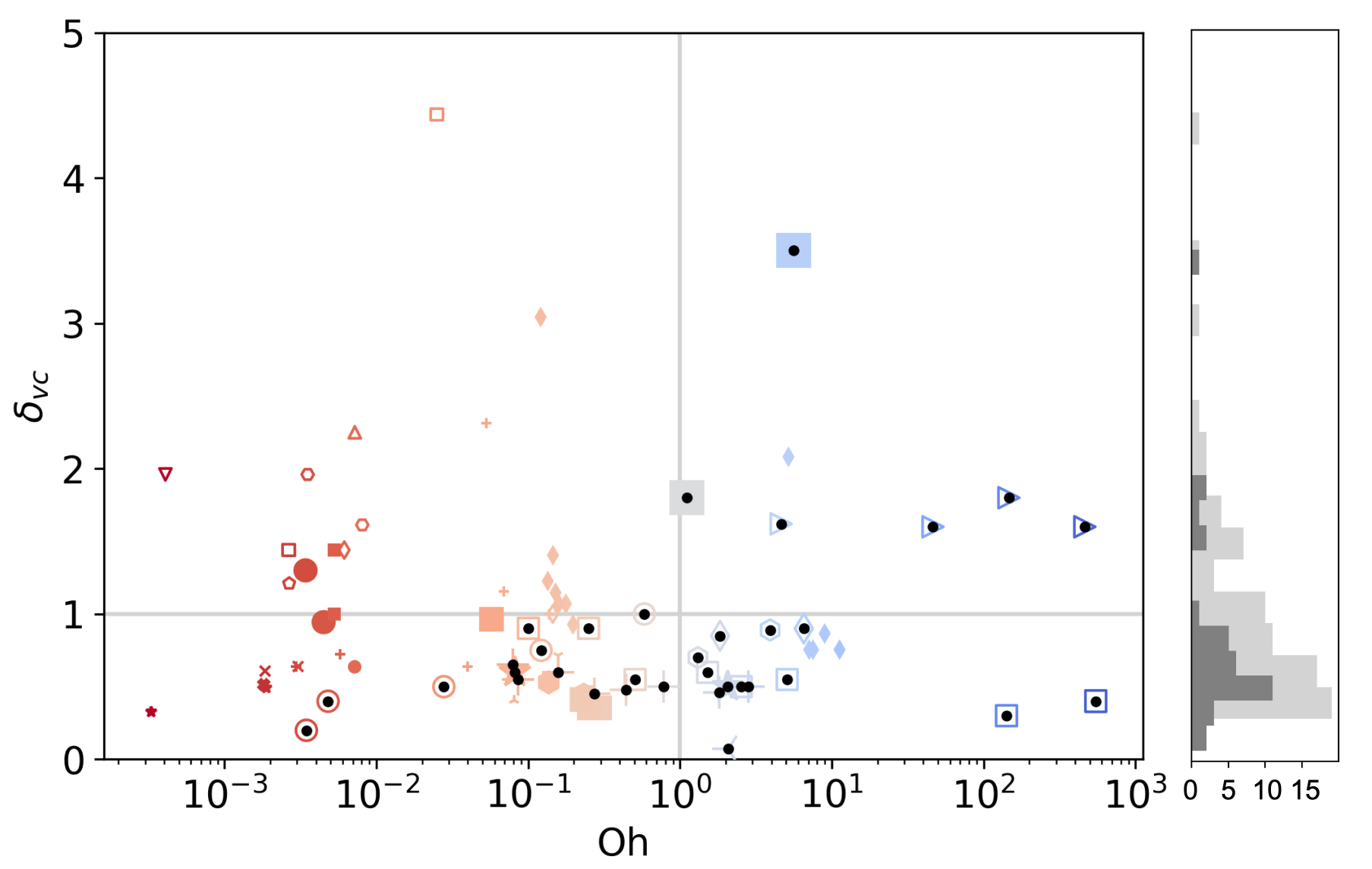}
\caption{Measurements of the values of the purely visco-capillary prefactors $\delta_{vc}$ as a function of the Ohnesorge number, for all experiments reproduced in the article. (Left) The biggest symbols with the added black dots correspond to direct measurements of $\delta_{vc}$ from fitting a linear regime present on the data. The medium-size symbols correspond to indirect measurements of $\delta_{vc}$ from two other prefactors (see section \ref{algebra}). The smallest symbols correspond to indirect measurements from a single other prefactor (least reliable). (Right) Distribution of the values of $\delta_{vc}$, for direct measurements (dark gray) and all measurements (light gray). 
\label{figSIdvc}}
\end{figure} 
All scaling regimes introduced in the article include dimensionless prefactors. For instance, the simple visco-capillary regime is $d=\delta_{vc} \Gamma t/\eta$. The prefactor $\delta_{vc}$ is expected to be a `constant of order 1', which more rigorously means that the variations of $\delta_{vc}$ with $d$, $t$, $\eta$ or $\Gamma$ can be at most logarithmic. This assumption that can be verified experimentally. For all experiments presenting a linear scaling $d\sim t$, we computed the value of $\delta_{vc}$ that better fitted the data. The values of $\delta_{vc}$ in all cases where the linear regime was present are given for spreading, coalescence and pinching in Tables~\ref{table4s}, \ref{table4c} and \ref{table4p}. The value of $\delta_{vc}$ appears to be independent of the Ohnesorge number, as shown in SI-Fig.~\ref{figSIdvc}. Only considering direct measurements the most frequent value (mode) is $\delta_{vc}\simeq 0.5$  and the mean is $\delta_{vc}\simeq 0.9$ (see Fig.~\ref{figSIdvc} for full distribution). We will show in section \ref{algebra} how additional values of $\delta_{vc}$ can be inferred from the prefactors of other scaling laws. 

\begin{figure}
\centering
\begin{overpic}[abs,unit=1mm,width=5cm]{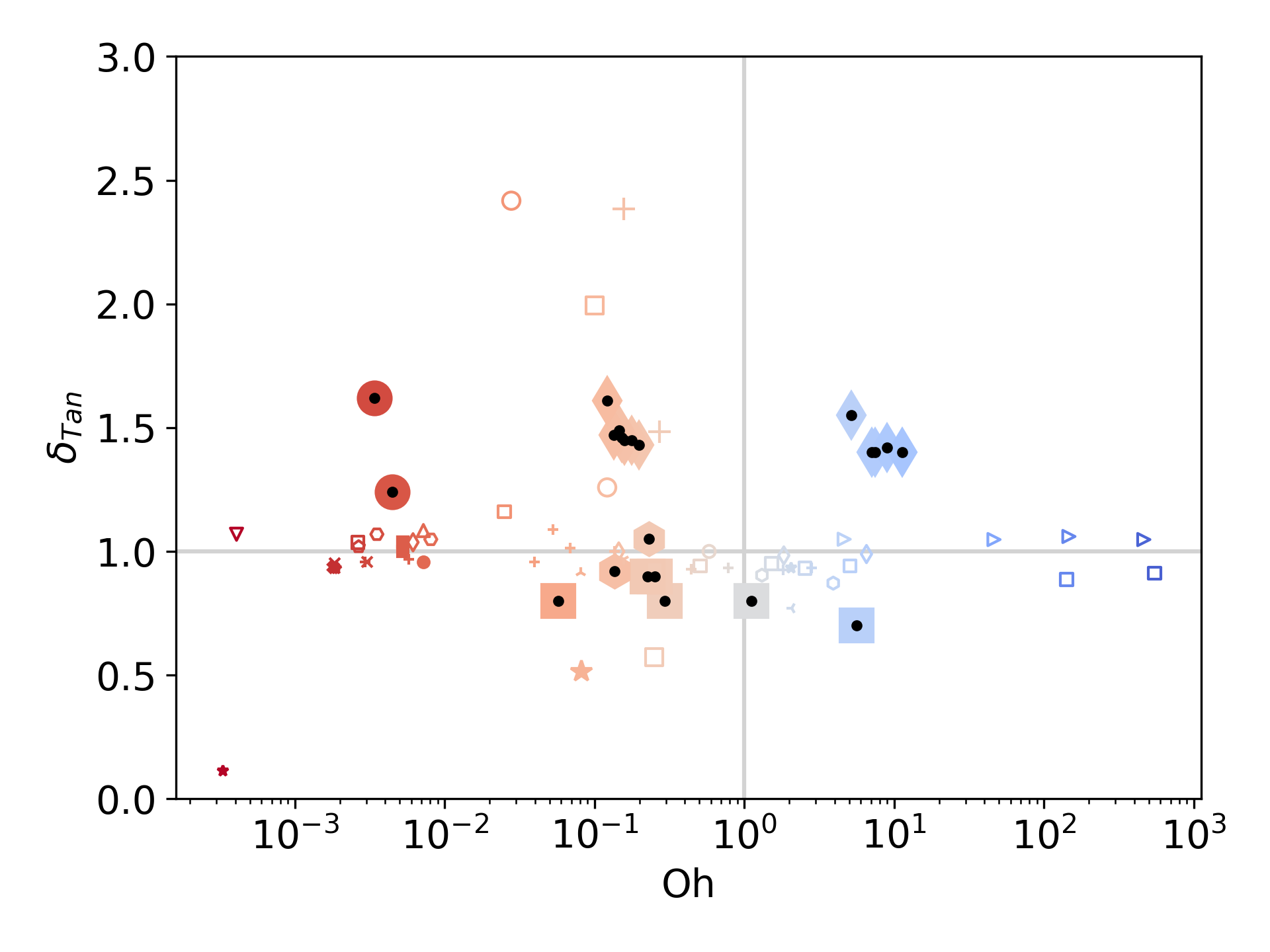}
\put(10,30){(a)}
\end{overpic}
\begin{overpic}[abs,unit=1mm,width=5cm]{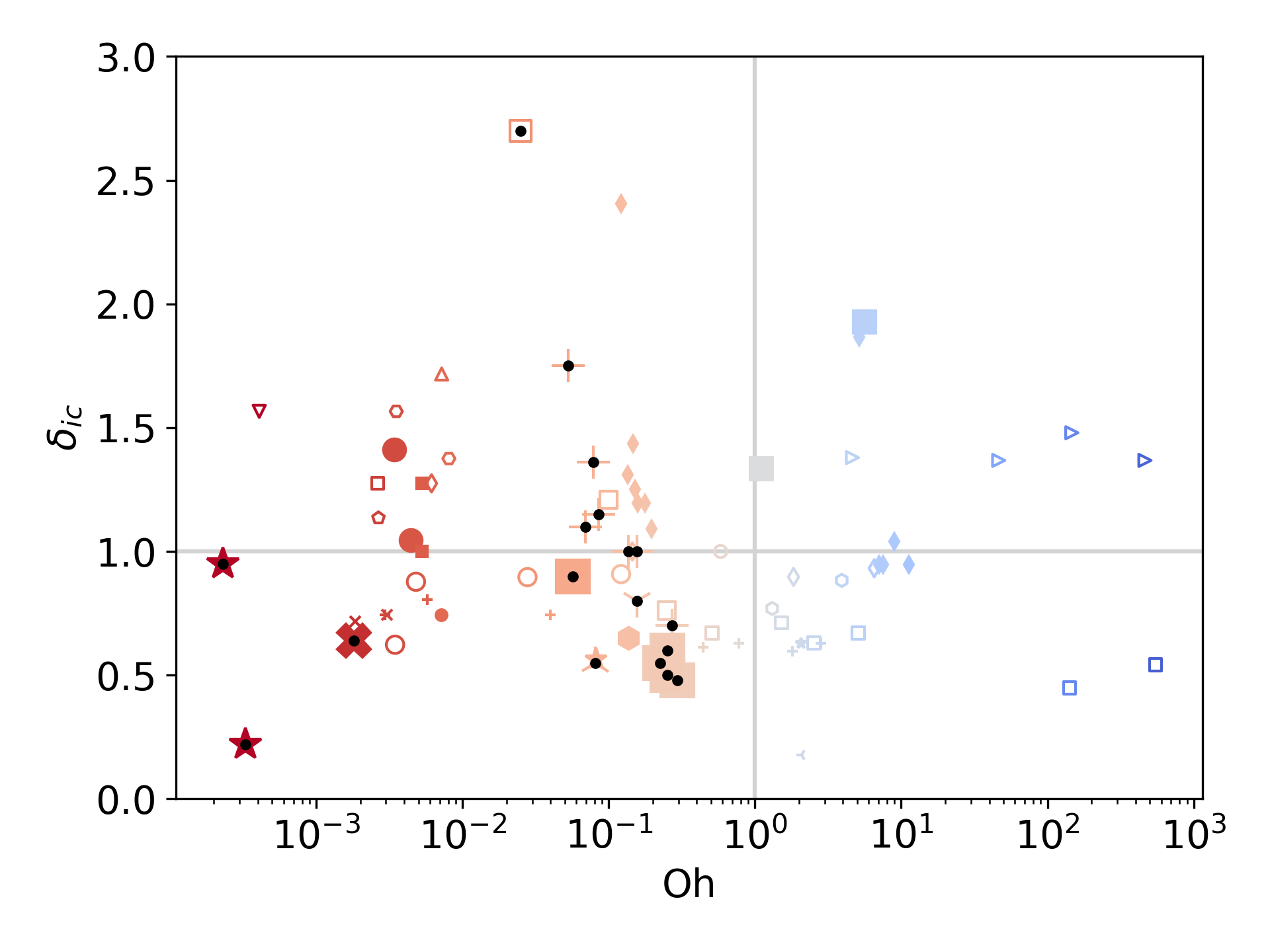}
\put(10,30){(b)}
\end{overpic}
\begin{overpic}[abs,unit=1mm,width=5cm]{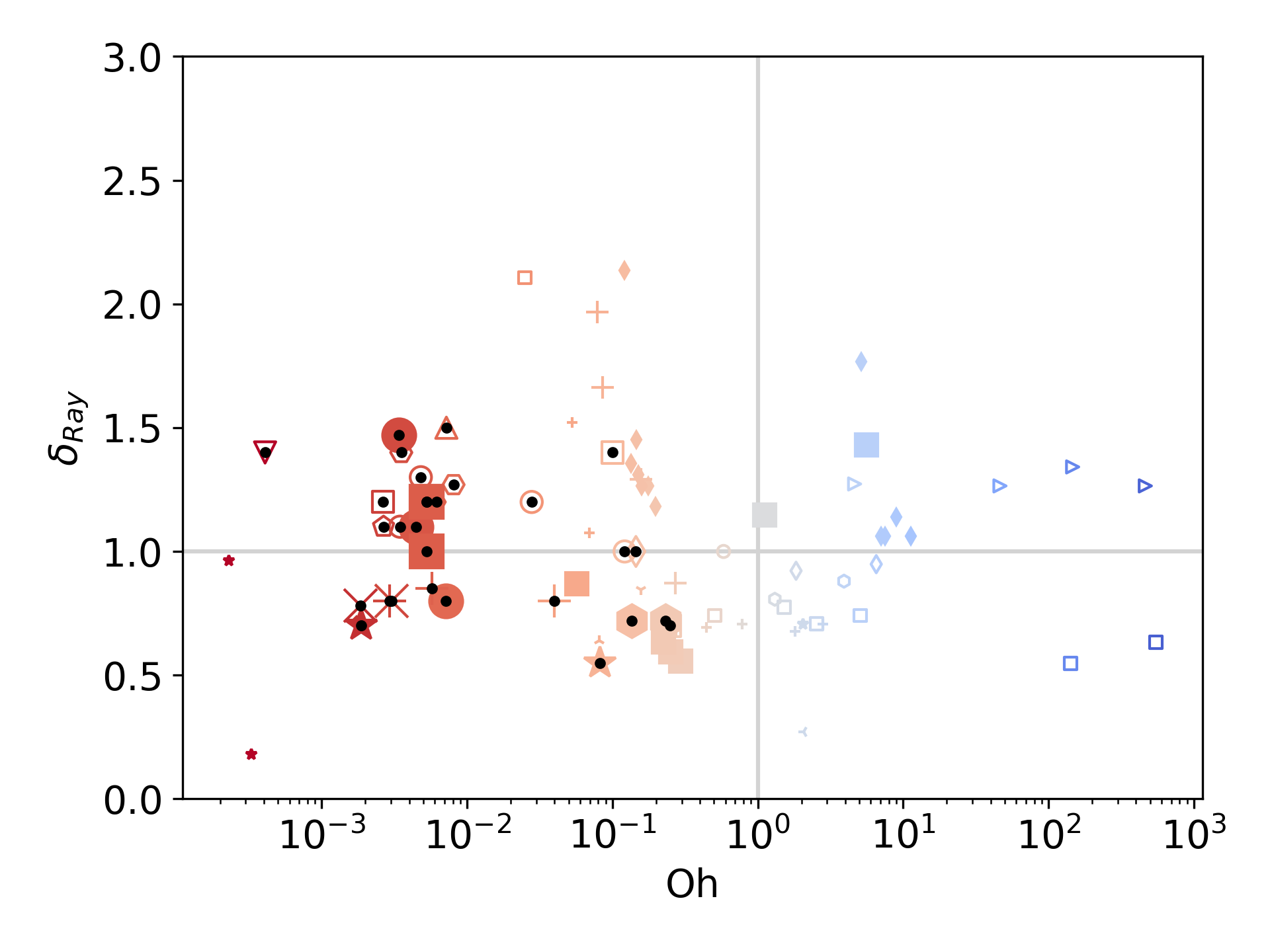}
\put(10,30){(c)}
\end{overpic}
\caption{ Values of dimensionless constants $\delta_{Tan}$, $\delta_{ic}$ and $\delta_{Ray}$, from direct fits of experimental data presenting the appropriate scalings (big symbols with black dots), and from indirect measurements (smaller symbols; see section~\ref{algebra}). 
\label{figSI2}}
\end{figure}

The values of $\delta_{ic}$, $\delta_{Tan}$ and $\delta_{Ray}$ obtained from experiments are shown in SI-Fig.~\ref{figSI2}. All values are indeed of order 1 and seem quite independent of the Ohnesorge number. When possible, the values of the measured prefactors are given in Tables~\ref{table4s}, \ref{table4c} and \ref{table4p}. 

\subsection{Unit prefactors\label{prefact}}
\begin{figure}
\centering
\begin{overpic}[abs,unit=1mm,width=8cm]{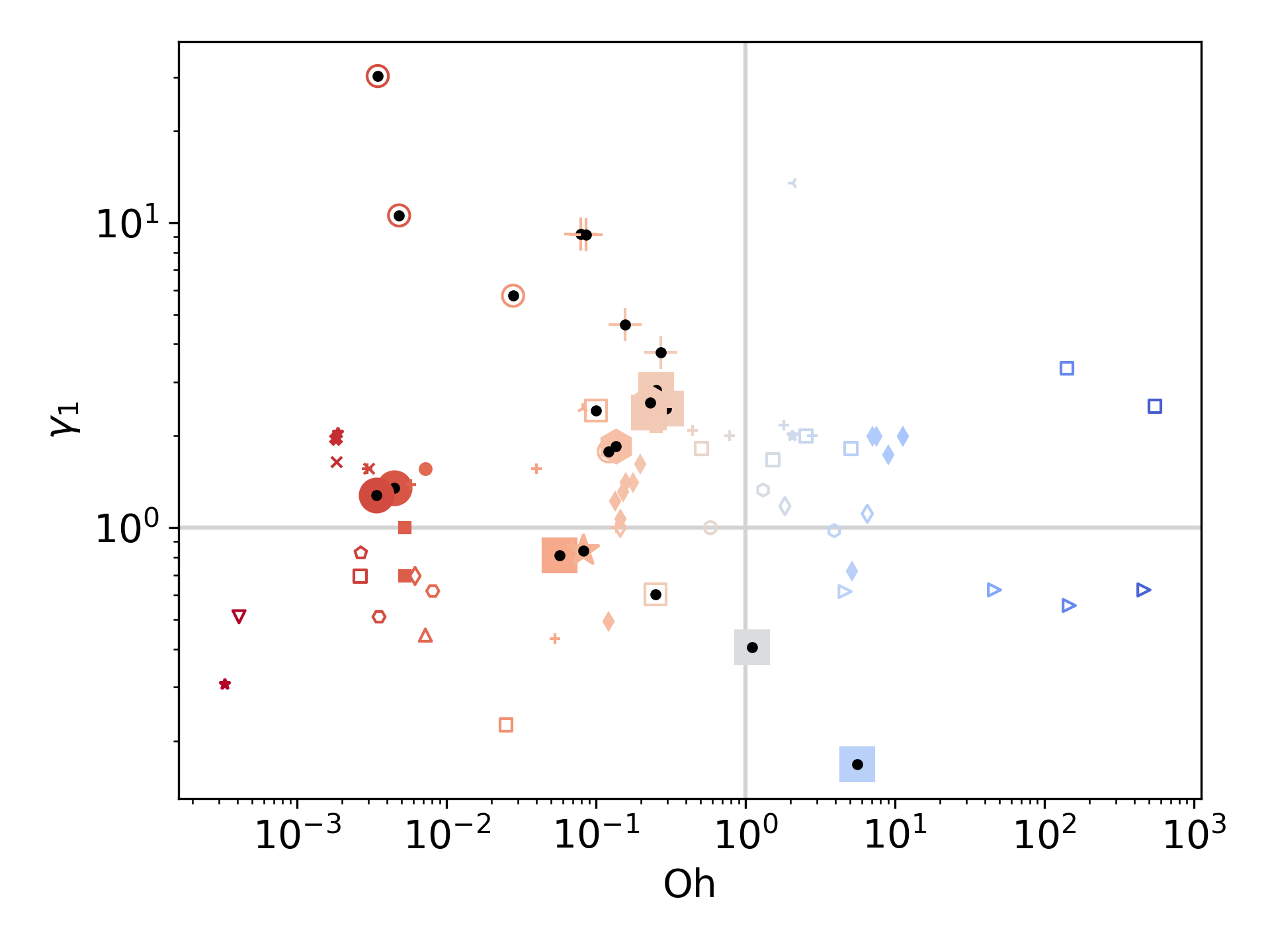}
\put(15,53){(a)}
\end{overpic}
\begin{overpic}[abs,unit=1mm,width=8cm]{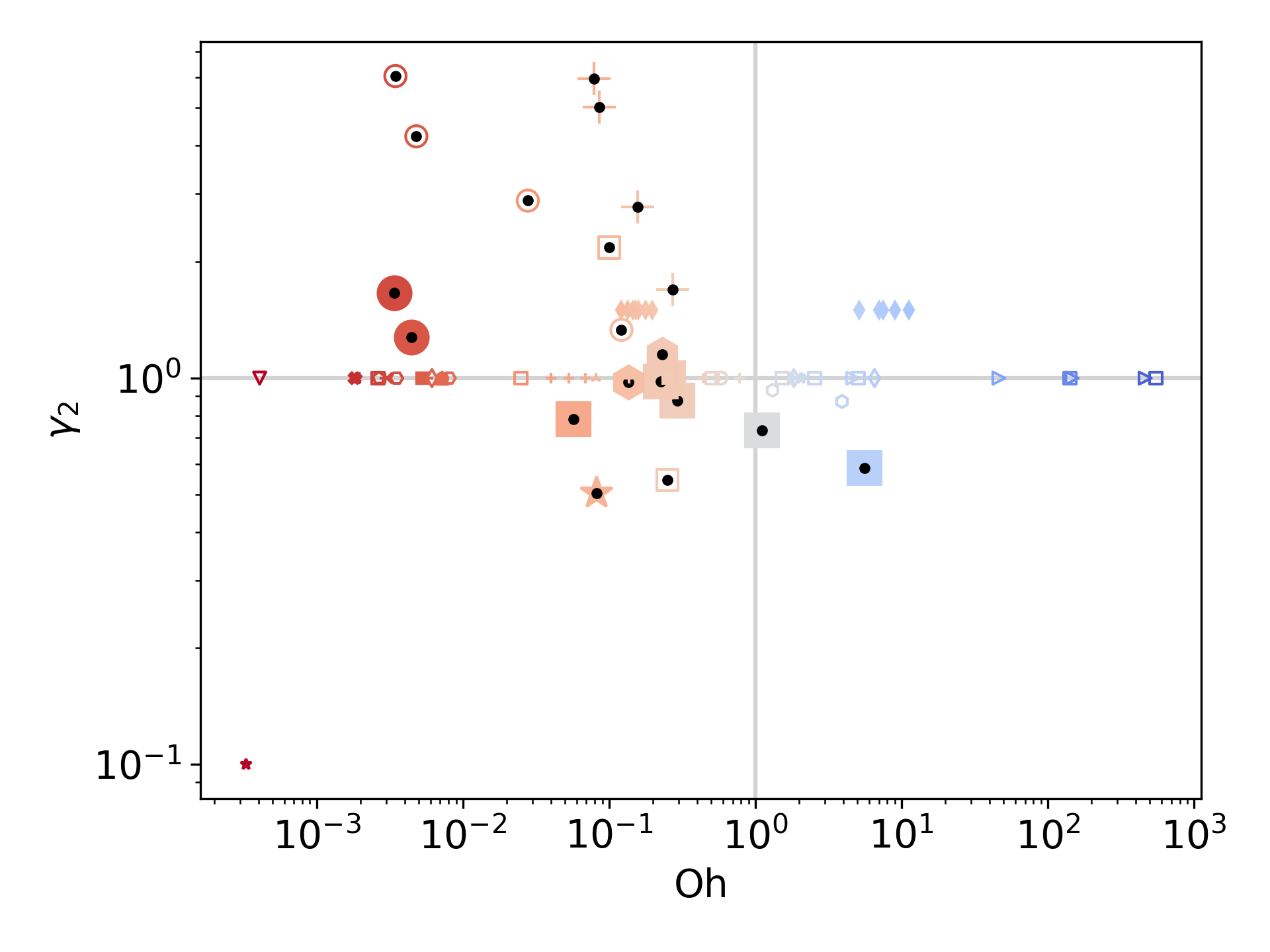}
\put(15,53){(b)}
\end{overpic}
\caption{ Values of the dimensionless prefactors $\gamma_1$ and $\gamma_2$ used for any of the four systems of units. The big data points with black dots correspond to direct measurements from fitting two consecutive regimes. The smaller data points correspond to values inferred from fitting a single regime.  
\label{figSI3}}
\end{figure} 
In all figures of the article, the units of length and time take into account the dimensionless constants ($\delta_{vc}$, $\delta_{ic}$ etc) included in the various scaling laws. For instance, in Fig. 5 giving the dynamics in Ohnesorge units, the axes are $t/\tau_o^*$ and $d/\ell_o^*$, where $\tau_o^*\equiv \gamma_1 \tau_o$ and $\ell_o^*\equiv \gamma_2 \ell_o$. In all figures associated with the four systems of units, the values of $\gamma_1$ and $\gamma_2$ for a particular data set are identical. The values of $\gamma_1$, $\gamma_2$ are given for spreading, coalescence and pinching data in Tables~\ref{table4s}, \ref{table4c} and \ref{table4p} and in Fig.~\ref{figSI3}. In this section we discuss how the values of $\gamma_1$ and $\gamma_2$ are computed.   

Let us take as an example a case of dynamics abiding to the simple visco-capillary scaling:  
\begin{equation}
d=\delta_{vc} \frac{\Gamma}{\eta} t  \leftrightarrow \frac{d}{\ell_{vi}} = \delta_{vc} \text{Oh}^{-2} \frac{t}{\tau_{vi}} \leftrightarrow \frac{d}{\ell_{vc}}=  \delta_{vc} \frac{t}{\tau_{vc}} \leftrightarrow \frac{d}{\ell_{ic}}  = \delta_{vc} \text{Oh}^{-1} \frac{t}{\tau_{ic}}  \leftrightarrow \frac{d}{\ell_{o}}=  \delta_{vc} \frac{t}{\tau_{o}}
\end{equation}
The goal of the coefficients $\gamma_1$ and $\gamma_2$ is to factor in the value of $\delta_{vc}$, such that the regime can be written as 
\begin{equation}
\frac{d}{\ell_{vi}^*} = \text{Oh}^{-2} \frac{t}{\tau_{vi}^*} \leftrightarrow \frac{d}{\ell_{vc^*}}=  \frac{t}{\tau_{vc}^*} \leftrightarrow \frac{d}{\ell_{ic}^*}  =\text{Oh}^{-1} \frac{t}{\tau_{ic}^*}  \leftrightarrow \frac{d}{\ell_{o}^*}=  \frac{t}{\tau_{o}^*}
\end{equation}
Regardless of the system of units, the constraint on $\gamma_1$ and $\gamma_2$ is the same: 
\begin{equation}
\frac{\delta_{vc}\gamma_1}{\gamma_2}=1 \label{cons1}
\end{equation}

Let us now assume that the data set first follows the simple visco-capillary regime and then displays Rayleigh's regime. In that case, the second regime provides an additional constraint: 
\begin{equation}
\frac{\delta_{Ray}\gamma_1^\frac{1}{2}}{\gamma_2}=1  \label{cons2}
\end{equation}
Solving the system made of Eq.~\ref{cons1} and Eq.~\ref{cons2} would lead to: 
\begin{align}
\gamma_1 &=\Big(\frac{\delta_{Ray}}{\delta_{vc}}\Big)^2\\
\gamma_2 &=\frac{\delta_{Ray}^2}{\delta_{vc}}
\end{align}

For all experimental data sets presenting two consecutive regimes, we followed the procedure outlined in the example above in order to obtain the values of $\gamma_1$ and $\gamma_2$. We encountered five cases summarized here: 
\begin{align}
&\text{- input:  } \delta_{vc} \text{  and  }  \delta_{Tan}  \rightarrow \gamma_1= \Big(\frac{\delta_{Tan}}{\delta_{vc}}\Big)^\frac{10}{9} \text{  and  } \gamma_2=\delta_{vc} \gamma_1\\
&\text{- input:  } \delta_{vc} \text{  and  }  \delta_{ic}  \rightarrow \gamma_1= \Big(\frac{\delta_{ic}}{\delta_{vc}}\Big)^3 \text{  and  } \gamma_2=\delta_{vc} \gamma_1\\
&\text{- input:  } \delta_{vc} \text{  and  }  \delta_{Ray} \rightarrow \gamma_1= \Big(\frac{\delta_{Ray}}{\delta_{vc}}\Big)^2 \text{  and  } \gamma_2=\delta_{vc} \gamma_1\\
&\text{- input:  } \delta_{Tan} \text{  and  }  \delta_{ic}  \rightarrow \gamma_1= \Big(\frac{\delta_{Tan}}{\delta_{ic}}\Big)^\frac{30}{17} \text{  and  } \gamma_2=\delta_{ic} \gamma_1^{\frac{2}{3}}\\
&\text{- input:  } \delta_{Tan} \text{  and  }  \delta_{Ray}  \rightarrow \gamma_1= \Big(\frac{\delta_{Tan}}{\delta_{Ray}}\Big)^\frac{5}{2} \text{  and  } \gamma_2=\delta_{Ray} \gamma_1^{\frac{1}{2}}\\
\end{align} 

In some experiments only a single regime is available, in which case we chose to set $\gamma_2=1$ and use the prefactor of the regime to obtain a value of $\gamma_1$. For instance, for the spreading experiment labeled `Eddi2013 Fig6 water', the data only show Rayleigh's regime, giving $\delta_{Ray}=1.2$, and we set $\gamma_2=1$ to get $\gamma_1=1/\delta_{Ray}^2\simeq 0.69$. In all these cases presenting a single regime, the values of $\gamma_1$ and $\gamma_2$ must be understood as quite approximate. In two cases we chose slightly different values of $\gamma_2$. In coalescence experiments by Yokota \textit{et al.} the confinement between two plates leads to a late spreading abiding to $d/\tau_{vc}\propto  (t/\tau_{vc})^\frac{1}{4}$, and the values of $\gamma_2$ are chosen to comply with this regime. The second example where we chose $\gamma_2\neq 1$ concerns the Tanner regime studied by Cazabat \textit{et al.}. In that case, the sole constraint is: 
\begin{equation}
\gamma_1 =\Big(\frac{\gamma_2}{\delta_{Tan}} \Big)^{10} \rightarrow \frac{\Delta \gamma_1}{\gamma_1} = 10 \frac{\Delta \gamma_2}{\gamma_2} +10 \frac{\Delta \delta_{Tan}}{\delta_{Tan}} 
\end{equation} 
In the right-hand side we express the relative uncertainty on $\gamma_1$ based on the uncertainty on $\gamma_2$ and $\delta_{Tan}$. Even if we assume that the uncertainty on $\delta_{Tan}$ obtained from fitting data is negligible, small variations of $\gamma_2$ can lead to substantial variations of $\gamma_1$. For Cazabat's data, choosing $\gamma_2=1$ led to unreasonable values of $\gamma_1$ so we took the liberty of slightly tuning $\gamma_2$ in order to recover values in accordance with the other data sets. This was done by deriving the inferred value of $\delta_{vc}$, as we shall explain now.    

\clearpage
\begin{table}
\csvreader[tabular=|l|c|c|c|c|c|c|c|,respect all,
    table head=\hline \textbf{Label} & \textbf{Oh}&  $\bm{\delta_{vc}}$ & $\bm{\delta_{Tan}}$ & $\bm{\delta_{ic}}$ & $\bm{\delta_{Ray}}$ & $\bm{\gamma_{1}}$ & $\bm{\gamma_{2}}$ \\ \hline,
    late after line=\\\hline,
    filter=\equal{\Type}{Spreading}
    ]%
{./FiguresArxivFinal/FIGS/Summary.csv}{1=\Label, 2=\Type, 8=\Oh, 9=\dvc, 10=\dtan, 11=\dic, 12=\dray, 13=\ct, 14=\cd }%
{\Label & \Oh & \dvc & \dtan & \dic & \dray & \ct & \cd}
\caption{Summary of the values of the dimensionless prefactors for all spreading experiments reproduced in the article. The content of this table is available in the supplementary file `\textbf{DataSummary.csv}'.  
\label{table4s}}
\end{table}

\begin{table}
\csvreader[tabular=|l|c|c|c|c|c|c|c|,respect all,
    table head=\hline \textbf{Label} & \textbf{Oh}&  $\bm{\delta_{vc}}$ & $\bm{\delta_{Tan}}$ & $\bm{\delta_{ic}}$ & $\bm{\delta_{Ray}}$ & $\bm{\gamma_{1}}$ & $\bm{\gamma_{2}}$ \\ \hline,
    late after line=\\\hline,
    filter=\equal{\Type}{Coalescence}
    ]%
{./FiguresArxivFinal/FIGS/Summary.csv}{1=\Label, 2=\Type, 8=\Oh, 9=\dvc, 10=\dtan, 11=\dic, 12=\dray, 13=\ct, 14=\cd }%
{\Label & \Oh & \dvc & \dtan & \dic & \dray & \ct & \cd}
\caption{Summary of the values of the dimensionless prefactors for all coalescence experiments reproduced in the article. The content of this table is available in the supplementary file `\textbf{DataSummary.csv}'.  
\label{table4c}}
\end{table}

\begin{table}
\csvreader[tabular=|l|c|c|c|c|c|c|c|,respect all,
    table head=\hline \textbf{Label} & \textbf{Oh}&  $\bm{\delta_{vc}}$ & $\bm{\delta_{Tan}}$ & $\bm{\delta_{ic}}$ & $\bm{\delta_{Ray}}$ & $\bm{\gamma_{1}}$ & $\bm{\gamma_{2}}$ \\ \hline,
    late after line=\\\hline,
    filter=\equal{\Type}{Pinching}
    ]%
{./FiguresArxivFinal/FIGS/Summary.csv}{1=\Label, 2=\Type, 8=\Oh, 9=\dvc, 10=\dtan, 11=\dic, 12=\dray, 13=\ct, 14=\cd }%
{\Label & \Oh & \dvc & \dtan & \dic & \dray & \ct & \cd}
\caption{Summary of the values of the dimensionless prefactors for all pinching experiments reproduced in the article. The content of this table is available in the supplementary file `\textbf{DataSummary.csv}'.  
\label{table4p}}
\end{table}

\clearpage
\subsection{Consistency relations\label{algebra}}
In all data we collected on spreading, coalescence and pinching, we never encountered a set presenting three consecutive regimes. However, such dynamics are absolutely possible. For instance, a spreading with a low value of Ohnesorge number could first follow the linear visco-capillary regime, then follow Rayleigh's regime and finally reach Tanner's regime. The data from Biance \textit{et al.} encompass the last two regimes ($\frac{1}{2}$ and $\frac{1}{10}$) but lack enough time resolution to describe the initial visco-capillary regime. If future experiments manage to resolve three consecutive regimes, one could check consistency relationships between the different prefactors. For instance, in the case outlined above, the three consecutive regimes would lead to the following constraints on $\gamma_1$ and $\gamma_2$: 
\begin{equation}
\begin{cases}
    \gamma_2 &=  \delta_{vc}  \gamma_1\\
    \gamma_2 &=  \delta_{Ray}  \gamma_1^\frac{1}{2}   \\
    \gamma_2 &=  \delta_{Tan}  \gamma_1^\frac{1}{10}   
 \end{cases} \rightarrow \delta_{Ray}^9=\delta_{vc}^4 \delta_{Tan}^5
\end{equation}
The constraints impose that the three constant prefactors are not independent and must respect the equation on the right-hand side. We have used such consistency equations to infer values of prefactors inaccessible in experiments. For instance, from Biance's data providing values for $\delta_{Ray}$ and $\delta_{Tan}$, we computed the predicted values of $\delta_{vc}=(\delta_{Ray}^9/\delta_{Tan}^5)^\frac{1}{4}$. These inferred values are present as the two red dots in SI-Fig.~\ref{figSIdvc}, corresponding to `Biance2004 Fig3 0p7' and `Biance2004 Fig3 1p2'.  

In general, if values of $\gamma_1$ and $\gamma_2$ are provided, the values of the constant prefactors can be inferred using the following equations: 
\begin{align}
\delta_{vc} &=\frac{\gamma_2}{\gamma_1}\\
\delta_{Tan} &=\frac{\gamma_2}{\gamma_1^\frac{1}{10}}\\
\delta_{ic} &=\frac{\gamma_2}{\gamma_1^\frac{2}{3}}\\
\delta_{Ray} &=\frac{\gamma_2}{\gamma_1^\frac{1}{2}}\\
\end{align}
These relations allow to infer values of the prefactors in all cases, even if a particular experiment is not expected to display a given regime. For instance, for a coalescence experiment, Tanner's regime is irrelevant, but a hypothetical value of $\delta_{Tan}$ can be computed anyway. All indirect measurements of prefactors in SI-Fig.~\ref{figSIdvc} and \ref{figSI2} are obtained using the equations given above.

\section{Categories of dimensionless numbers}
The article aims at analyzing dynamics based on four dimensional parameters: three materials parameters ($\rho$, $\eta$ and $\Gamma$) and one geometric parameter $D$. We call these quantities `parameters' because they are expected to be constant for a particular experiment, in contrast to the variables $d$ and $t$. These four dimensional parameters and the associated two dimensional variables can combine to produce a variety of quantities without dimensions. According to dimensional analysis, the dynamics can be fully characterized by 3 dimensionless numbers (6 quantities -3 dimensions)~\cite{Buckingham1914}. In the article we favored descriptions based on four different choices of dimensionless times and lengths in conjunction with the Ohnesorge number. In this section, we entertain different approaches and try to categorize the different kinds of dimensionless numbers in the hope to facilitate comparison with different viewpoints. 

In the article, we encounter a few kinds of quantities without dimensions, which we shall summarize here and comment thereafter. 
\begin{itemize}
\item[-] Scaling exponents: any scaling regime $d=K t^\alpha$ is associated with a dimensionless exponent $\alpha$.
\item[-] Simple dynamic dimensionless numbers: Re, Ca, We. 
\item[-] Variable geometric ratio: $\Lambda$. 
\item[-] Dimensionless constants: $\delta_{vi}$, $\delta_{vc}$, $\delta_{ic}$, $\delta_{Tan}$, $\delta_{Ray}$.  
\item[-] Ohnesorge and Laplace numbers: Oh, La. 
\item[-] Dimensionless lengths: $d/\ell_{vi}$, $d/\ell_{vc}$, $d/\ell_{ic}$, $d/\ell_{o}$. 
\item[-] Dimensionless times: $t/\tau_{vi}$, $d/\tau_{vc}$, $d/\tau_{ic}$, $d/\tau_{o}$. 
\item[-] Dimensionless unit prefactors: $\gamma_1$, $\gamma_2$. 
\end{itemize}

\subsection{Scaling exponents}
In the article we discussed five main scaling laws, associated with the exponents $1$ (vc), $\frac{1}{2}$ (vi and Rayleigh), $\frac{2}{3}$ (ic) and $\frac{1}{10}$ (Tanner). The values of these exponents are connected to the dimensions of the components of the prefactor $K$ in $d=Kt^\alpha$. For instance, $\alpha=\frac{2}{3}$ for the inertio-capillary regime because $[K]=([\Gamma]/[\rho])^\frac{1}{3}=\mathcal{L}.\mathcal{T}^{-\frac{2}{3}}$. Since physical quantities tend to have dimensions $\mathcal{M}^a.\mathcal{L}^b.\mathcal{T}^c$, with $a$, $b$ and $c$ small integers, then scaling exponents with a mechanical underpinning can only be rational numbers built from ratios of small integers. However, if one adds logarithmic corrections as in $d\propto Kt^\alpha \log(d/D)$, then the apparent scaling exponent may deviate from simple fractions. In that case, one can define the apparent exponent as the logarithmic derivative~\cite{Eddi2013}: 
\begin{equation}
\alpha_* \equiv \frac{\partial \log d}{\partial \log t}
\end{equation}
In theory one may derive the evolution of the apparent exponent $\alpha_*(t)$ for all dynamics reported in the article. In practice, the use of a derivative can introduce a substantial source of error. In SI-Fig.~\ref{Fig_wecare}a we give the computed apparent exponent for an example of spreading from Eddi \textit{et al.}~\cite{Eddi2013}, obtained with different methods of derivation.

\subsection{Simple dynamic dimensionless numbers}
\begin{figure}
\centering
\includegraphics[width=15cm,clip]{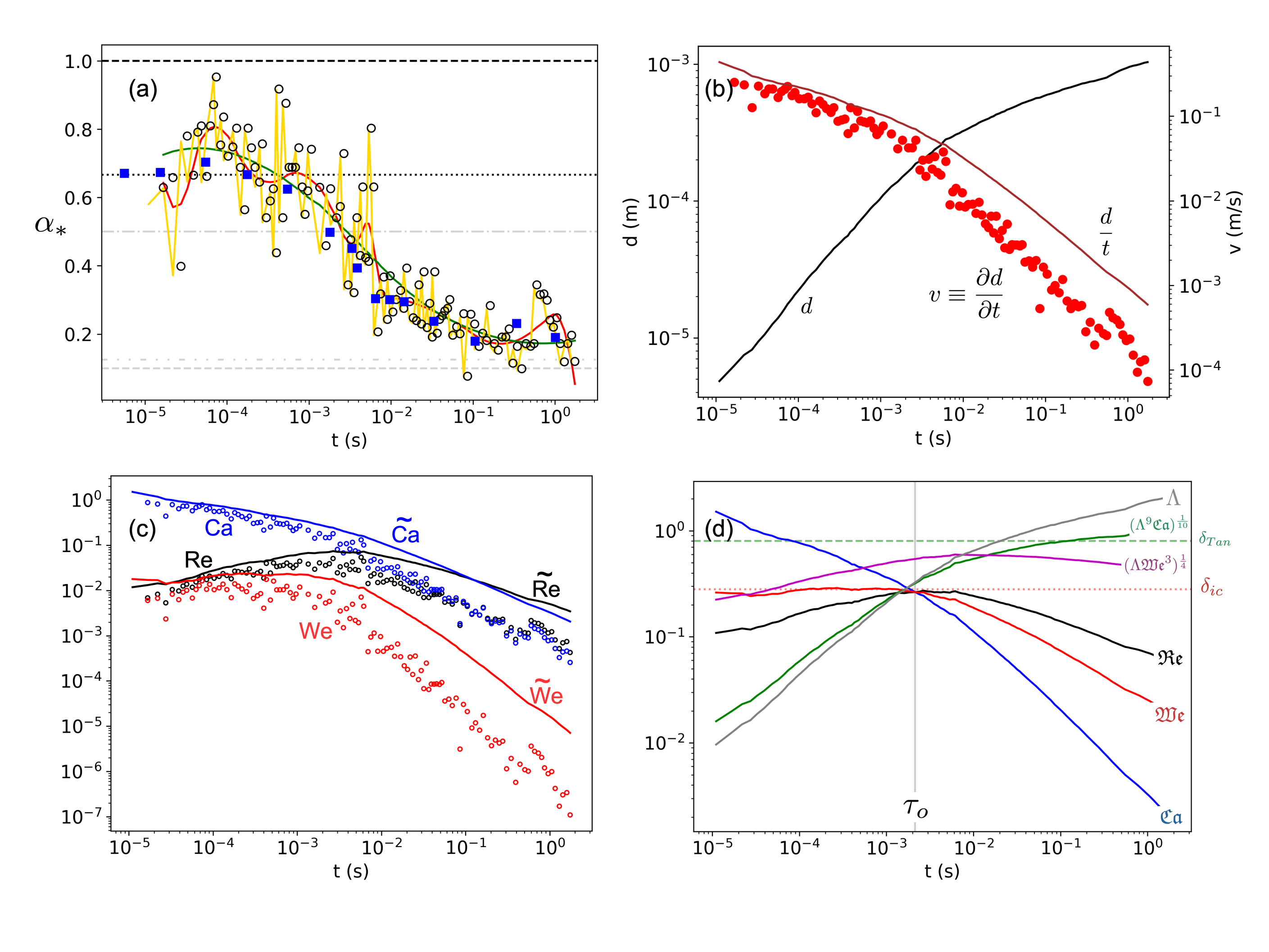}
\caption{Alternative dimensionless representations for the data set `Eddi2013 Fig6 220'~\cite{Eddi2013}. (a) Apparent scaling exponent $\alpha_*$ obtained using different methods: difference between adjacent points on logarithmic scale (open circles), logarithmic derivative after cubic interpolations of the data points obtained with the method interpolate.splprep on python with two different smoothing parameters (red and green), by computing the ratio $vt/d$ (yellow), and as provided in the original article~\cite{Eddi2013} (blue squares). (b) Comparison between the speed $v\equiv \partial d/\partial t$ (red dots) and the speed $d/t$ (dark red line). The black curve is $d(t)$. (c) Comparison between the traditional dimensionless numbers Re, Ca and We (open circles), and the alternate numbers $\tilde{\text{Re}}$, $\tilde{\text{Ca}}$ and $\tilde{\text{We}}$. (d) Evolution of the values of the dimensionless numbers based directly on the form of the scaling laws. The numbers $\mathfrak{Re}$, $\mathfrak{Ca}$ and $\mathfrak{We}$ are identical for $t=\tau_o$. The constants $\delta_{ic}$ and $\delta_{Tan}$ identify plateaus in the evolution of $\mathfrak{We}$ and $(\Lambda^9\mathfrak{Ca})^\frac{1}{10}$. 
\label{Fig_wecare}}
\end{figure}   
Although without dimensions, the scaling exponents are usually not considered as dimensionless numbers per say. Traditional dimensionless numbers are understood as ratios of dimensional quantities. What we call a `simple dimensionless number' is a ratio built from two material parameters with dimensions of the form $\mathcal{M}.\mathcal{L}^b.\mathcal{T}^c$, and a kinematic quantity with dimensions $\mathcal{L}^{b'}.\mathcal{T}^{c'}$. Following standard definitions, the article introduced three such simple dimensionless numbers: 
\begin{align}
\text{Re} &\equiv \frac{\rho d v}{\eta} = \frac{d v}{\nu} \\
\text{Ca} &\equiv \frac{\eta v}{\Gamma}= \frac{v}{c} \\
\text{We} &\equiv \frac{\rho d v^2}{\Gamma}= \frac{d v^2}{\kappa}
\end{align}
These definitions are rooted in steady hydrodynamics, where $d$ and $v$ are usually understood as control parameters imposed by the experimenter. Here $d$ and $v$ are the variable extent and speed of the spreading/coalescing/pinching object. Hence, the dimensionless numbers Re, Ca and We are dynamical, in the sense that their values can change over the course of an experiment. Computation of the values of the simple dimensionless numbers is contingent on deriving the leading edge speed:
\begin{equation}
v\equiv\frac{\partial d}{\partial t} = \alpha_* \frac{d}{t} 
\end{equation}
In SI-Fig.~\ref{Fig_wecare}b we give the edge speed computed from simple differences without any additional processing, the ratio $d/t$ and the curve $d(t)$ for the example `Eddi2013 Fig6 220'~\cite{Eddi2013}. From this perspective, the apparent scaling exponent can be defined as a ratio of two speeds: $\alpha_* = vt/d$. The result of this definition is given as the yellow curve in SI-Fig.~\ref{Fig_wecare}a. 

For a given data set $d(t)$, the values of Re, Ca and We can be computed at any time after deriving the speed $v$, as shown on an example in SI-Fig.~\ref{Fig_wecare}. If the Reynolds number is constant it means that $d\sim t^\frac{1}{2}$. If the Capillary number is constant it means that $d\sim t$. If the Weber number is constant it means that $d\sim t^\frac{2}{3}$. In this example, the dynamics first seem to display a constant Weber number, and indeed $\alpha_*\simeq \frac{2}{3}$, as show in SI-Fig.~\ref{Fig_wecare}a. The value of the constant will be connected to the dimensionless numbers $\delta_{ic}$ in section~\ref{dimconsseq}.  

The derivation of the data $d(t)$ involved in the computation of the speed $v$ usually introduces a source of error. To circumvent this issue, one may define alternate simple dimensionless numbers based on $d$ and $t$ only:
\begin{align}
\tilde{\text{Re}} &\equiv \frac{\rho d^2}{\eta t} = \frac{\text{Re}}{\alpha_*} \\
\tilde{\text{Ca}} &\equiv \frac{\eta d}{\Gamma t}=\frac{\text{Ca}}{\alpha_*} \\
\tilde{\text{We}} &\equiv \frac{\rho d ^3}{\Gamma t^2}= \frac{\text{We}}{\alpha_*^2}
\end{align}
The values of these alternate numbers are given in SI-Fig.~\ref{Fig_wecare}c.  

As mentioned in the article, dimensionless numbers are defined modulo an overall power. Usually, this power is chosen such that the material parameters appearing in the number have integer exponents. This historical habit may not be best suited to our purpose. In the context of a comparative study of multiple scaling regimes, it may be more judicious to define alternate numbers based directly on the form of the scaling laws:  
\begin{align}
\mathfrak{Re} &\equiv d\Big(\frac{\rho}{\eta t}\Big)^\frac{1}{2} =  \tilde{\text{Re}}^\frac{1}{2} =\Big(\frac{\text{Re}}{\alpha_*}\Big)^\frac{1}{2} \\
\mathfrak{Ca} &\equiv d\frac{\eta}{\Gamma t} = \tilde{\text{Ca}} = \frac{\text{Ca}}{\alpha_*}\\
\mathfrak{We} &\equiv d\Big(\frac{\rho}{\Gamma t^2}\Big)^\frac{1}{3}=\tilde{\text{We}}^\frac{1}{3} = \frac{\text{We}^\frac{1}{3}}{\alpha_*^\frac{2}{3}}
\end{align}
The values of these alternate numbers are given in SI-Fig.~\ref{Fig_wecare}d.  

The three simple dimensionless numbers are related by the `we care' identity, which can be expressed in a few ways: 
\begin{align}
\text{We} &= \text{Ca}\text{Re}\\
\tilde{\text{We}} &= \tilde{\text{Ca}}\tilde{\text{Re}}\\
\mathfrak{We}^3 &=\mathfrak{Ca} \mathfrak{Re}^2\\
\kappa &= c \nu\
\end{align}

\subsection{Variable geometric ratio}
In addition to Re, Ca and We, we also introduced a dynamic geometric size ratio: 
\begin{equation}
\Lambda\equiv\frac{d}{D}
\end{equation}
For a given experiment, the value of this number evolves as the spreading, coalescence or pinching proceeds. The size ratio can be connected to a number of shape descriptors. Taken spreading as an example, the size ratio can be related to the apparent contact angle or the curvature. The actual relationship between these geometric measures and the size ratio can vary depending on context, but a few examples can help as an illustration. 

If the spreading geometry follows a pancake with cylindrical symmetry, then conservation of volume imposes that $D^3\propto h d^2$, where $h$ is the height of the pancake. The size ratio of the pancake is then $h/d\propto \Lambda^{-3}$. 

If the spreading follows a spherical cap geometry, then the conservation of volume is expressed as $D^3\propto \theta d^3$, where $\theta\propto\Lambda^{-3}$ is the apparent contact angle.  

\subsection{Dimensionless constants\label{dimconsseq}}
In the article, we introduced five dimensionless constants: $\delta_{vi}$, $\delta_{vc}$, $\delta_{ic}$, $\delta_{Tan}$ and $\delta_{Ray}$. Let us first discuss the three constants based on simple scaling laws associated with Re, Ca and We. The constants $\delta_{vi}$, $\delta_{vc}$, $\delta_{ic}$ can be understood as special values of the dynamical numbers Re, Ca and We. How these special values are defined is somewhat arbitrary, but usually one seeks time intervals where the dynamical numbers are constant and `close to 1'. Using the traditional or alternate simple dynamical numbers, the three simple scaling laws can be expressed as: 
\begin{align}
d=\delta_{vi} \Big(\frac{\eta}{\rho}\Big)^\frac{1}{2} t^\frac{1}{2} & \leftrightarrow \delta_{vi} = \mathfrak{Re}=\tilde{\text{Re}}^\frac{1}{2} = \Big(\frac{\text{Re}}{\alpha_*}\Big)^\frac{1}{2}  \label{Redvi} \\
d=\delta_{vc} \frac{\Gamma}{\eta} t & \leftrightarrow \delta_{vc} = \mathfrak{Ca}=\tilde{\text{Ca}} = \frac{\text{Ca}}{\alpha_*}  \label{Cadvc}\\
d=\delta_{ic} \Big(\frac{\Gamma}{\rho}\Big)^\frac{1}{3} t^\frac{2}{3} & \leftrightarrow \delta_{ic} =\mathfrak{We}= \tilde{\text{We}}^\frac{1}{3} = \frac{\text{We}^\frac{1}{3}}{\alpha_*^\frac{2}{3}} \label{Wedic}
\end{align} 
The numbers $\delta_{vi}$, $\delta_{vc}$ and $\delta_{ic}$ are understood as constant, which are meant to identify the values of plateaus exhibited by the time-series $\mathfrak{Re}(t)$,$\mathfrak{Ca}(t)$ and $\mathfrak{We}(t)$. Thus, in the following we will use the symbols $\mathfrak{Re}_0$,$\mathfrak{Ca}_0$ and $\mathfrak{We}_0$ to respectively stand for $\delta_{vi}$, $\delta_{vc}$ and $\delta_{ic}$. For instance, in SI-Fig.~\ref{Fig_wecare}d, we can identify a roughly constant value of $\mathfrak{We}$ for $10^{-5}\text{ s}\lesssim t \lesssim 2~10^{-3}$~s, giving $\delta_{ic}=\mathfrak{We}_0\simeq 0.28$. Then, as stated in the article, the special value of the traditional Weber number is $\text{We}_0=\mathfrak{We}_0^3 \alpha_0^2$, with $\alpha_0=\frac{2}{3}$. 

In contrast to the three simple scaling laws, the two-size dependent regimes discussed in the article cannot be derived by simple dimensional analysis. In simple scaling laws, the length $d$ depends on three quantities, the variable $t$ and a choice of two material parameters. In contrast, the size-dependent regimes also depend on $D$, which does not lead to a unique possible scaling. When expressed in terms of dimensionless numbers, the size-dependent regimes are connected to products between simple dynamical numbers and the size ratio: 
\begin{align}
d = {\delta}_{Tan} \Big(\frac{\Gamma}{\eta}\Big)^\frac{1}{10} D^\frac{9}{10} t^\frac{1}{10}
 & \leftrightarrow \delta_{Tan} = (\Lambda^9 \mathfrak{Ca})^\frac{1}{10} =(\Lambda^9 \tilde{\text{Ca}})^\frac{1}{10} = \Big(\Lambda^9 \frac{\text{Ca}}{\alpha_*}\Big)^\frac{1}{10}\\
 d ={\delta}_{Ray} \Big(\frac{\Gamma D}{\rho}\Big)^\frac{1}{4}  t^\frac{1}{2}& \leftrightarrow \delta_{Ray} =(\Lambda \mathfrak{We}^3)^\frac{1}{4}= (\Lambda \tilde{\text{We}})^\frac{1}{4} = \Big(\Lambda \frac{\text{We}}{\alpha_*^2}\Big)^\frac{1}{4}
\end{align}
The curves corresponding to $(\Lambda^9 \tilde{\text{Ca}})^\frac{1}{10}$ and $(\Lambda \tilde{\text{We}})^\frac{1}{4}$ are given in SI-Fig.~\ref{Fig_wecare}d, on which we indicate the value of $\delta_{Tan}$ corresponding to a plateau of $(\Lambda^9 \tilde{\text{Ca}})^\frac{1}{10}$. Note that this plateau is rather limited due to the effect of gravity, which brings $\alpha$ closer to $\frac{1}{8}$ rather than $\frac{1}{10}$~\cite{Cazabat1986}. 

In practice, the values of the dimensionless constants used in the article were obtained by directly fitting the data $d(t)$ to a particular exponent in a prescribe time range, and are most often equal to the values provided in the original papers. See section~\ref{general} for a discussion on a fitting procedure letting the exponent free to adopt values beyond the fives regimes discussed in the article. 

\subsection{Ohnesorge and Laplace numbers}
In contrast to the three simple dimensionless numbers, the Ohnesorge number depends on three rather than two material parameters. In addition, like the size ratio, the Ohnesorge number depends on the extrinsic size $D$. As stated in the article, the Ohnesorge number can be expressed from the simple dimensionless numbers as:
\begin{equation}
\text{Oh}^2\equiv \frac{\eta^2}{\Gamma\rho D}=\frac{\text{Ca}}{\text{Re}}\Lambda = \frac{\text{We}}{\text{Re}^2}\Lambda = \frac{\text{Ca}^2}{\text{We}}\Lambda
\label{ohcare}
\end{equation} 
The different decompositions are built in such a way that the product is independent of the variables $d$ and $t$. Thus, the Ohnesorge number is a constant in each experiment. 

The Ohnesorge number can also be written in terms of the kinematic ratios associated with the material parameters: 
\begin{equation}
\text{Oh}=\frac{\nu}{(\kappa D)^\frac{1}{2}} = \Big(\frac{\nu}{c D}\Big)^\frac{1}{2} = \frac{\kappa^\frac{1}{2}}{c D^\frac{1}{2}}
\label{ohkin}
\end{equation} 
\noindent where we recall that $\nu\equiv \eta/\rho$, $\kappa\equiv\Gamma/\rho$ and $c\equiv\Gamma/\eta$.

The Laplace number $\text{La}=\text{Oh}^{-2}$ gives an alternative definition of the dimensionless combination present in the Ohnesorge number. Again, dimensionless numbers are defined modulo an overall power. 

\subsection{Dimensionless lengths}
Of the four systems of units discussed in the article, three share the same dimensionless length, which is none other than the size ratio: 
\begin{equation}
\frac{d}{\ell_{vi}}=\frac{d}{\ell_{vc}}=\frac{d}{\ell_{ic}}=\frac{d}{D}=\Lambda
\end{equation} 
In contrast, the dimensionless length of the Ohnesorge units compares the variable $d$ to the intrinsic Ohnesorge length $\ell_o$, which can be expressed from $D$ by using the Ohnesorge number: 
\begin{equation}
\frac{d}{\ell_{o}}=\frac{d}{D}\frac{D}{\ell_o}=\Lambda\text{La} =\frac{\Lambda}{\text{Oh}^2} = \frac{\text{Re}}{\text{Ca}}=\frac{\text{Re}^2}{\text{We}}=  \frac{\text{We}}{\text{Ca}^2}
\end{equation}

\subsection{Dimensionless times}
In the article, we used four different time scales related by the Ohnesorge number: 
\begin{align}
&\tau_{vi} \xrightarrow{\times \text{Oh}} \tau_{ic} \xrightarrow{\times \text{Oh}} \tau_{vc} \xrightarrow{\times \text{Oh}^2} \tau_{o}\\
&\frac{\rho D^2}{\eta} \xrightarrow{\times \text{Oh}} \Big(\frac{\rho D^3}{\Gamma}\Big)^\frac{1}{2} \xrightarrow{\times \text{Oh}} \frac{\eta D}{\Gamma} \xrightarrow{\times \text{Oh}^2} \frac{\eta^3}{\Gamma^2 \rho}
\label{Ohorder2}
\end{align}
Using these time scales in conjunction with the variable $t$, we built four dimensionless times that can be expressed from the simple dimensionless numbers. The three size-dependent time scales can be obtained from the associated simple dimensionless numbers as follows: 
\begin{align}
\frac{t}{\tau_{vi}}&=\Big(\frac{\Lambda}{\mathfrak{Re}}\Big)^2=\frac{\Lambda^2}{\tilde{\text{Re}}}=\alpha_* \frac{\Lambda^2}{\text{Re}}\\
\frac{t}{\tau_{ic}}&=\Big(\frac{\Lambda}{\mathfrak{We}}\Big)^\frac{3}{2}=\frac{\Lambda^\frac{3}{2}}{\tilde{\text{We}^\frac{1}{2}}}=\alpha_* \frac{\Lambda^\frac{3}{2}}{\text{We}^\frac{1}{2}}\\
\frac{t}{\tau_{vc}}&=\frac{\Lambda}{\mathfrak{Ca}}=\frac{\Lambda}{\tilde{\text{Ca}}}=\alpha_* \frac{\Lambda}{\text{Ca}}
\end{align}
Note that the ratios $\frac{\Lambda}{\mathfrak{Re}}$, $\frac{\Lambda}{\mathfrak{We}}$ and $\frac{\Lambda}{\mathfrak{Ca}}$ effectively produce inverse dimensionless numbers where the varying length scale $d$ has been replaced by the constant $D$, such that the only variability in the ratios come from $t$. 

The fourth dimensionless time can be expressed in a few equivalent ways illustrating different properties: 
\begin{equation}
 \frac{t}{\tau_{o}} \equiv \frac{t\Gamma^2 \rho}{\eta^3} = \frac{\Lambda}{\tilde{\text{Ca}} \text{Oh}^2}= \frac{1}{\mathfrak{Ca}} \frac{d}{\ell_{o}} = \alpha_* \frac{\text{Re}}{\text{Ca}^2} = \alpha_* \frac{\text{We}}{\text{Ca}^3} = \alpha_* \frac{\text{Re}^2}{\text{WeCa}}
\end{equation}
In Ohnesorge's units the alternate dimensionless numbers can be understood as the dimensionless equivalents of the kinematic quantities $c$, $\nu^\frac{1}{2}$ and $\kappa^\frac{1}{3}$, which are now understood as variable rather than constant: 
\begin{equation}
 \frac{d}{\ell_{o}}  = \mathfrak{Ca}(t) \frac{t}{\tau_{o}}= \mathfrak{Re}(t) \Big(\frac{t}{\tau_{o}}\Big)^\frac{1}{2} =\mathfrak{We}(t) \Big(\frac{t}{\tau_{o}}\Big)^\frac{2}{3}
\end{equation}
This polymorphism of the Ohnesorge units is exhibited quite visually by the intersections of the three curves $\mathfrak{Re}(t)$, $\mathfrak{Ca}(t)$ and $\mathfrak{We}(t)$, marking the instant with $t=\tau_o$, as shown on an example in SI-Fig.~\ref{Fig_wecare}d. 

\subsection{Dimensionless unit prefactors}
In sections~\ref{prefact} and \ref{algebra} we described how the actual units used in the article include prefactors taking into account the dimensionless constants. For instance, in the case of dynamics decomposed into an early visco-capillary regime and a late inertio-capillary regime, then $\gamma_1=(\delta_{ic}/\delta_{vc})^3$ and $\gamma_2=\delta_{vc} (\delta_{ic}/\delta_{vc})^3$.  Putting everything together and choosing the Ohnesorge units as an example, one has: 
\begin{equation}
 \frac{d}{\ell_{o}^*}  = \mathcal{C} \frac{t}{\tau_{o}^*}= \mathcal{R} \Big(\frac{t}{\tau_{o}^*}\Big)^\frac{1}{2} =\mathcal{W} \Big(\frac{t}{\tau_{o}^*}\Big)^\frac{2}{3}
\end{equation}
\noindent where we have defined renormalized versions of the simple dimensionless numbers: 
\begin{align}
\mathcal{C}&\equiv \frac{\mathfrak{Ca}(t)}{\mathfrak{Ca}_0}\\
\mathcal{R}&\equiv \frac{\mathfrak{Re}(t)}{\mathfrak{Re}_0}\\
\mathcal{W}&\equiv \frac{\mathfrak{We}(t)}{\mathfrak{We}_0}
\end{align}
With these new definitions, we have $\mathcal{W}^3=\mathcal{C}\mathcal{R}^2$. 

\begin{figure}
\centering
\includegraphics[width=15cm,clip]{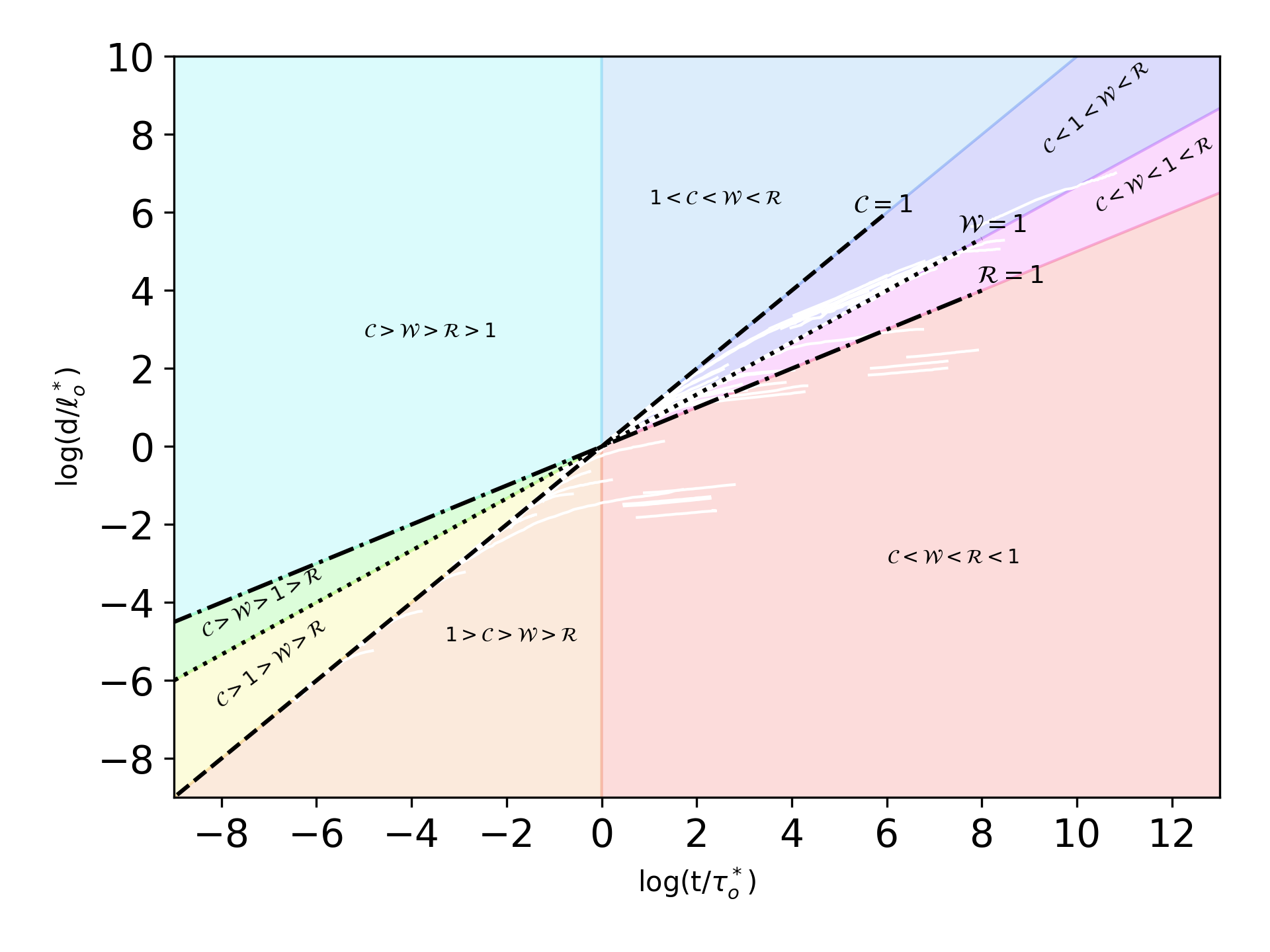}
\caption{The different dynamical domains in Ohnesorge units. Each colored domain corresponds to a different inequality between the three simple dimensionless numbers and the threshold $1$. Note that because of the `we care' relationship, \textit{i.e.} $\mathcal{W}^3=\mathcal{C}\mathcal{R}^2$, the only two possible inequalities between the three simple dimensionless numbers are $\mathcal{C}<\mathcal{W}<\mathcal{R}$ (if $t>\tau_o$) or $\mathcal{C}>\mathcal{W}>\mathcal{R}$ (if $t<\tau_o$). The different domains are then defined by the placement of $1$ within these inequalities.   
\label{dimdomains}}
\end{figure}

Overall, the dimensionless plots in the four systems of units provided in the article can be understood as representations of choices of different ratios of the simple dimensionless numbers normalized by their constant values. This mapping from dimensionless time and space to simple dimensionless numbers can be illustrated in Ohnesorge units, where the various sectors of the plot are associated with different inequalities between the dimensionless numbers, as illustrate in SI-Fig.~\ref{dimdomains}. 

\subsection{Quasi-properties and generalized dynamical number\label{general}}
So far we have assumed that the five scaling laws described in the article provided a good description of most of the regimes encountered in the analyzed experiments on spreading, coalescence and pinching. Indeed, the various representations of the data in the four systems of units largely supported this hypothesis. Although close to the reference exponents provided by the five scalings, the actual values may be slightly different. For instance, data may display an exponent $\alpha\simeq 0.45$ instead of $\alpha=\frac{1}{2}$ for Rayleigh's law. The dimensionless constants $\delta_{vc}$, $\delta_{ic}$ etc. provided in Tables~\ref{table4s}-\ref{table4p} were all obtained by assuming one of the standard exponents and fitting the experimental data with the associated regime. Instead, we can relax the constraint on the value of the exponent and fit the data to obtain both the exponent and the prefactor. 

If $d=Kt^\alpha$, dimensional analysis imposes that $[K]=\mathcal{L}.\mathcal{T}^{-\alpha}$. We know that if $\alpha$ is a rational number constructed from small integers, then the kinematic coefficient $K$ can be connected to ratios of traditional material parameters, like $\Gamma/\eta$, $(\Gamma/\rho)^\frac{1}{3}$ etc. If the value of $\alpha$ cannot be expressed as a simple fraction, one may say that $K$ is a `quasi-property'~\cite{Jaishankar2013}. In this case, the different systems of units can still be used to estimate an expected value of $K$. For instance, if one assumes than a spreading regime fitted by $d=Kt^\alpha$ can be well represented by the Ohnesorge units, then the expected value of $K$ is $K_o=\ell_o.\tau_o^{-\alpha}$, irrespective of the value of $\alpha$, \textit{i.e.} even if it is beyond 1, $\frac{1}{2}$ or $\frac{2}{3}$. For any particular system of units with length $\ell$ and time $\tau$, one can define a generalized dynamical dimensionless number: 
\begin{equation}
\mathfrak{N}(\alpha) \equiv \frac{dt^{-\alpha}}{\ell \tau^{-\alpha}}
\end{equation} 
For instance, for the Ohnesorge units we have: 
\begin{equation}
\mathfrak{N}_o(\alpha) \equiv \frac{d \Gamma^{1-2\alpha} \rho^{1-\alpha} }{ \eta^{2-3\alpha} t^{\alpha}} = \begin{cases}
    \mathfrak{C}       & \quad \text{if } \alpha=1\\
   \mathfrak{W}  & \quad \text{if } \alpha=\frac{2}{3}\\
   \mathfrak{R}  & \quad \text{if } \alpha=\frac{1}{2}\\
  \end{cases}
\end{equation} 

\begin{figure}
\centering
\begin{overpic}[abs,unit=1mm,width=8cm]{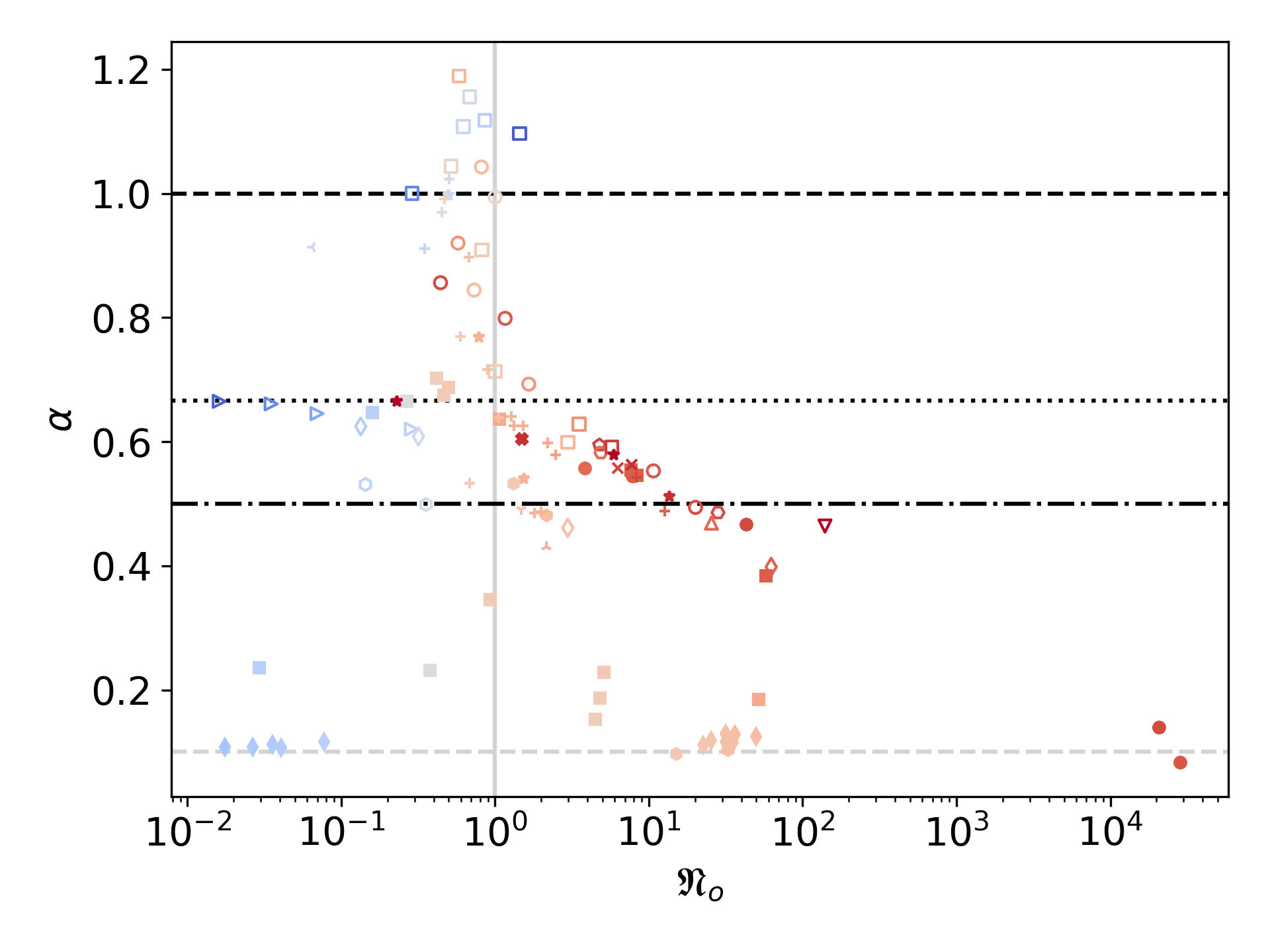}
\put(17,52){(a)}
\end{overpic}
\begin{overpic}[abs,unit=1mm,width=8cm]{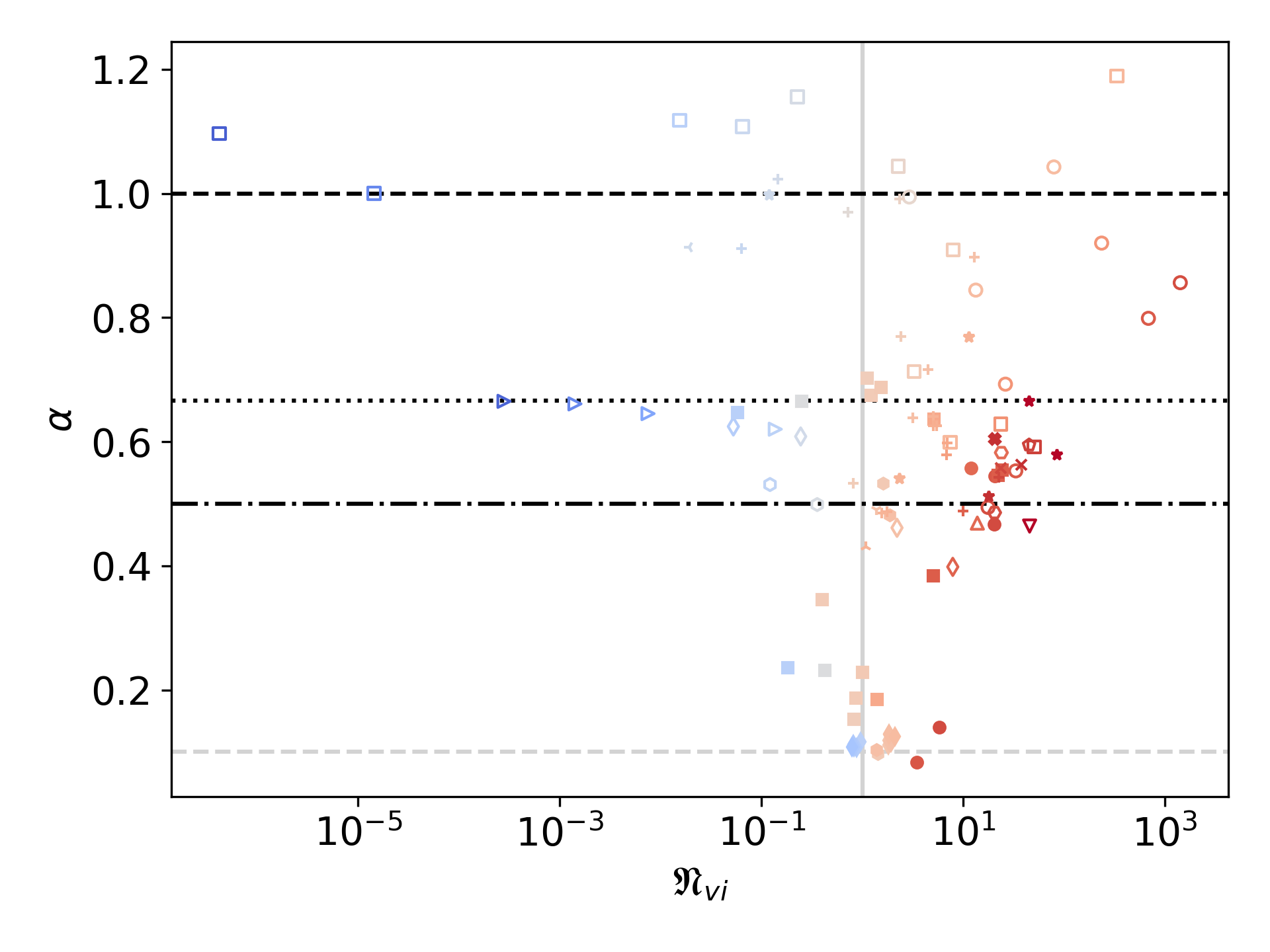}
\put(17,52){(b)}
\end{overpic}
\begin{overpic}[abs,unit=1mm,width=8cm]{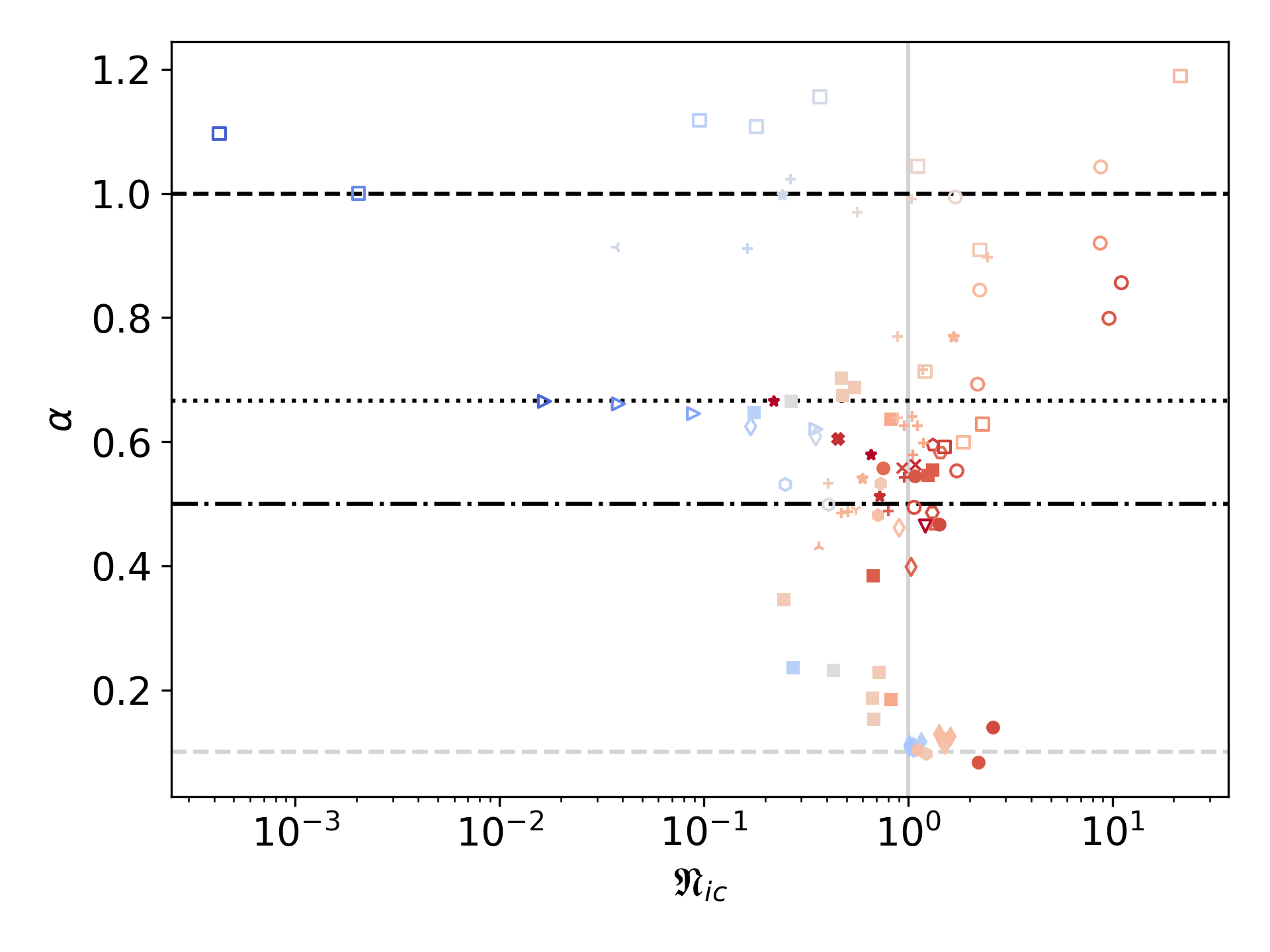}
\put(17,52){(c)}
\end{overpic}
\begin{overpic}[abs,unit=1mm,width=8cm]{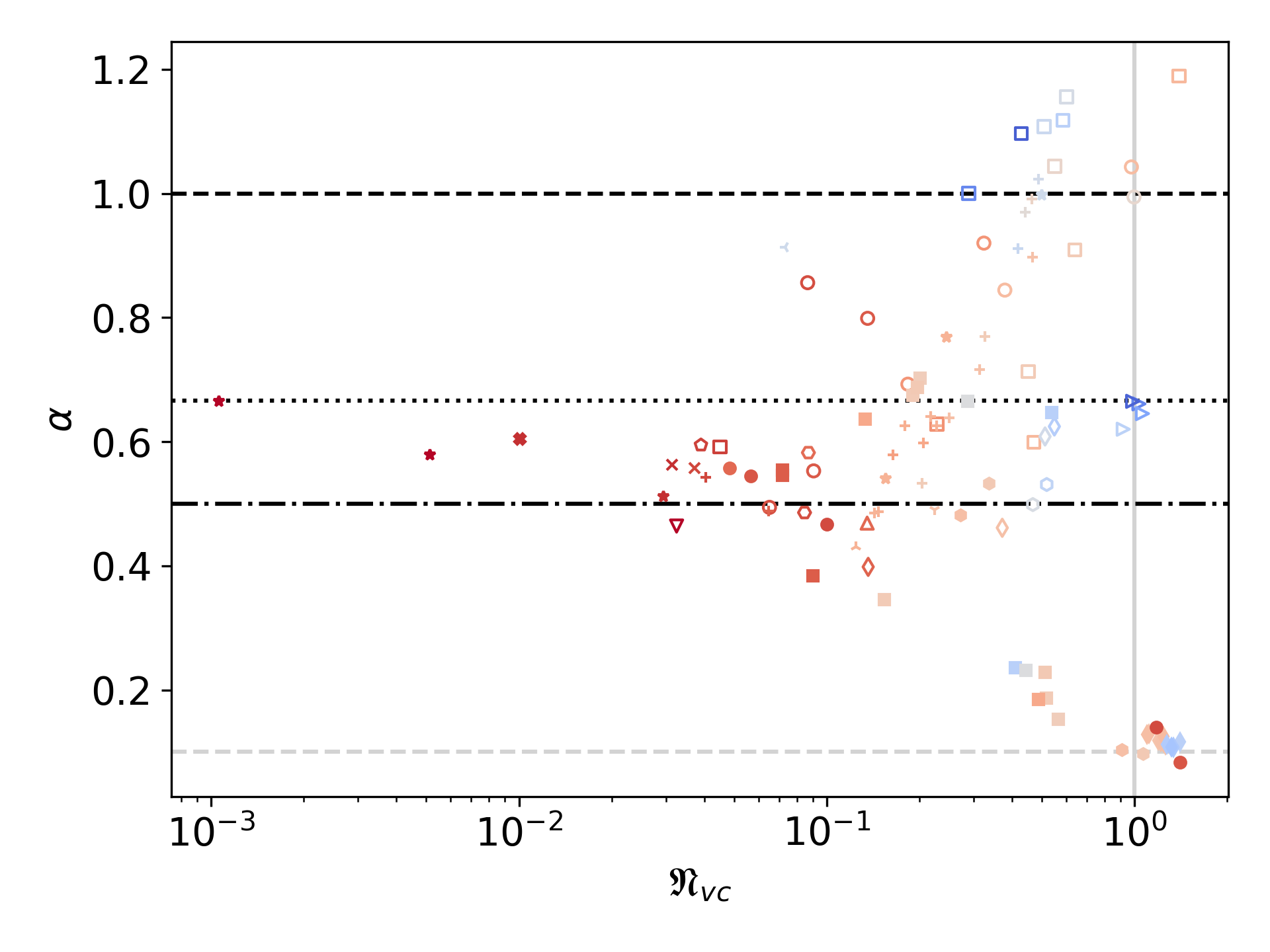}
\put(17,52){(d)}
\end{overpic}
\caption{Spreading exponent and generalized dynamical dimensionless numbers in Ohnesorge, visco-inertial, visco-capillary and inertio-capillary units. Agreement with a particular system of units corresponds $\mathfrak{N}\propto 1$ (vertical line). 
\label{AlphaK}}
\end{figure}

The values of $\mathfrak{N}_o$ and $\alpha$ obtained by fitting both the prefactor and the exponent on all the regimes of the data shown in the article are given in SI-Fig.~\ref{AlphaK}. Regimes exhibiting a value of $\mathfrak{N}_o$ close to 1 are in good agreement with the Ohnesorge units. In contrast, dynamics associated with very small or large values of $\mathfrak{N}_o$ indicate departure from the Ohnesorge units. This is particularly true when the extrinsic size $D$ starts to have an impact, as in the Tanner or Rayleigh regimes. In SI-Fig.~\ref{AlphaK} we also provide plots for the values of $\mathfrak{N}_{vi}$, $\mathfrak{N}_{ic}$ and $\mathfrak{N}_{vc}$, by using the respective visco-inertial, inertio-capillary and visco-capillary units to define the $\mathfrak{N}$. 

\clearpage

\section{Open questions}
In this section, we highlight some of the issues raised in the main article by providing additional plots focusing on sub-sets of the experimental data. 

\subsection{Size-dependence of the inertio-capillary regimes for Oh$<1$}
\begin{figure}[h]
\centering
\includegraphics[width=15cm,clip]{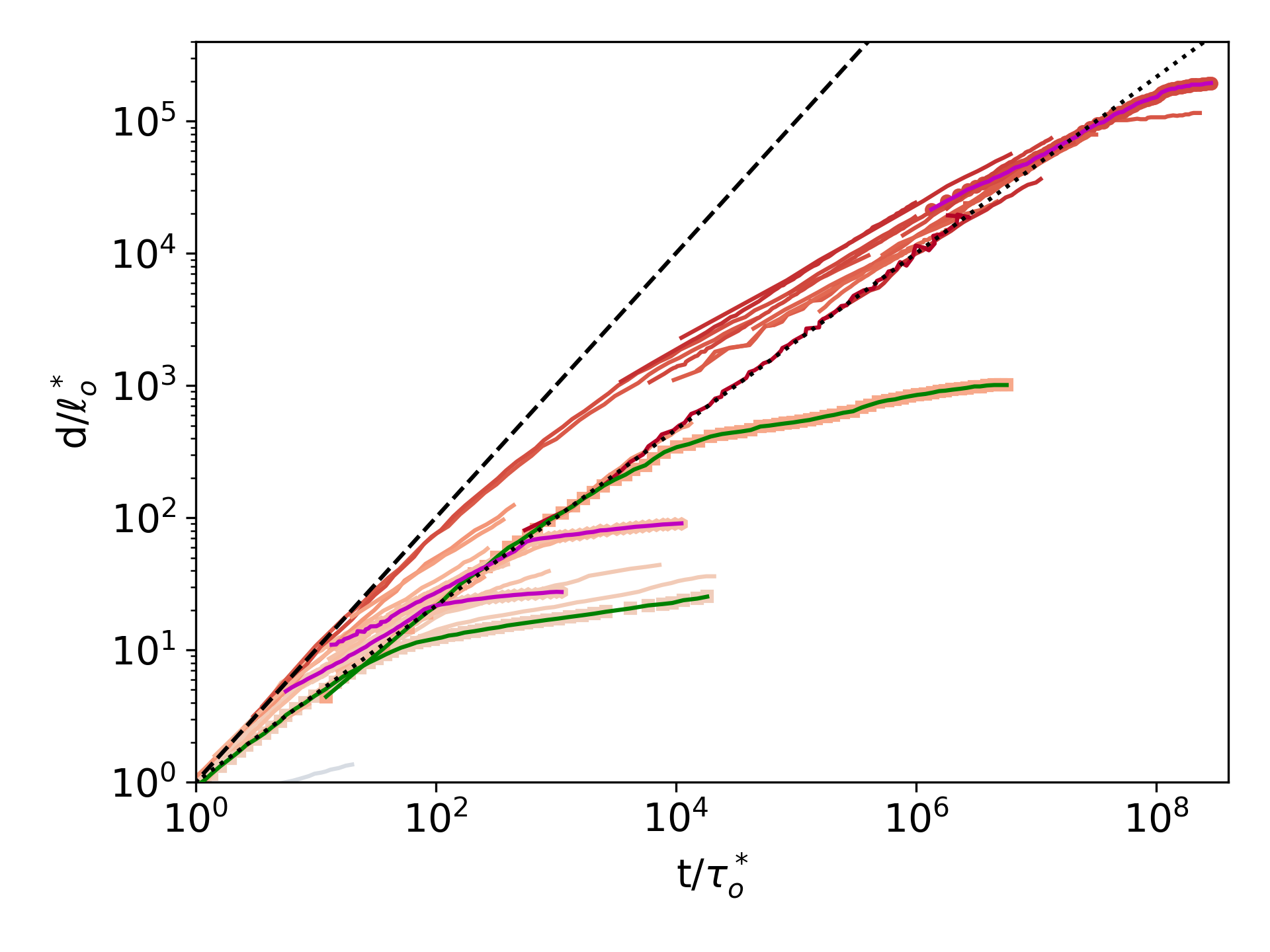}
\caption{Spreading, coalescence and pinching dynamics for $d/\ell_o^*>1$ and $t/\tau_o^*>1$. The highlighted spreading curves have $\text{Oh}=2.31~10^{-1}$ (`Chen2014 Fig3b 60cP'), $\text{Oh}=2.95~10^{-1}$ (`Eddi2013 Fig4 0p37'), $\text{Oh}=5.71~10^{-2}$ (`Eddi2013 Fig6 11') and $\text{Oh}=5.71~10^{-2}$ (`Biance2004 Fig3 1p2').
\label{sizesmallOh}}
\end{figure}   
In SI-Fig.~\ref{sizesmallOh} we provide a close-up on the region $d/\ell_o^*>1$ and $t/\tau_o^*>1$. In this sector of the operating space, we have $\mathcal{C}<\mathcal{W}<\mathcal{R}$. Data from spreading, coalescence and pinching exhibit two possible trends. Starting at $d/\ell_o^*=1$ and $t/\tau_o^*=1$, some dynamics directly follow the inertio-capillary regime ($\alpha=\frac{2}{3}$, dotted line), while other remain on the visco-capillary regime ($\alpha=1$, dashed line), until $d/\ell_o^*=t/\tau_o^*=\text{Oh}^{-1}$, after which they follow Rayleigh's regime ($\alpha=\frac{1}{2}$), reaching $d/\ell_o^*=D$ when crossing the dotted line. To put it bluntly: why do some dynamics take a short cut via the $\frac{2}{3}$ regime, while other take the long route through Rayleigh's regime? Let us take spreading as an example. Highlighted in SI-Fig.~\ref{sizesmallOh} are four spreading curves with $\text{Oh}=2.31~10^{-1}$ (`Chen2014 Fig3b 60cP'), $\text{Oh}=2.95~10^{-1}$ (`Eddi2013 Fig4 0p37'), $\text{Oh}=5.71~10^{-2}$ (`Eddi2013 Fig6 11') and $\text{Oh}=5.71~10^{-2}$ (`Biance2004 Fig3 1p2'). The data from Eddi \textit{et al.} abide to the $\frac{2}{3}$ scaling, while the data from Chen \textit{et al.} and Biance \textit{et al.} follow Rayleigh's regime. Some criterion beyond the value of the Ohnesorge number seems to dictate the course of events. Note that in experiments by Chen \textit{et al.}, the substrate was partially wetting ($\theta=63^\circ$). 

\subsection{Inertio-capillary regime for Oh$>1$}
\begin{figure}[h]
\centering
\begin{overpic}[abs,unit=1mm,width=8.5cm]{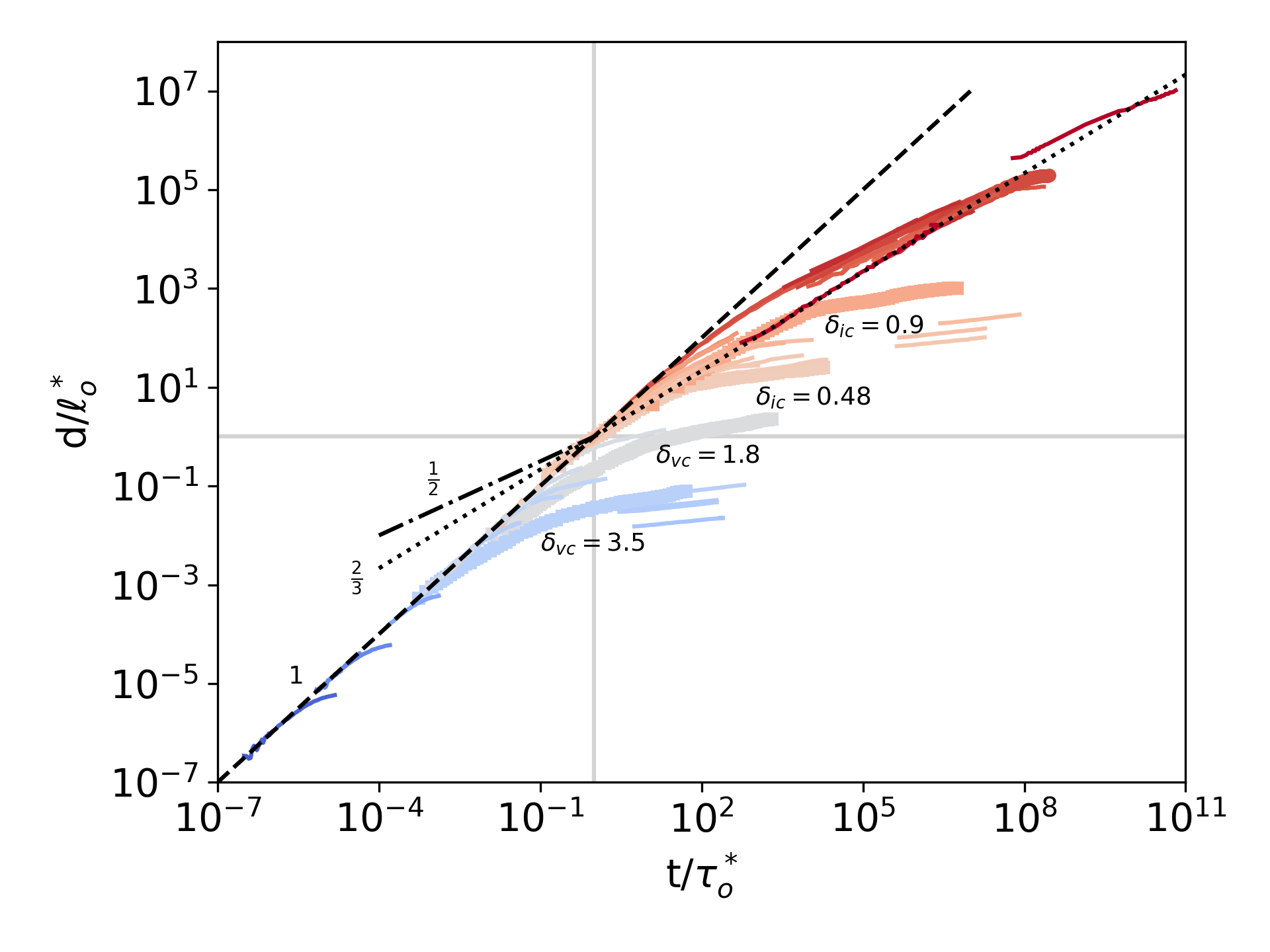}
\put(17,56){(a)}
\end{overpic}
\begin{overpic}[abs,unit=1mm,width=8.5cm]{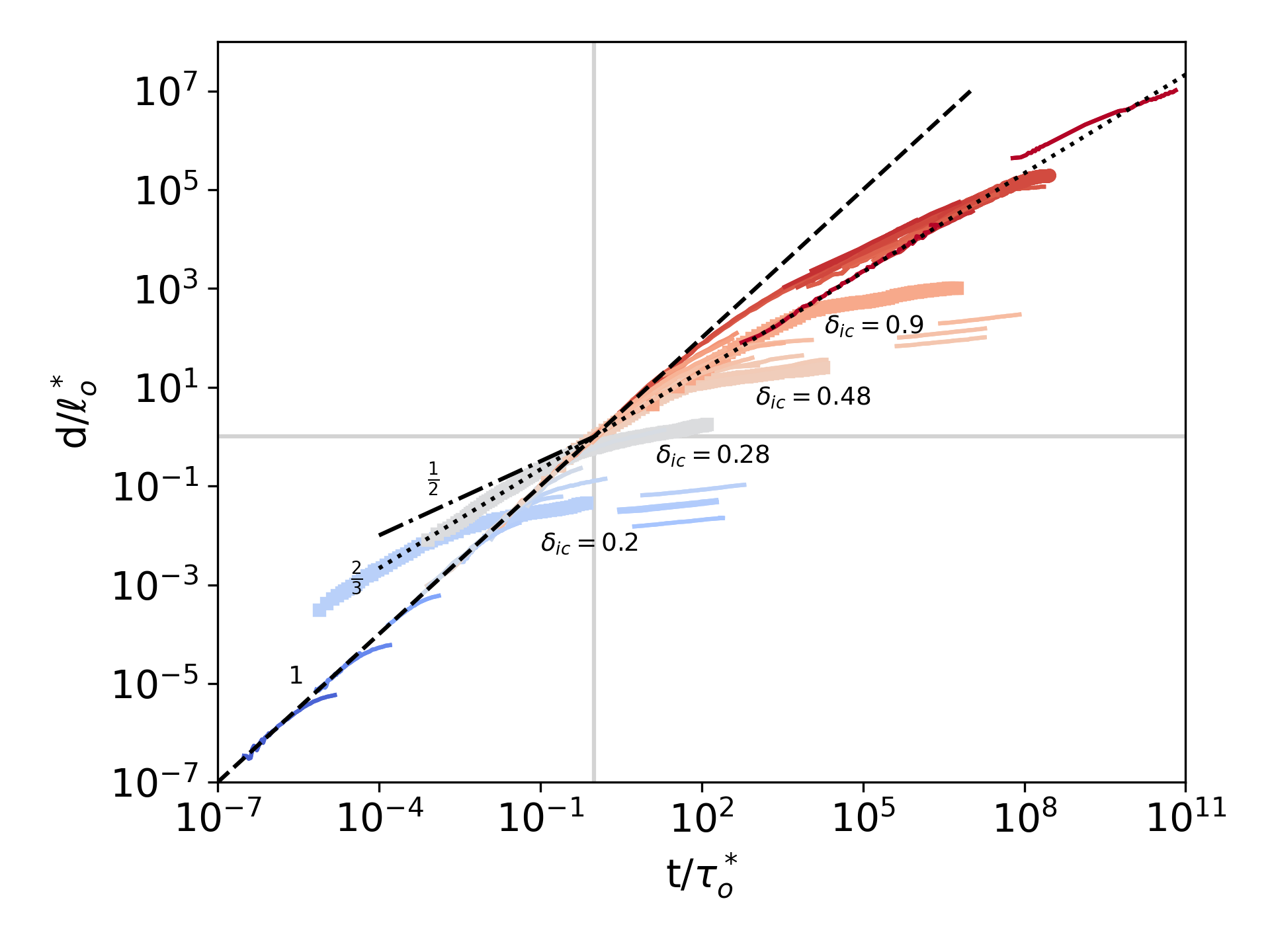}
\put(17,56){(b)}
\end{overpic}
\caption{Two ways to interpret the data sets `Eddi2013 Fig6 220' ($\text{Oh}=1.1$) and `Eddi2013 Fig6 1120' ($\text{Oh}=5.6$). (a) Visco-capillary interpretation: the first data points are interpreted as belonging to a visco-capillary regime, followed by a long intermediate range dominated by a logarithmic correction, finally reaching a Tanner regime in the late spreading. The fitted values of $\delta_{vc}$ and $\delta_ {Tan}$ are used to compute the unit prefactors $\gamma_1=(\delta_{Tan}/\delta_{vc})^\frac{10}{9}$ and $\gamma_2=\delta_{vc} \gamma_1$. (b) Inertio-capillary interpretation: the early spreading is interpreted as an inertio-capillary regime followed by a Tanner regime in the late spreading. The fitted values of $\delta_{ic}$ and $\delta_ {Tan}$ are used to compute the unit prefactors $\gamma_1=(\delta_{Tan}/\delta_{ic})^\frac{30}{17}$ and $\gamma_2=\delta_{ic} \gamma_1^\frac{2}{3}$. 
\label{Eddi23}}
\end{figure}

The existence of a $\frac{2}{3}$ regime of spreading for Oh$<1$ as exhibited above is not too surprising and has been discussed in the literature~\cite{}. While compiling the data reproduced in the article, we also noticed that some spreading experiments with Oh$\gtrsim1$ could also be interpreted as following a $\frac{2}{3}$ regime extending in the region $d/\ell_o^*<1$ and $t/\tau_o^*<1$, where $\mathcal{C}>1\gtrsim\mathcal{W}>\mathcal{R}$. These experiments are labeled `Eddi2013 Fig6 220' and `Eddi2013 Fig6 1120'. In the main article and in the Table~\ref{table4s} we chose to interpret these data from a visco-capillary perspective in accordance with the original paper~\cite{Eddi2013}. That is to say that the spreading was interpreted as displaying a visco-capillary regime with an intermediate regime dominated by a logarithmic correction. In this interpretation, the first few data points are used to obtain a value of $\delta_{vc}$, and the associated unit prefactors $\gamma_1$ and $\gamma_2$ computed with the addition of the late spreading prefactor $\delta_{Tan}$ produce the curves displayed in SI-Fig.~\ref{Eddi23}a. The values of $\delta_{vc}$ derived in this way are sensibly larger than the average, as can be seen in SI-Fig.~\ref{figSIdvc}. A different way to interpret these two data sets is to consider than the early regime abides to a $\frac{2}{3}$ regime, and to build the unit prefactors from $\delta_{ic}$ and $\delta_{Tan}$. The example of `Eddi2013 Fig6 220' was treated in details in SI-Fig.~\ref{Fig_wecare}, where it was quite convincing than for $t/\tau_o^*<1$ the dynamics displayed a constant Weber number, and so followed the $\frac{2}{3}$ regime. In SI-Fig.~\ref{Eddi23}b we give the result of interpreting the data sets `Eddi2013 Fig6 220' and `Eddi2013 Fig6 1120' in this alternate way. From this perspective, the inferred values of $\delta_{ic}$ are consistent but seem to decrease with increasing Ohnesorge number. We expect that investigating the dynamics at yet earlier times would help deciphering the respective roles of inertia and viscosity for the onset of spreading. More broadly for values of the Ohnesorge number close to 1, we can expect a broader diversity of dynamics than reflected in the main paper. For instance, recent data on pinching near $\text{Oh}\propto 1$ suggest a variety of rich behaviors~\cite{Castrejon2015}. Unfortunately, the information on the values of the material parameters used in the experiments of this reference was too scarce to allow us to reproduce the plots.  

%\subsection{Liquid-on-liquid spreading}
%
%\subsection{Generalized integral scale}

\section{Animated figures}
The files `Fig3{\_}VC.gif', `Fig3{\_}IC.gif', `Fig4{\_}VI.gif' and `Fig5{\_}Oh.gif' give animated versions of Fig.~3, 4 and 5 of the main article, showing how every data set is included in the graph. The images used to create the animated files are given in the zip archives of the same names. Note that the values of $\eta$, $\rho$, $\Gamma$ and $D$ given on each image have standard units, \textit{i.e.} Pa.s for viscosity, kg/m$^3$ for density, N/m for surface tension and m for the extrinsic size. 

The file `TowardOhUnits.gif' gives an animated figure built from the succession of three images with different units for the data sets shown in Fig.~5 of the main article: standard units ($d$ in meters and $t$ in seconds), `bare' Ohnesorge units $d/\ell_o$ and $t/\tau_o$, and normalized Ohnesorge units $d/\ell_o^*$ and $t/\tau_o^*$. 

\clearpage

\section{Details on experimental data\label{datacomments}}
In this section we provide additional details on the experiments reproduced in the article. The readers are referred to the original publications for the full context. The horizontal lines on some of the plots give the value of $D$, when it is within the range of the data. The vertical plain, dashed, dotted and dotted-dashed lines respectively give the values of $\tau_o^*$, $\tau_{vc}^*$, $\tau_{ic}^*$ and $\tau_{vi}^*$, when within the range of the data. 

\begin{table}[h]
  \centering
  \begin{tabular}{ | p{9cm} | p{9cm} | }
    \hline
    \textbf{Cazabat1986 Fig1a 0p78} & \textbf{Cazabat1986 Fig1a 1p5}  \\
    \begin{minipage}{.5\textwidth}
      \includegraphics[width=\linewidth]{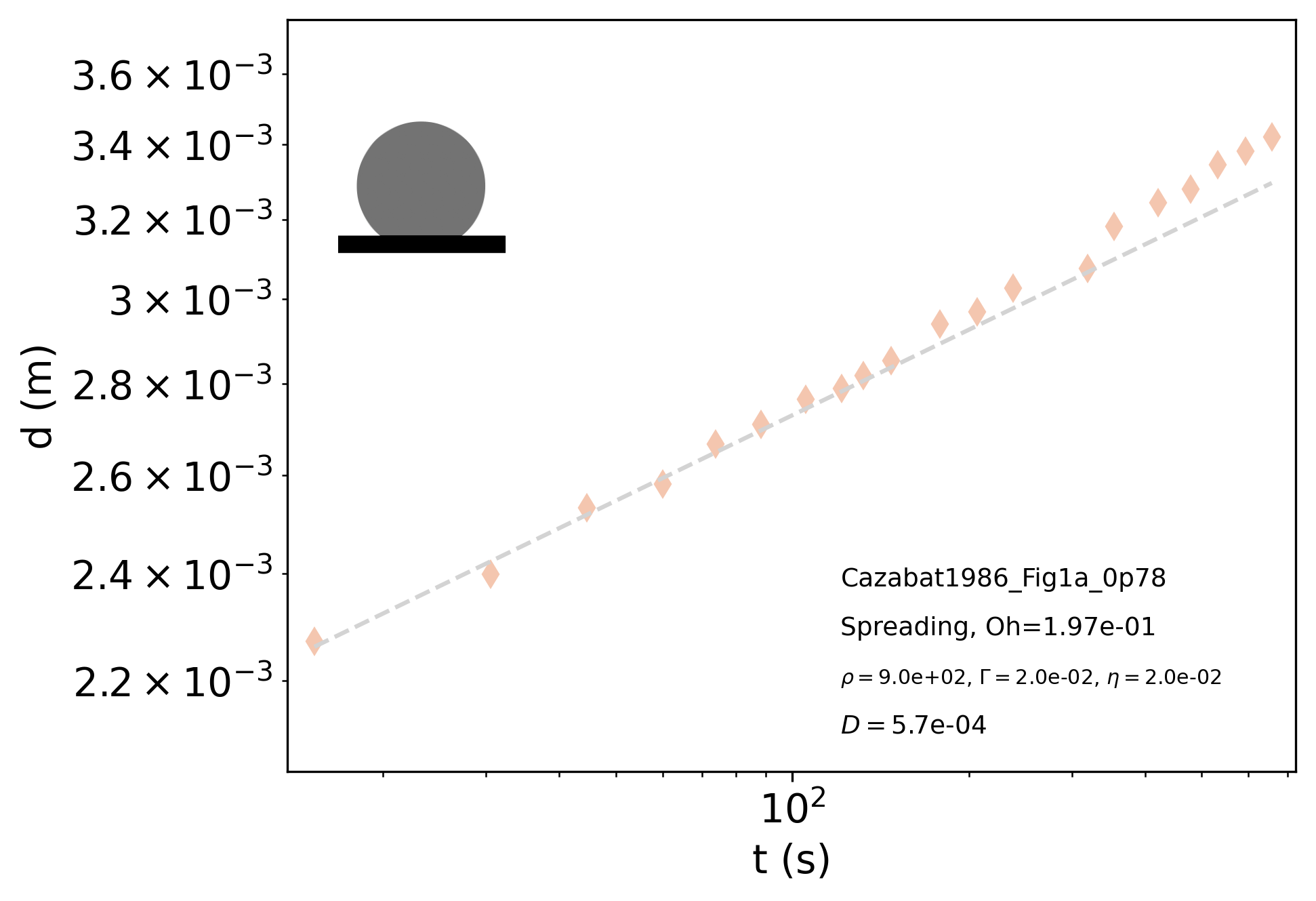}
    \end{minipage}
    &
    \begin{minipage}{.5\textwidth}
      \includegraphics[width=\linewidth]{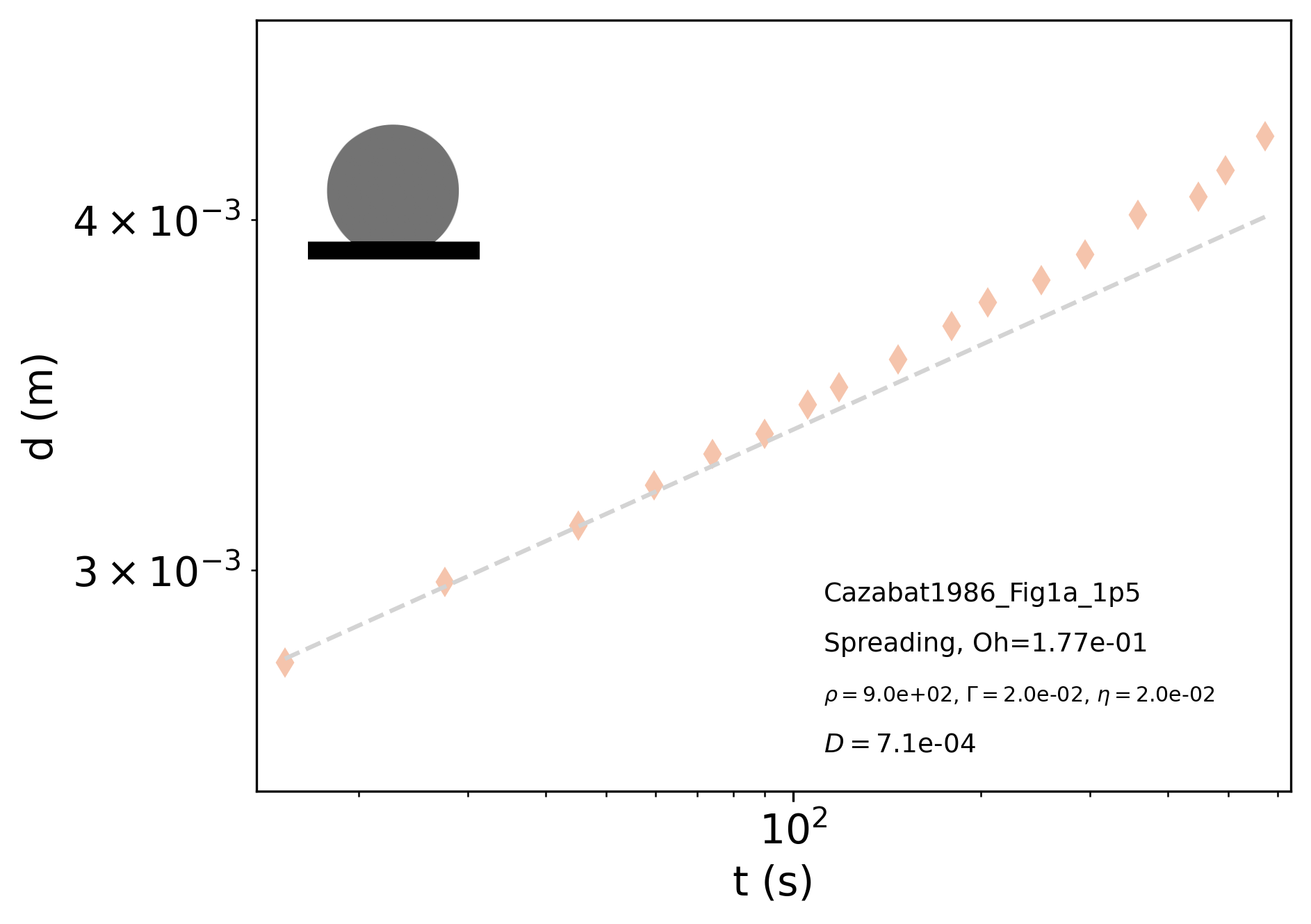}
    \end{minipage}
    \\
  Silicone oil in ambient air, on smooth hydrophilic glass. \newline Effect of gravity for $d\gtrsim (\Gamma/\rho g)^\frac{1}{2}$. & Silicone oil in ambient air, on smooth hydrophilic glass. \newline Effect of gravity for $d\gtrsim (\Gamma/\rho g)^\frac{1}{2}$.  \\
      \hline \hline
    \textbf{Cazabat1986 Fig1a 2p9} & \textbf{Cazabat1986 Fig1a 3p8}  \\ 
    \begin{minipage}{.5\textwidth}
      \includegraphics[width=\linewidth]{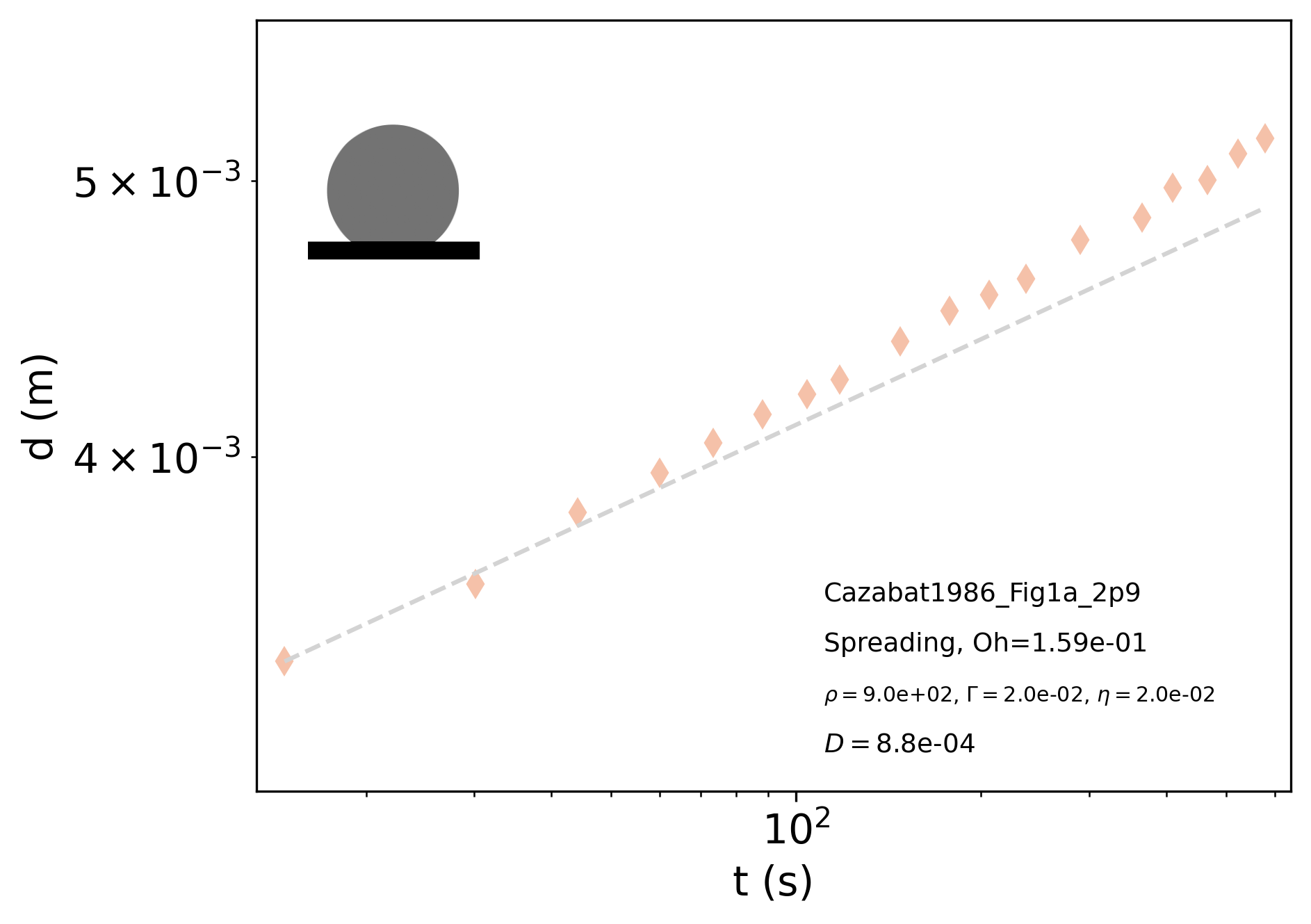}
    \end{minipage}
    &
    \begin{minipage}{.5\textwidth}
      \includegraphics[width=\linewidth]{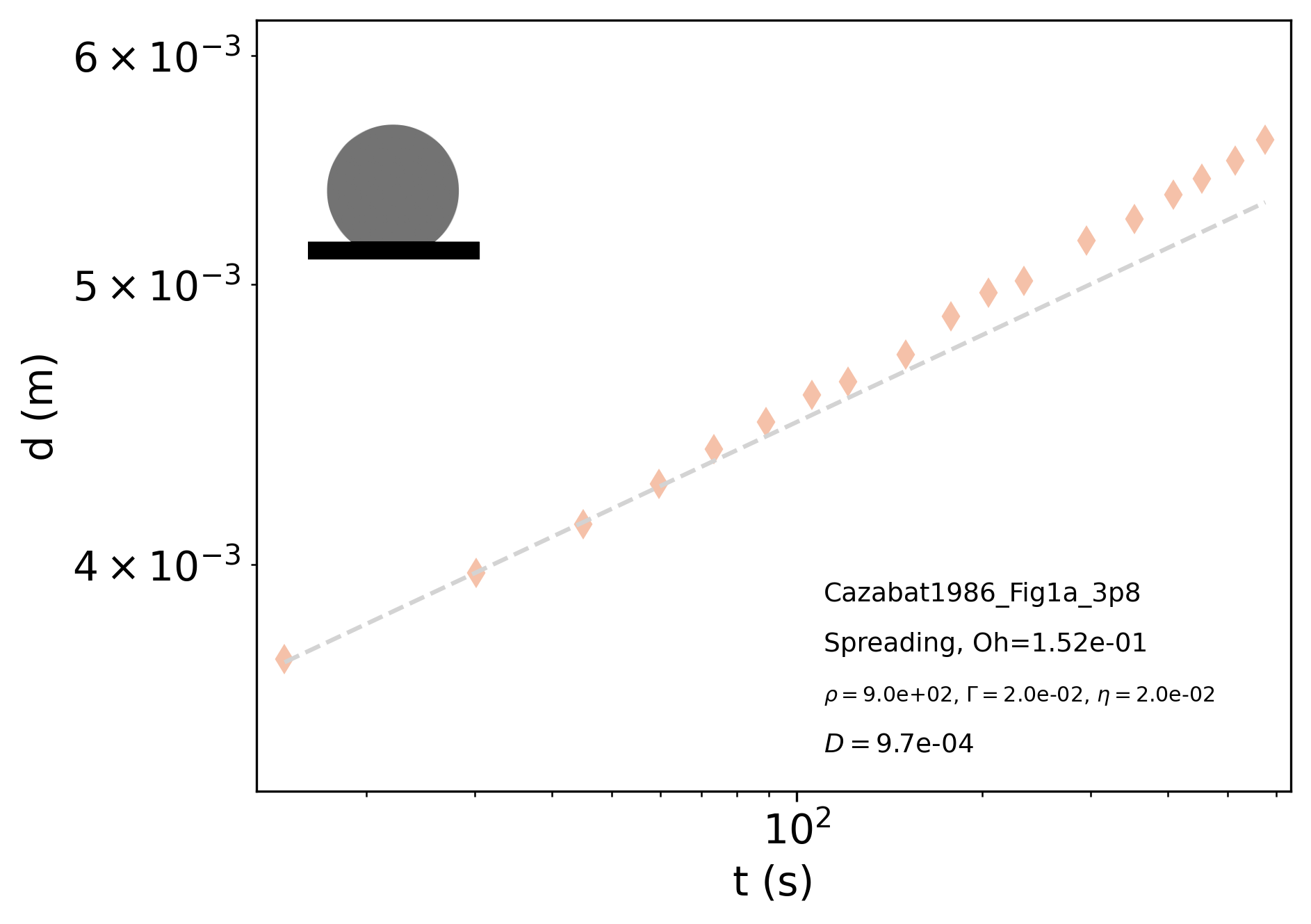}
    \end{minipage}
    \\ 
  Silicone oil in ambient air, on smooth hydrophilic glass. \newline Effect of gravity for $d\gtrsim (\Gamma/\rho g)^\frac{1}{2}$. & Silicone oil in ambient air, on smooth hydrophilic glass. \newline Effect of gravity for $d\gtrsim (\Gamma/\rho g)^\frac{1}{2}$.  \\ \hline
  \end{tabular}
\end{table}

\begin{table} 
 \centering 
 \begin{tabular}{ | p{9cm} | p{9cm} | } 
 \hline 
 \textbf{Cazabat1986 Fig1a 4p7} & \textbf{Cazabat1986 Fig1a 7p8}  \\ 
 \begin{minipage}{.5\textwidth} 
 \includegraphics[width=\linewidth]{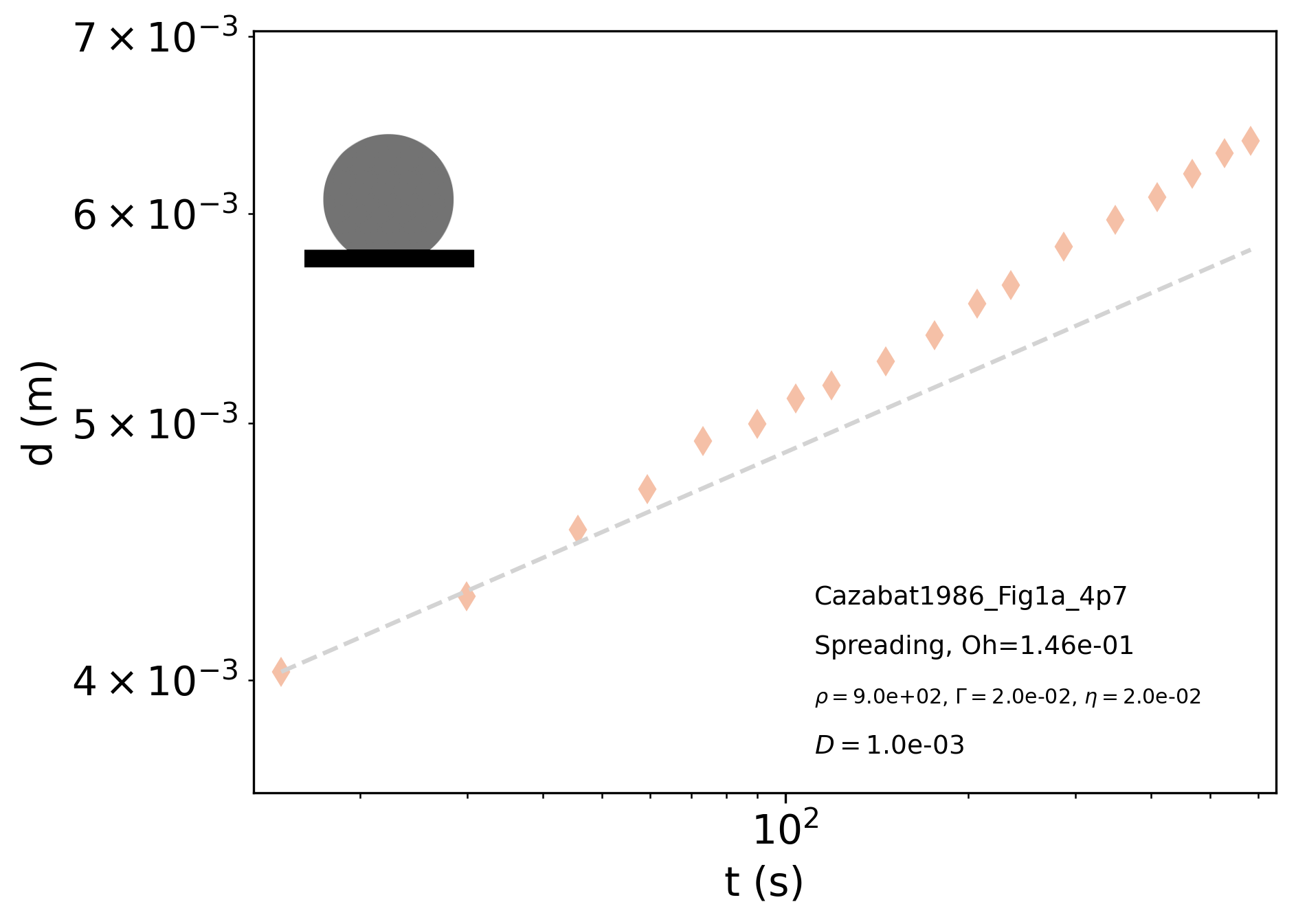} 
 \end{minipage}
 & 
 \begin{minipage}{.5\textwidth} 
 \includegraphics[width=\linewidth]{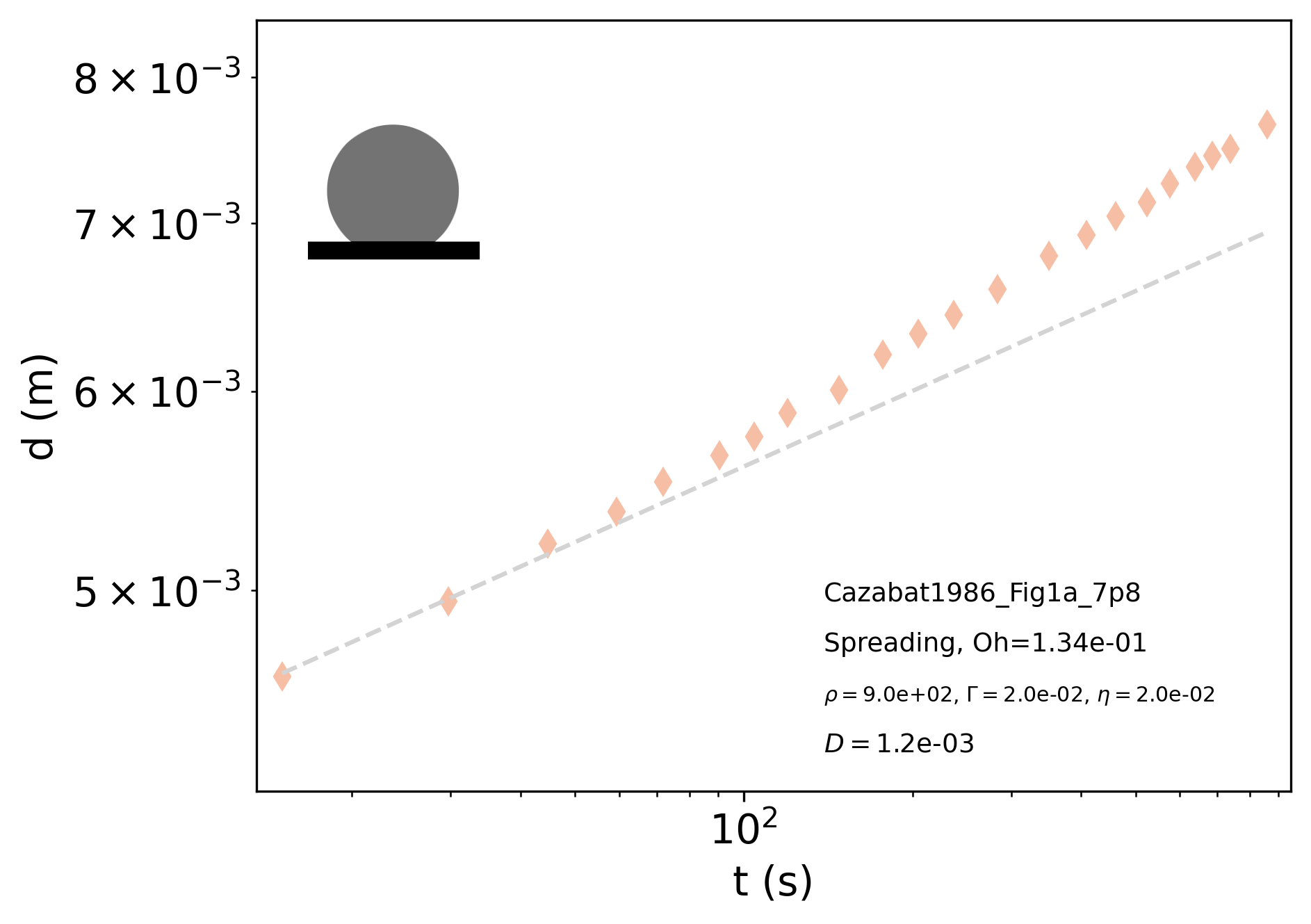} 
 \end{minipage} 
 \\ 
Silicone oil in ambient air, on smooth hydrophilic glass. \newline Effect of gravity for $d\gtrsim (\Gamma/\rho g)^\frac{1}{2}$. & Silicone oil in ambient air, on smooth hydrophilic glass. \newline Effect of gravity for $d\gtrsim (\Gamma/\rho g)^\frac{1}{2}$.\\ \hline \hline 
\textbf{Cazabat1986 Fig1a 14p4} & \textbf{Cazabat1986 Fig1b 37p9}  \\ 
 \begin{minipage}{.5\textwidth} 
 \includegraphics[width=\linewidth]{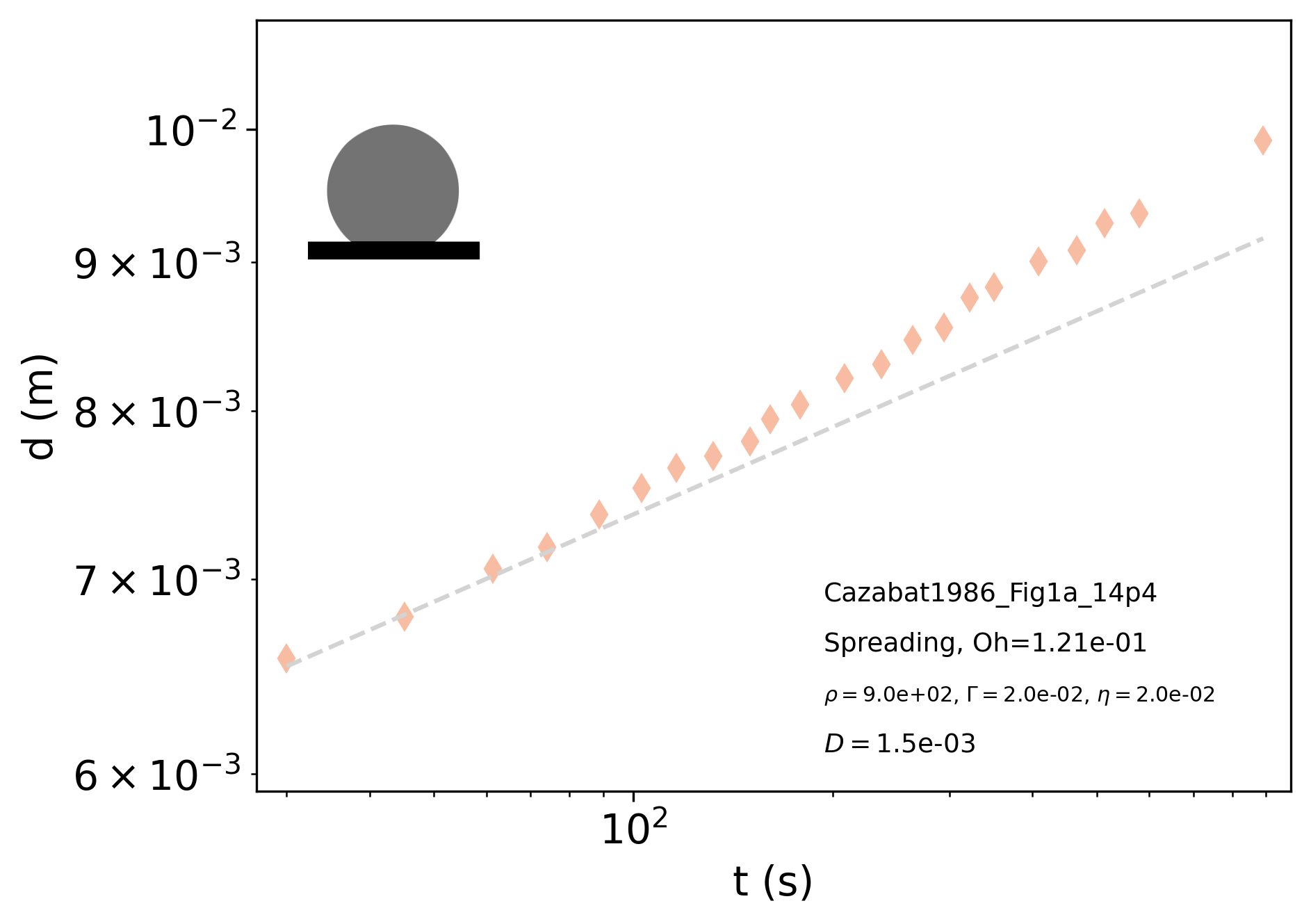} 
 \end{minipage}
 & 
 \begin{minipage}{.5\textwidth} 
 \includegraphics[width=\linewidth]{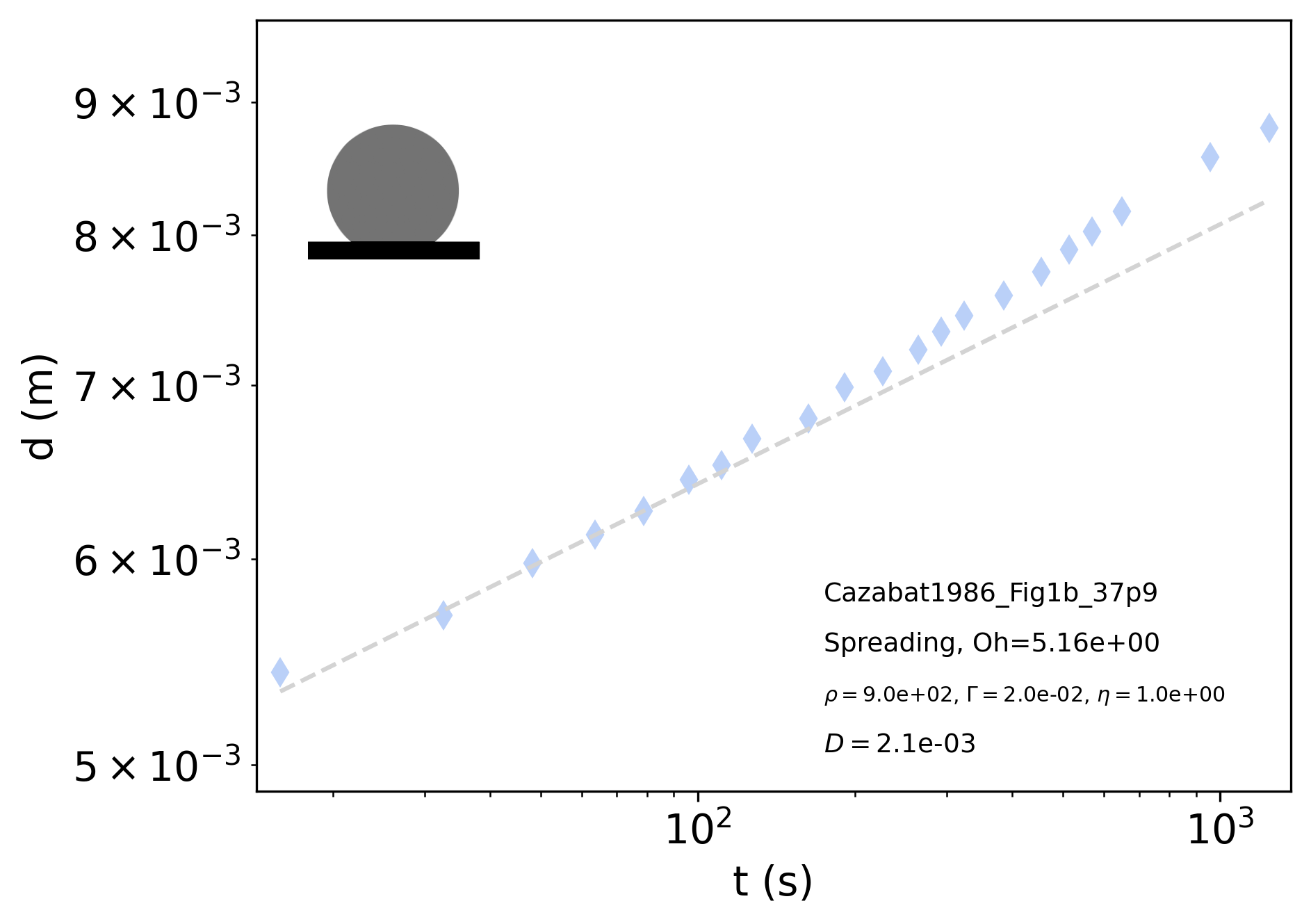} 
 \end{minipage} 
 \\ 
Silicone oil in ambient air, on smooth hydrophilic glass. \newline Effect of gravity for $d\gtrsim (\Gamma/\rho g)^\frac{1}{2}$. & Silicone oil in ambient air, on smooth hydrophilic glass. \newline Effect of gravity for $d\gtrsim (\Gamma/\rho g)^\frac{1}{2}$.\\ \hline \hline 
\textbf{Cazabat1986 Fig1b 5p8} & \textbf{Cazabat1986 Fig1b 4p03}  \\ 
 \begin{minipage}{.5\textwidth} 
 \includegraphics[width=\linewidth]{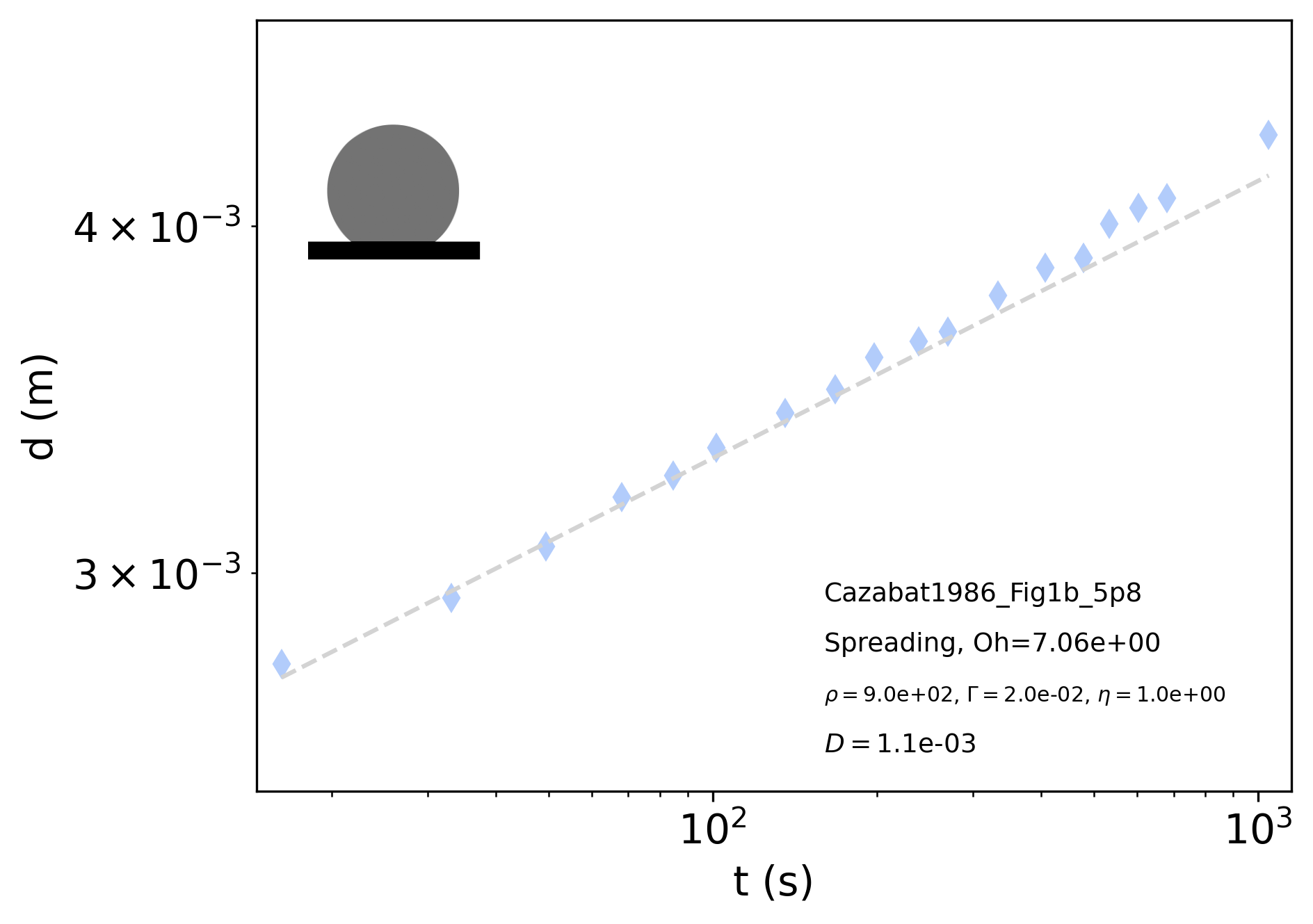} 
 \end{minipage}
 & 
 \begin{minipage}{.5\textwidth} 
 \includegraphics[width=\linewidth]{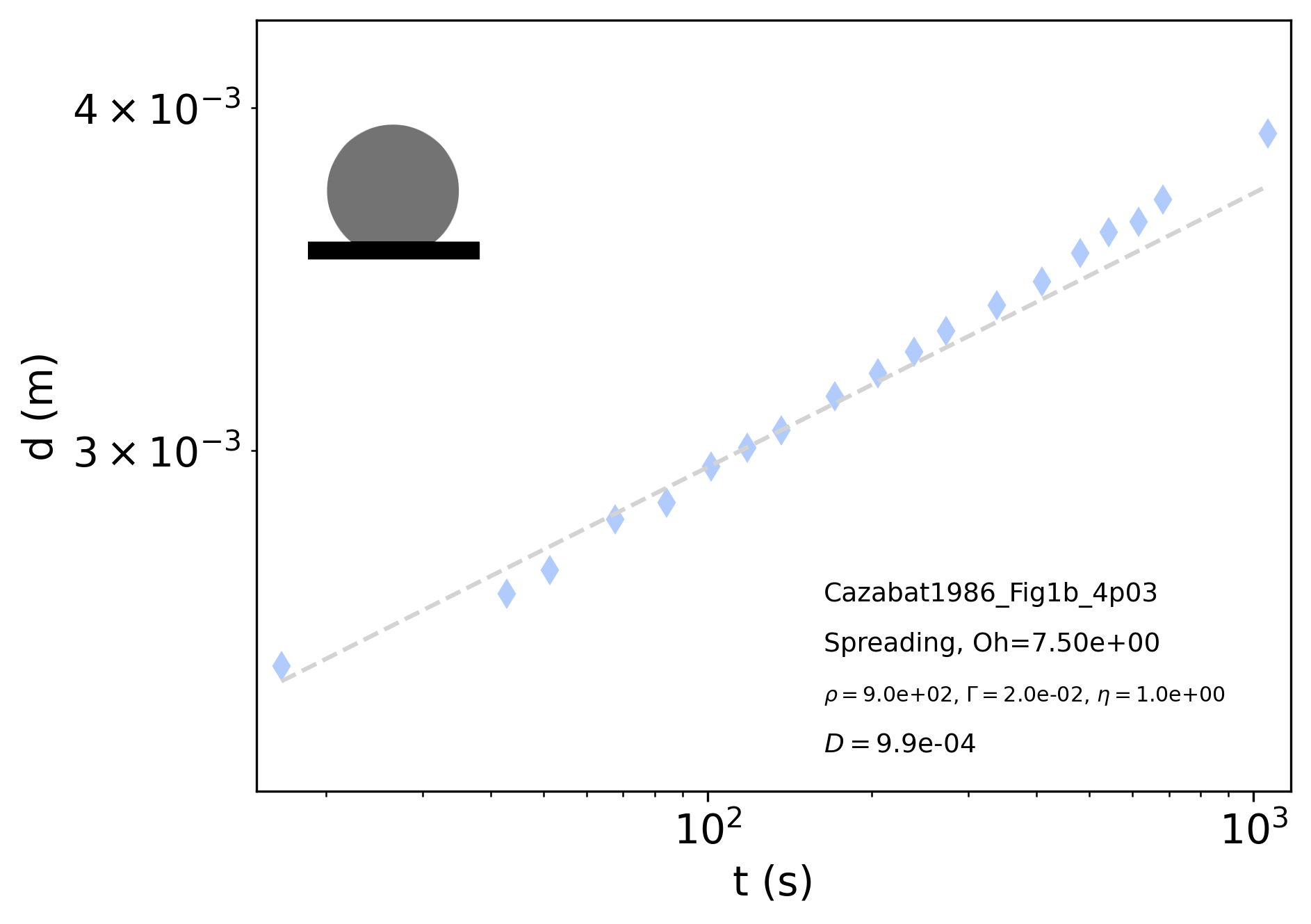} 
 \end{minipage} 
 \\ 
Silicone oil in ambient air, on smooth hydrophilic glass. \newline Effect of gravity for $d\gtrsim (\Gamma/\rho g)^\frac{1}{2}$. & Silicone oil in ambient air, on smooth hydrophilic glass. \newline Effect of gravity for $d\gtrsim (\Gamma/\rho g)^\frac{1}{2}$.\\ \hline \hline 
\end{tabular} 
 \end{table} 
\begin{table} 
 \centering 
 \begin{tabular}{ | p{9cm} | p{9cm} | } 
 \hline 
 \textbf{Cazabat1986 Fig1b 1p35} & \textbf{Cazabat1986 Fig1b 0p35}  \\ 
 \begin{minipage}{.5\textwidth} 
 \includegraphics[width=\linewidth]{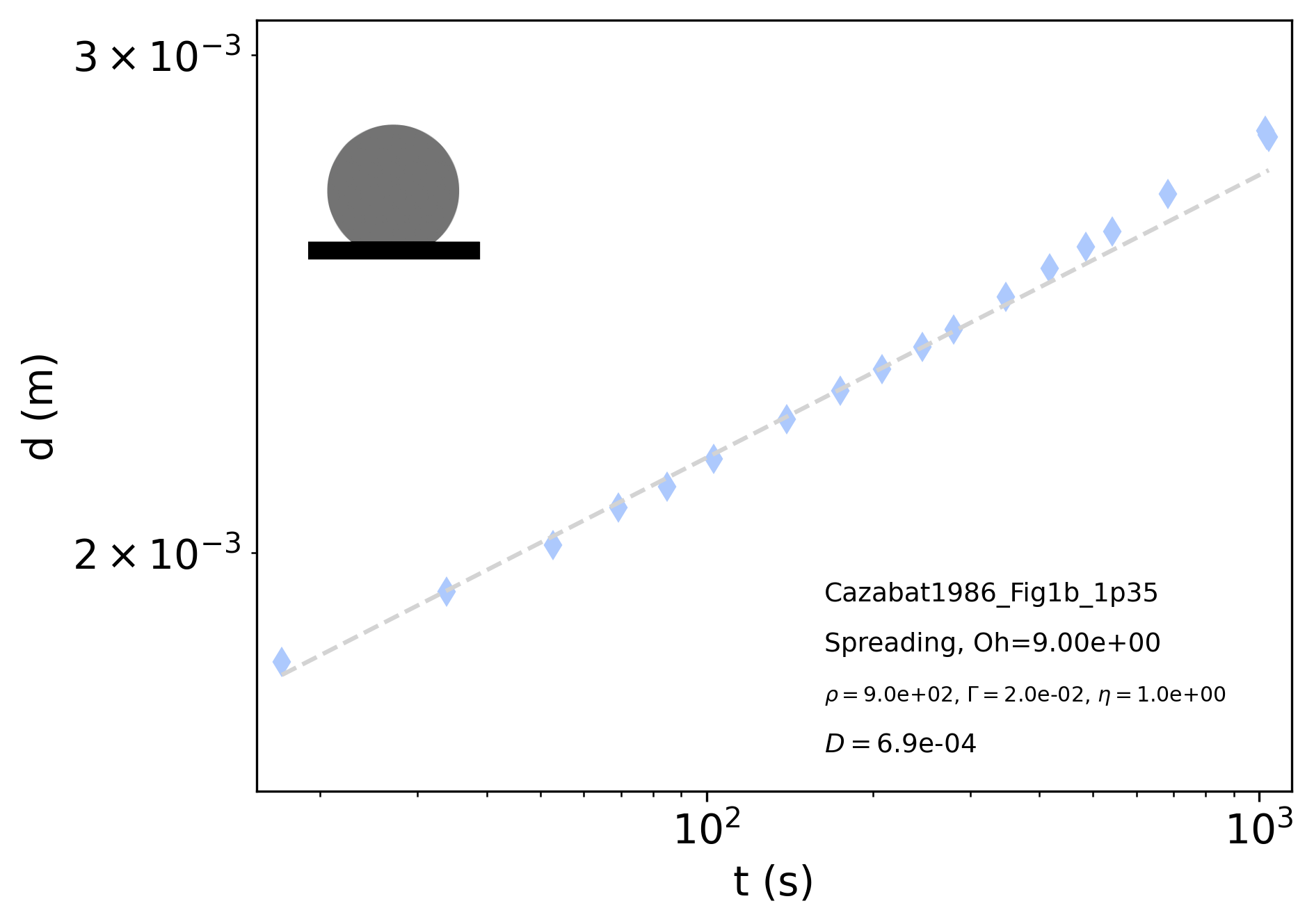} 
 \end{minipage}
 & 
 \begin{minipage}{.5\textwidth} 
 \includegraphics[width=\linewidth]{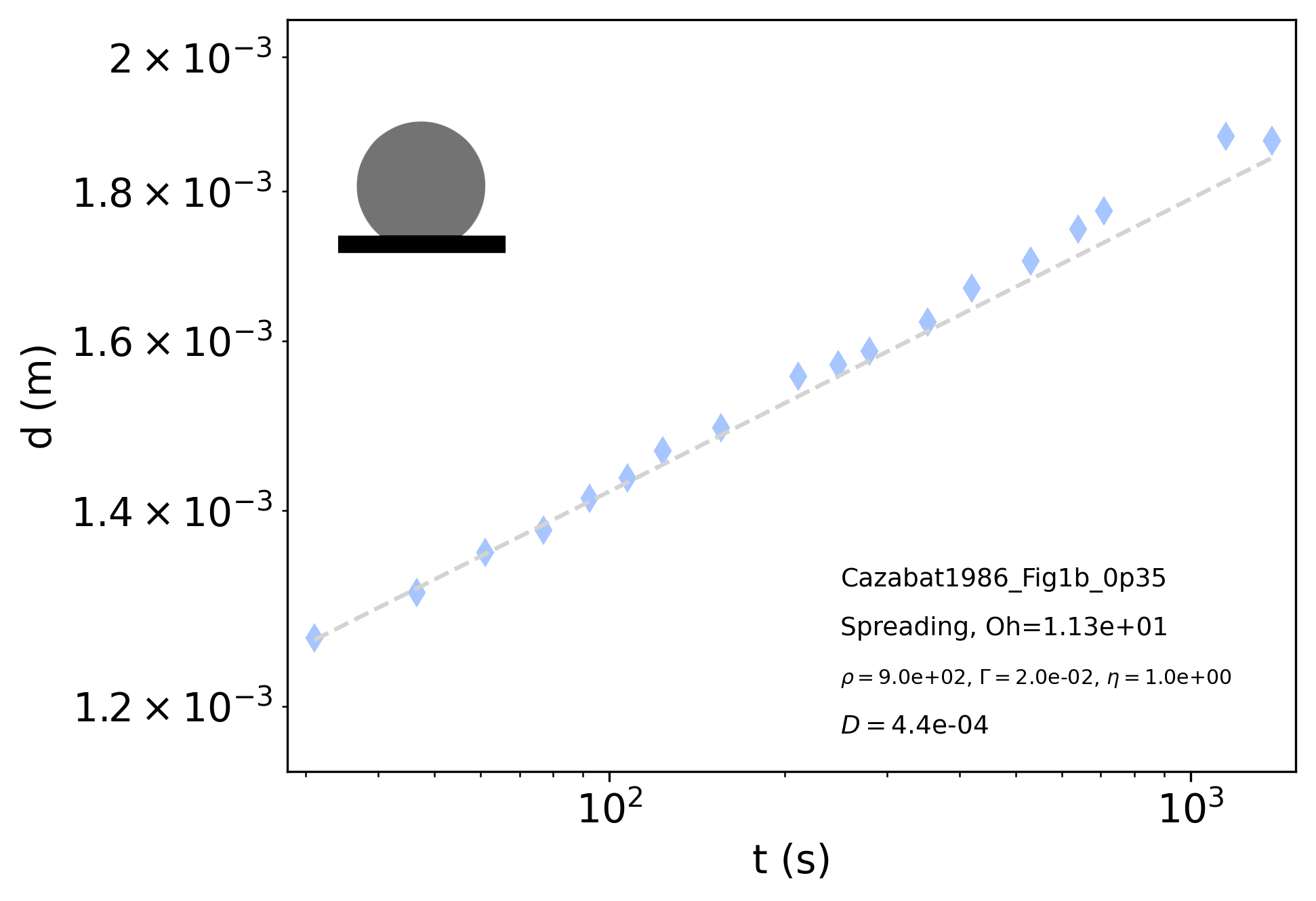} 
 \end{minipage} 
 \\ 
Silicone oil in ambient air, on smooth hydrophilic glass. \newline Effect of gravity for $d\gtrsim (\Gamma/\rho g)^\frac{1}{2}$. & Silicone oil in ambient air, on smooth hydrophilic glass. \newline Effect of gravity for $d\gtrsim (\Gamma/\rho g)^\frac{1}{2}$.\\ \hline \hline 
\textbf{Biance2004 Fig3 0p27} & \textbf{Biance2004 Fig3 0p7}  \\ 
 \begin{minipage}{.5\textwidth} 
 \includegraphics[width=\linewidth]{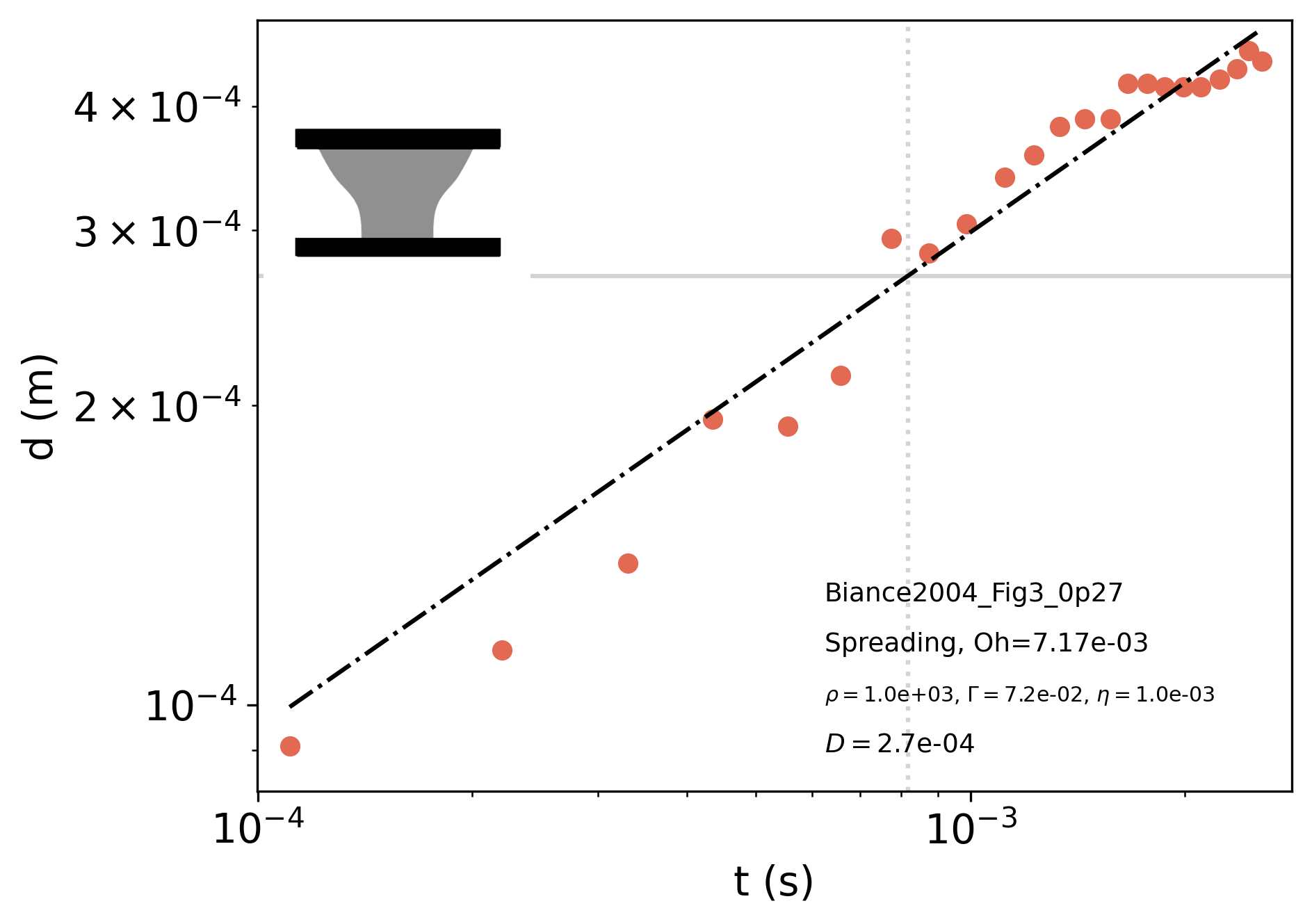} 
 \end{minipage}
 & 
 \begin{minipage}{.5\textwidth} 
 \includegraphics[width=\linewidth]{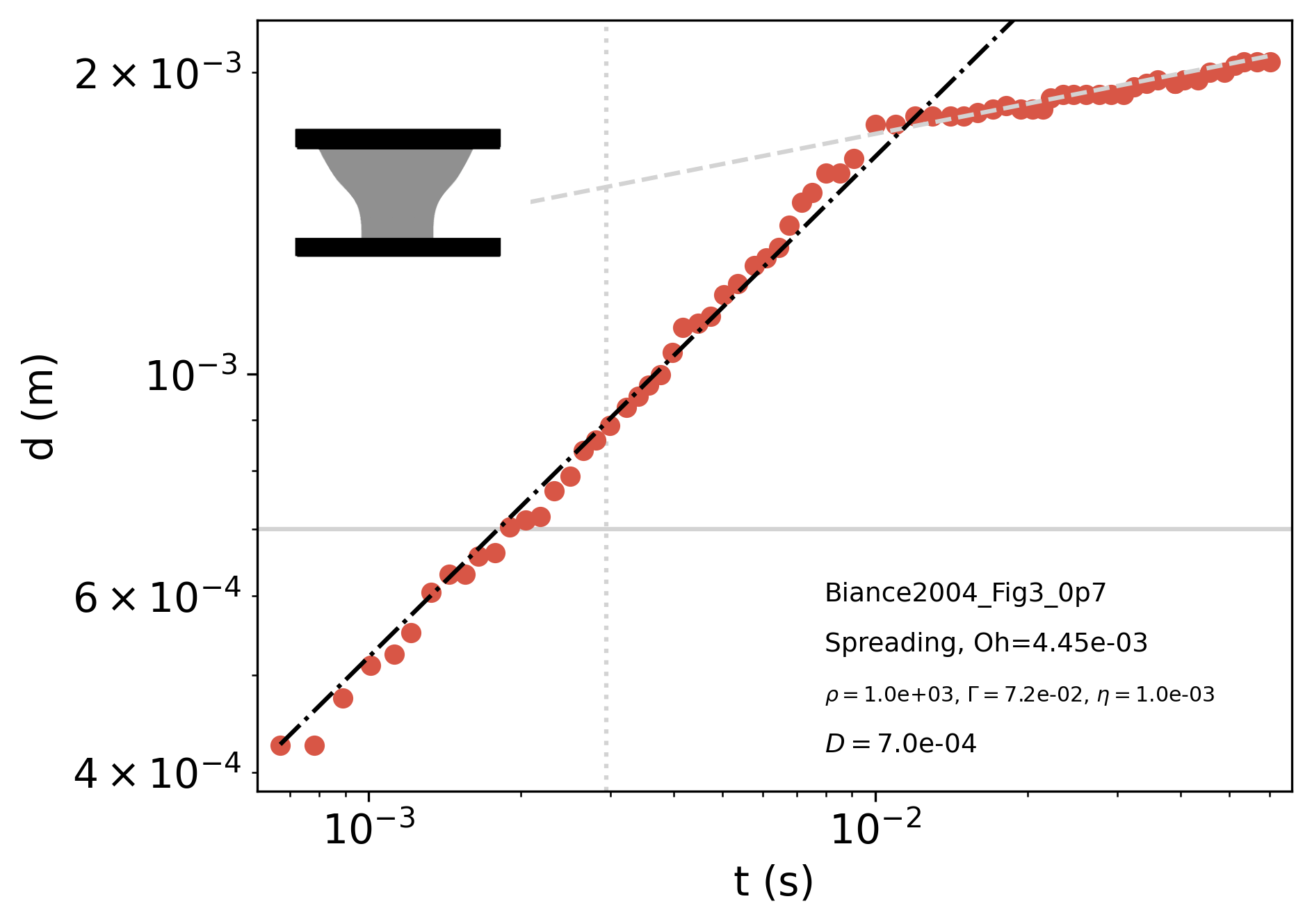} 
 \end{minipage} 
 \\ 
Water in ambient air, on inverted hydrophilic glass & Water in ambient air, on inverted hydrophilic glass\\ \hline \hline 
\textbf{Biance2004 Fig3 1p2} & \textbf{Eddi2013 Fig4 0p37}  \\ 
 \begin{minipage}{.5\textwidth} 
 \includegraphics[width=\linewidth]{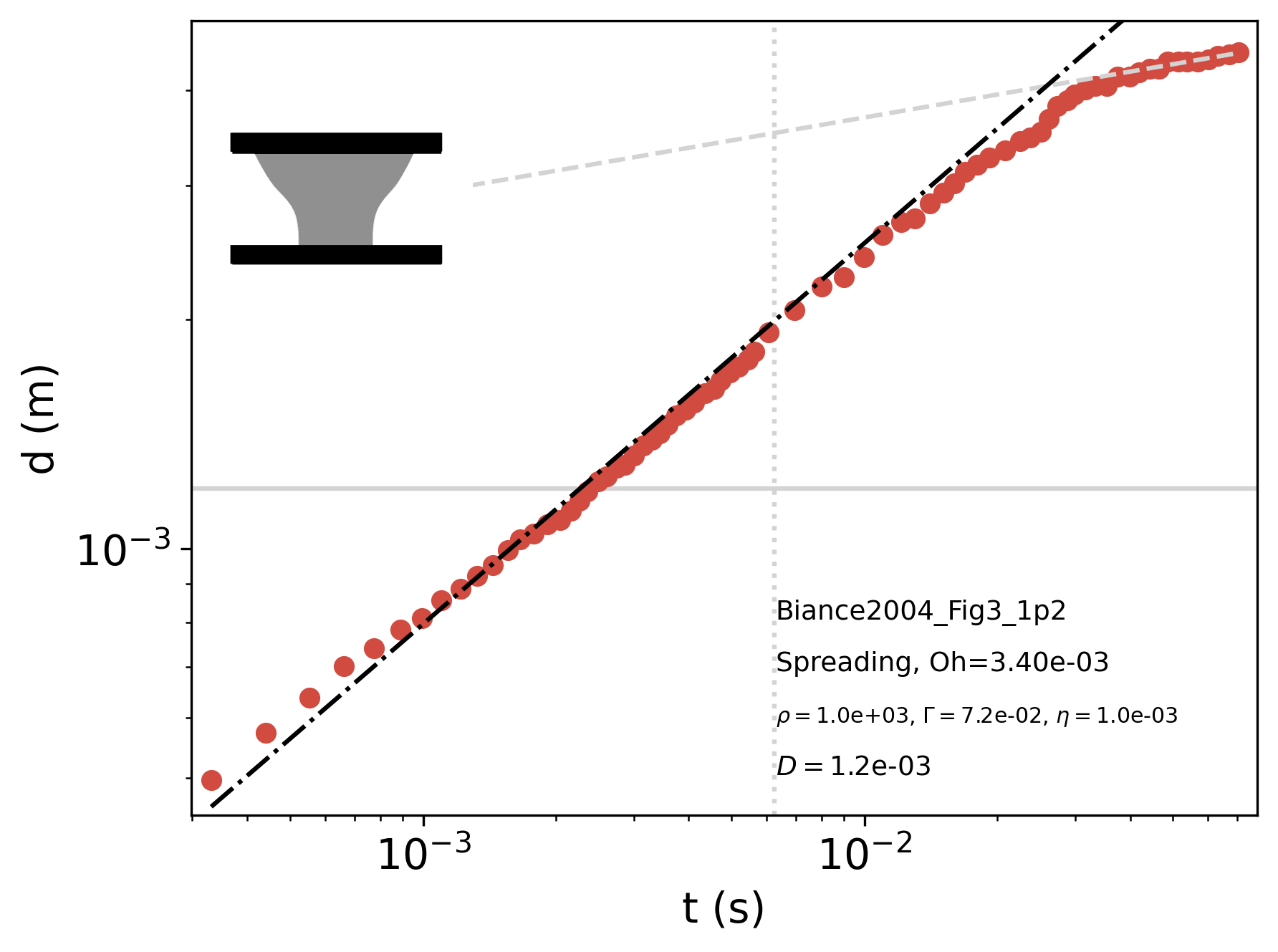} 
 \end{minipage}
 & 
 \begin{minipage}{.5\textwidth} 
 \includegraphics[width=\linewidth]{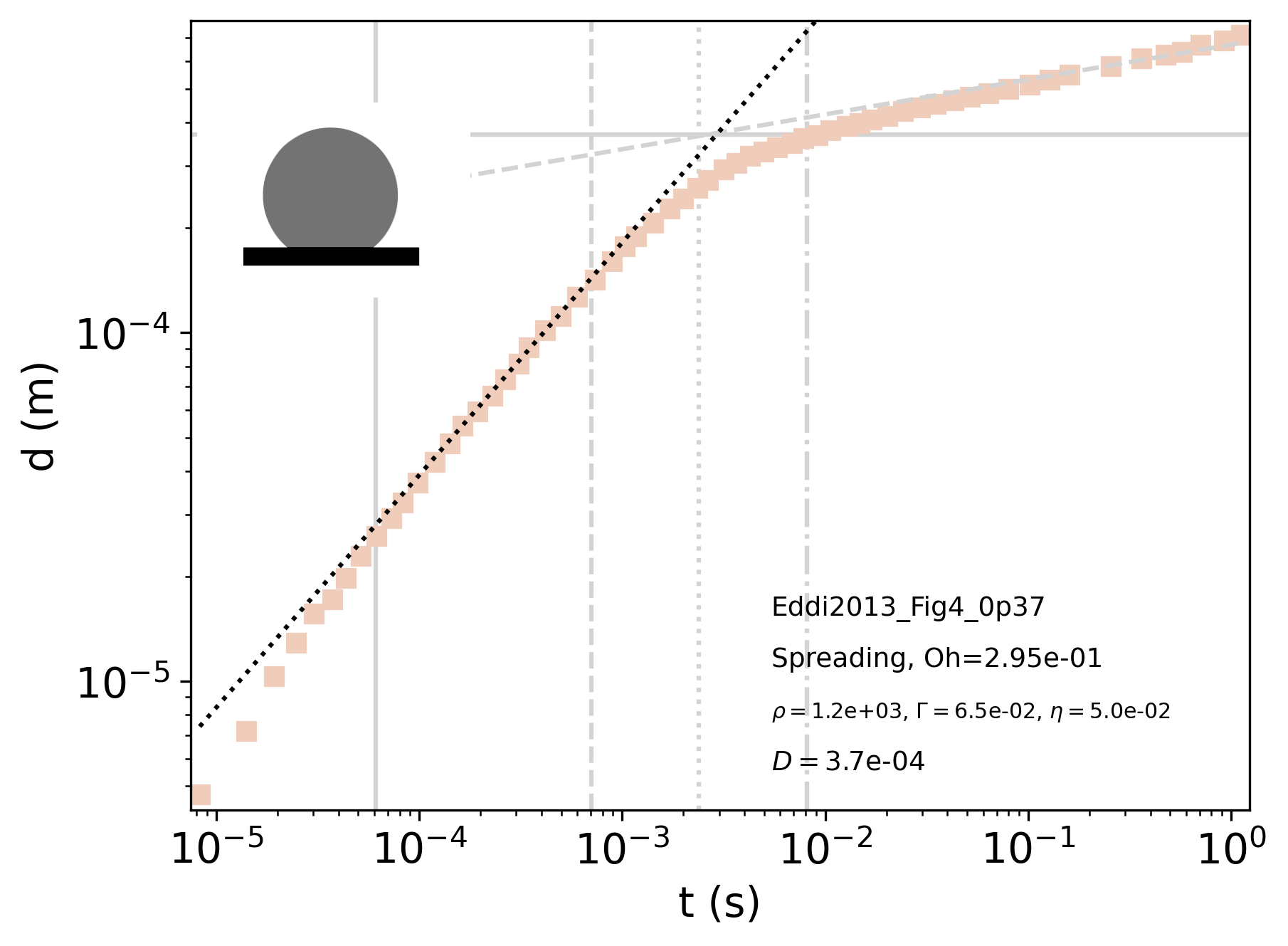} 
 \end{minipage} 
 \\ 
Water in ambient air, on inverted hydrophilic glass & Water-glycerol mixture in ambient air, on hydrophilic glass\\ \hline \hline 
\end{tabular} 
 \end{table} 
\begin{table} 
 \centering 
 \begin{tabular}{ | p{9cm} | p{9cm} | } 
 \hline 
 \textbf{Eddi2013 Fig4 0p5} & \textbf{Eddi2013 Fig4 0p63}  \\ 
 \begin{minipage}{.5\textwidth} 
 \includegraphics[width=\linewidth]{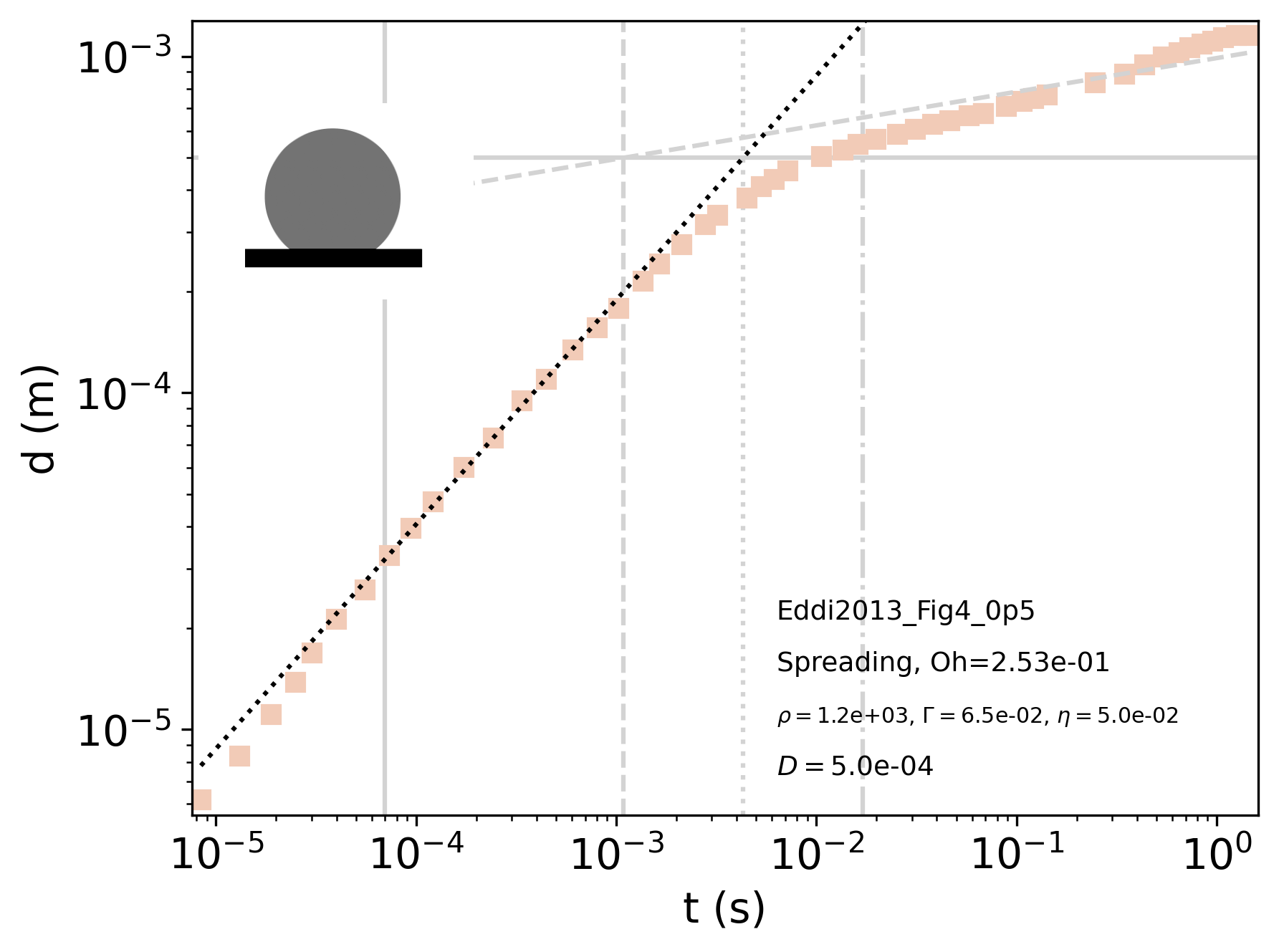} 
 \end{minipage}
 & 
 \begin{minipage}{.5\textwidth} 
 \includegraphics[width=\linewidth]{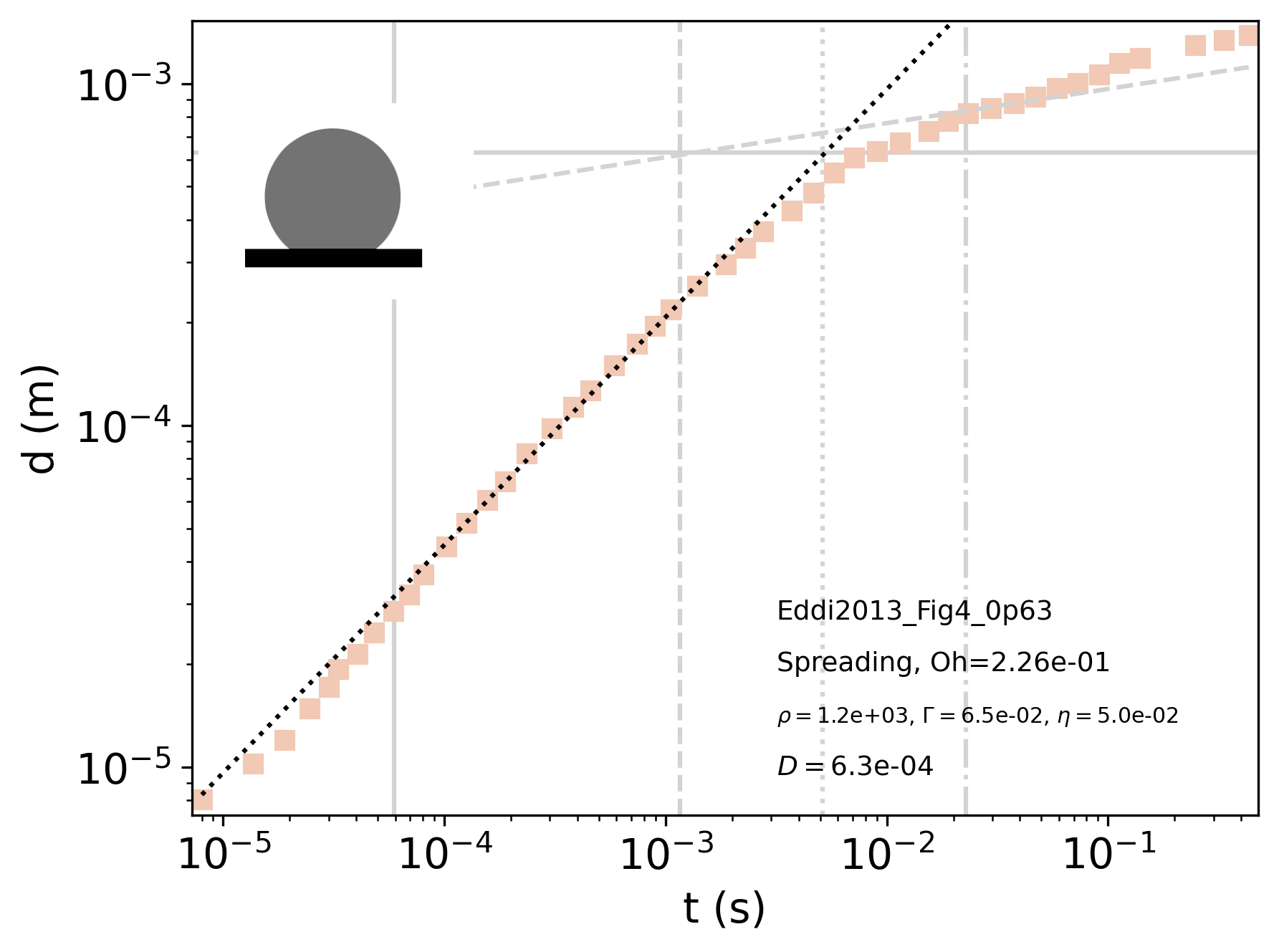} 
 \end{minipage} 
 \\ 
Water-glycerol mixture in ambient air, on hydrophilic glass & Water-glycerol mixture in ambient air, on hydrophilic glass\\ \hline \hline 
\textbf{Eddi2013 Fig5a 105deg} & \textbf{Eddi2013 Fig5b 0}  \\ 
 \begin{minipage}{.5\textwidth} 
 \includegraphics[width=\linewidth]{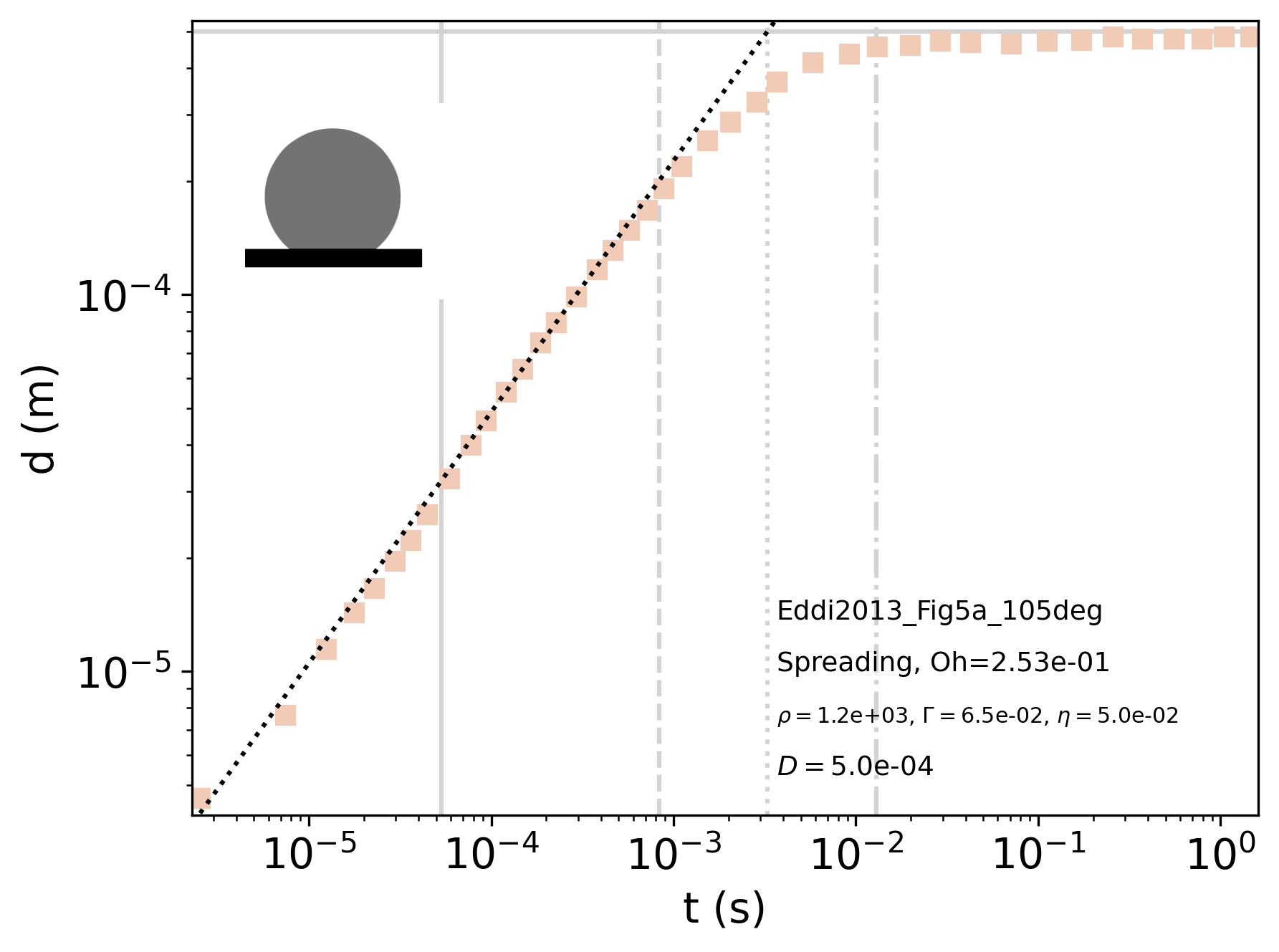} 
 \end{minipage}
 & 
 \begin{minipage}{.5\textwidth} 
 \includegraphics[width=\linewidth]{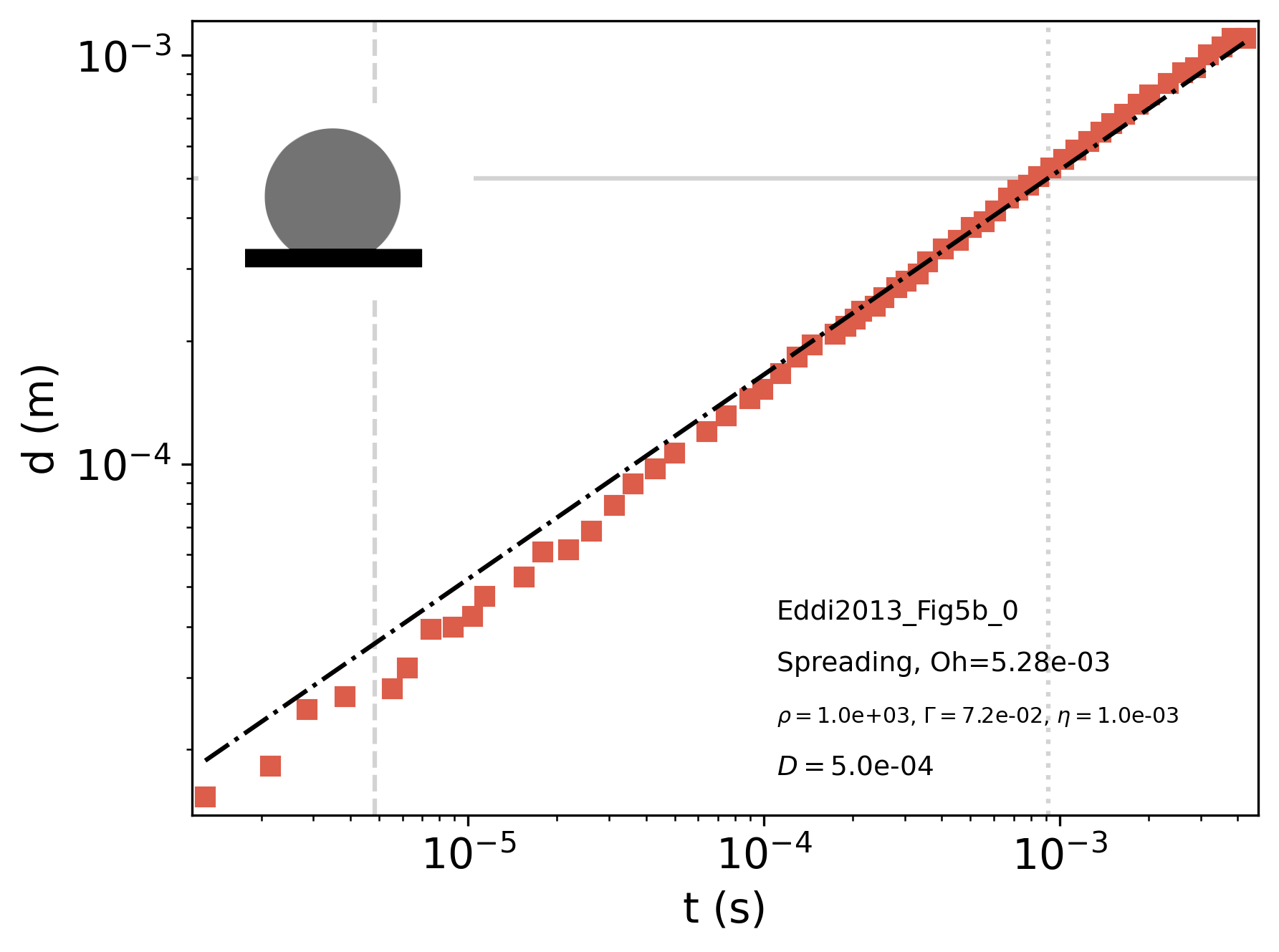} 
 \end{minipage} 
 \\ 
Water-glycerol mixture in ambient air, \newline on hydrophobic fluoropolymer-coated glass. & Water in ambient air, on hydrophilic glass.\\ \hline \hline 
\textbf{Eddi2013 Fig5b 115} & \textbf{Eddi2013 Fig6 water}  \\ 
 \begin{minipage}{.5\textwidth} 
 \includegraphics[width=\linewidth]{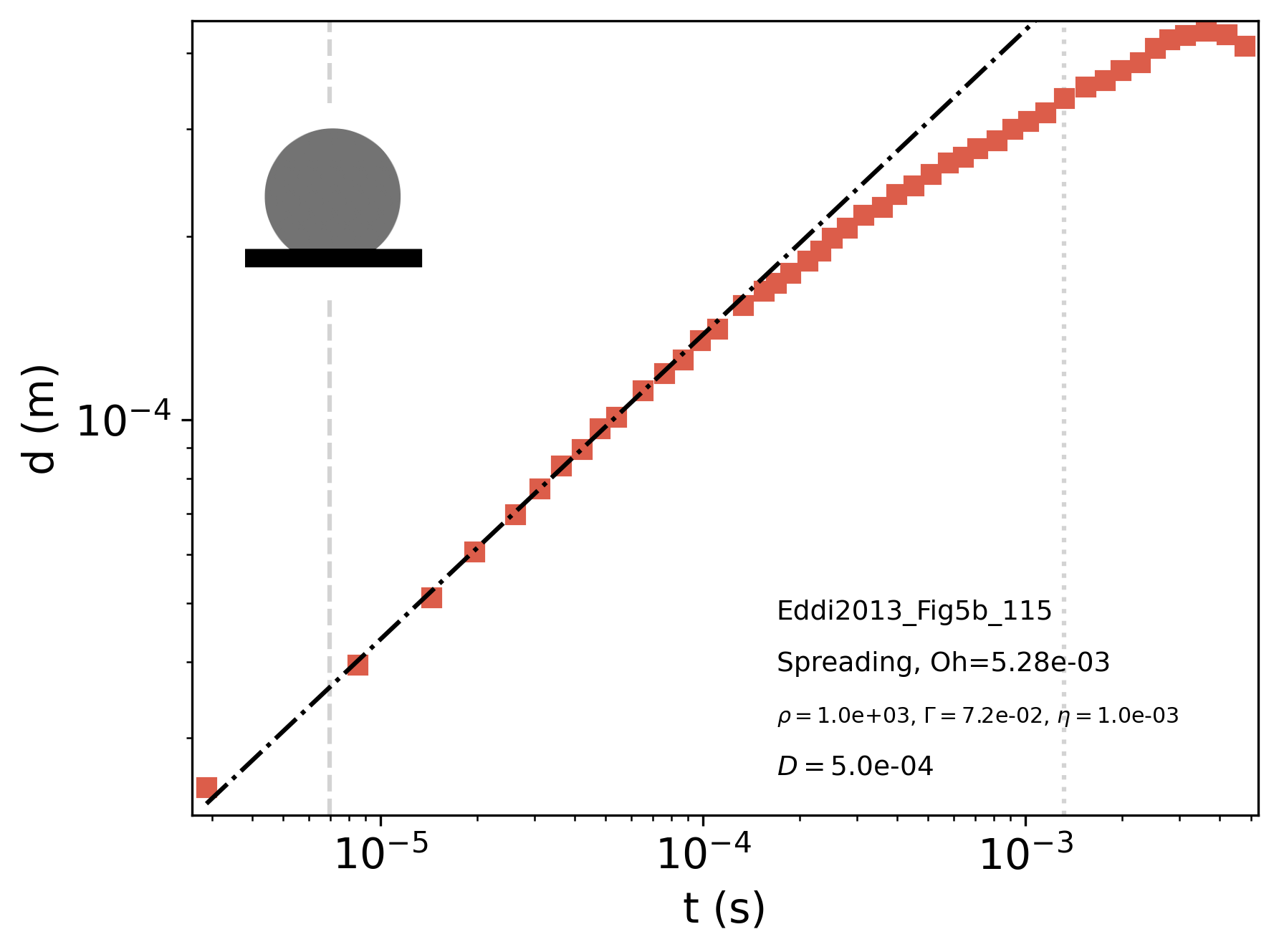} 
 \end{minipage}
 & 
 \begin{minipage}{.5\textwidth} 
 \includegraphics[width=\linewidth]{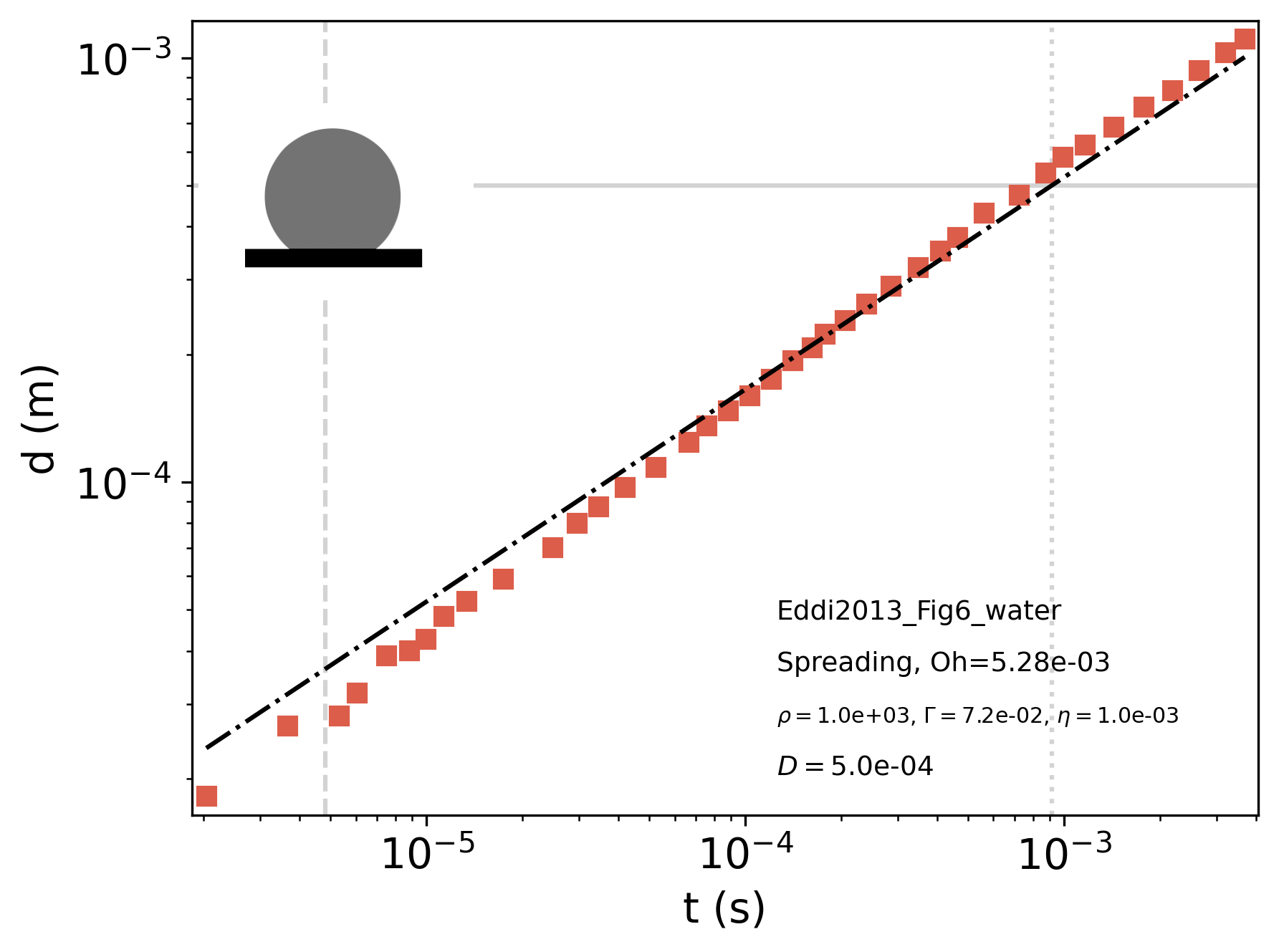} 
 \end{minipage} 
 \\ 
Water in ambient air, on hydrophobic Teflon-coated glass. & Water in ambient air, on hydrophilic glass.\\ \hline \hline 
\end{tabular} 
 \end{table} 
\begin{table} 
 \centering 
 \begin{tabular}{ | p{9cm} | p{9cm} | } 
 \hline 
 \textbf{Eddi2013 Fig6 11} & \textbf{Eddi2013 Fig6 1120}  \\ 
 \begin{minipage}{.5\textwidth} 
 \includegraphics[width=\linewidth]{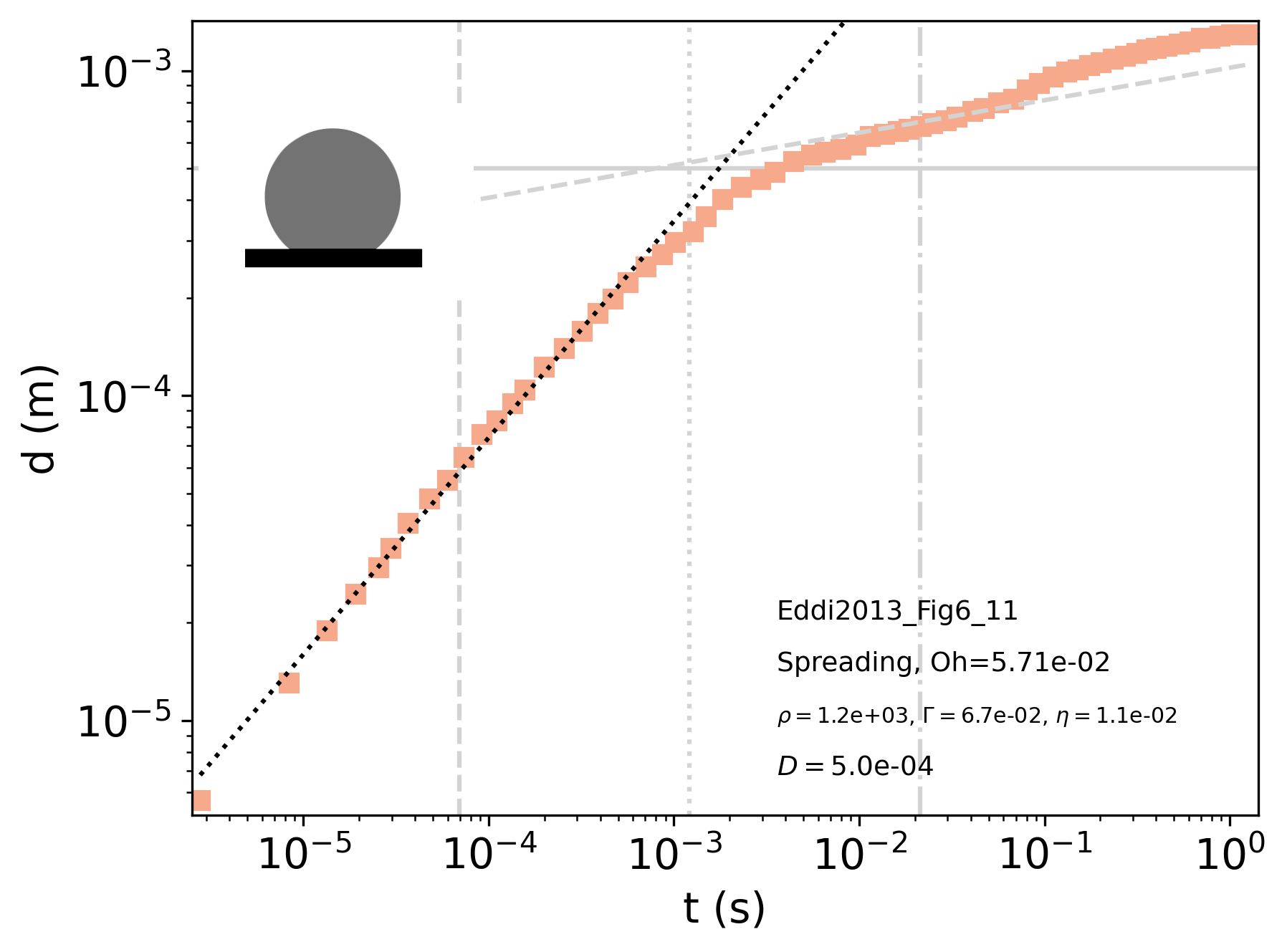} 
 \end{minipage}
 & 
 \begin{minipage}{.5\textwidth} 
 \includegraphics[width=\linewidth]{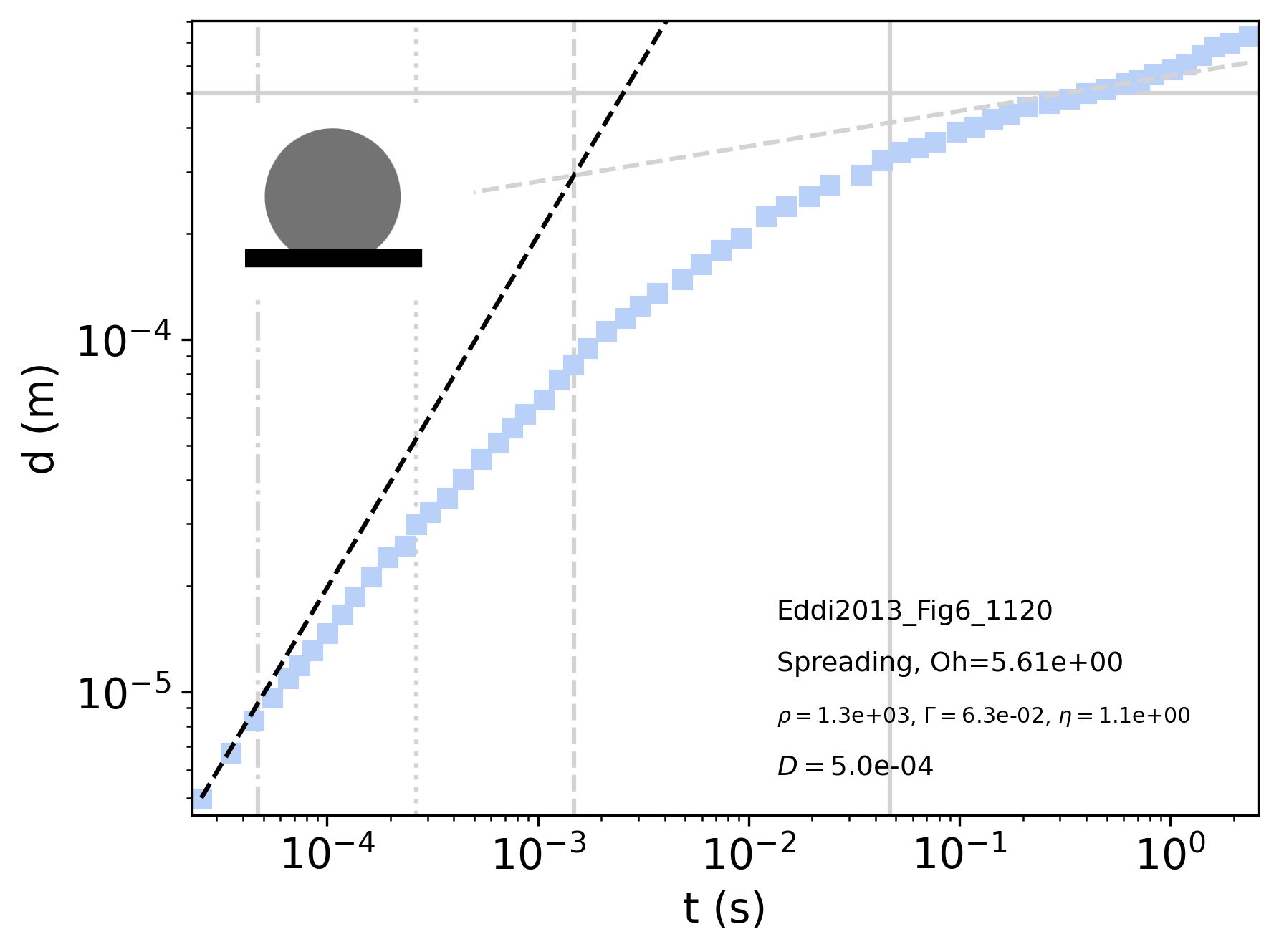} 
 \end{minipage} 
 \\ 
Water-glycerol mixture in ambient air, on hydrophilic glass & Water-glycerol mixture in ambient air, on hydrophilic glass\\ \hline \hline 
\textbf{Eddi2013 Fig6 220} & \textbf{Chen2014 Fig3b 60cP}  \\ 
 \begin{minipage}{.5\textwidth} 
 \includegraphics[width=\linewidth]{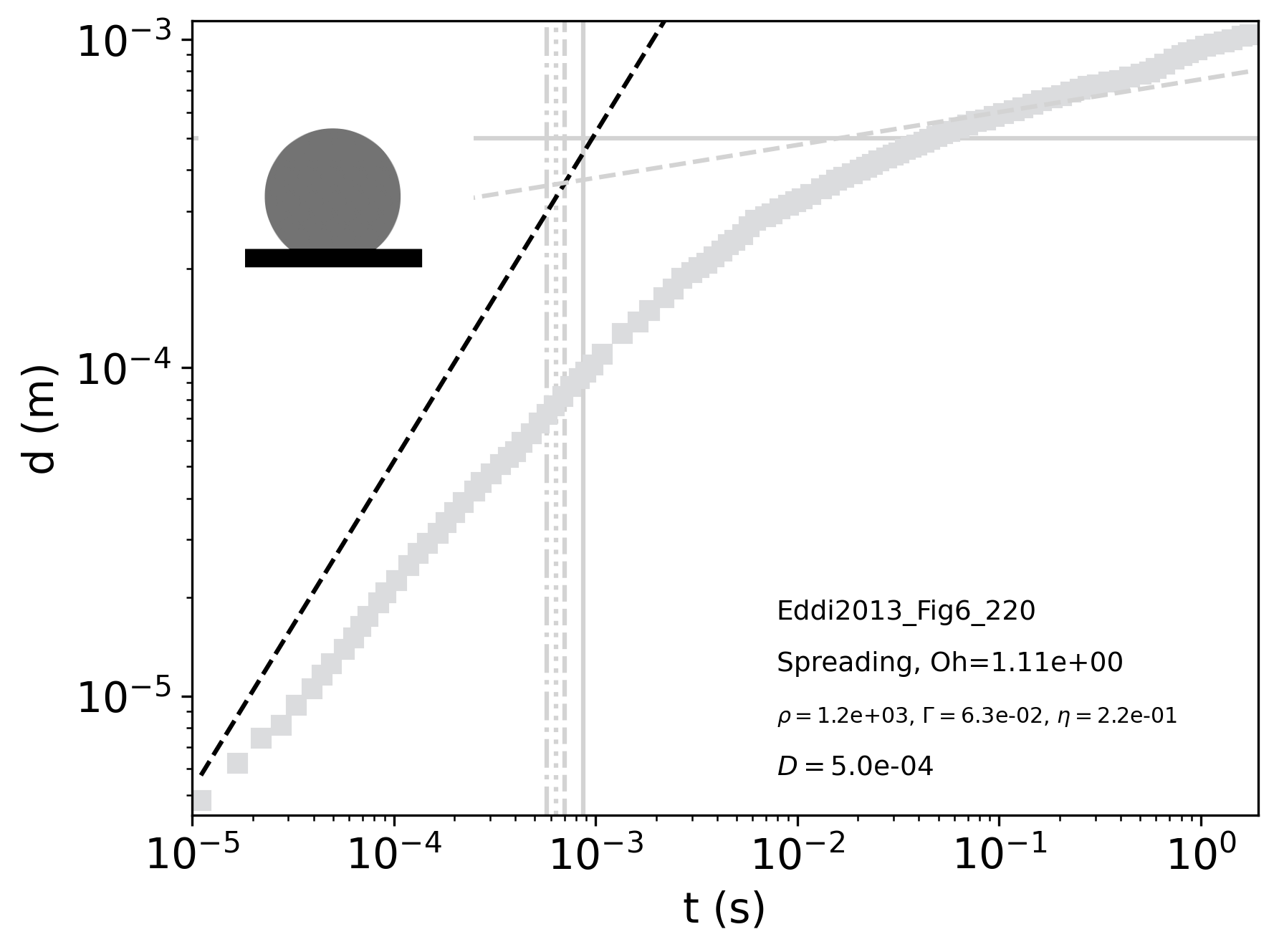} 
 \end{minipage}
 & 
 \begin{minipage}{.5\textwidth} 
 \includegraphics[width=\linewidth]{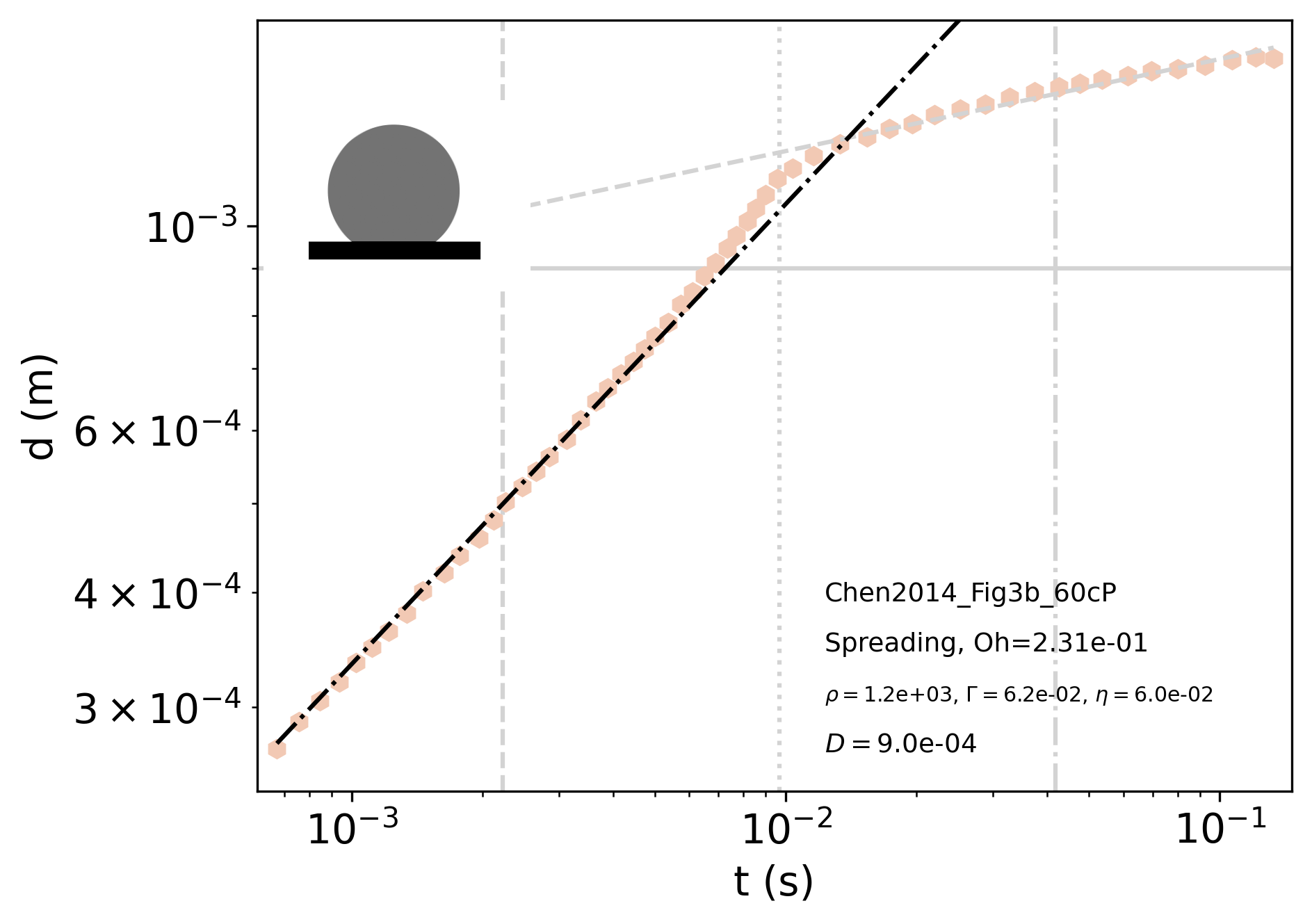} 
 \end{minipage} 
 \\ 
Water-glycerol mixture in ambient air, on hydrophilic glass & Water-glycerol mixture in ambient air, on partial wetting substrate ($\theta=63^\circ$).\\ \hline \hline 
\textbf{Chen2014 Fig3b 35p5cP} & \textbf{Menchaca2001 Fig9}  \\ 
 \begin{minipage}{.5\textwidth} 
 \includegraphics[width=\linewidth]{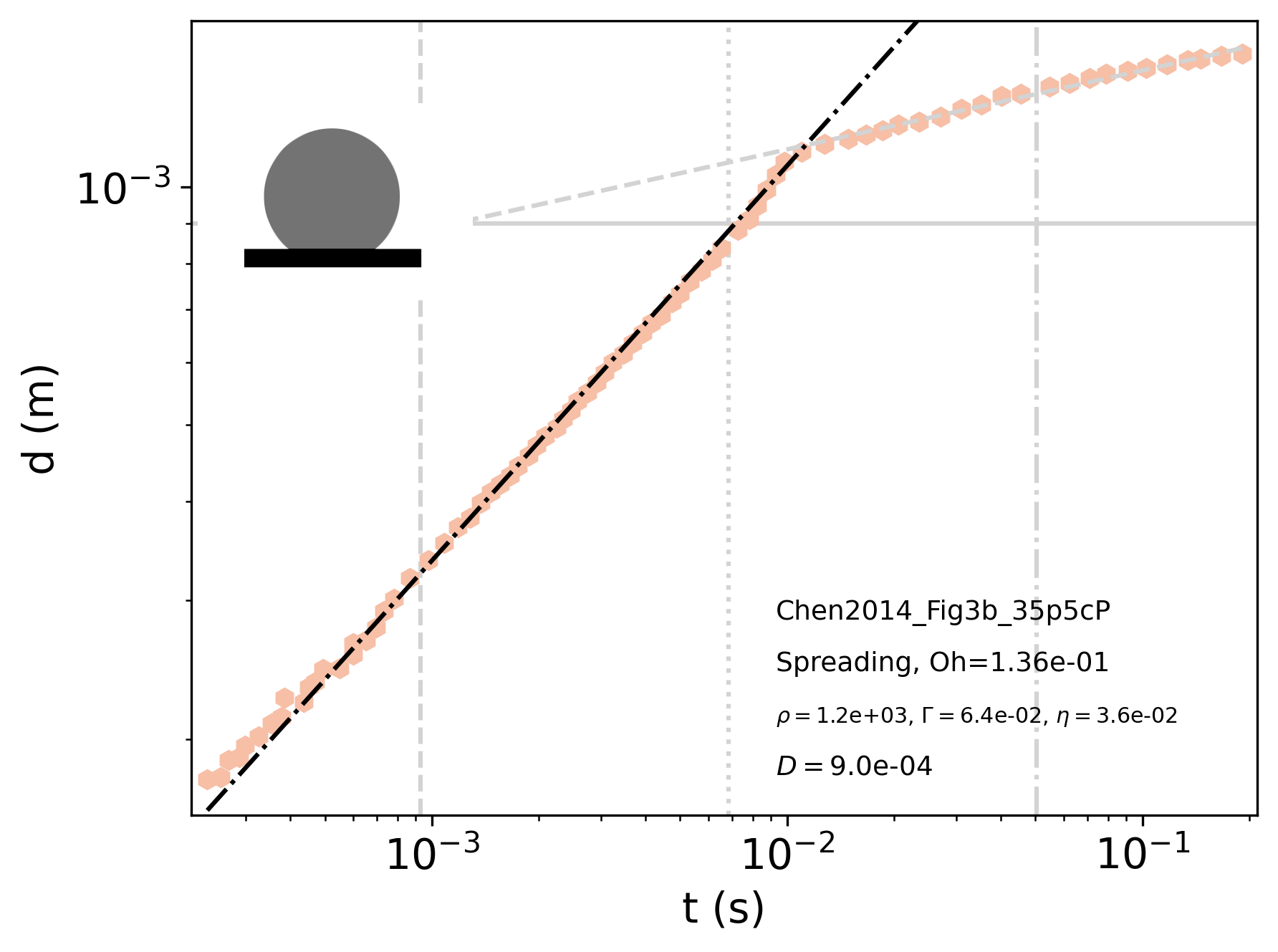} 
 \end{minipage}
 & 
 \begin{minipage}{.5\textwidth} 
 \includegraphics[width=\linewidth]{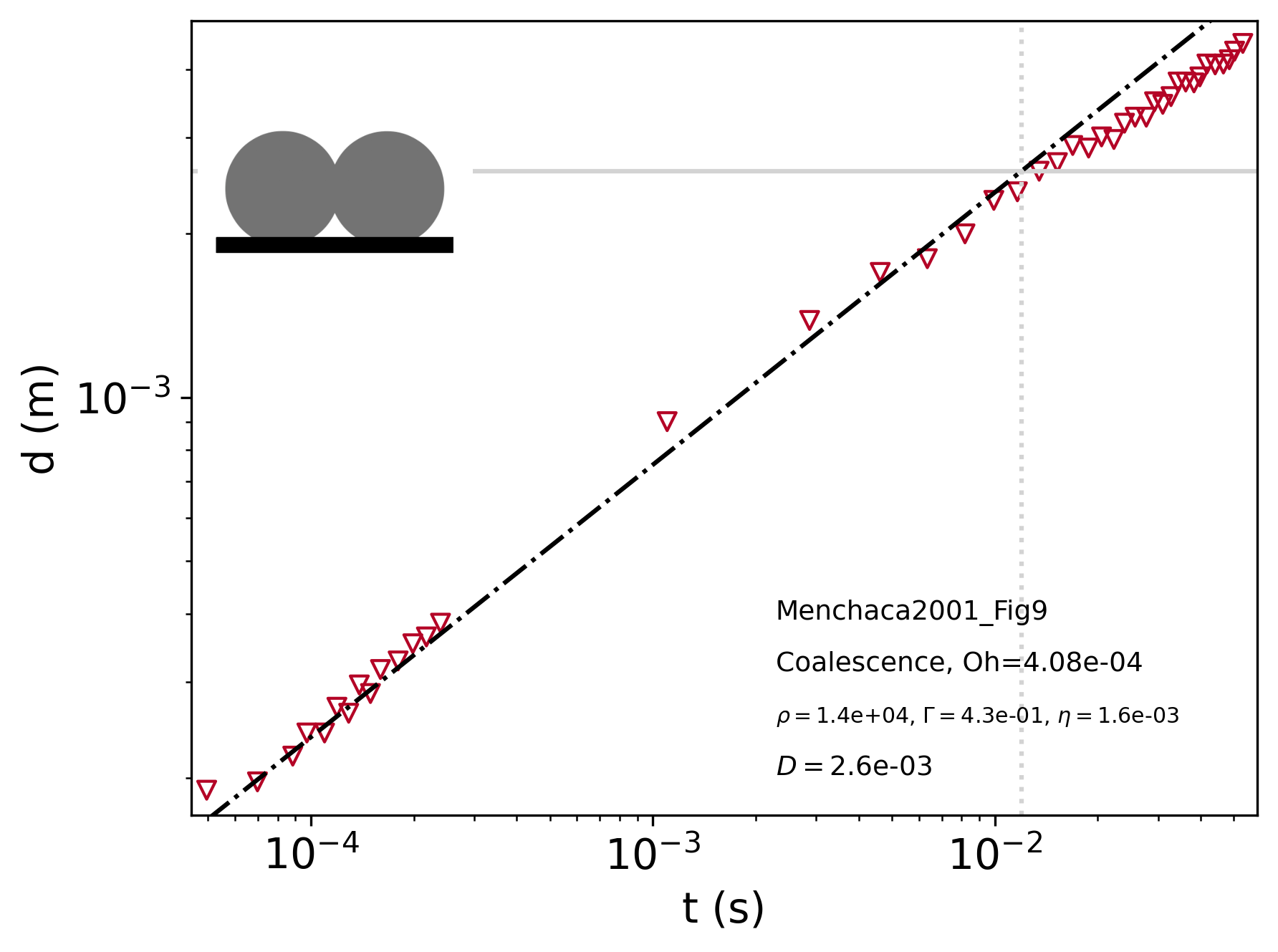} 
 \end{minipage} 
 \\ 
Water-glycerol mixture in ambient air, on partial wetting substrate ($\theta=63^\circ$). & Mercury in ambient air.\\ \hline \hline 
\end{tabular} 
 \end{table} 
\begin{table} 
 \centering 
 \begin{tabular}{ | p{9cm} | p{9cm} | } 
 \hline 
 \textbf{Wu2004 Fig4 1p94} & \textbf{Yao2005 100000cS 5cm}  \\ 
 \begin{minipage}{.5\textwidth} 
 \includegraphics[width=\linewidth]{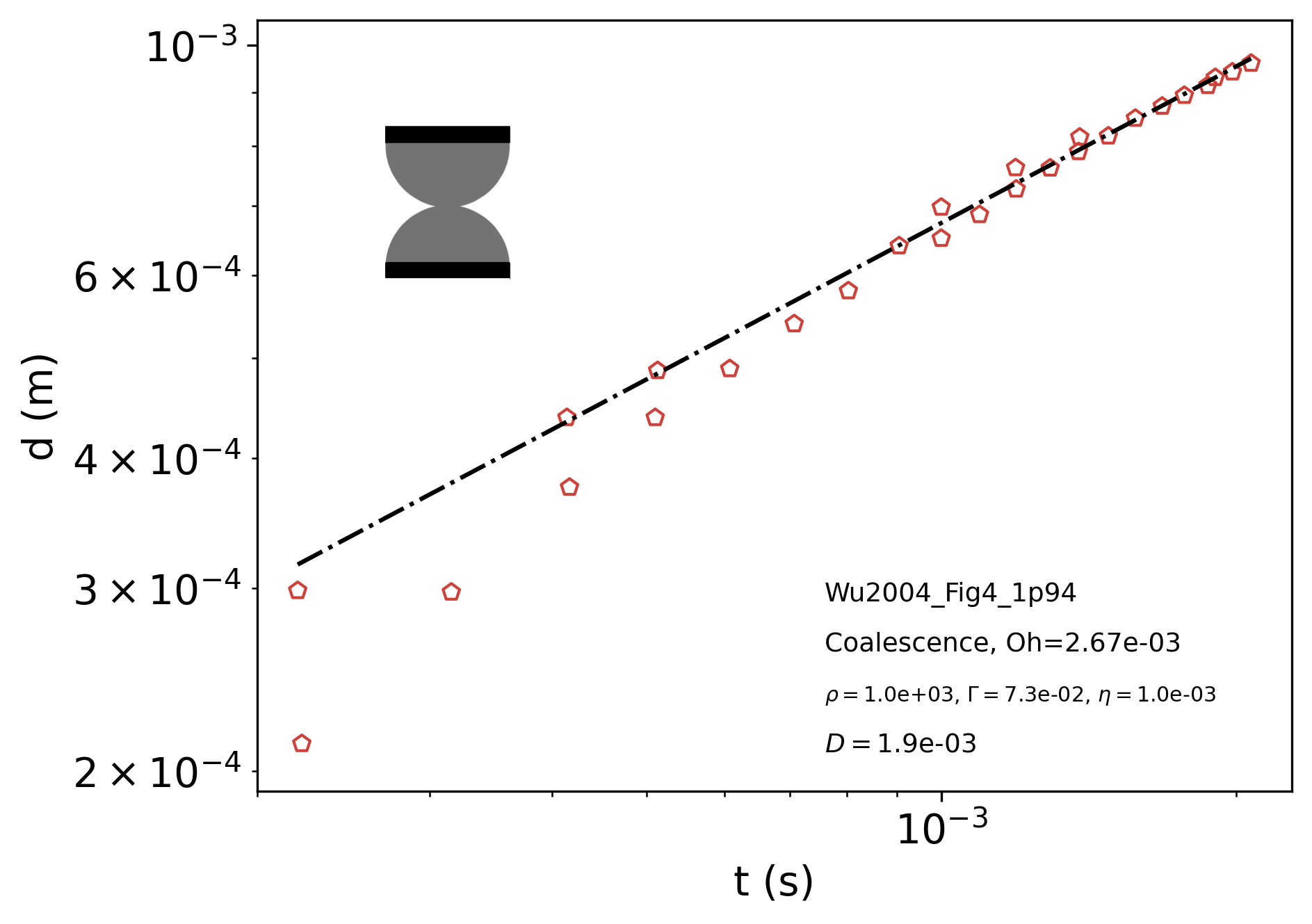} 
 \end{minipage}
 & 
 \begin{minipage}{.5\textwidth} 
 \includegraphics[width=\linewidth]{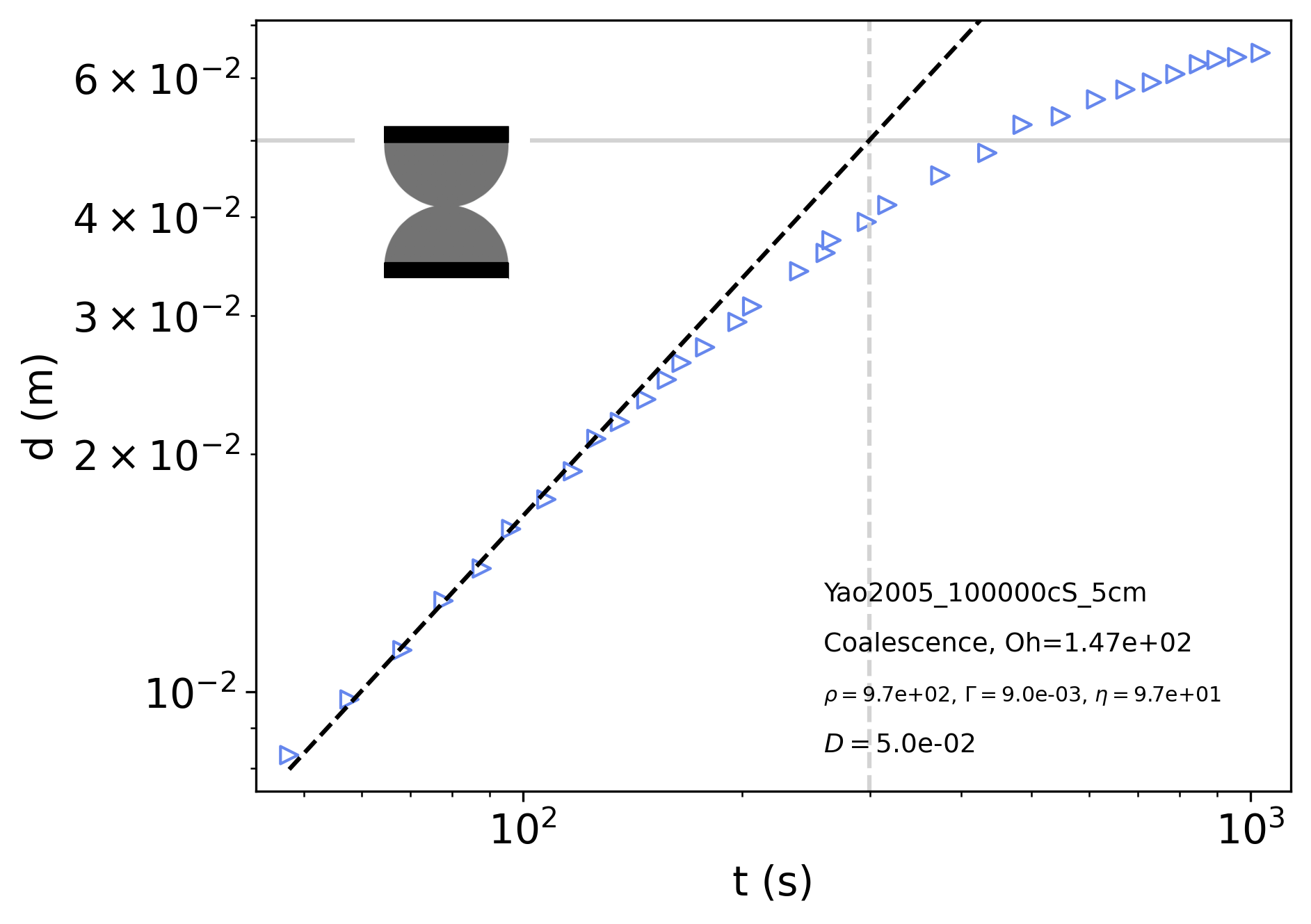} 
 \end{minipage} 
 \\ 
Water in ambient air. & Siliconee oil in density-matched water-alcohol mixture.\\ \hline \hline 
\textbf{Yao2005 100000cS 0p5cm} & \textbf{Yao2005 10000cS 0p5cm}  \\ 
 \begin{minipage}{.5\textwidth} 
 \includegraphics[width=\linewidth]{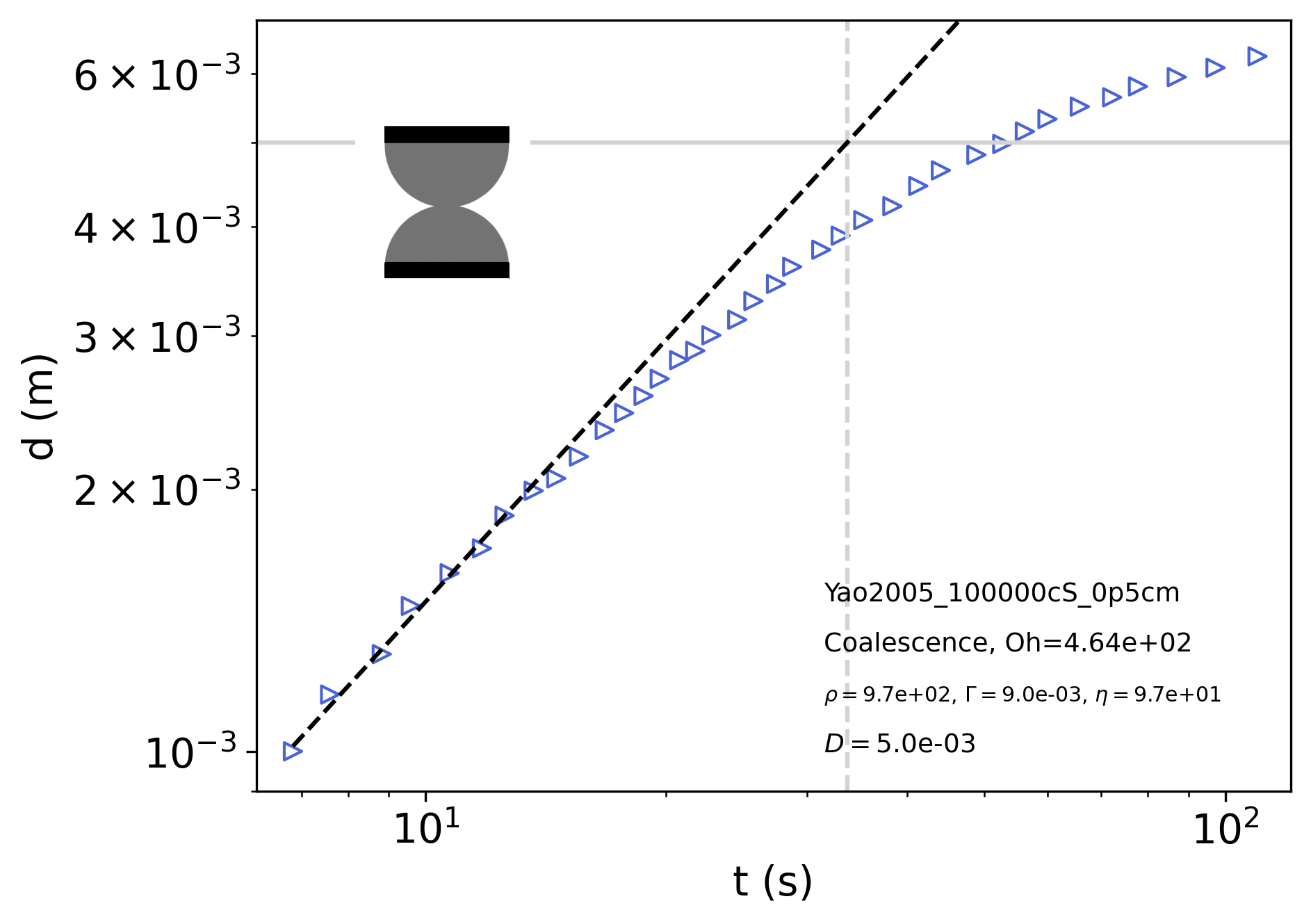} 
 \end{minipage}
 & 
 \begin{minipage}{.5\textwidth} 
 \includegraphics[width=\linewidth]{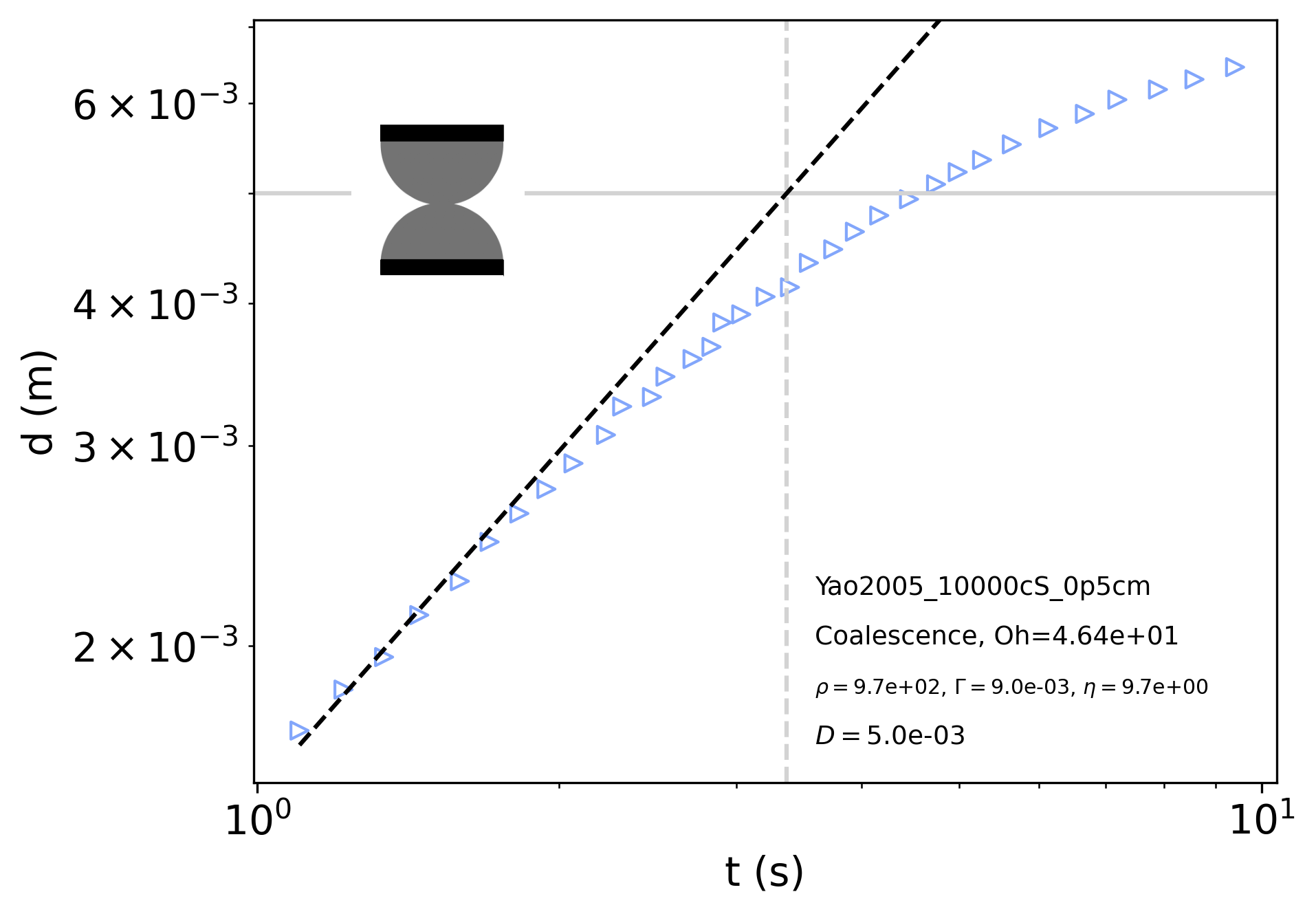} 
 \end{minipage} 
 \\ 
Siliconee oil in density-matched water-alcohol mixture. & Siliconee oil in density-matched water-alcohol mixture.\\ \hline \hline 
\textbf{Yao2005 1000cS 0p5cm} & \textbf{Thoroddsen2005 Fig6}  \\ 
 \begin{minipage}{.5\textwidth} 
 \includegraphics[width=\linewidth]{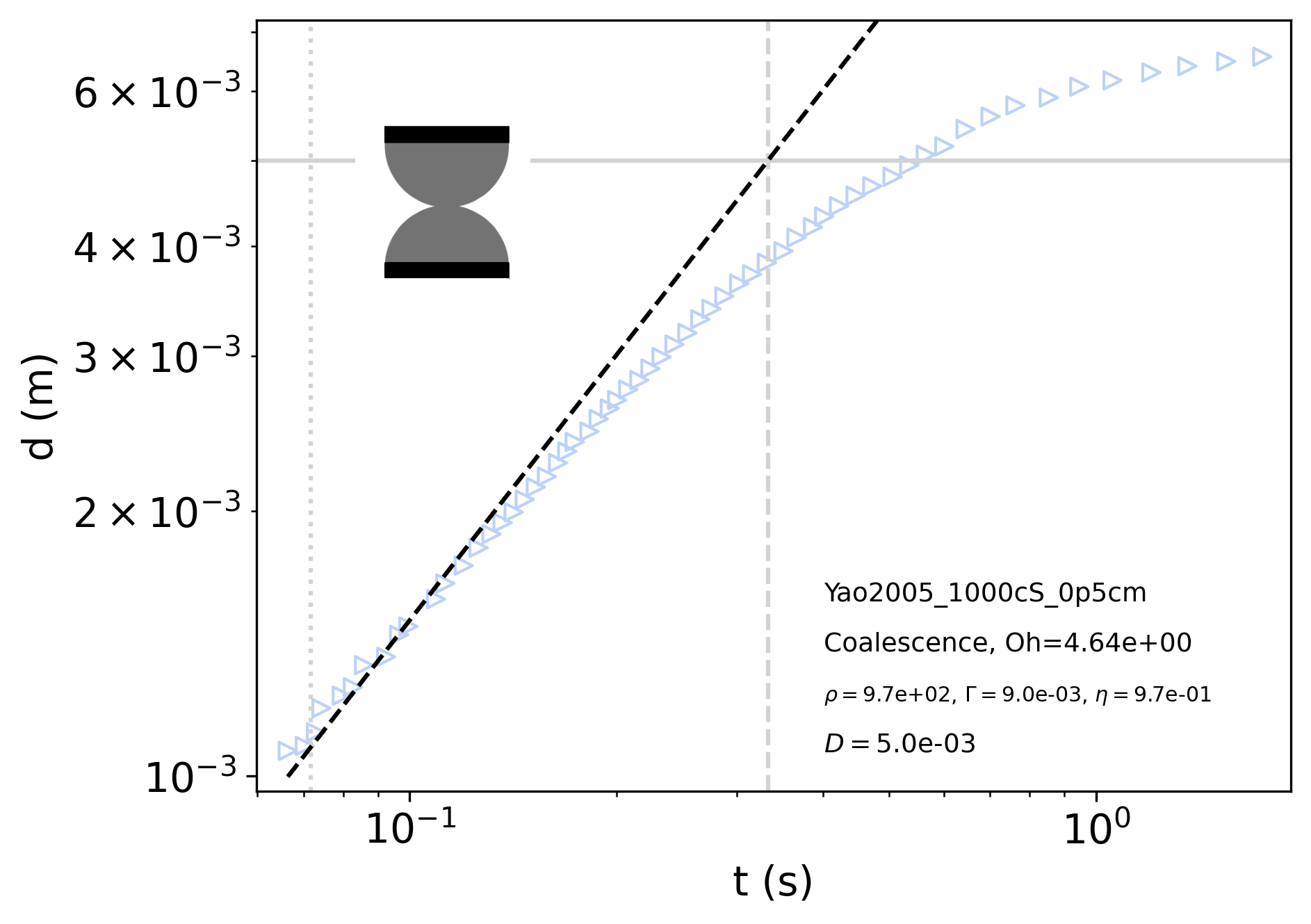} 
 \end{minipage}
 & 
 \begin{minipage}{.5\textwidth} 
 \includegraphics[width=\linewidth]{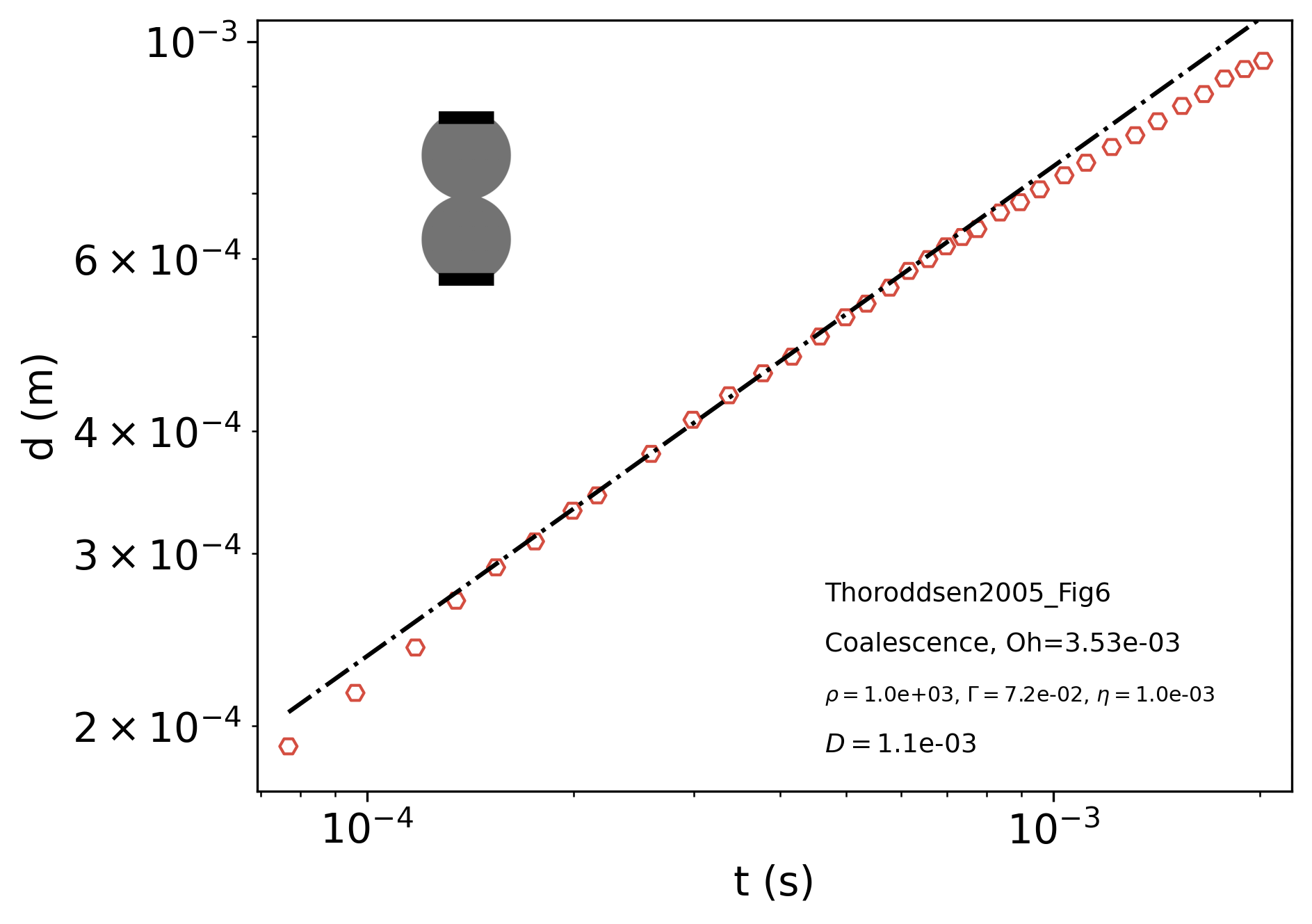} 
 \end{minipage} 
 \\ 
Siliconee oil in density-matched water-alcohol mixture. & Water in ambient air.\\ \hline \hline 
\end{tabular} 
 \end{table} 
\begin{table} 
 \centering 
 \begin{tabular}{ | p{9cm} | p{9cm} | } 
 \hline 
 \textbf{Thoroddsen2005b Fig4} & \textbf{Aarts2005 Fig2 100mPas}  \\ 
 \begin{minipage}{.5\textwidth} 
 \includegraphics[width=\linewidth]{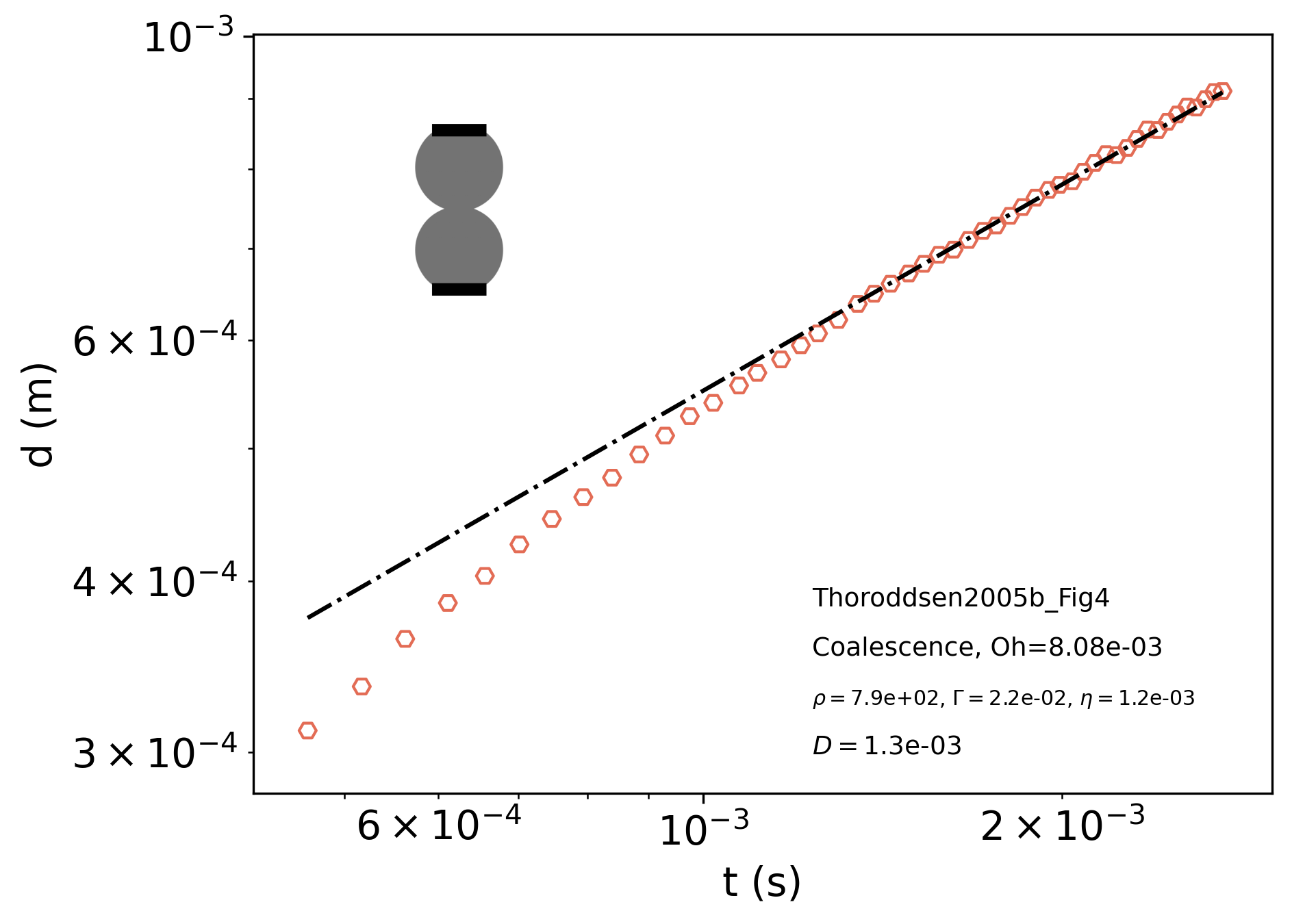} 
 \end{minipage}
 & 
 \begin{minipage}{.5\textwidth} 
 \includegraphics[width=\linewidth]{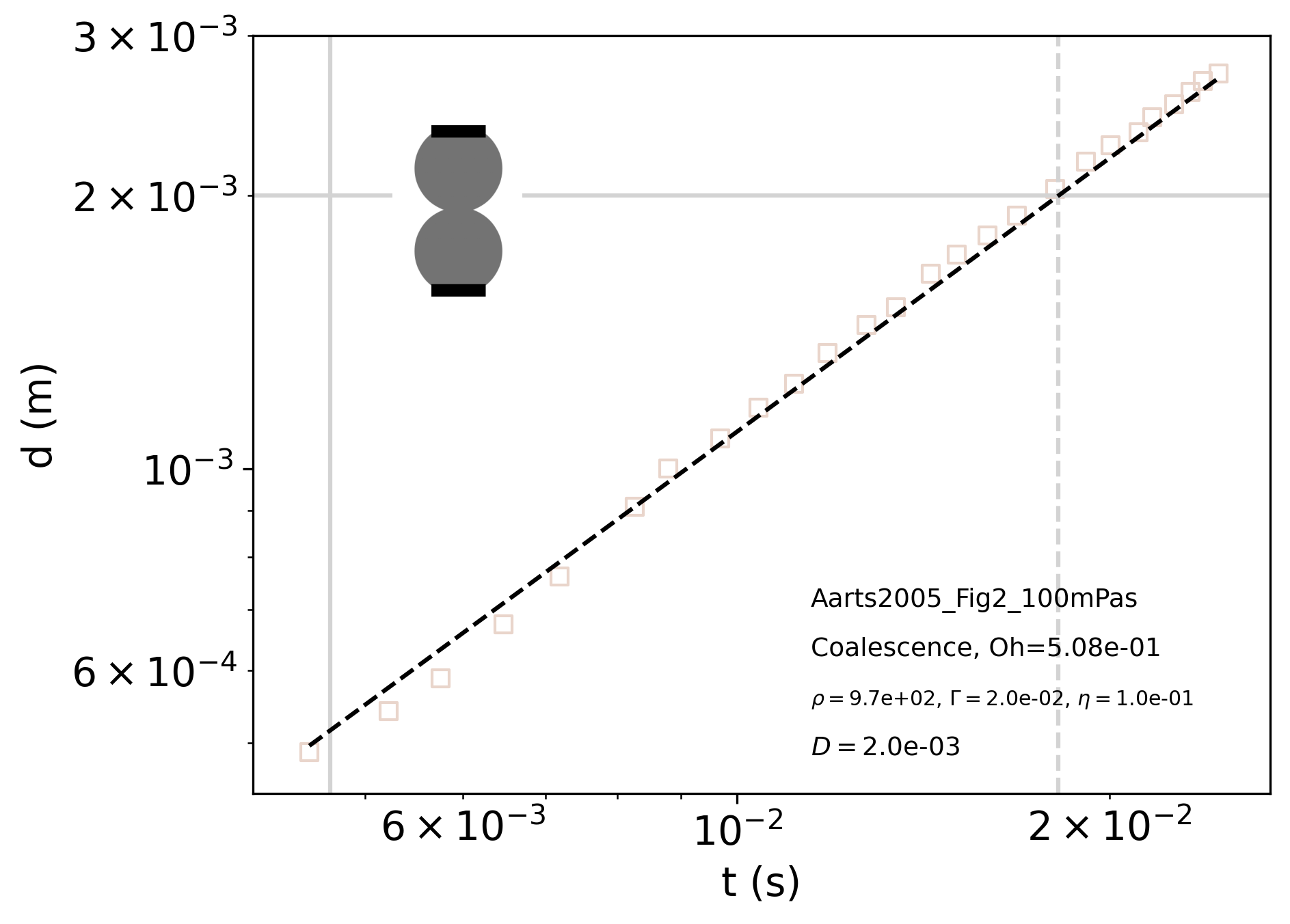} 
 \end{minipage} 
 \\ 
Air bubbles in ethyl alcohol.  \newline Density and viscosity are that of the outer fluid.  & Silicone oil in ambient air.\\ \hline \hline 
\textbf{Aarts2005 Fig2 1Pas} & \textbf{Aarts2005 Fig2 500mPas}  \\ 
 \begin{minipage}{.5\textwidth} 
 \includegraphics[width=\linewidth]{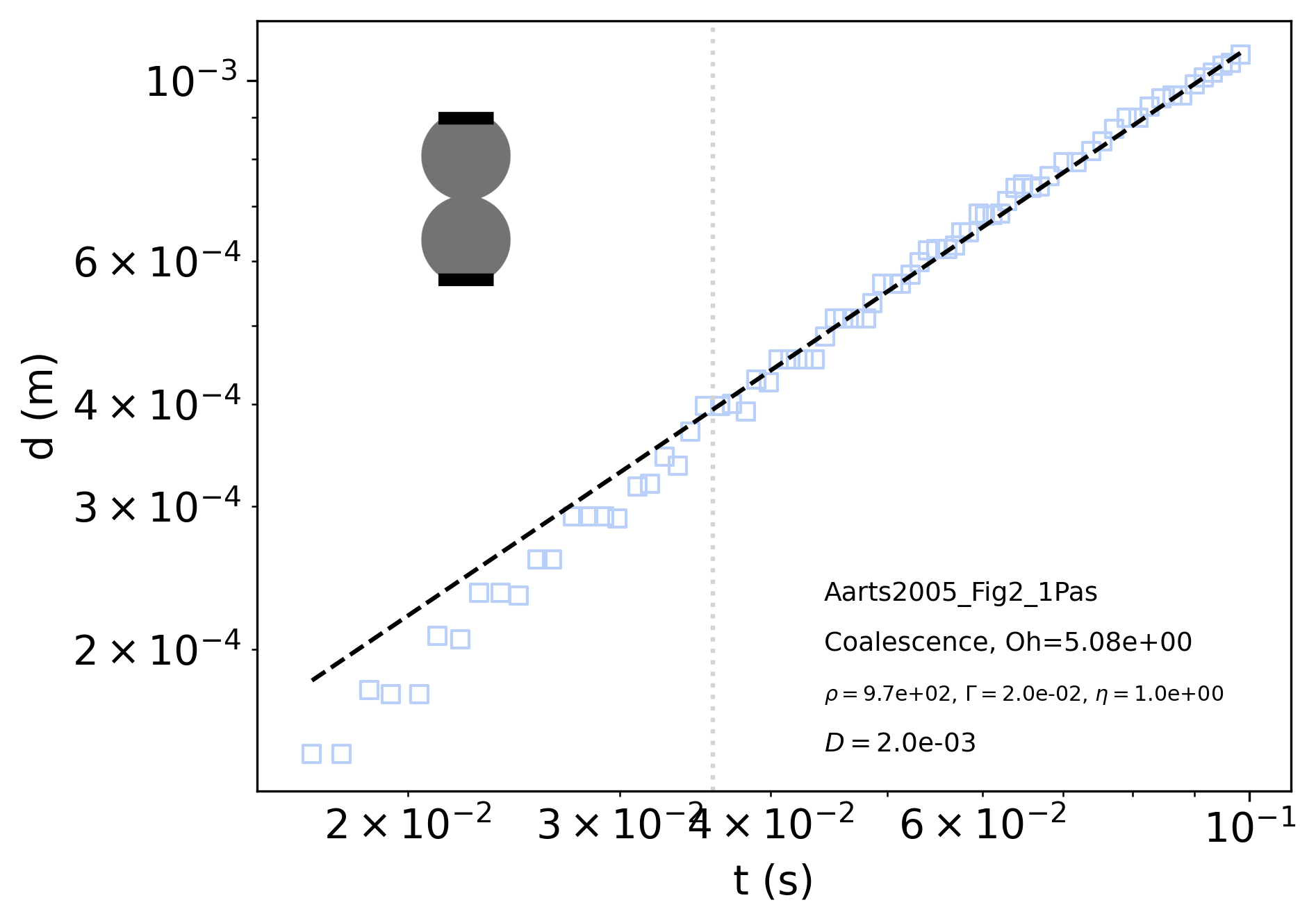} 
 \end{minipage}
 & 
 \begin{minipage}{.5\textwidth} 
 \includegraphics[width=\linewidth]{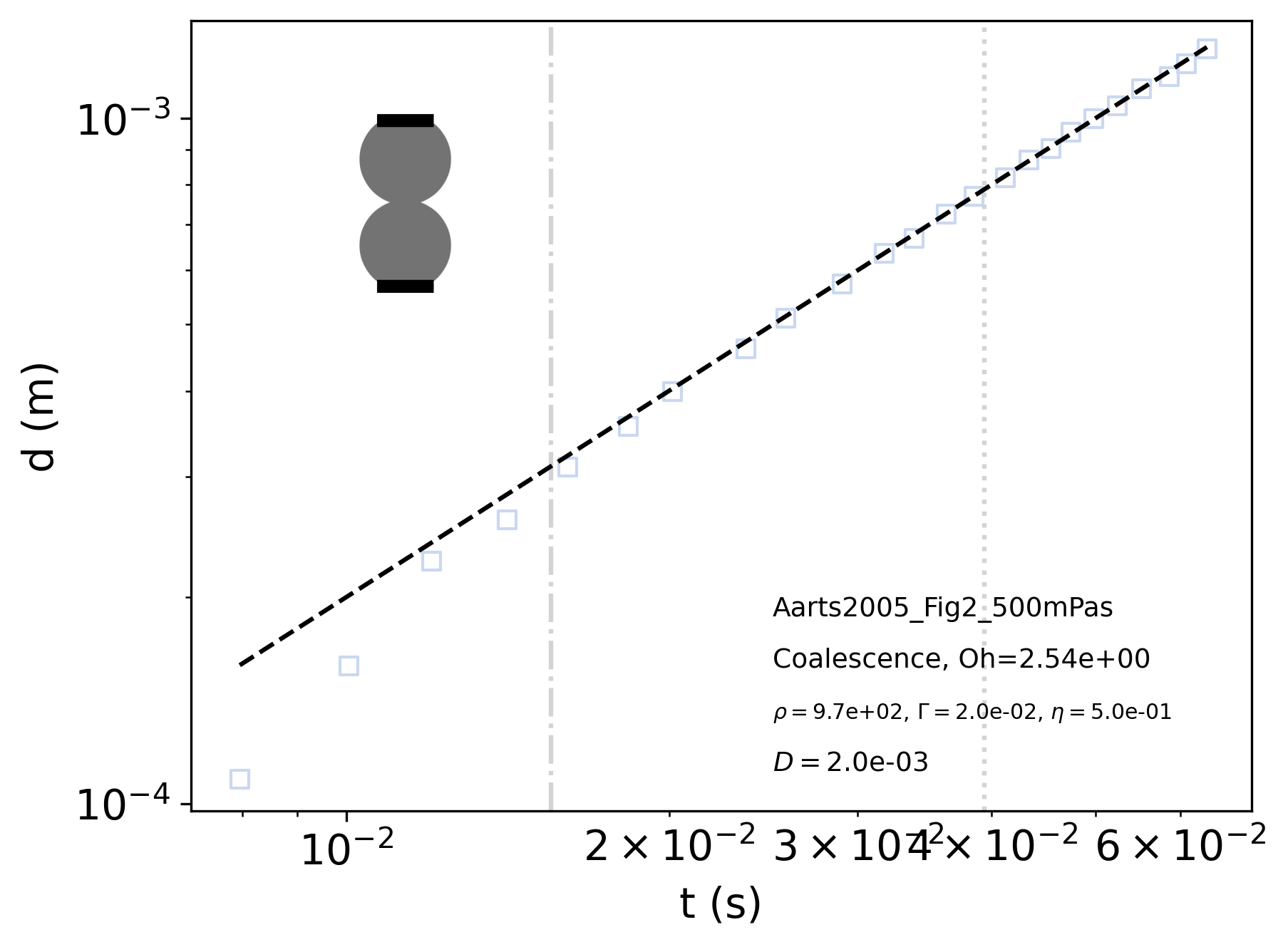} 
 \end{minipage} 
 \\ 
Silicone oil in ambient air. & Silicone oil in ambient air.\\ \hline \hline 
\textbf{Aarts2005 Fig2 300mPas} & \textbf{Aarts2005 Fig3 5mPas}  \\ 
 \begin{minipage}{.5\textwidth} 
 \includegraphics[width=\linewidth]{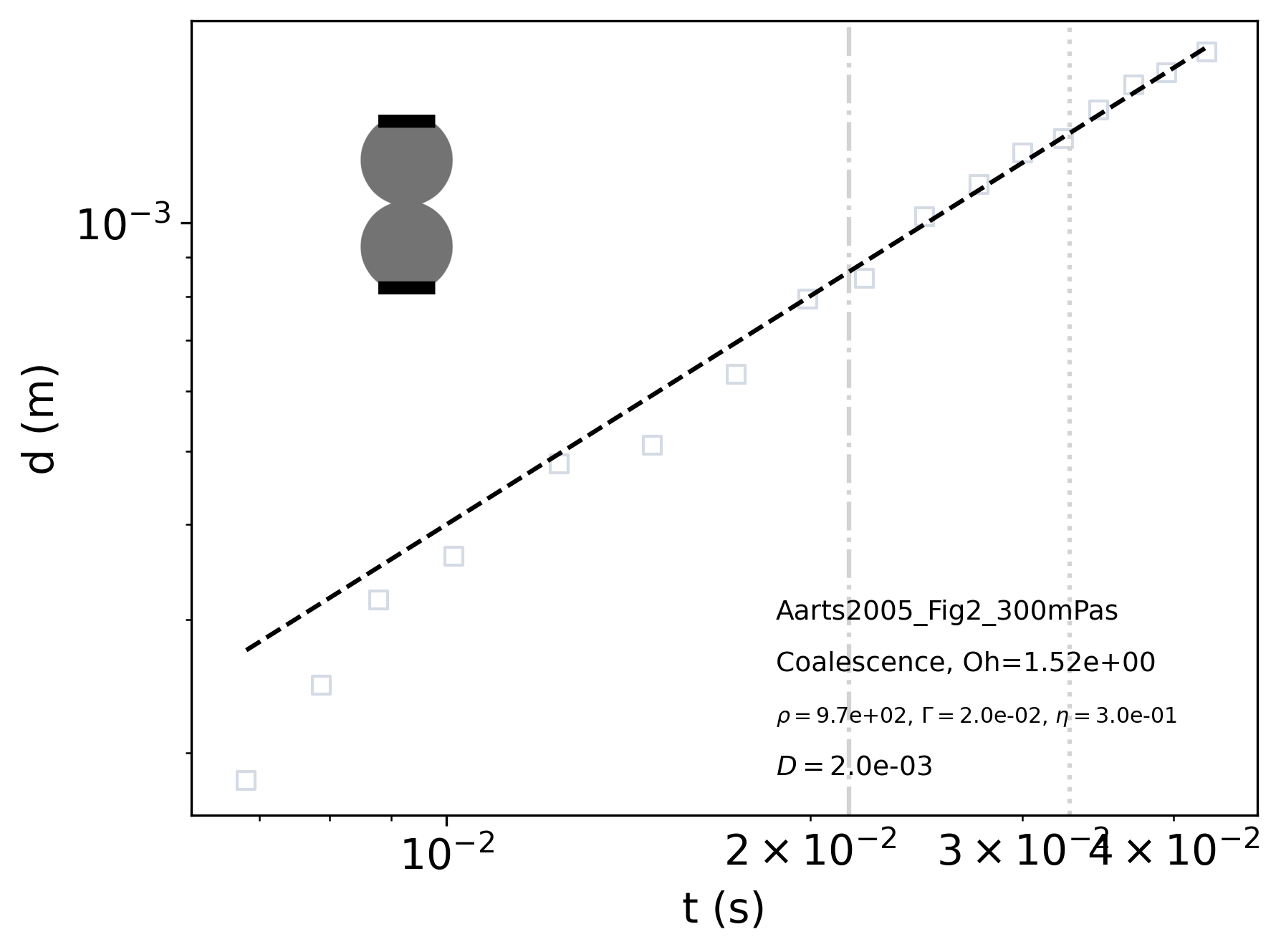} 
 \end{minipage}
 & 
 \begin{minipage}{.5\textwidth} 
 \includegraphics[width=\linewidth]{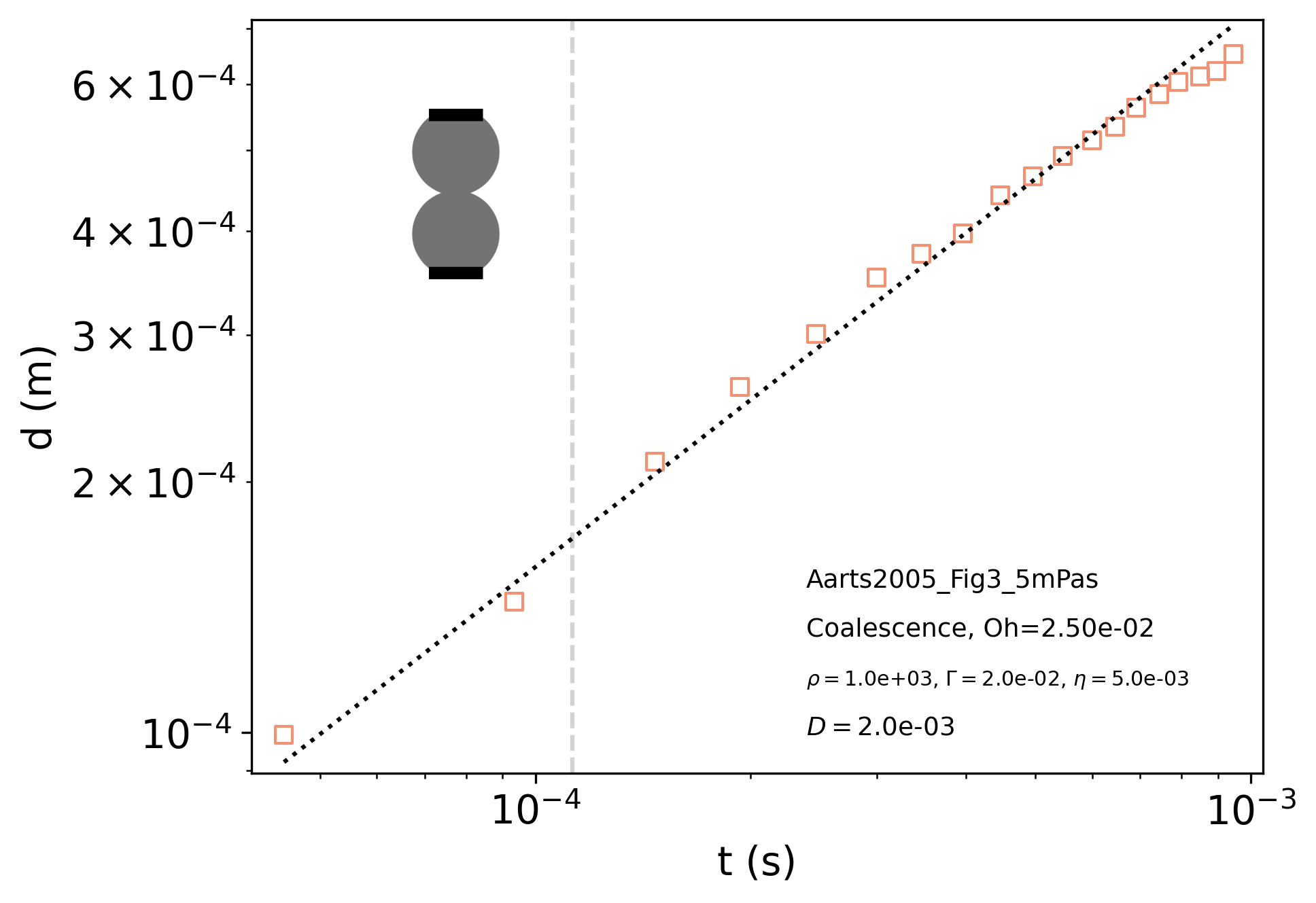} 
 \end{minipage} 
 \\ 
Silicone oil in ambient air. & Silicone oil in ambient air.\\ \hline \hline 
\end{tabular} 
 \end{table} 
\begin{table} 
 \centering 
 \begin{tabular}{ | p{9cm} | p{9cm} | } 
 \hline 
 \textbf{Aarts2005 Fig3 20mPas} & \textbf{Aarts2005 Fig3 50mPas}  \\ 
 \begin{minipage}{.5\textwidth} 
 \includegraphics[width=\linewidth]{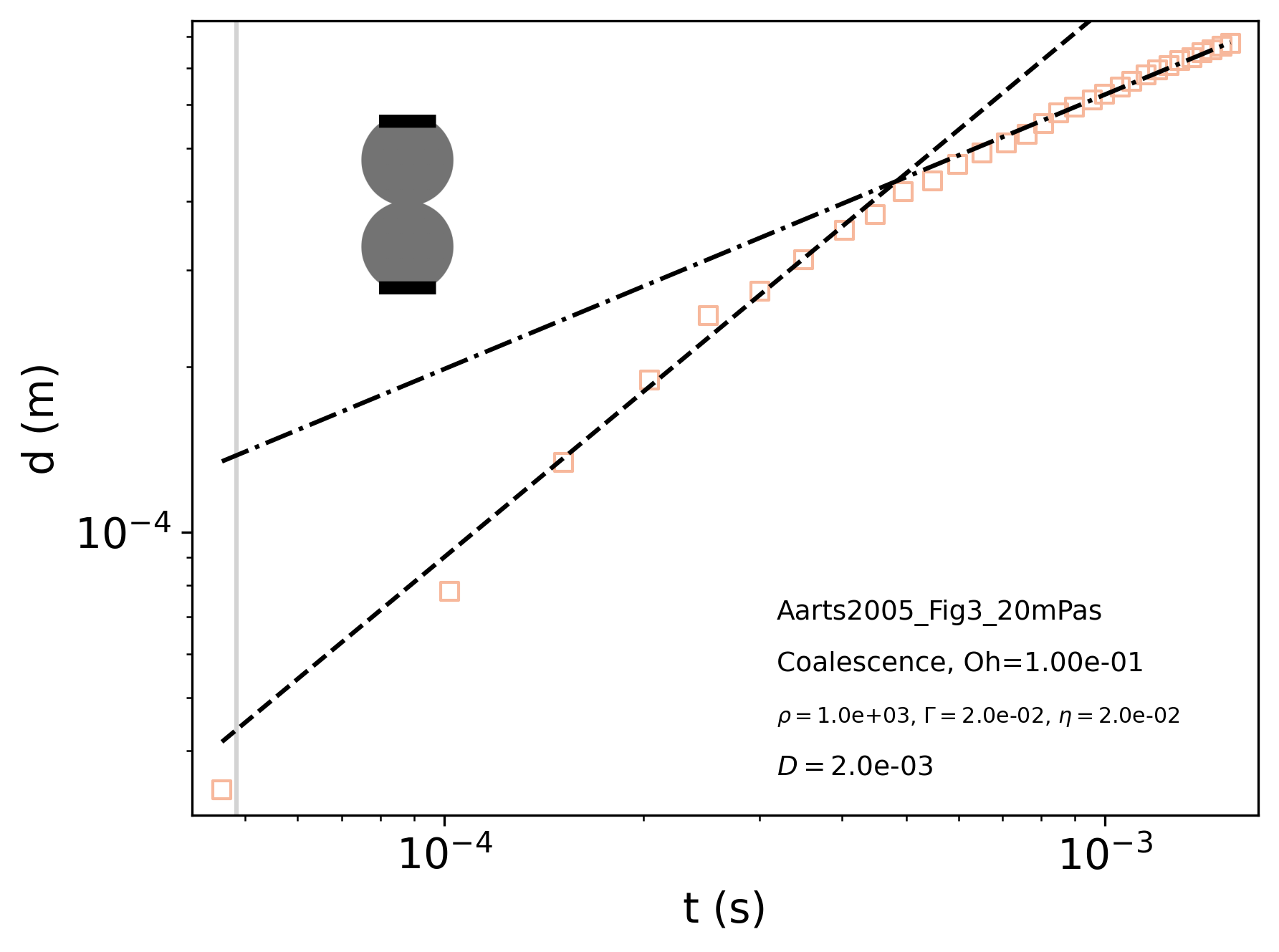} 
 \end{minipage}
 & 
 \begin{minipage}{.5\textwidth} 
 \includegraphics[width=\linewidth]{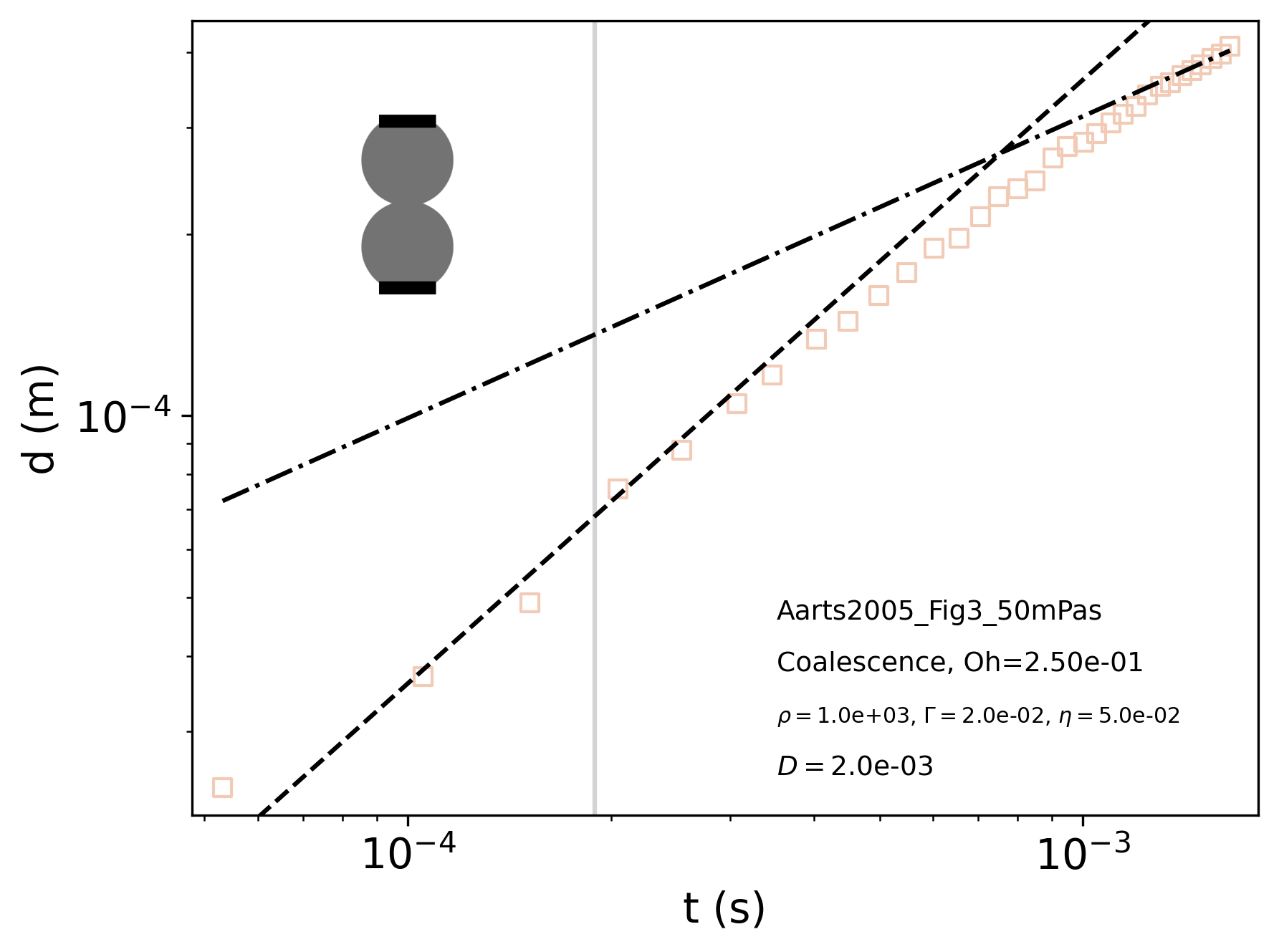} 
 \end{minipage} 
 \\ 
Silicone oil in ambient air. & Silicone oil in ambient air.\\ \hline \hline 
\textbf{Aarts2005 Fig3 1mPas} & \textbf{Aarts2008 Fig9 bubb17}  \\ 
 \begin{minipage}{.5\textwidth} 
 \includegraphics[width=\linewidth]{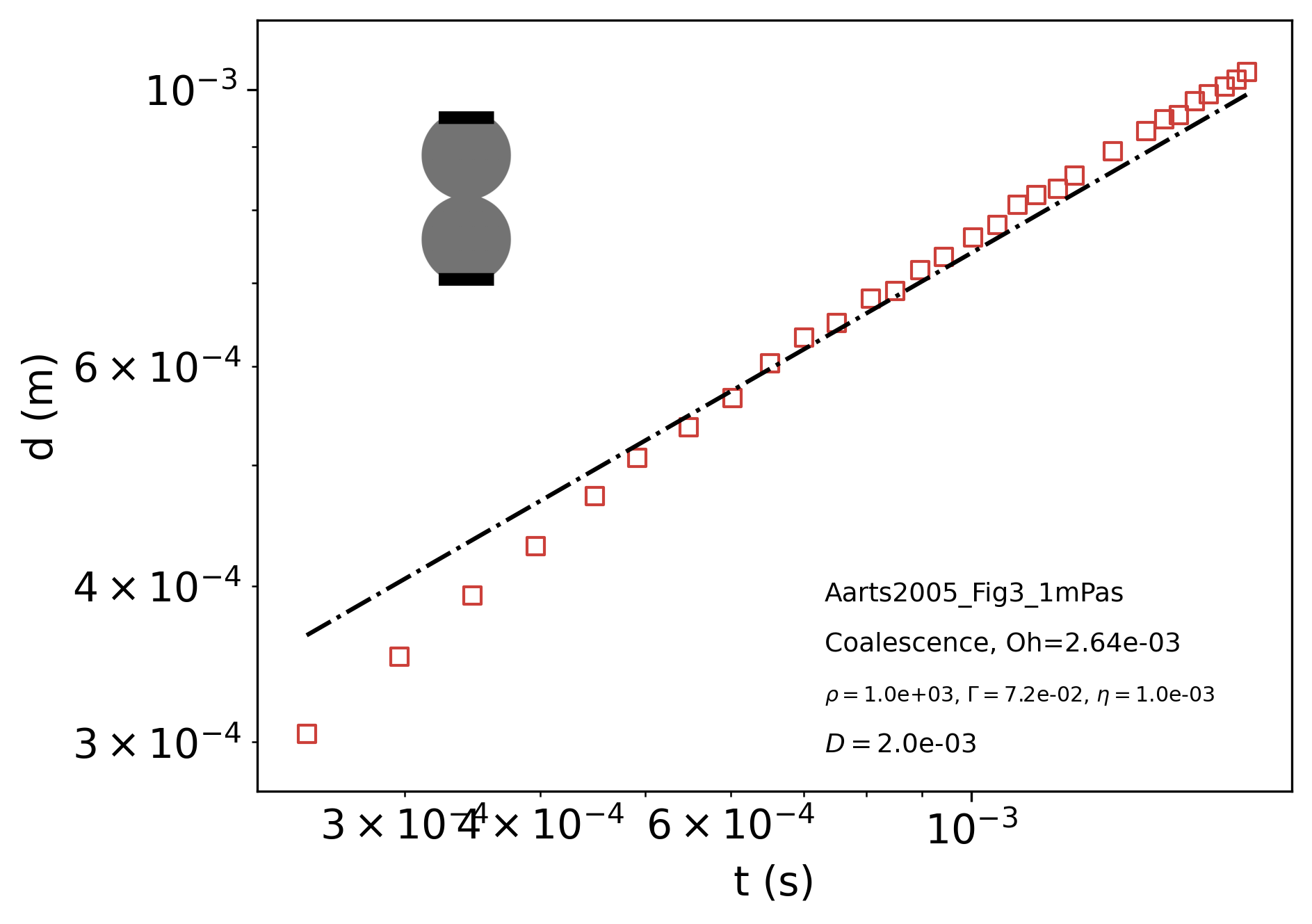} 
 \end{minipage}
 & 
 \begin{minipage}{.5\textwidth} 
 \includegraphics[width=\linewidth]{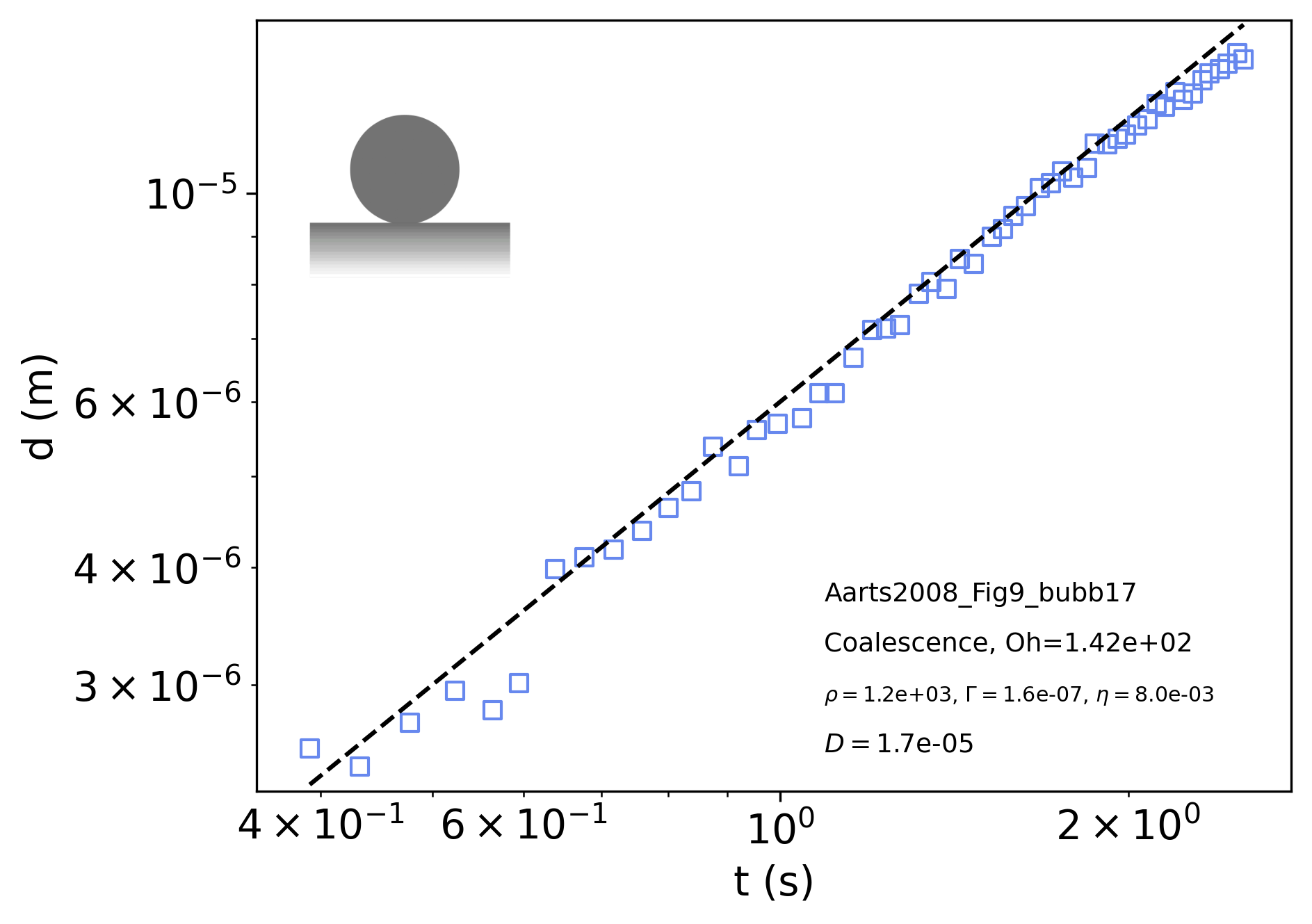} 
 \end{minipage} 
 \\ 
Water in ambient air. & PMMA Poly(styrene) in Decalin, gas phase. \newline The density used is that of the PMMA colloid.\\ \hline \hline 
\textbf{Aarts2008 Fig9 drop17} & \textbf{Yokota2011 Fig2 289}  \\ 
 \begin{minipage}{.5\textwidth} 
 \includegraphics[width=\linewidth]{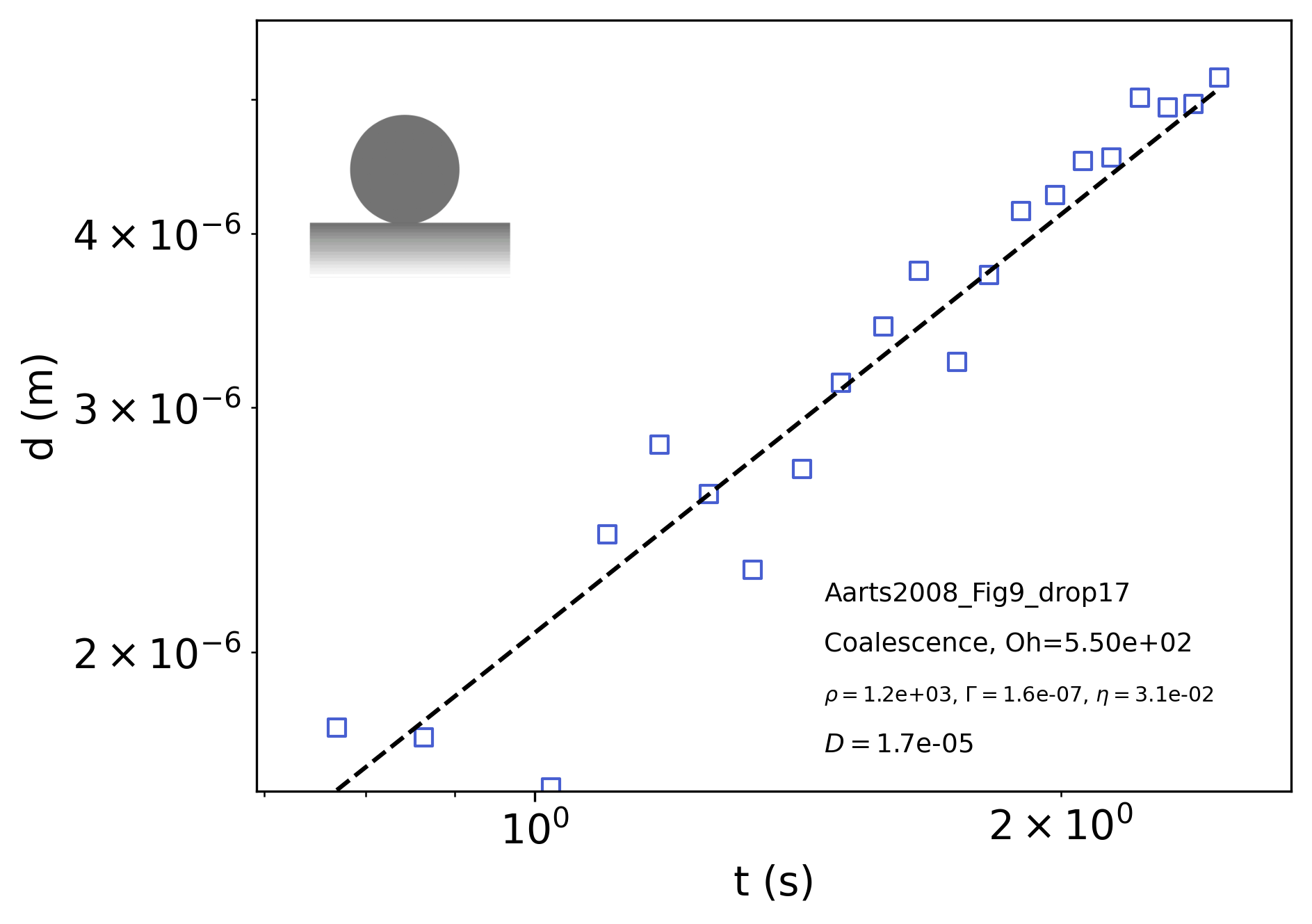} 
 \end{minipage}
 & 
 \begin{minipage}{.5\textwidth} 
 \includegraphics[width=\linewidth]{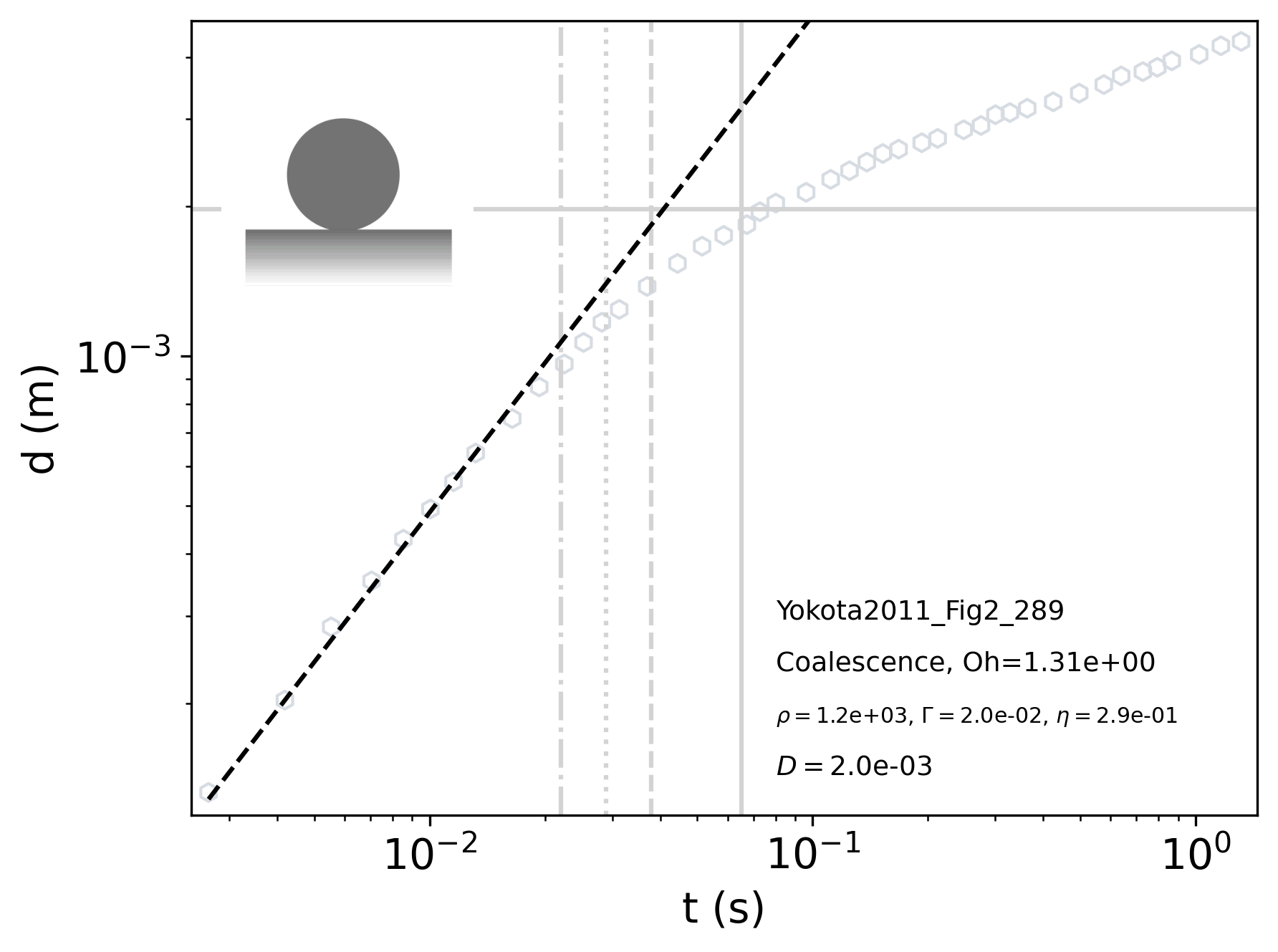} 
 \end{minipage} 
 \\ 
PMMA Poly(styrene) in Decalin, liquid phase. \newline The density used is that of the PMMA colloid. & Water-glycerol mixture in PDMS oil. \newline A precise value of density was not available.\newline In a Hele-Shaw cell $D=(R H)^\frac{1}{2}$, where $R$ and $H$ are respectively the drop radius and the cell thickness. \newline Long-time behavior follows $d/\ell_{vc}\propto (t/\tau_{vc})^\frac{1}{4}$.\\ \hline \hline 
\end{tabular} 
 \end{table} 
\begin{table} 
 \centering 
 \begin{tabular}{ | p{9cm} | p{9cm} | } 
 \hline 
 \textbf{Yokota2011 Fig2 888} & \textbf{Paulsen2011 Fig2 1p9}  \\ 
 \begin{minipage}{.5\textwidth} 
 \includegraphics[width=\linewidth]{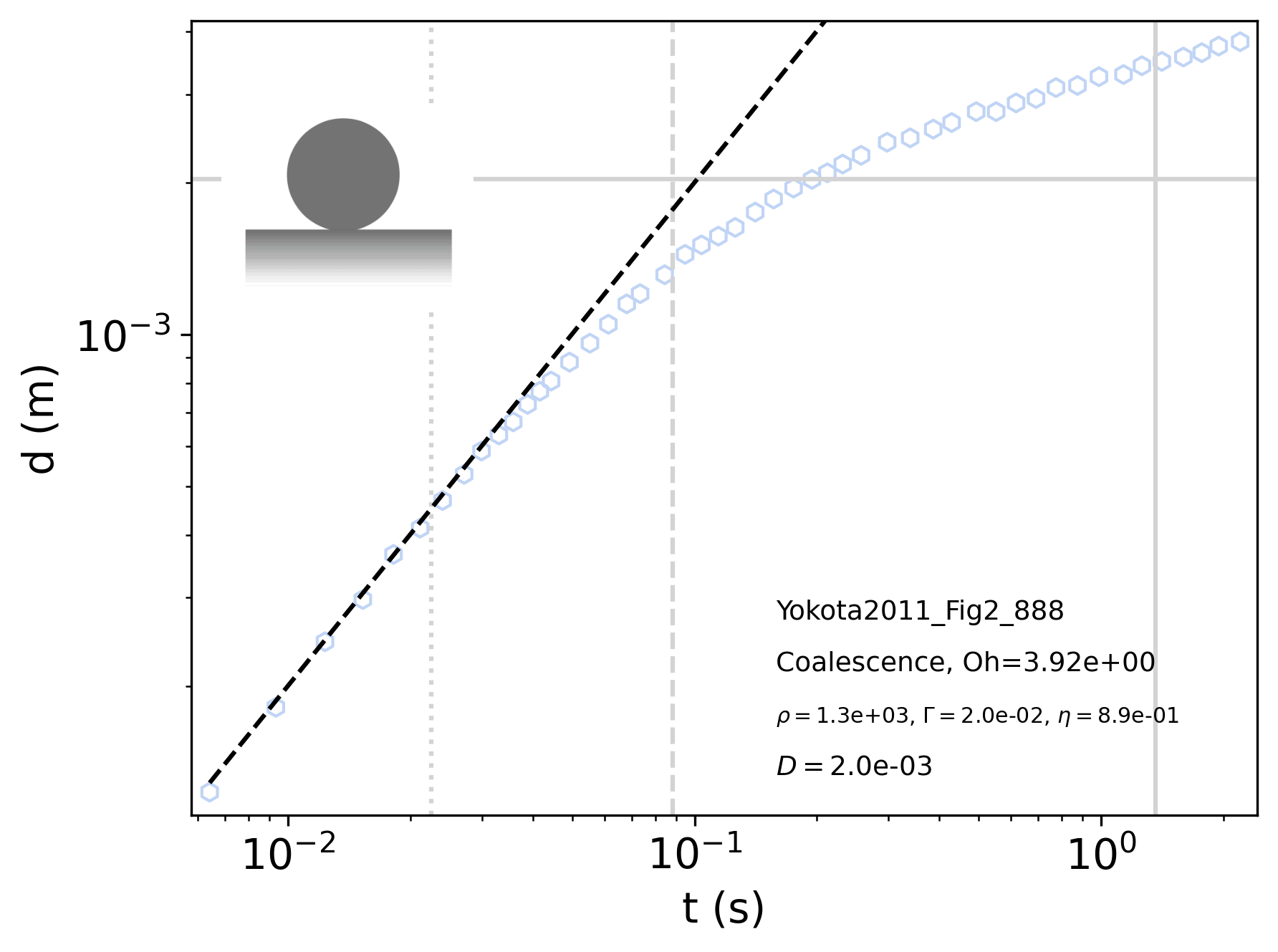} 
 \end{minipage}
 & 
 \begin{minipage}{.5\textwidth} 
 \includegraphics[width=\linewidth]{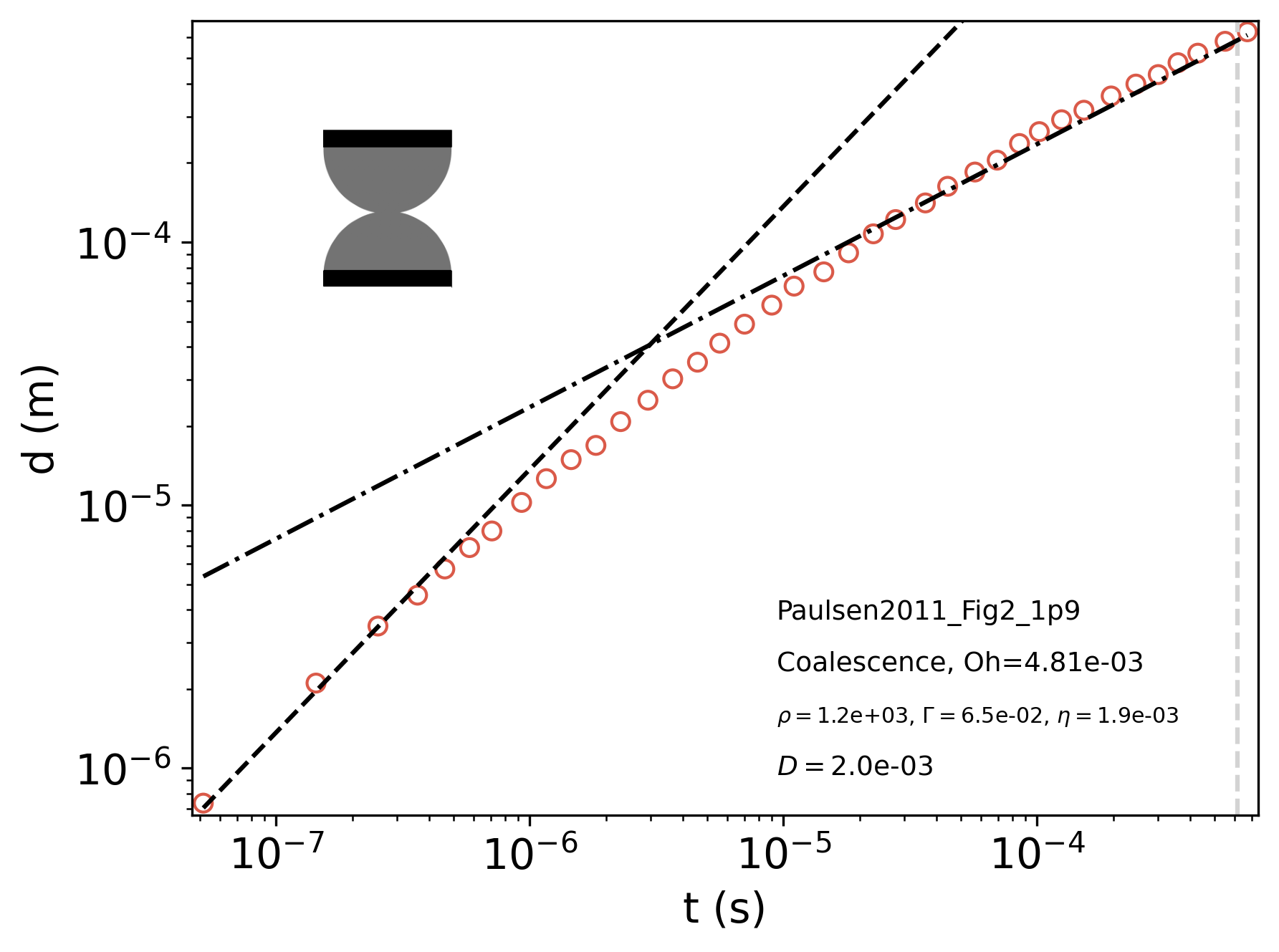} 
 \end{minipage} 
 \\ 
Water-glycerol mixture in PDMS oil. \newline A precise value of density was not available.\newline In a Hele-Shaw cell $D=(R H)^\frac{1}{2}$, where $R$ and $H$ are respectively the drop radius and the cell thickness. \newline Long-time behavior follows $d/\ell_{vc}\propto (t/\tau_{vc})^\frac{1}{4}$. & Glycerol-water-NaCl mixture in ambient air.\\ \hline \hline 
\textbf{Paulsen2011 Fig2 11} & \textbf{Paulsen2011 Fig2 48}  \\ 
 \begin{minipage}{.5\textwidth} 
 \includegraphics[width=\linewidth]{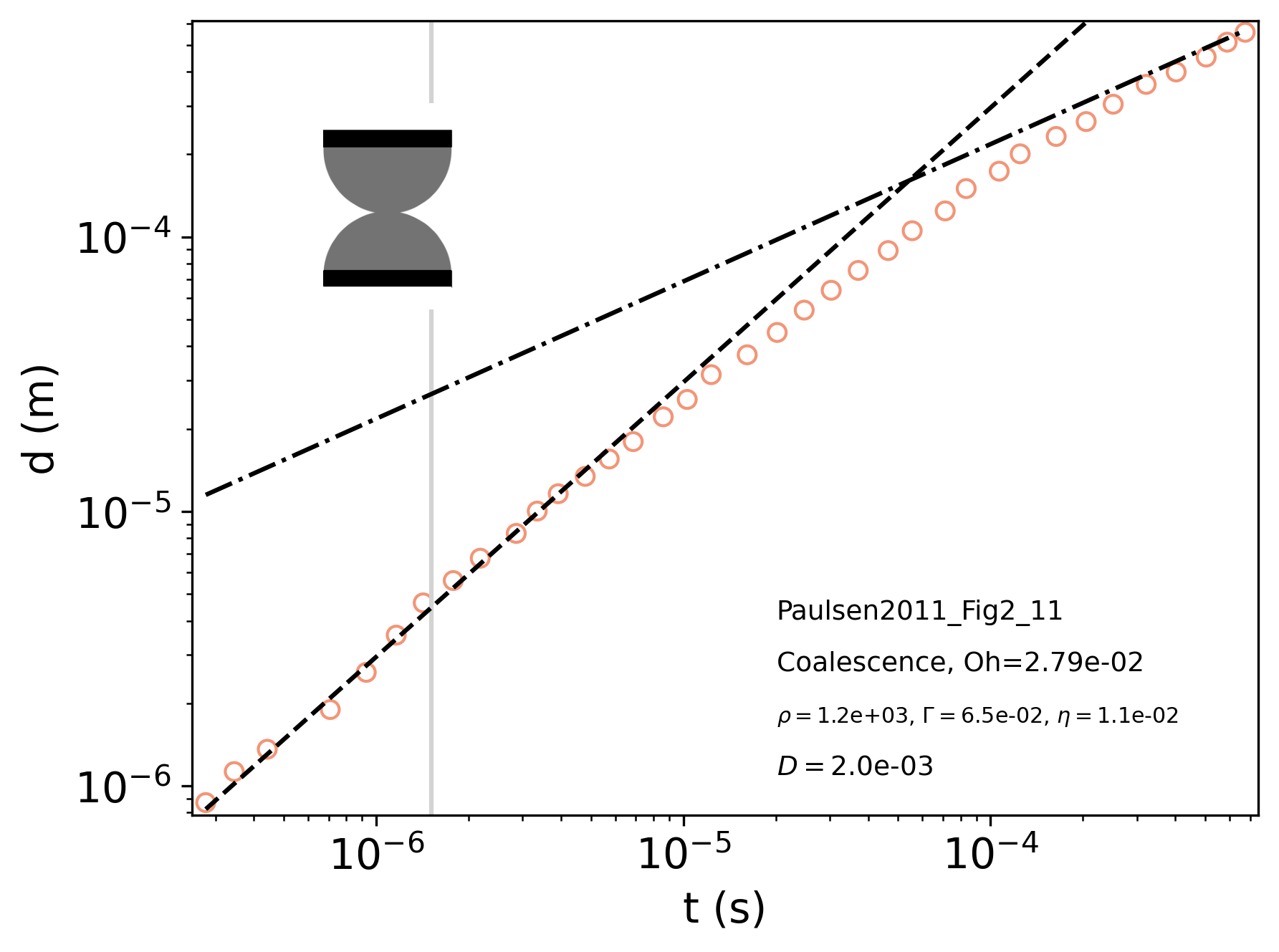} 
 \end{minipage}
 & 
 \begin{minipage}{.5\textwidth} 
 \includegraphics[width=\linewidth]{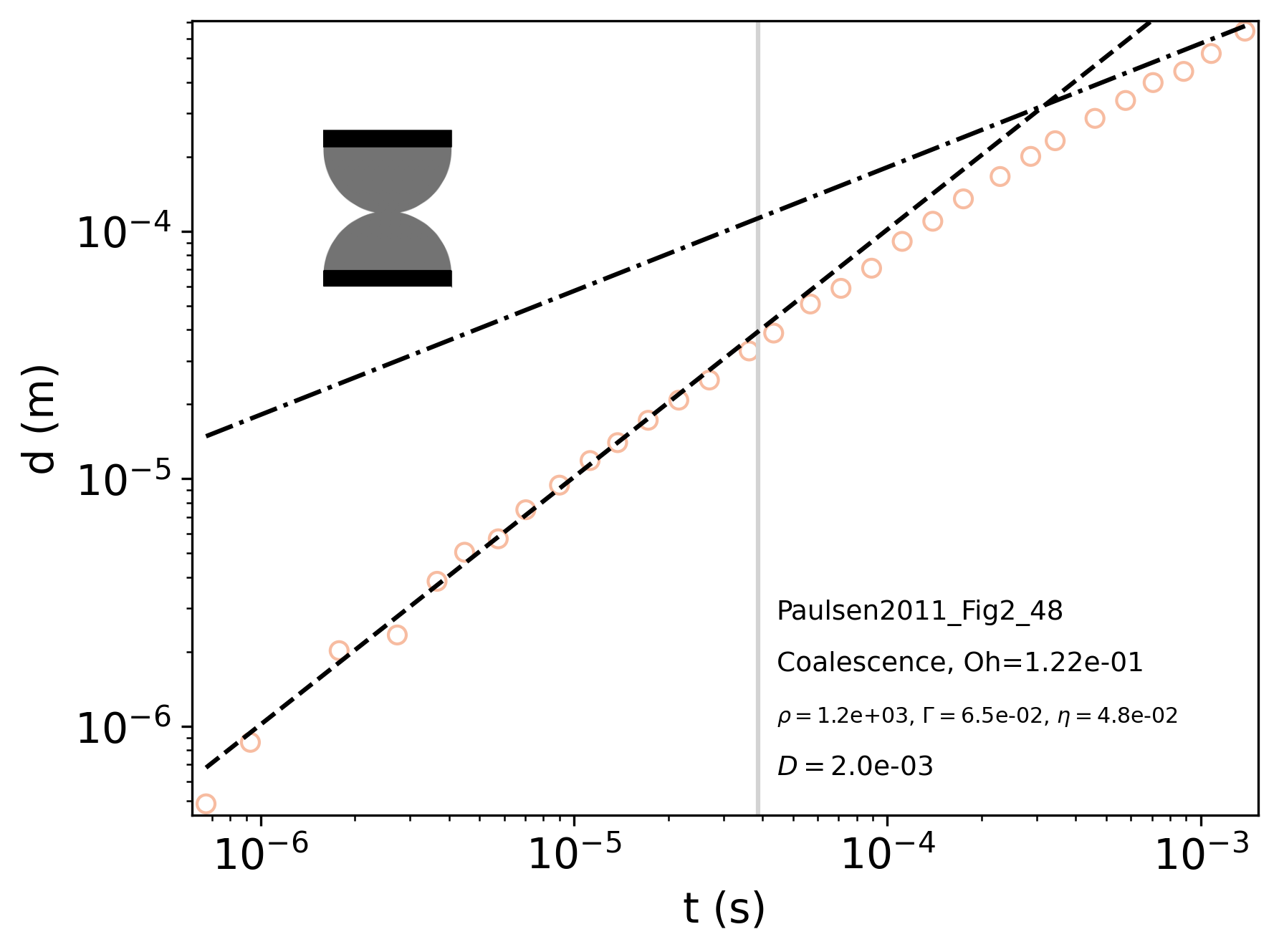} 
 \end{minipage} 
 \\ 
Glycerol-water-NaCl mixture in ambient air. & Glycerol-water-NaCl mixture in ambient air.\\ \hline \hline 
\textbf{Paulsen2011 Fig2 230} & \textbf{Paulsen2014 Fig1}  \\ 
 \begin{minipage}{.5\textwidth} 
 \includegraphics[width=\linewidth]{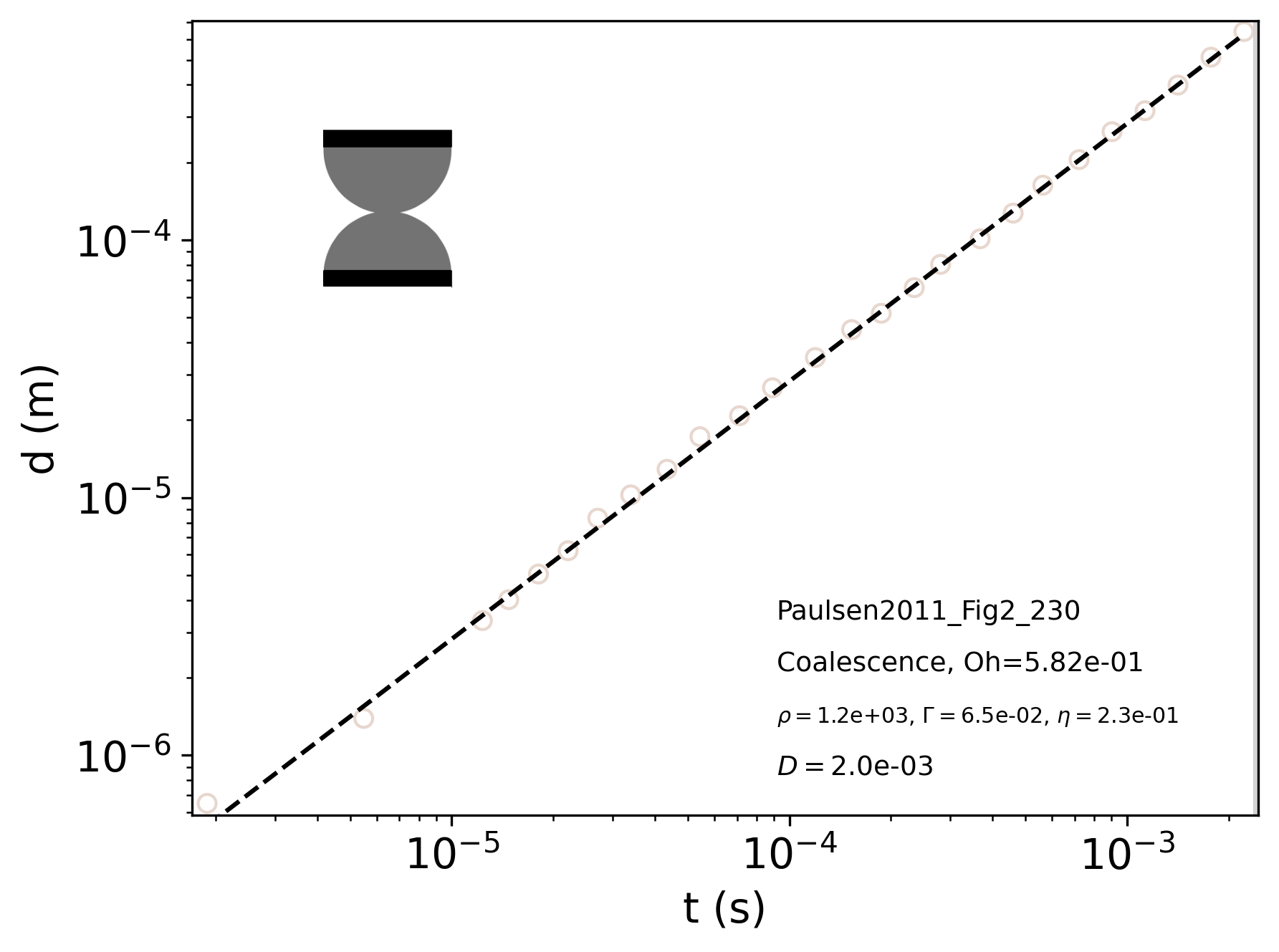} 
 \end{minipage}
 & 
 \begin{minipage}{.5\textwidth} 
 \includegraphics[width=\linewidth]{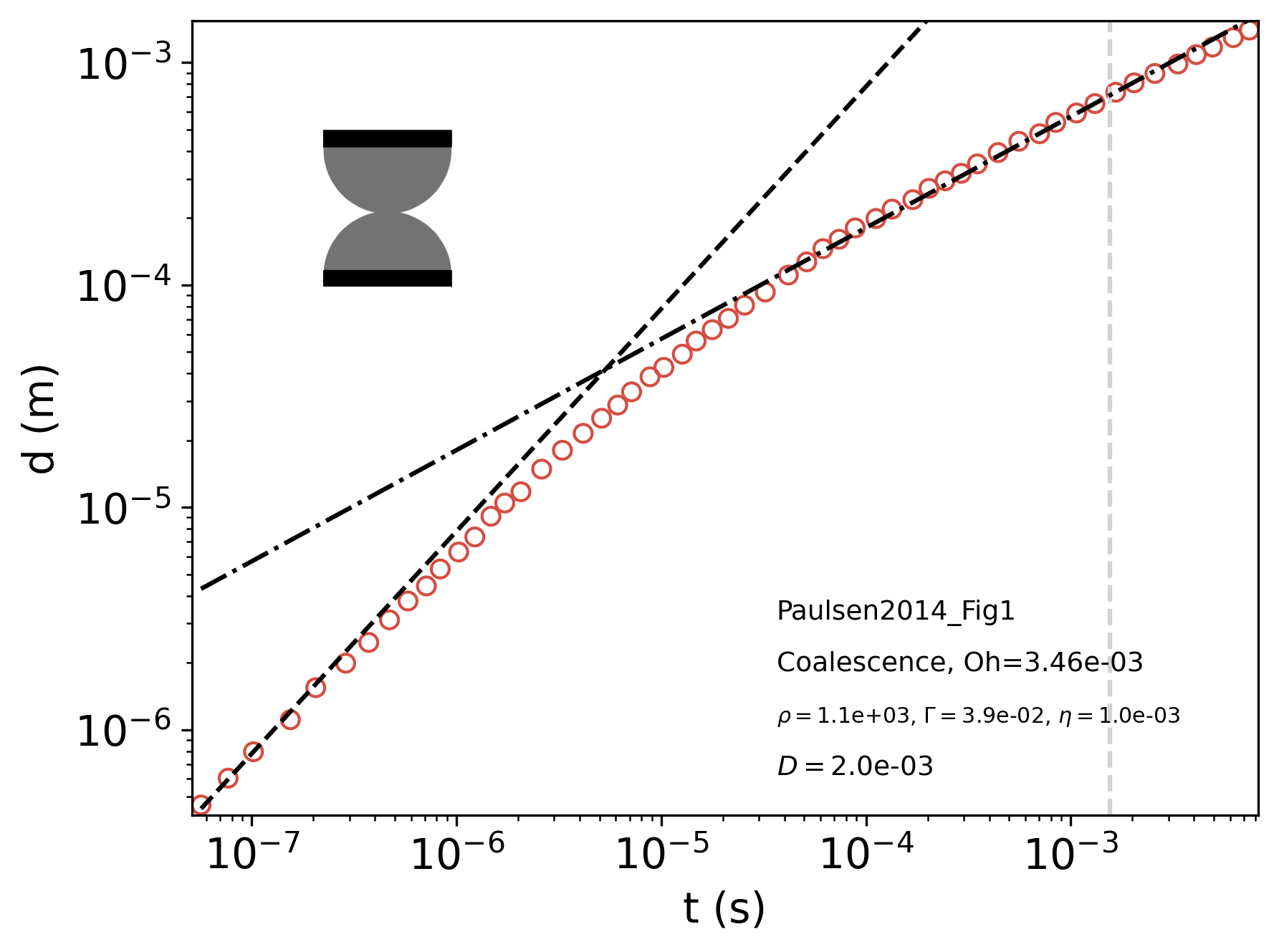} 
 \end{minipage} 
 \\ 
Glycerol-water-NaCl mixture in ambient air. & Salt water in silicone oil.\\ \hline \hline 
\end{tabular} 
 \end{table} 
\begin{table} 
 \centering 
 \begin{tabular}{ | p{9cm} | p{9cm} | } 
 \hline 
 \textbf{Soto2018 Fig5} & \textbf{Rahman2019 Fig6 6p65}  \\ 
 \begin{minipage}{.5\textwidth} 
 \includegraphics[width=\linewidth]{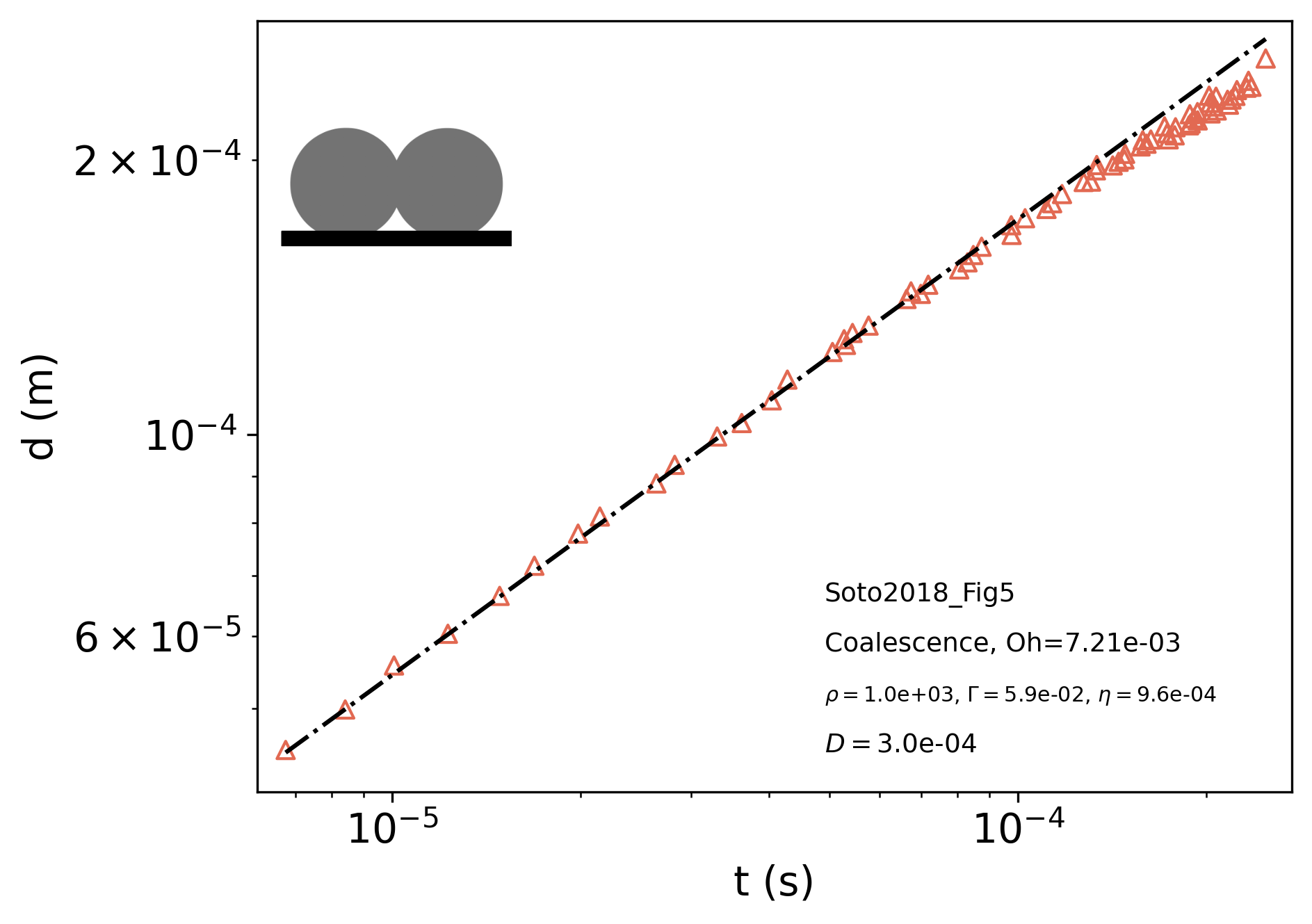} 
 \end{minipage}
 & 
 \begin{minipage}{.5\textwidth} 
 \includegraphics[width=\linewidth]{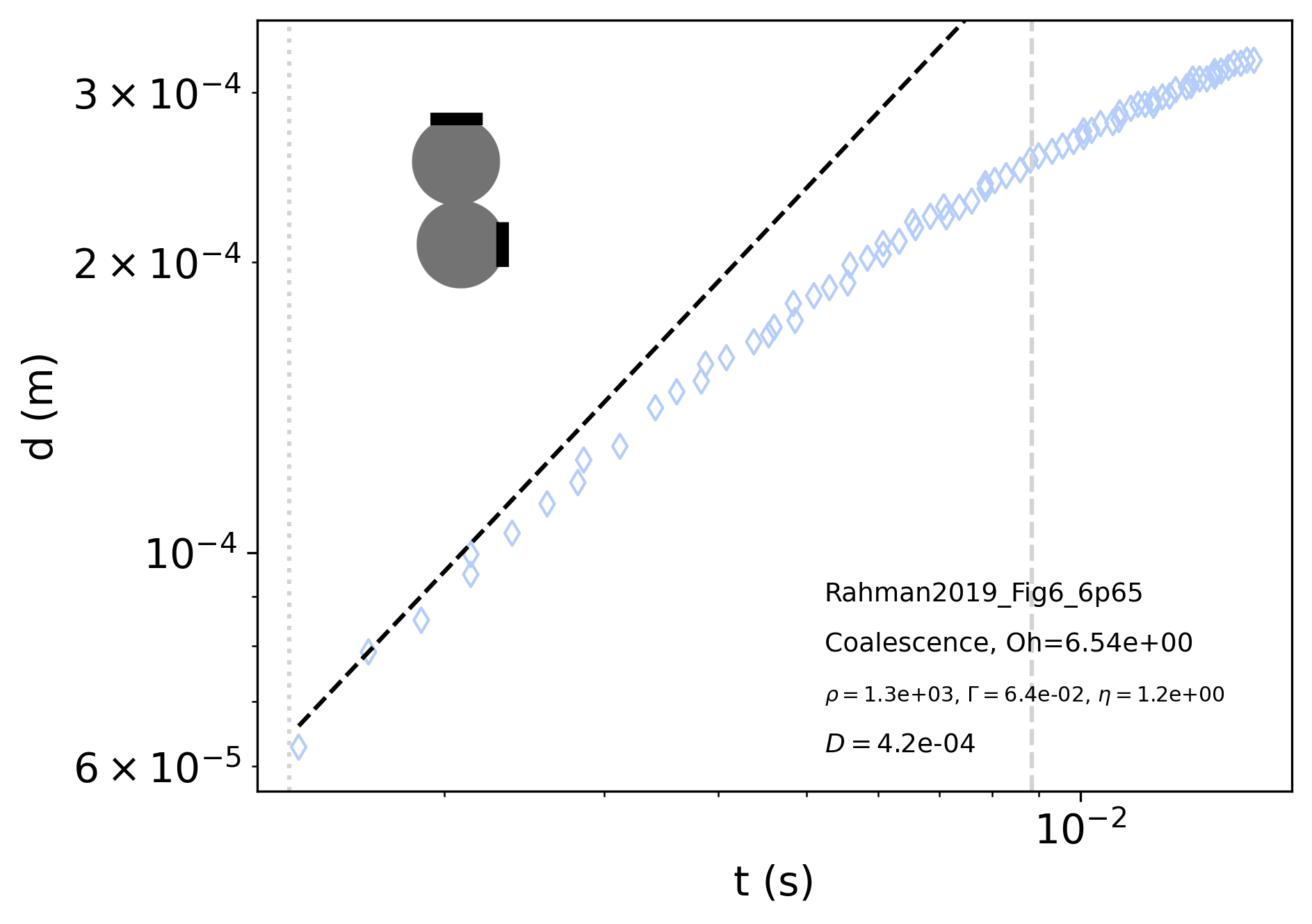} 
 \end{minipage} 
 \\ 
CO$_2$ bubble in water. \newline  Density and viscosity are that of the outer fluid. & Water-glycerol mixture in ambient air. \\ \hline \hline 
\textbf{Rahman2019 Fig6 1p84} & \textbf{Rahman2019 Fig6 0p146}  \\ 
 \begin{minipage}{.5\textwidth} 
 \includegraphics[width=\linewidth]{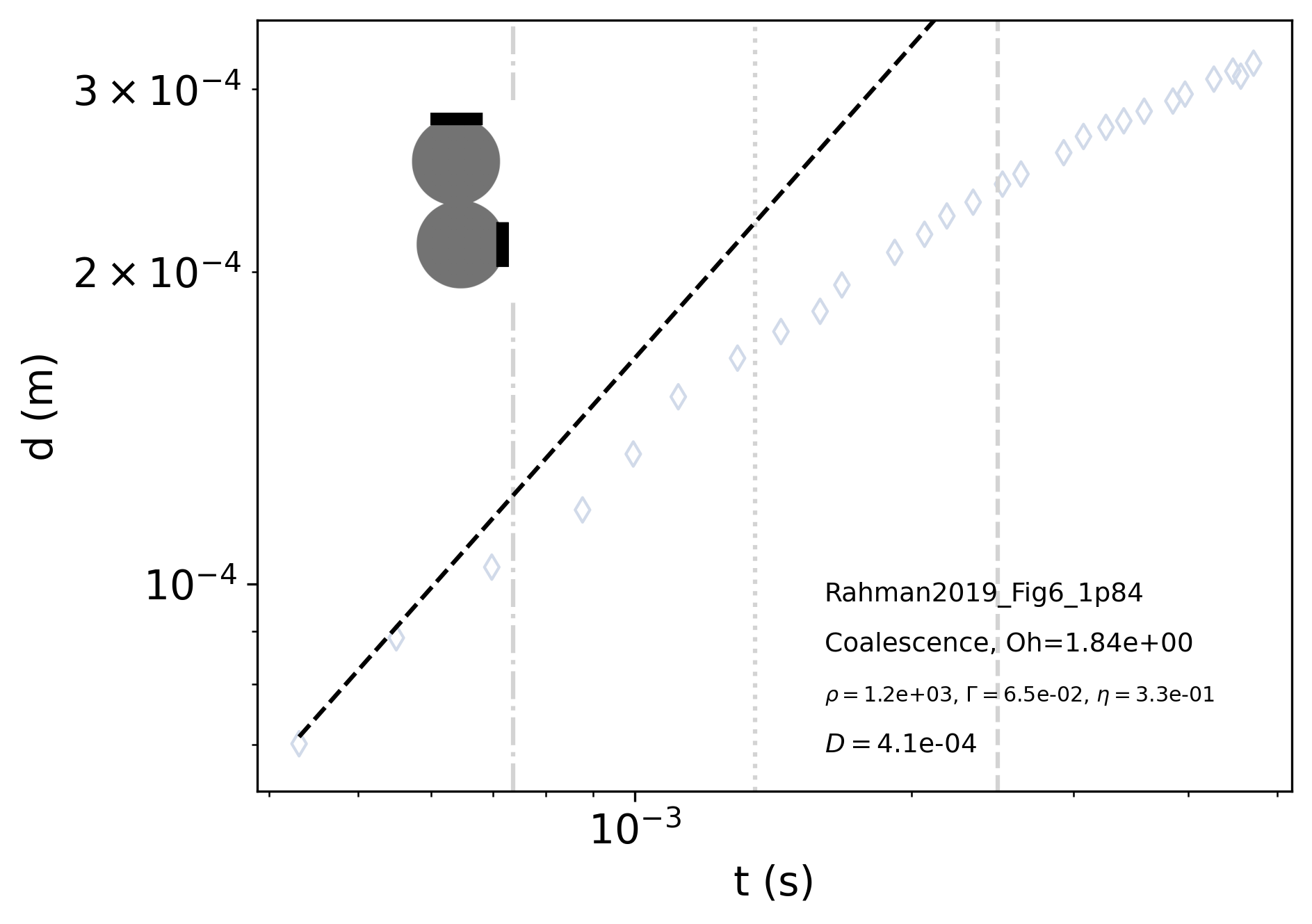} 
 \end{minipage}
 & 
 \begin{minipage}{.5\textwidth} 
 \includegraphics[width=\linewidth]{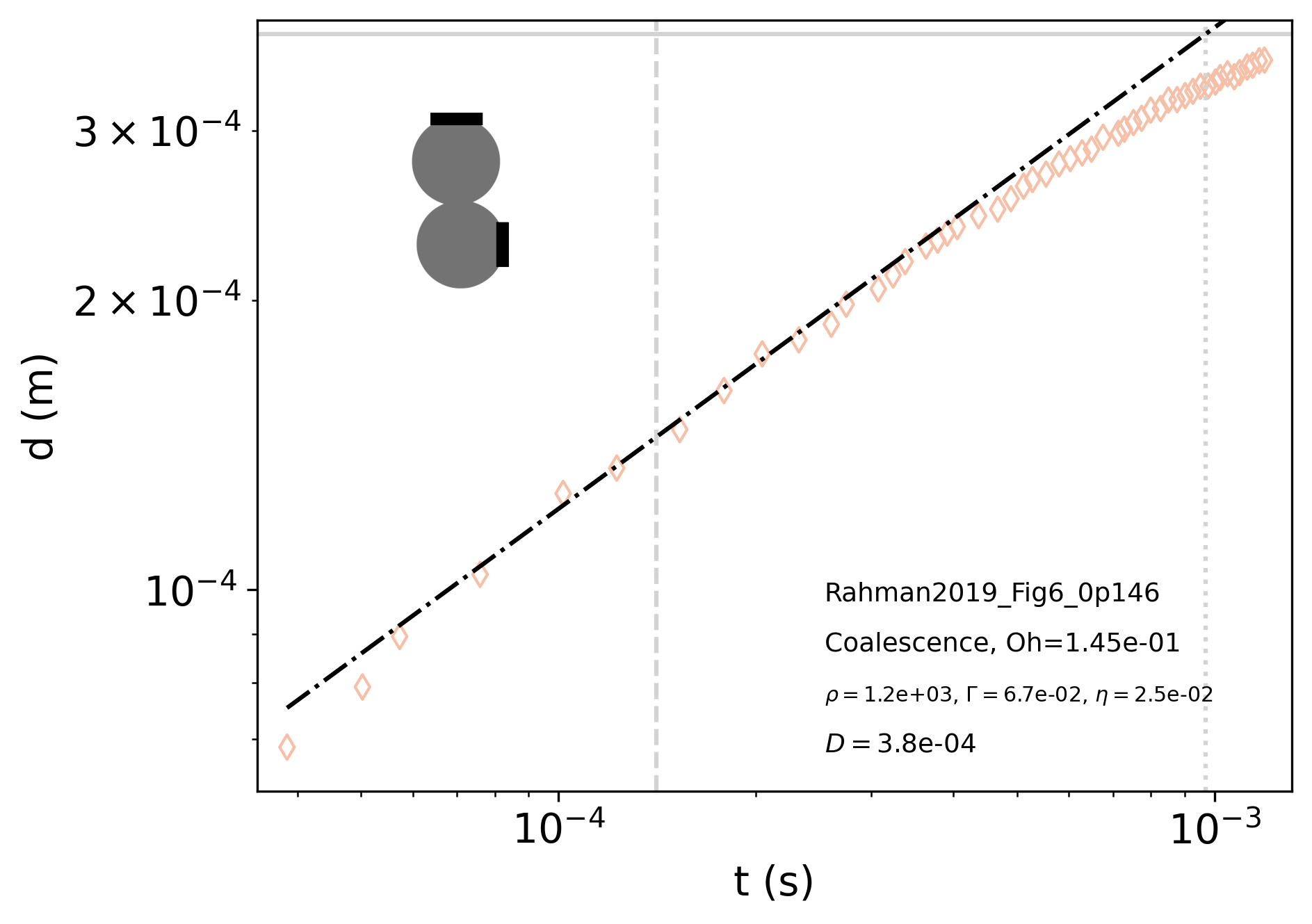} 
 \end{minipage} 
 \\ 
Water-glycerol mixture in ambient air.  & Water-glycerol mixture in ambient air. \\ \hline \hline 
\textbf{Rahman2019 Fig6 0p00692} & \textbf{Chen1997 Fig7}  \\ 
 \begin{minipage}{.5\textwidth} 
 \includegraphics[width=\linewidth]{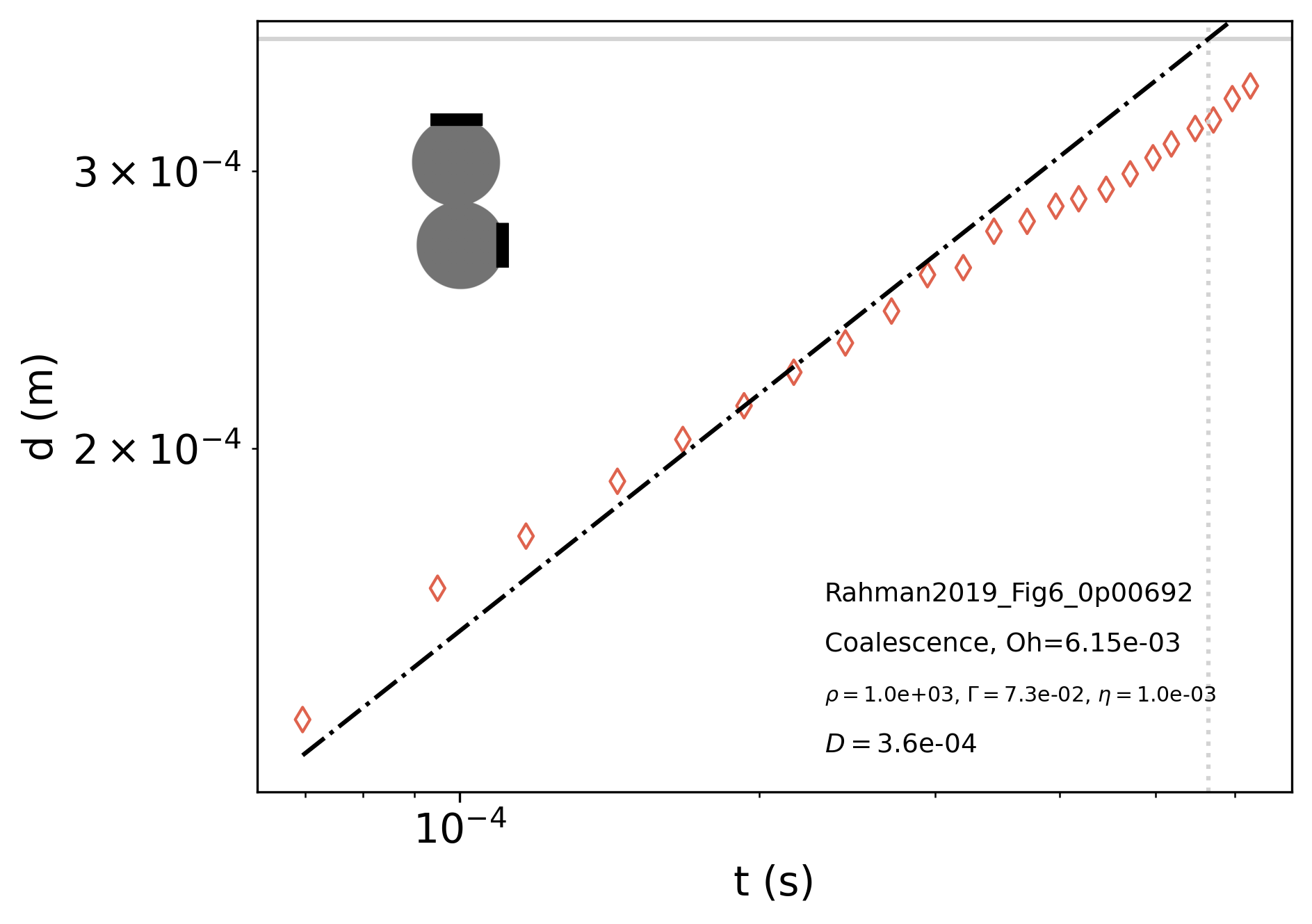} 
 \end{minipage}
 & 
 \begin{minipage}{.5\textwidth} 
 \includegraphics[width=\linewidth]{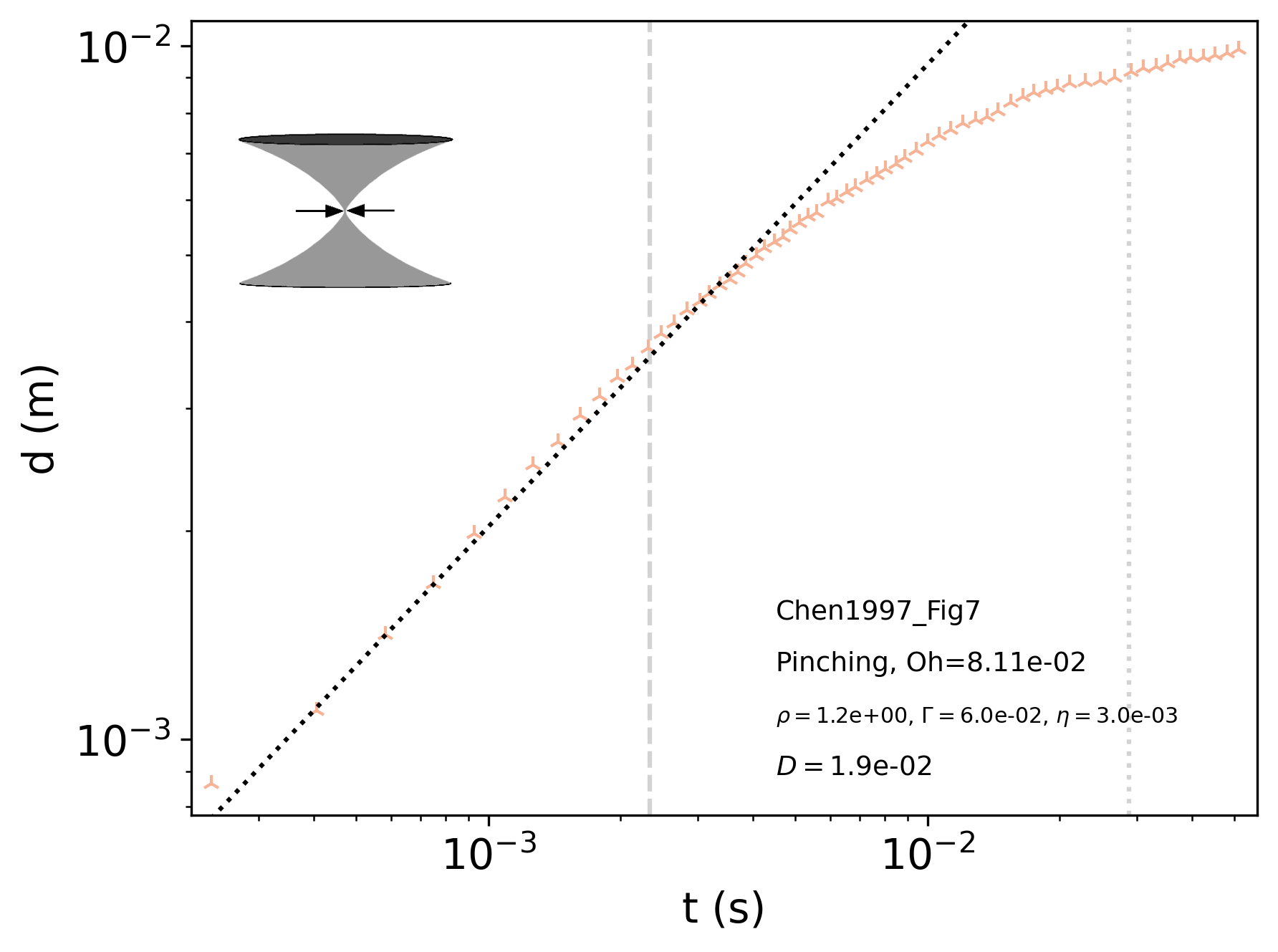} 
 \end{minipage} 
 \\ 
Water-glycerol mixture in ambient air.  & Catenoid soap film in ambient air.\newline Density is that of air, viscosity is that of the soap film. \\ \hline \hline 
\end{tabular} 
 \end{table} 
\begin{table} 
 \centering 
 \begin{tabular}{ | p{9cm} | p{9cm} | } 
 \hline 
 \textbf{McKinley2000 Fig4} & \textbf{Chen2002 Fig3}  \\ 
 \begin{minipage}{.5\textwidth} 
 \includegraphics[width=\linewidth]{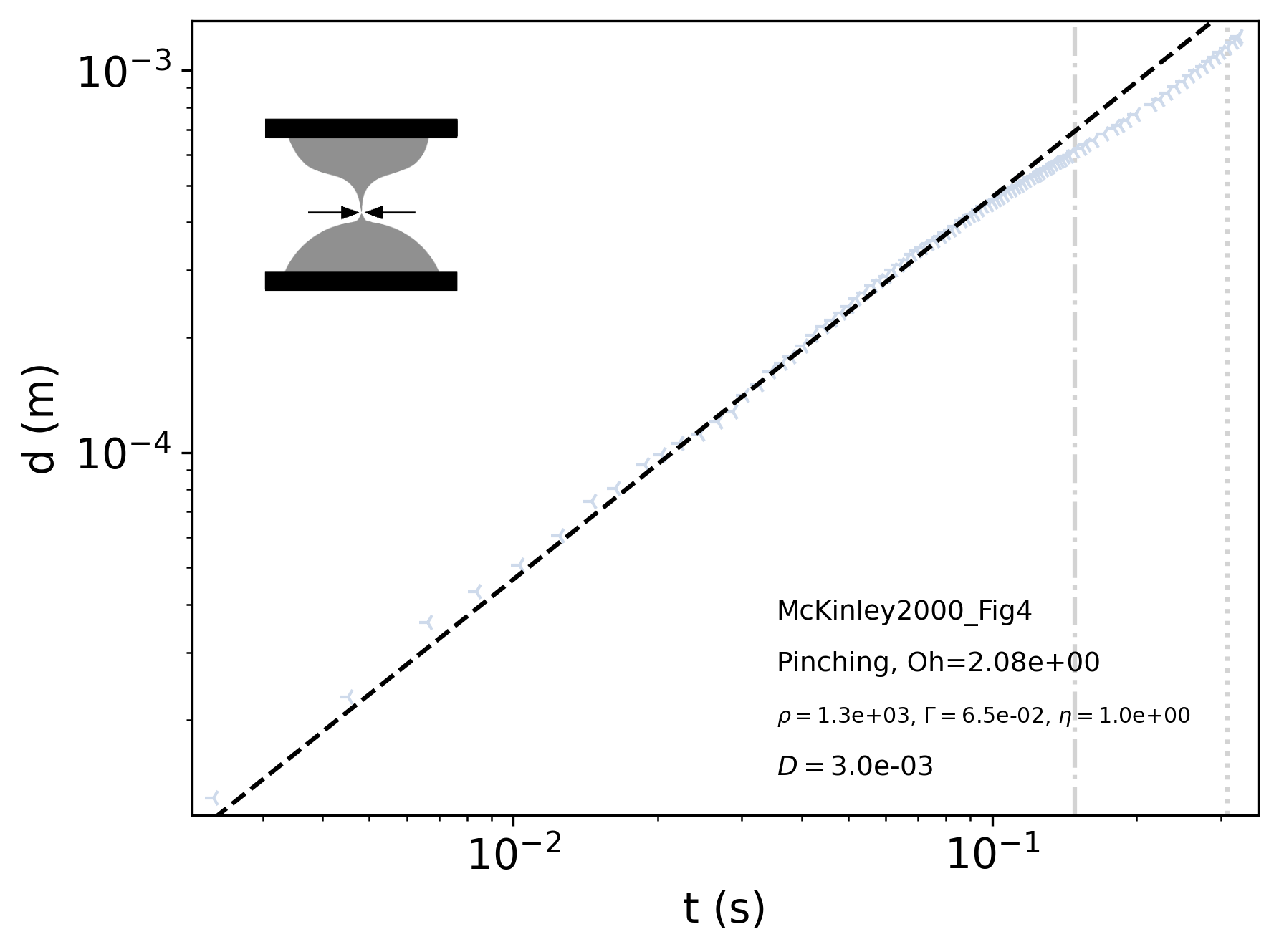} 
 \end{minipage}
 & 
 \begin{minipage}{.5\textwidth} 
 \includegraphics[width=\linewidth]{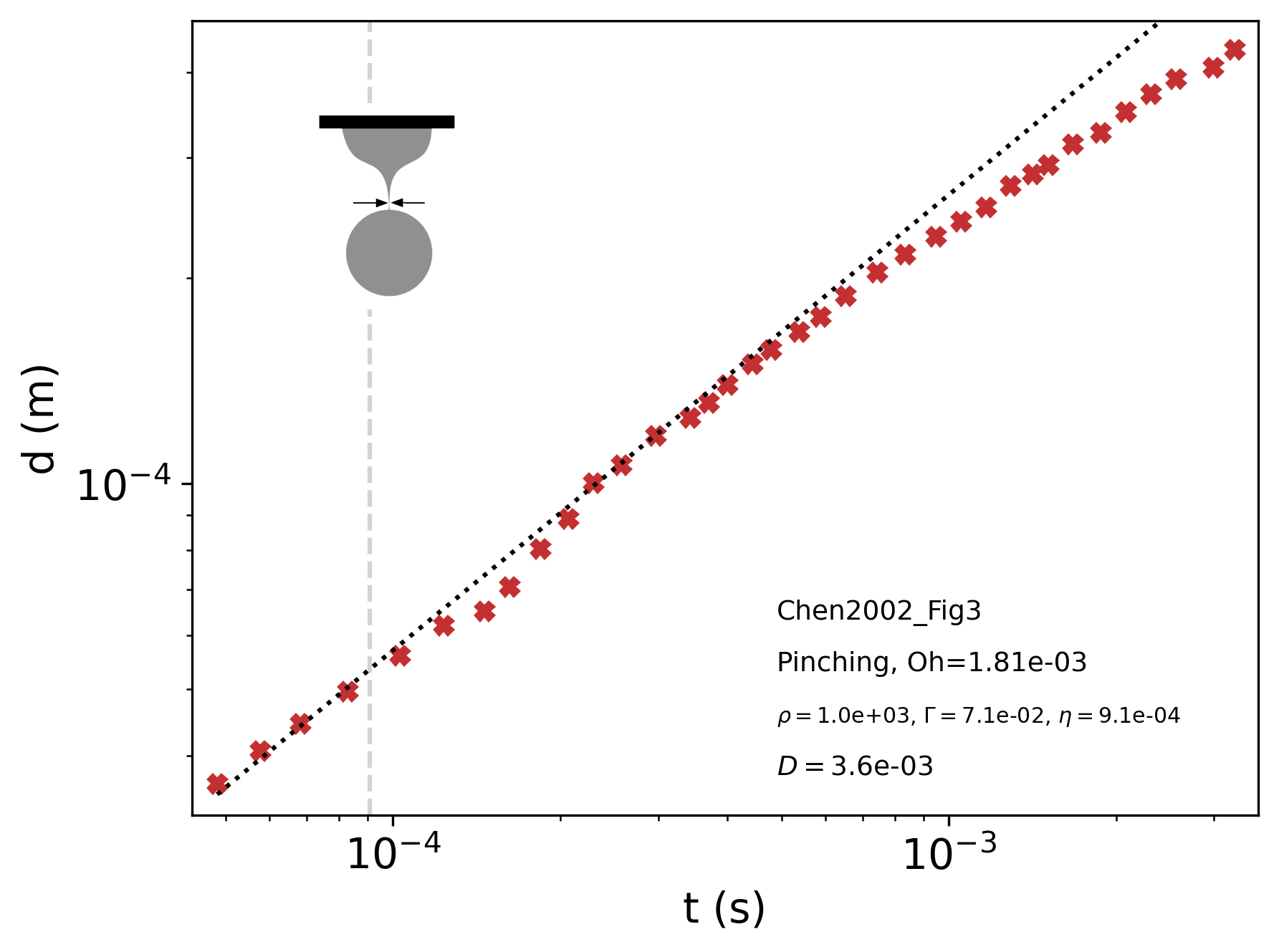} 
 \end{minipage} 
 \\ 
Glycerol in ambient air. & Water in ambient air.\\ \hline \hline 
\textbf{Burton2004 Fig5} & \textbf{Burton2005 0p9}  \\ 
 \begin{minipage}{.5\textwidth} 
 \includegraphics[width=\linewidth]{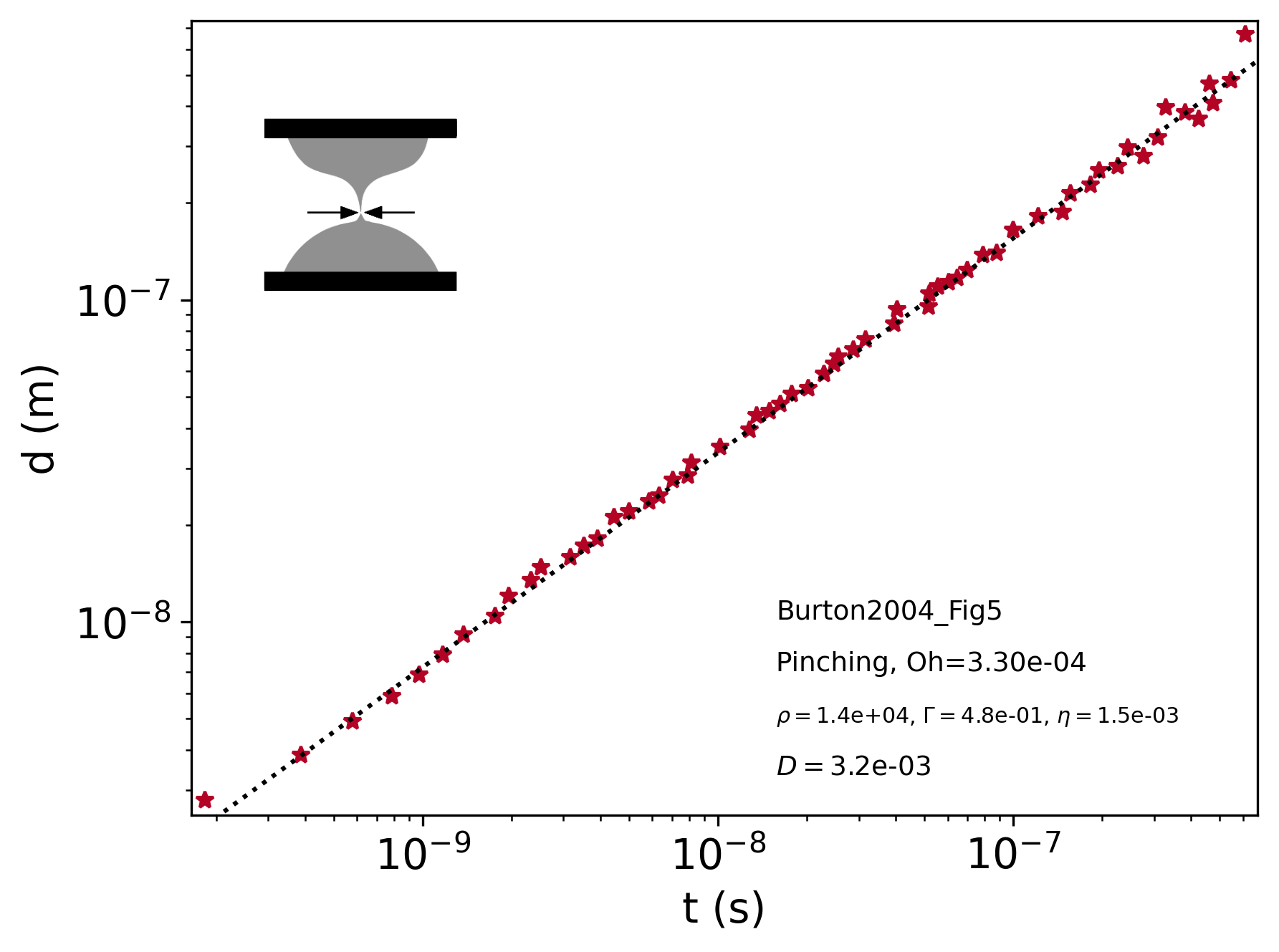} 
 \end{minipage}
 & 
 \begin{minipage}{.5\textwidth} 
 \includegraphics[width=\linewidth]{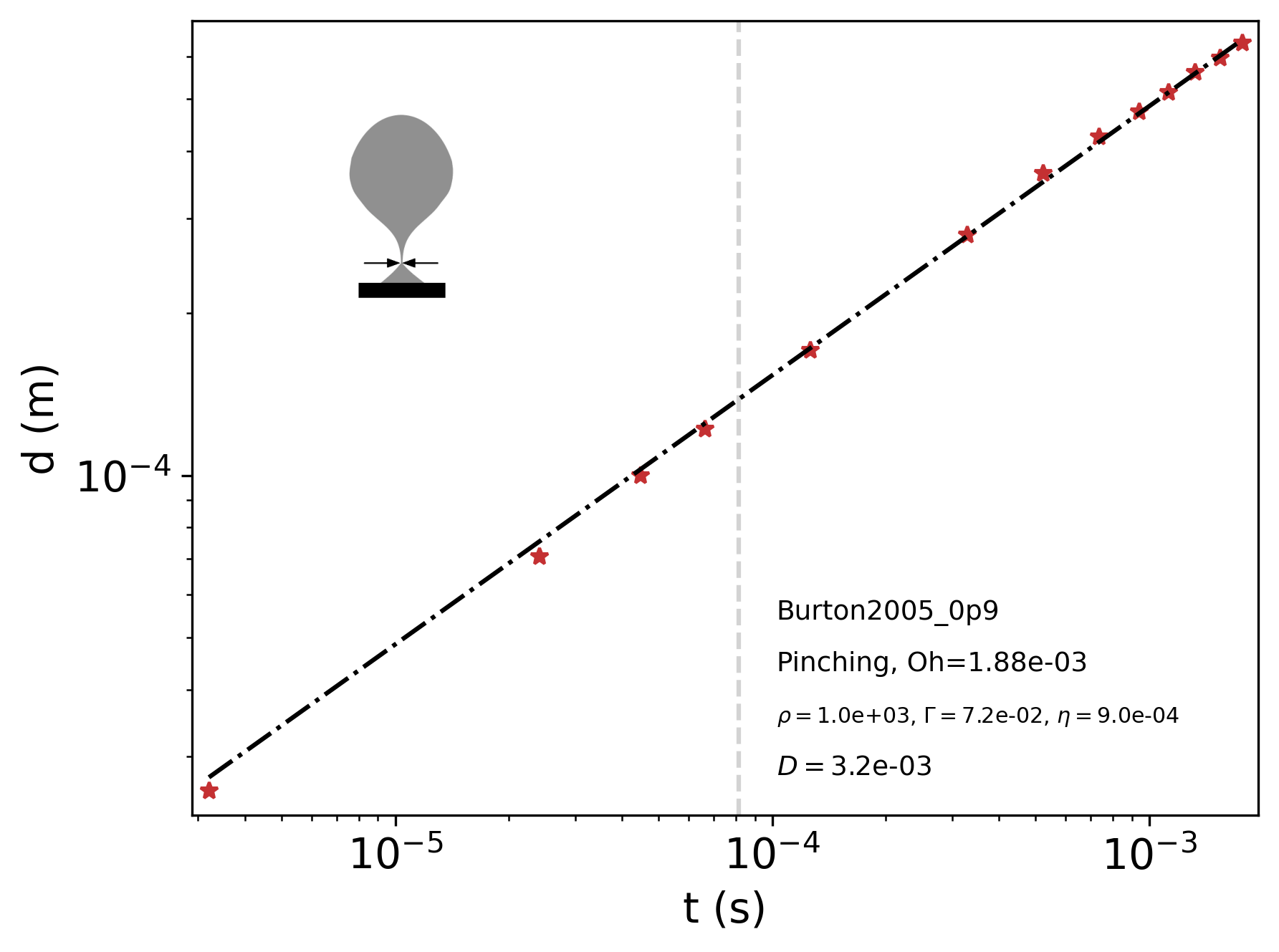} 
 \end{minipage} 
 \\ 
Mercury in ambient air. & Air bubble in water.\newline Density and viscosity are that of the outer fluid.\\ \hline \hline 
\textbf{Burton2005 37p4} & \textbf{Burton2005 1011}  \\ 
 \begin{minipage}{.5\textwidth} 
 \includegraphics[width=\linewidth]{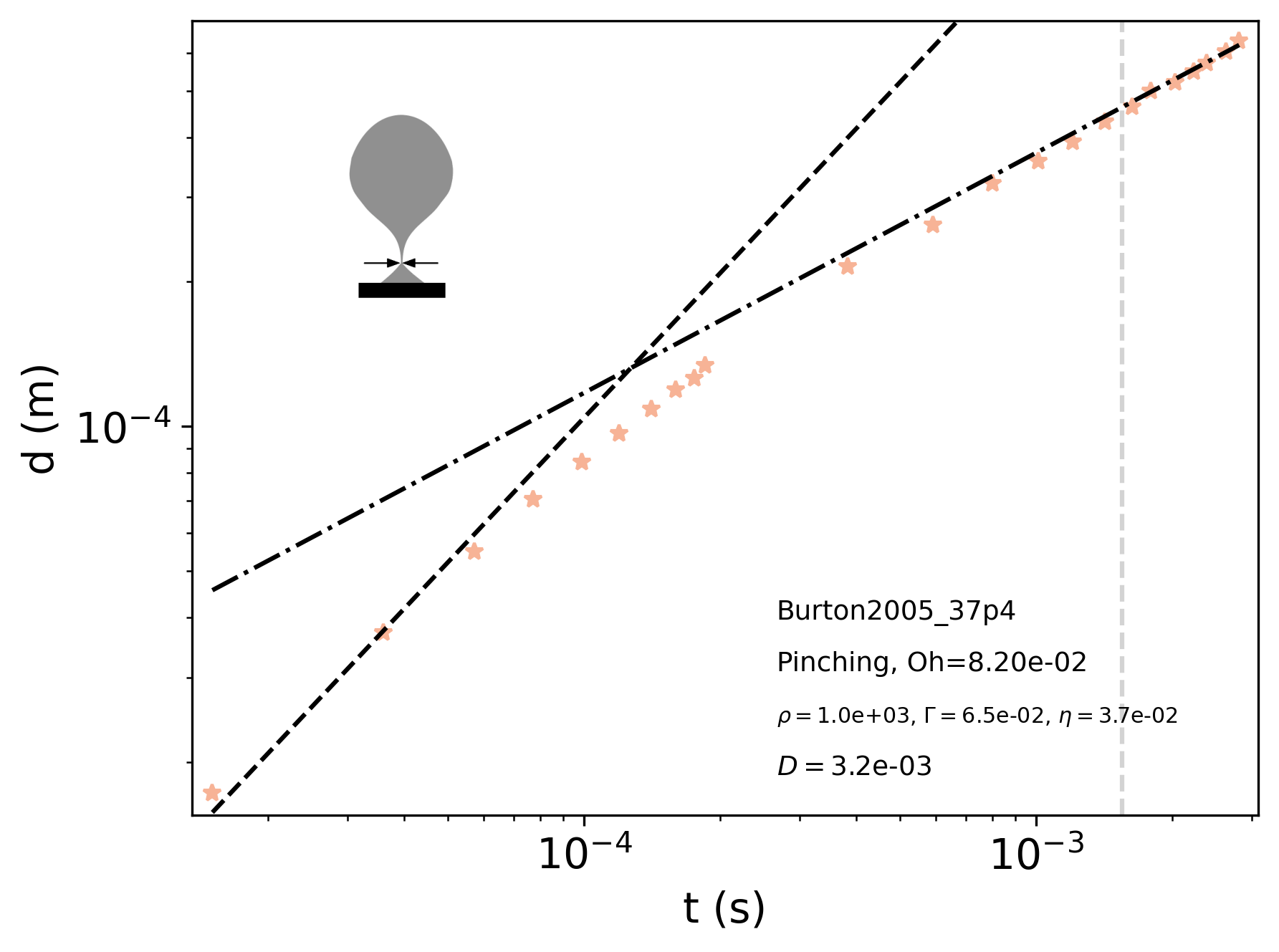} 
 \end{minipage}
 & 
 \begin{minipage}{.5\textwidth} 
 \includegraphics[width=\linewidth]{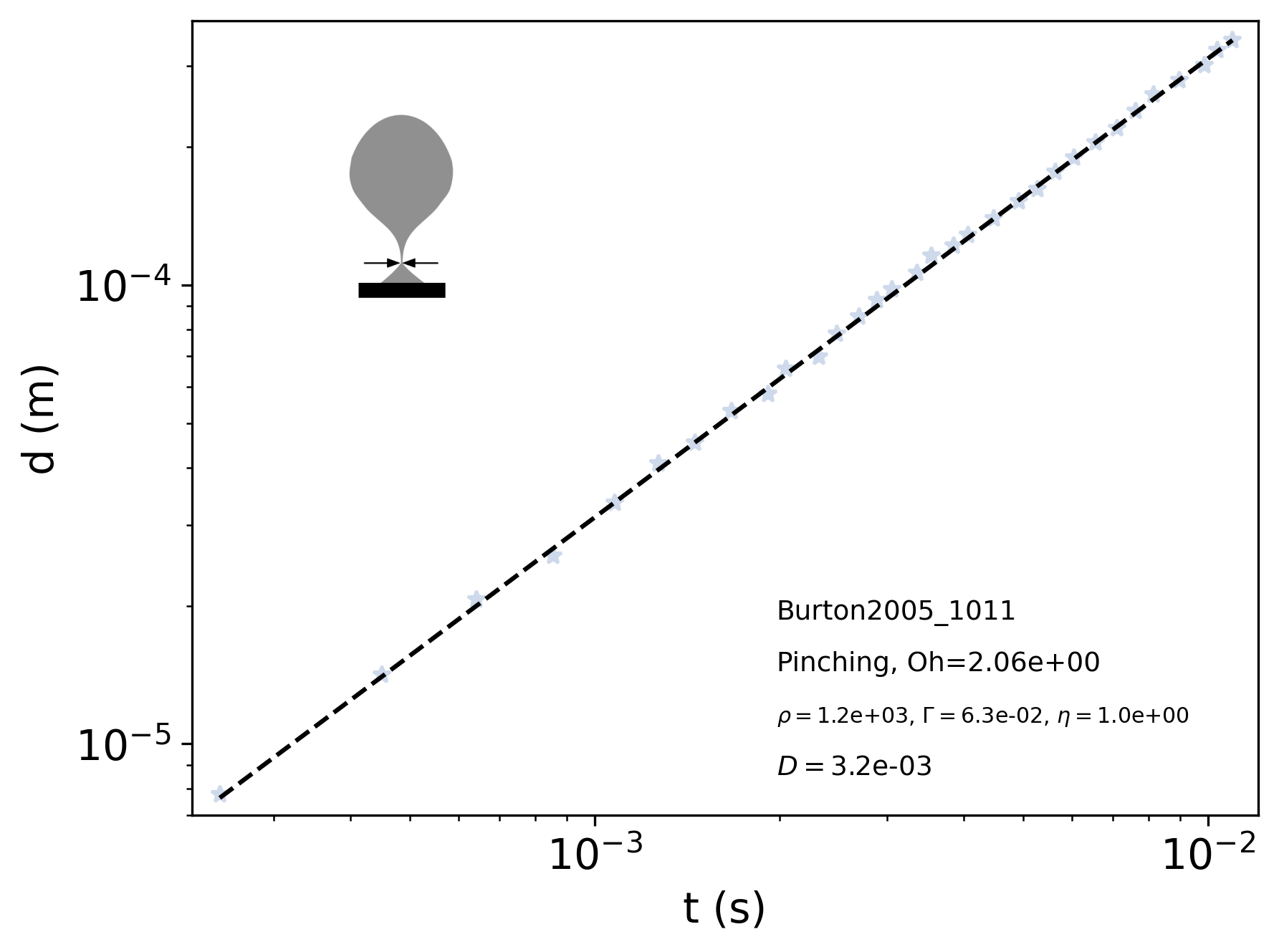} 
 \end{minipage} 
 \\ 
Air bubble in water-glycerol mixture.\newline Density and viscosity are that of the outer fluid. & Air bubble in water-glycerol mixture.\newline Density and viscosity are that of the outer fluid.\\ \hline \hline 
\end{tabular} 
 \end{table} 
\begin{table} 
 \centering 
 \begin{tabular}{ | p{9cm} | p{9cm} | } 
 \hline 
 \textbf{Burton2007 Fig7} & \textbf{Keim2006 Fig2 4p1}  \\ 
 \begin{minipage}{.5\textwidth} 
 \includegraphics[width=\linewidth]{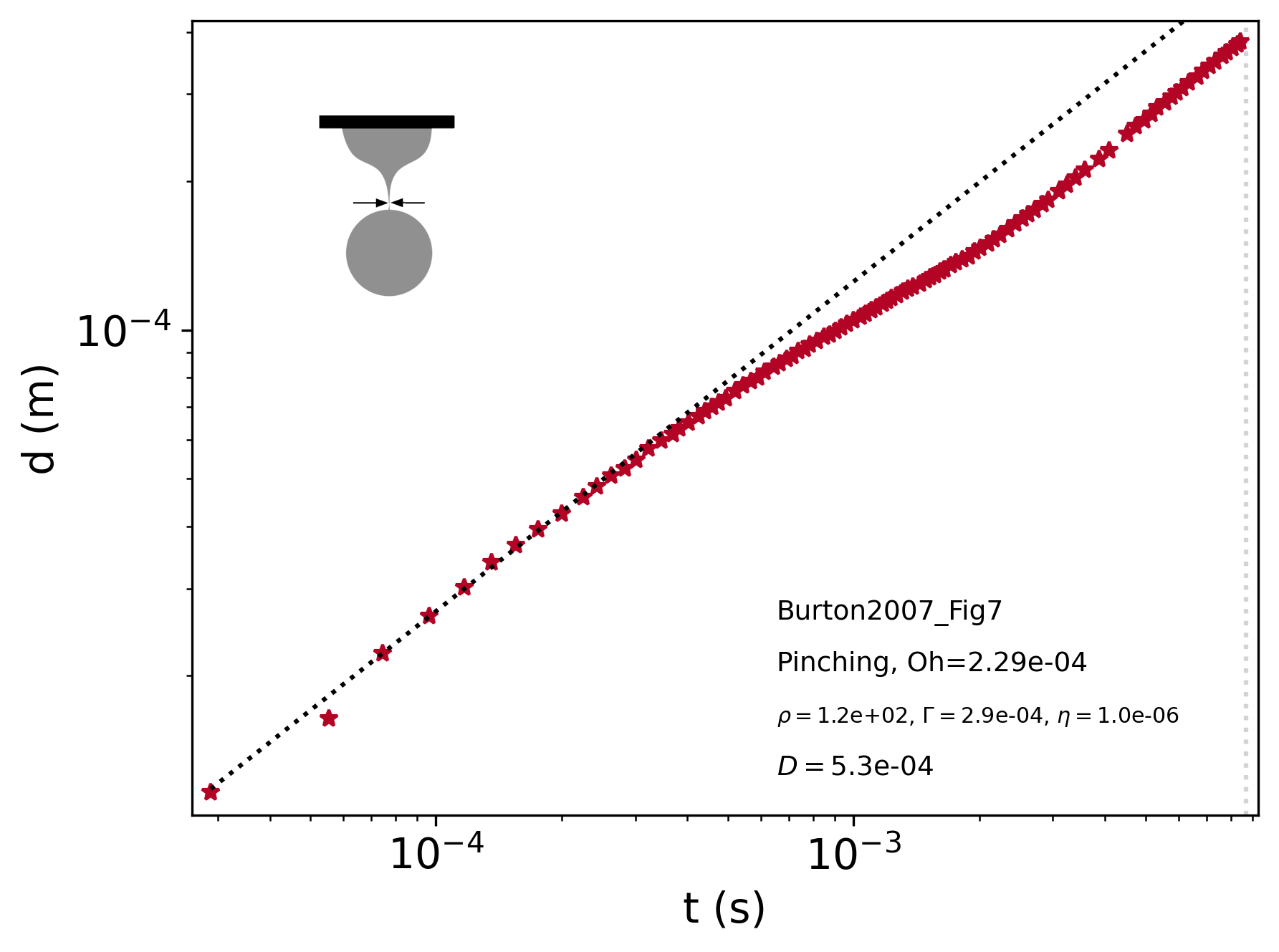} 
 \end{minipage}
 & 
 \begin{minipage}{.5\textwidth} 
 \includegraphics[width=\linewidth]{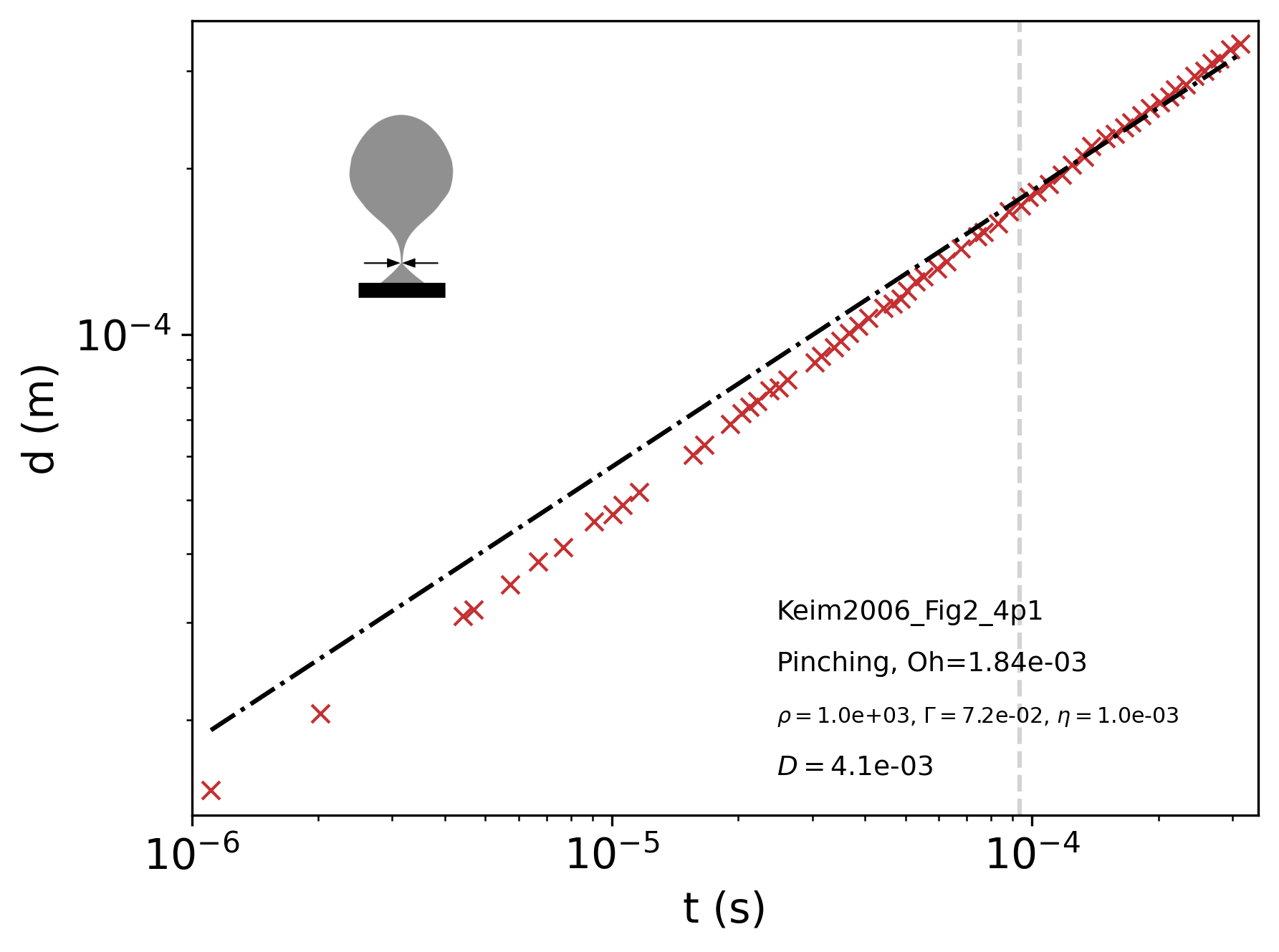} 
 \end{minipage} 
 \\ 
Superfluid $^4\text{He}$ at temperature T=1.34K. \newline No viscosity is available. The artificial value $\eta=10^{-6}$ Pa.s is used in SI-Fig. 2b. & Air bubble in water. \newline Density and viscosity are that of the outer fluid.\\ \hline \hline 
\textbf{Keim2006 Fig2 1p5} & \textbf{Bolanos2009 Fig6 water}  \\ 
 \begin{minipage}{.5\textwidth} 
 \includegraphics[width=\linewidth]{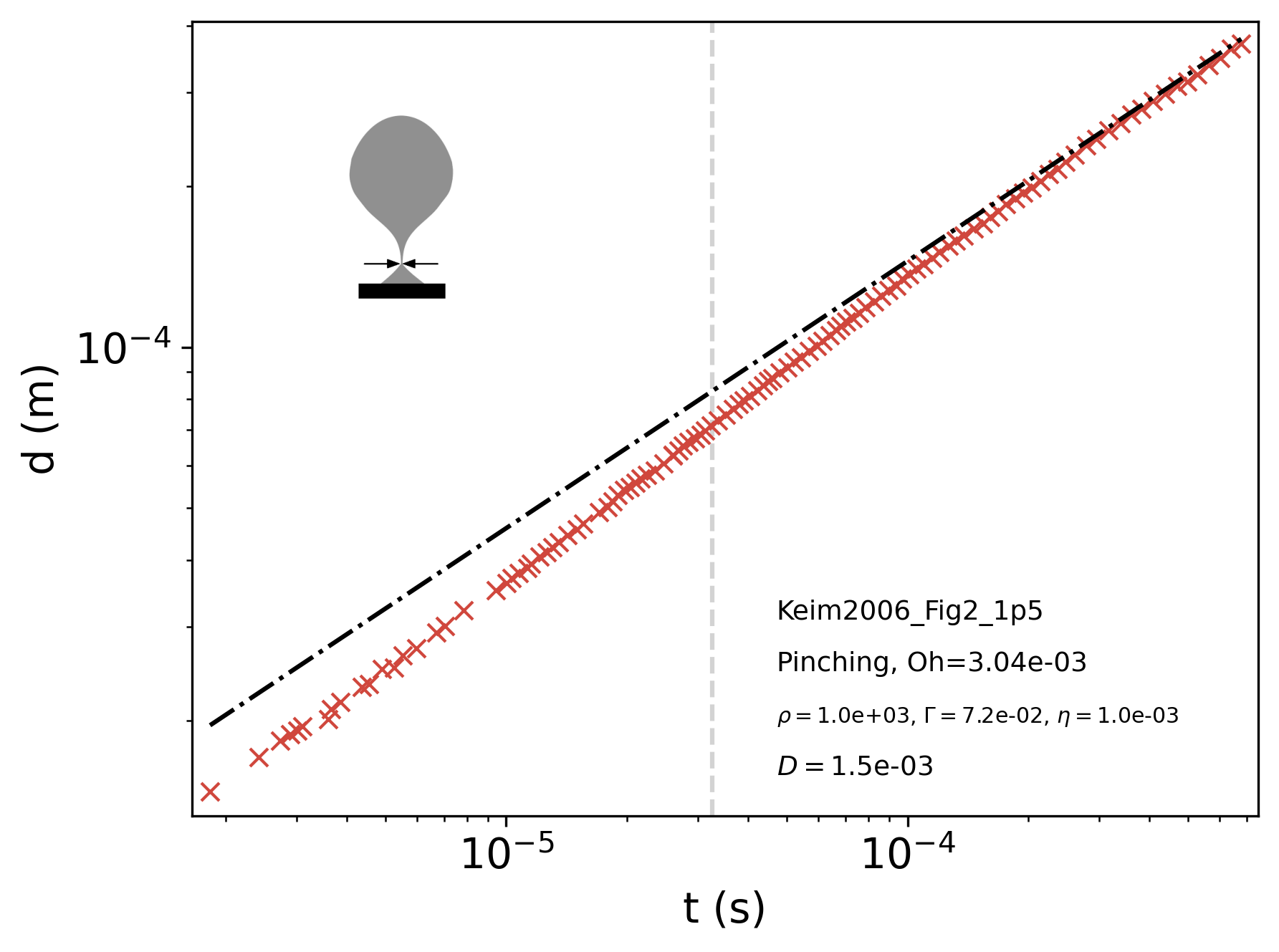} 
 \end{minipage}
 & 
 \begin{minipage}{.5\textwidth} 
 \includegraphics[width=\linewidth]{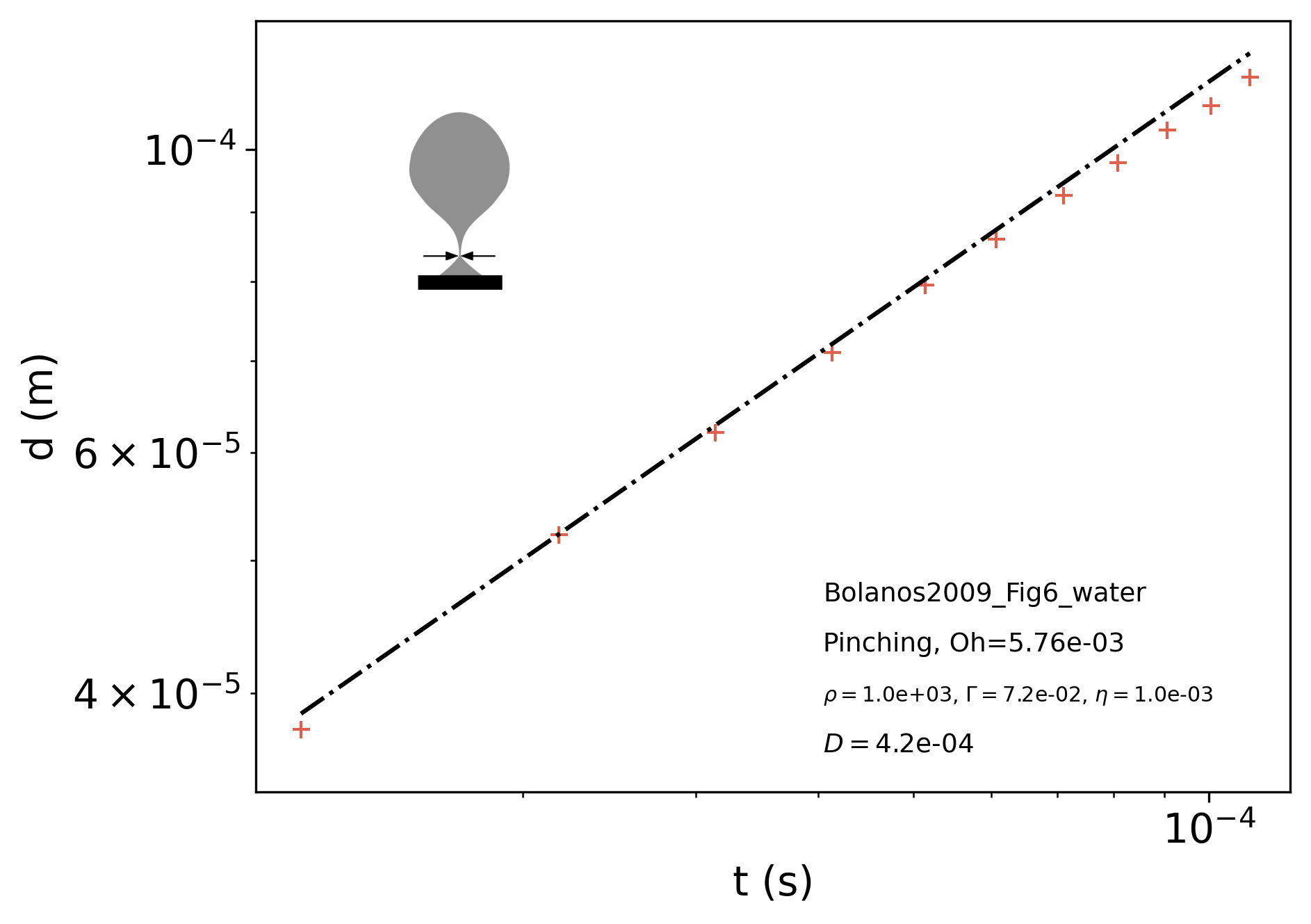} 
 \end{minipage} 
 \\ 
Air bubble in water. \newline Density and viscosity are that of the outer fluid. & Air bubble in water. \newline Density and viscosity are that of the outer fluid.\\ \hline \hline 
\textbf{Bolanos2009 Fig7 water} & \textbf{Bolanos2009 Fig6 O2}  \\ 
 \begin{minipage}{.5\textwidth} 
 \includegraphics[width=\linewidth]{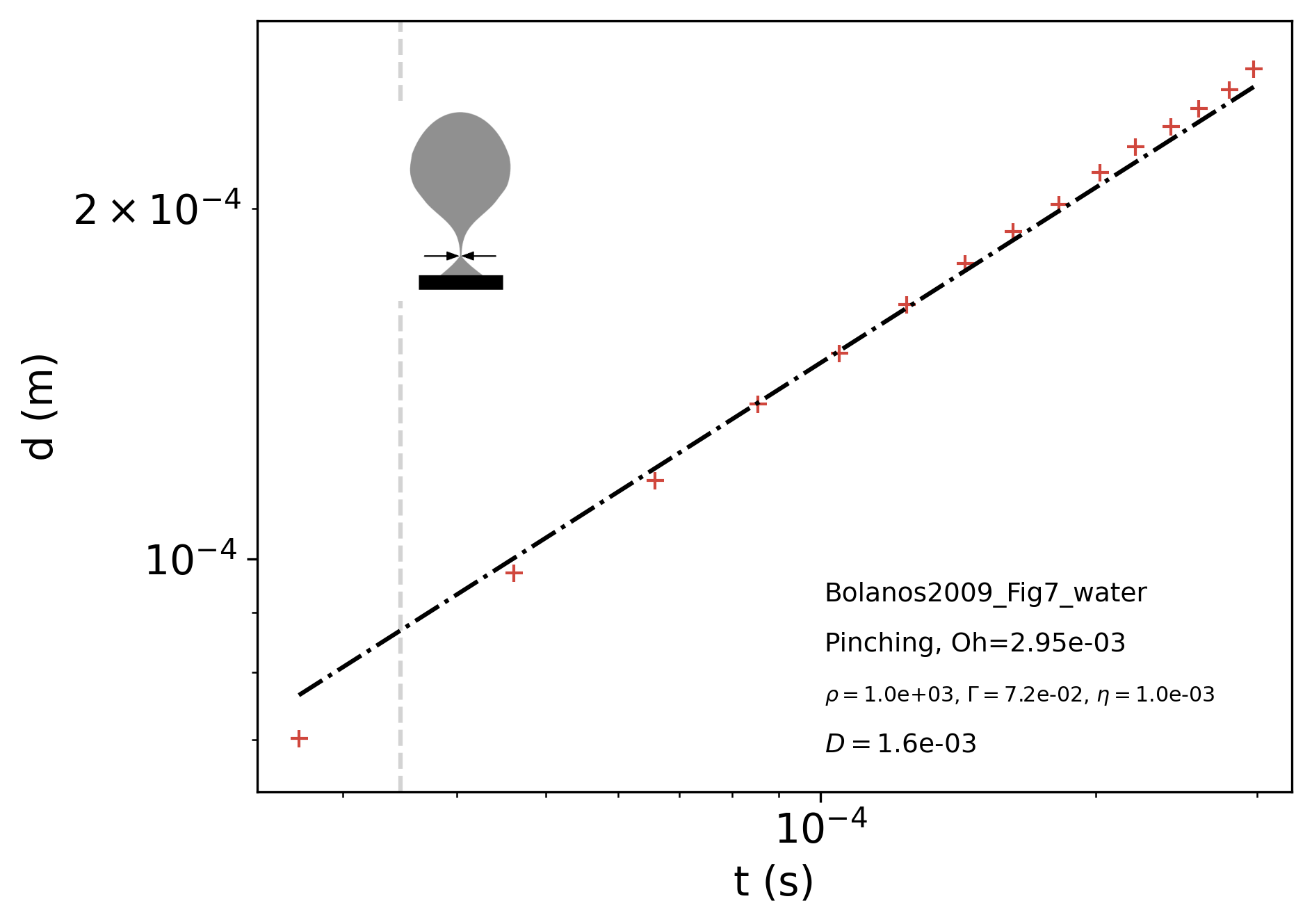} 
 \end{minipage}
 & 
 \begin{minipage}{.5\textwidth} 
 \includegraphics[width=\linewidth]{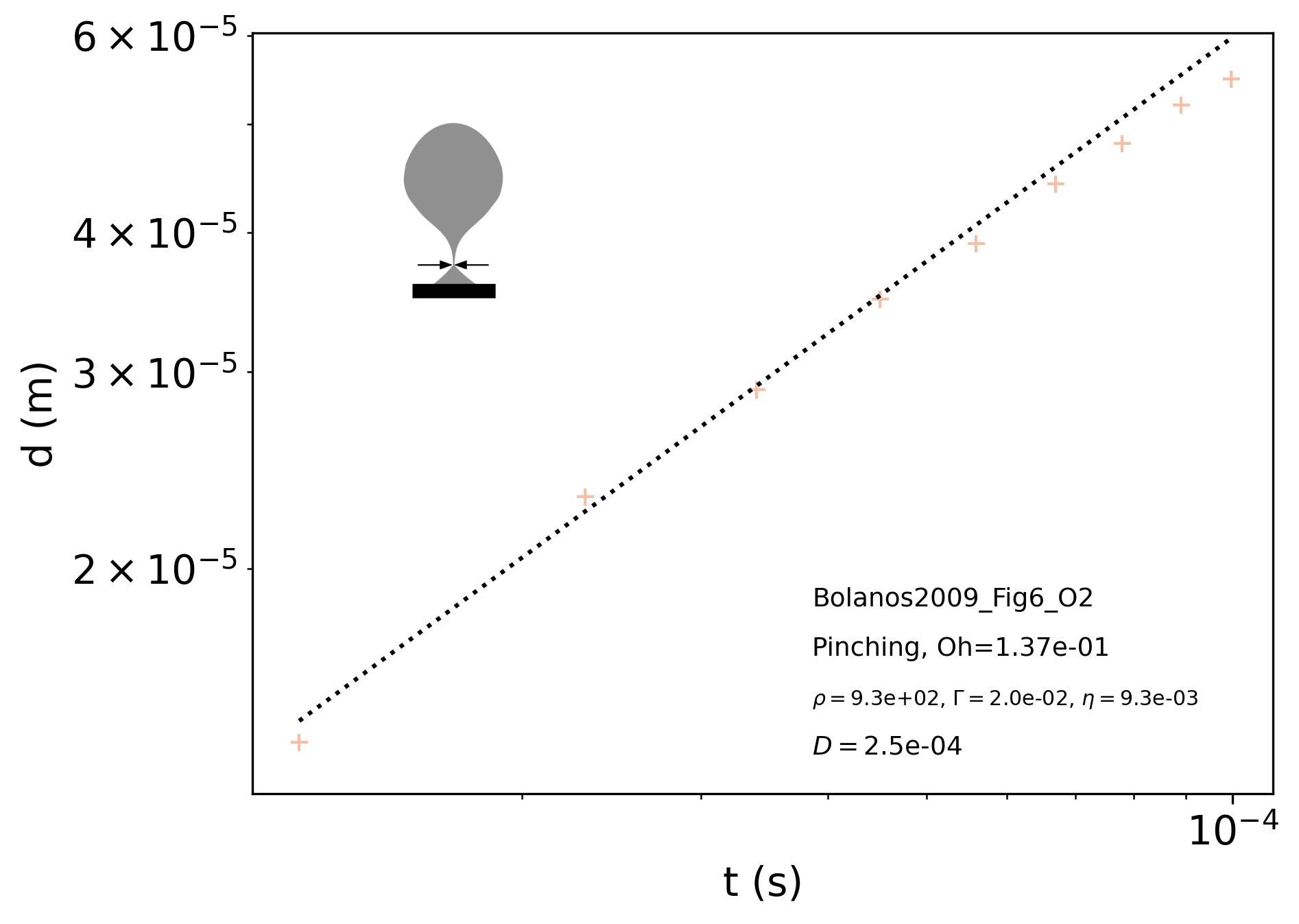} 
 \end{minipage} 
 \\ 
Air bubble in water. \newline Density and viscosity are that of the outer fluid. & Air bubble in silicone oil. \newline Density and viscosity are that of the outer fluid.\\ \hline \hline 
\end{tabular} 
 \end{table} 
\begin{table} 
 \centering 
 \begin{tabular}{ | p{9cm} | p{9cm} | } 
 \hline 
 \textbf{Bolanos2009 Fig7 O7} & \textbf{Bolanos2009 Fig7 O8}  \\ 
 \begin{minipage}{.5\textwidth} 
 \includegraphics[width=\linewidth]{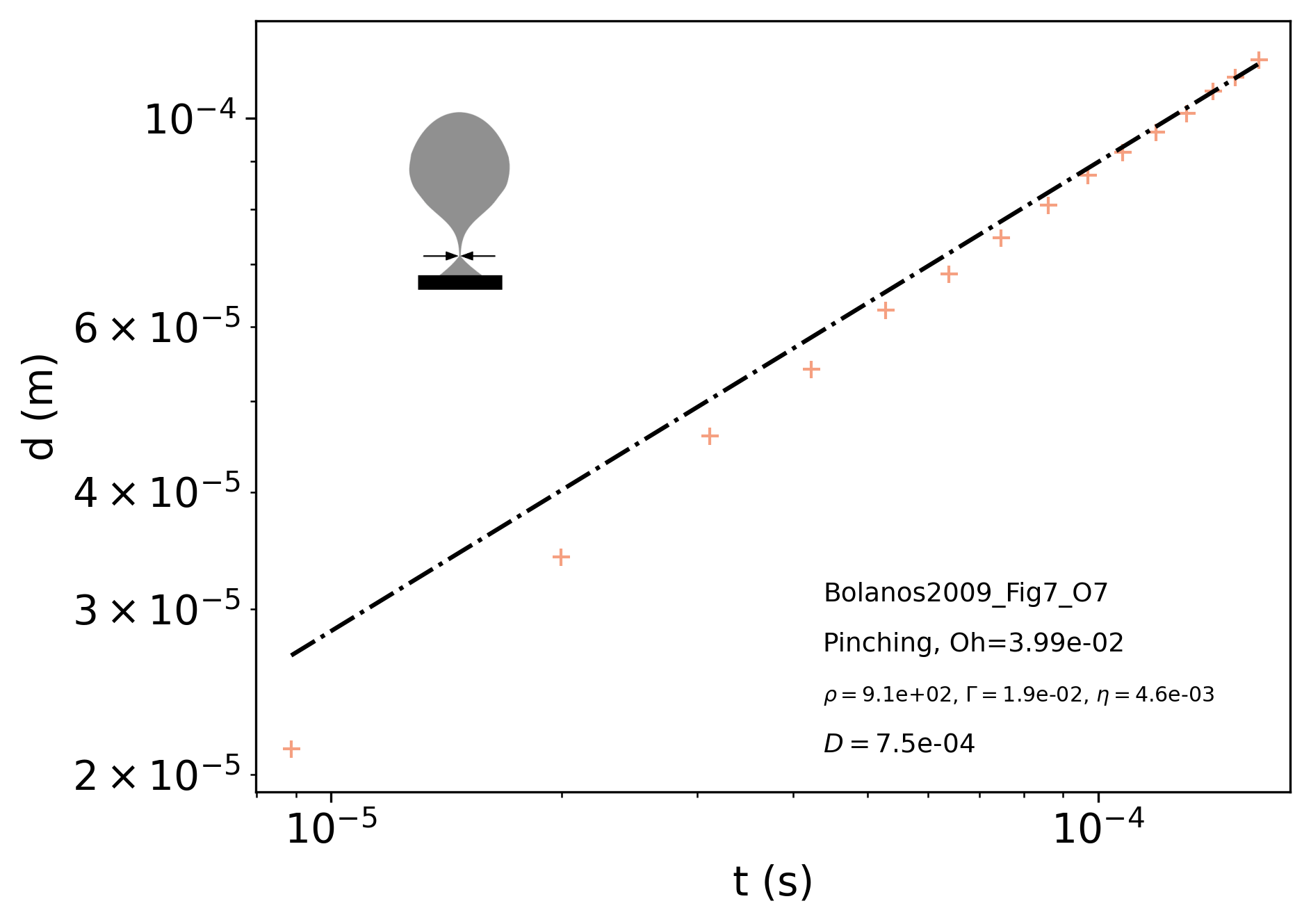} 
 \end{minipage}
 & 
 \begin{minipage}{.5\textwidth} 
 \includegraphics[width=\linewidth]{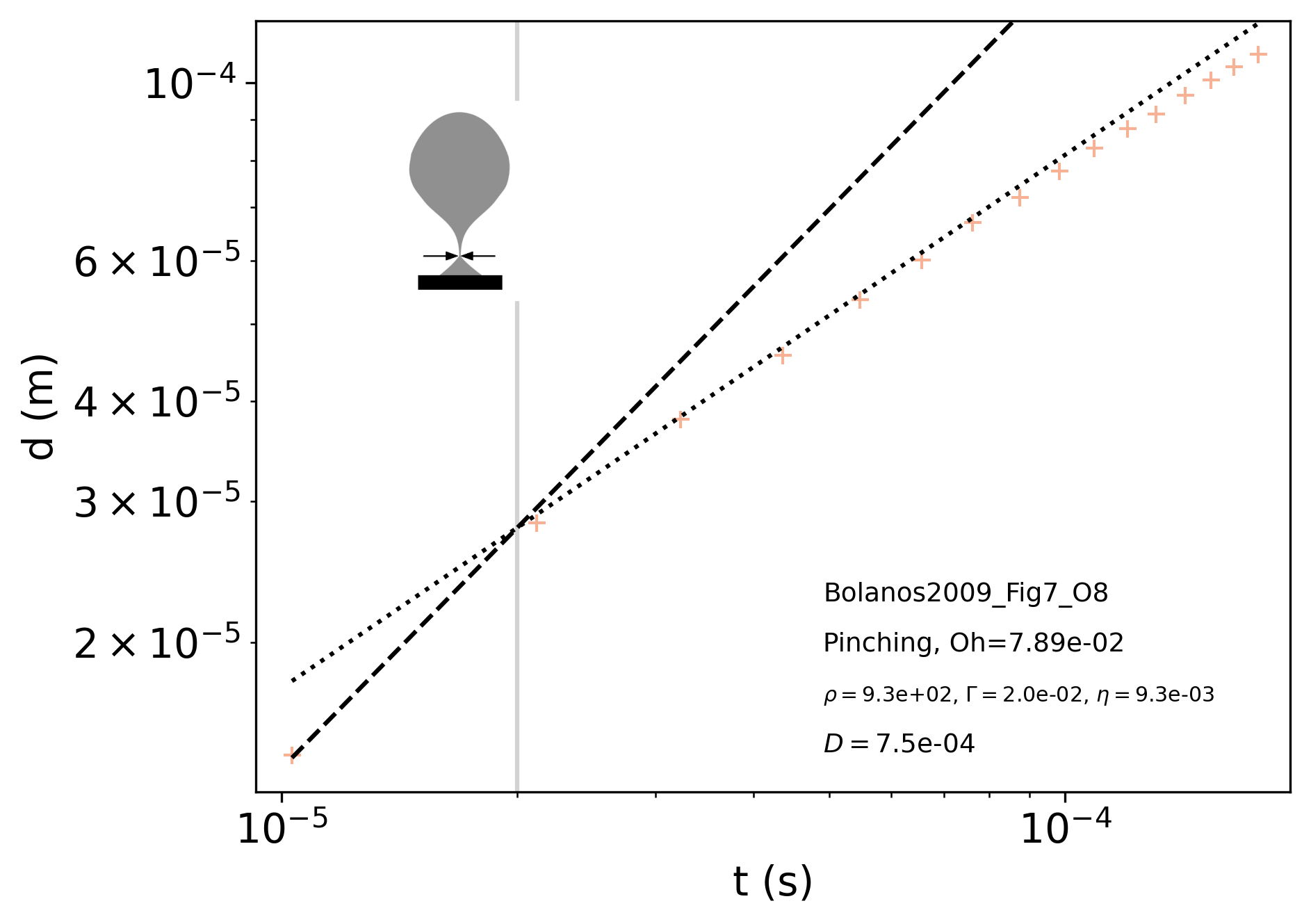} 
 \end{minipage} 
 \\ 
Air bubble in silicone oil. \newline Density and viscosity are that of the outer fluid. & Air bubble in silicone oil. \newline Density and viscosity are that of the outer fluid.\\ \hline \hline 
\textbf{Bolanos2009 Fig7 O9} & \textbf{Bolanos2009 Fig8 G1}  \\ 
 \begin{minipage}{.5\textwidth} 
 \includegraphics[width=\linewidth]{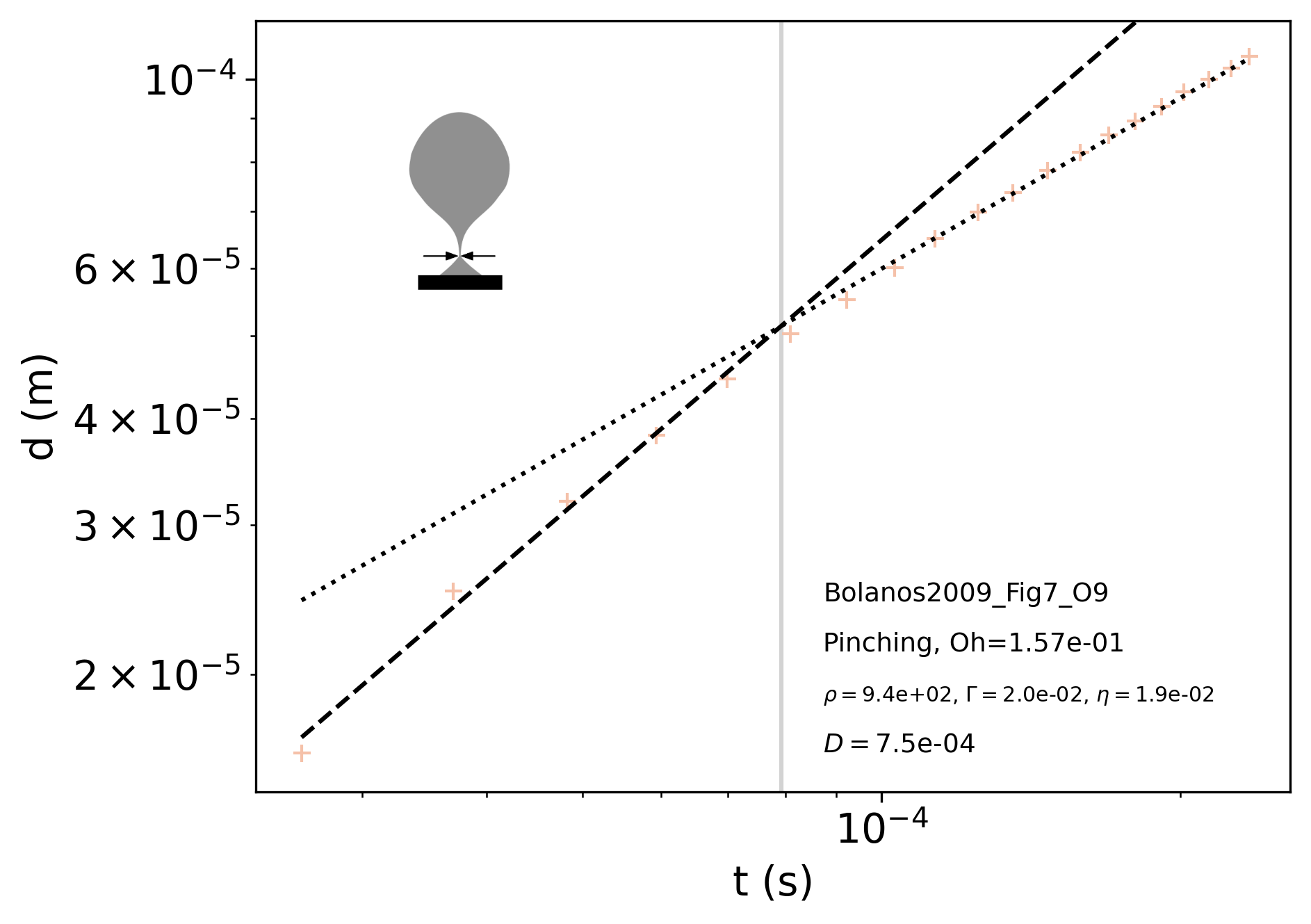} 
 \end{minipage}
 & 
 \begin{minipage}{.5\textwidth} 
 \includegraphics[width=\linewidth]{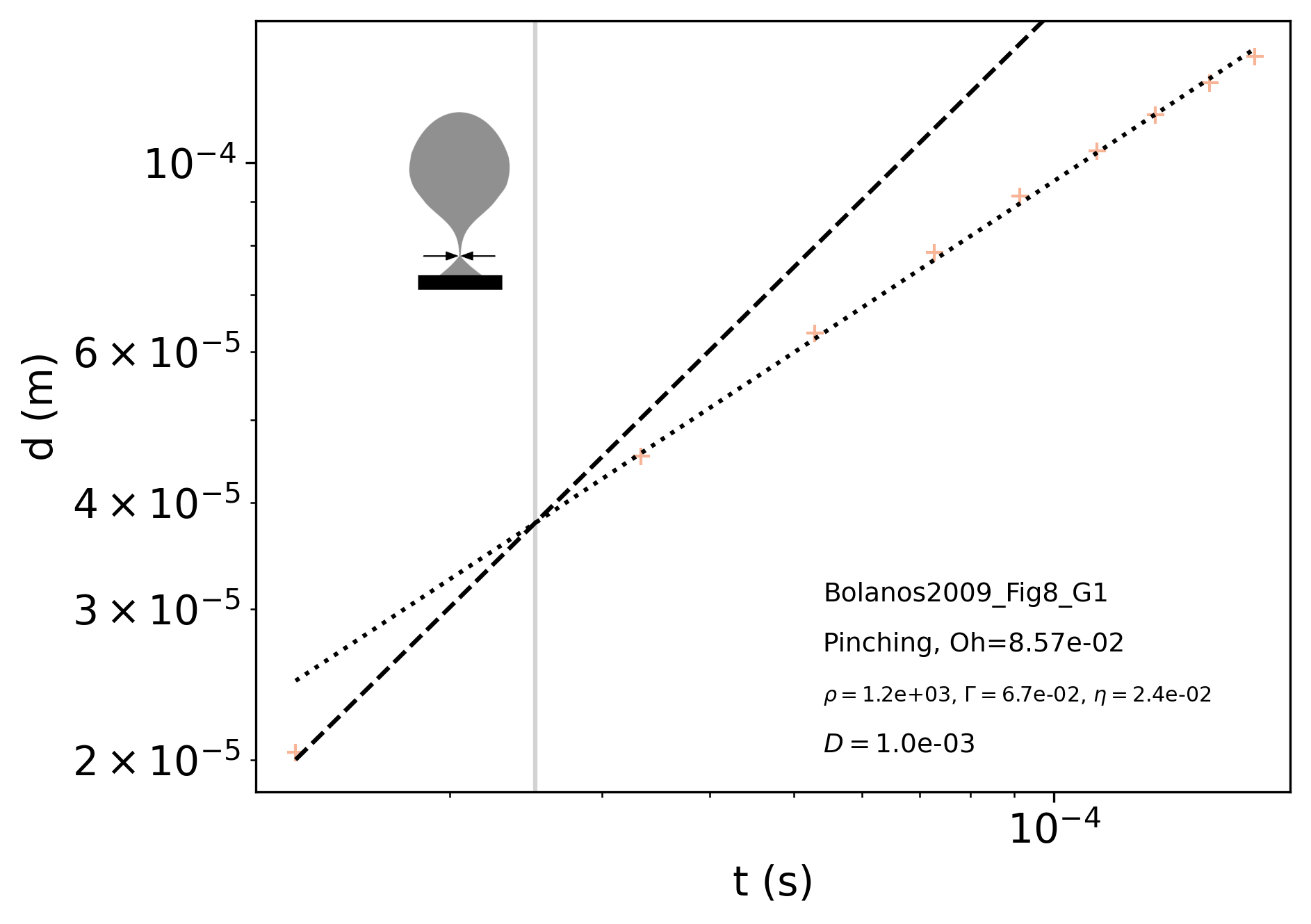} 
 \end{minipage} 
 \\ 
Air bubble in silicone oil. \newline Density and viscosity are that of the outer fluid. & Air bubble in water-glycerol mixture. \newline Density and viscosity are that of the outer fluid.\\ \hline \hline 
\textbf{Bolanos2009 Fig8 G2} & \textbf{Bolanos2009 Fig8 G4}  \\ 
 \begin{minipage}{.5\textwidth} 
 \includegraphics[width=\linewidth]{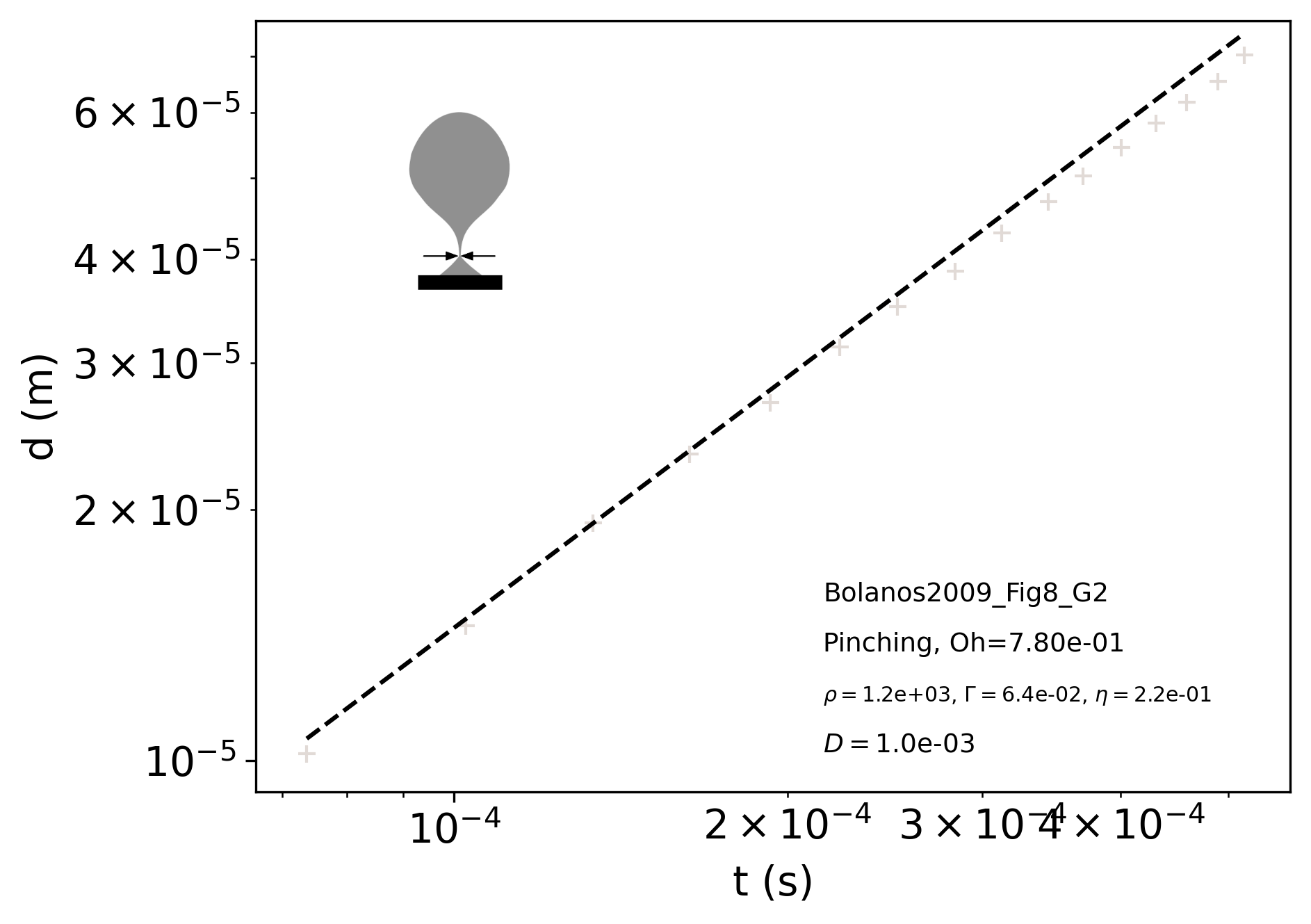} 
 \end{minipage}
 & 
 \begin{minipage}{.5\textwidth} 
 \includegraphics[width=\linewidth]{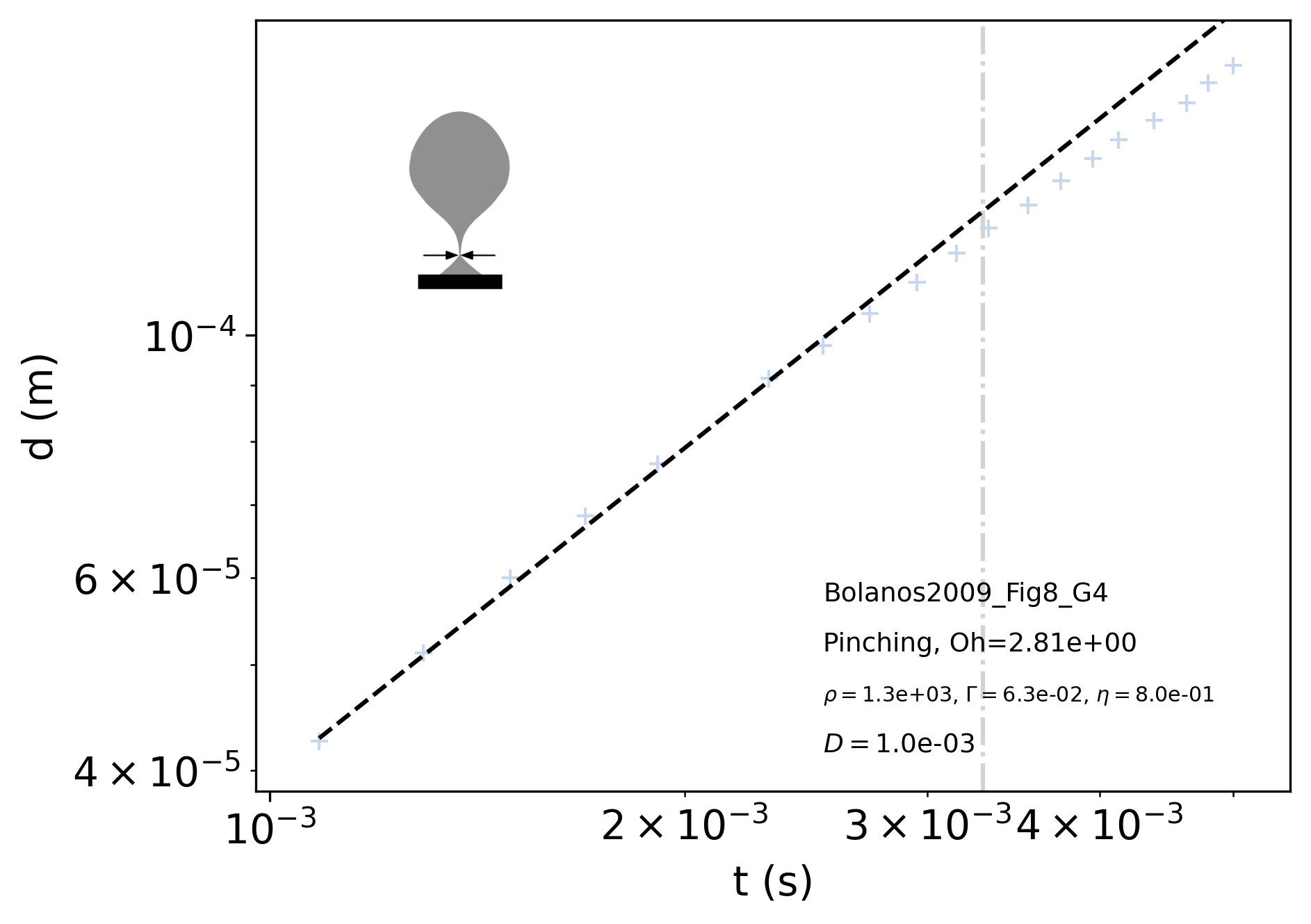} 
 \end{minipage} 
 \\ 
Air bubble in water-glycerol mixture. \newline Density and viscosity are that of the outer fluid. & Air bubble in water-glycerol mixture. \newline Density and viscosity are that of the outer fluid.\\ \hline \hline 
\end{tabular} 
 \end{table} 
\begin{table} 
 \centering 
 \begin{tabular}{ | p{9cm} | p{9cm} | } 
 \hline 
 \textbf{Bolanos2009 Fig9 G5} & \textbf{Bolanos2009 Fig9 G6}  \\ 
 \begin{minipage}{.5\textwidth} 
 \includegraphics[width=\linewidth]{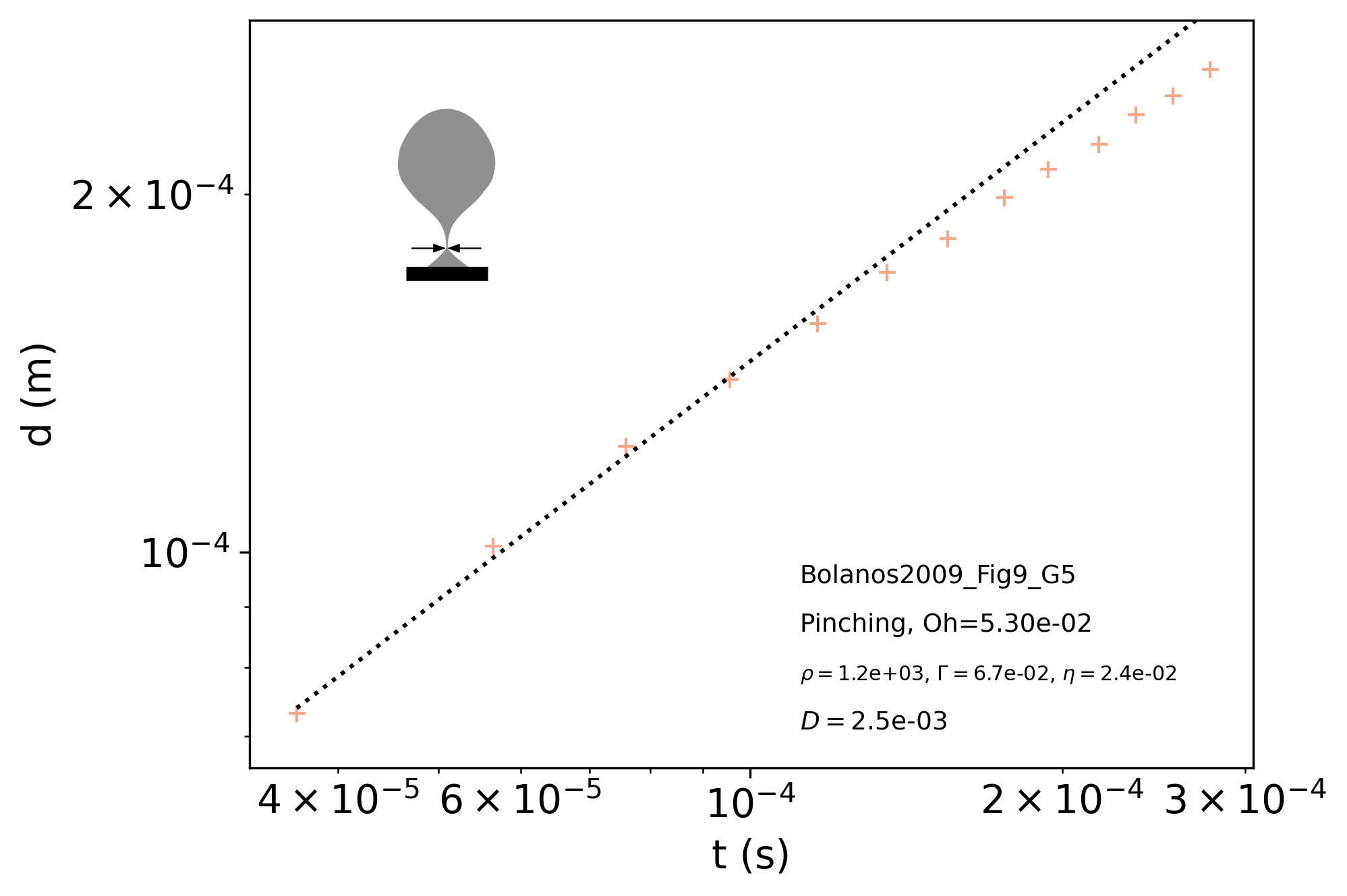} 
 \end{minipage}
 & 
 \begin{minipage}{.5\textwidth} 
 \includegraphics[width=\linewidth]{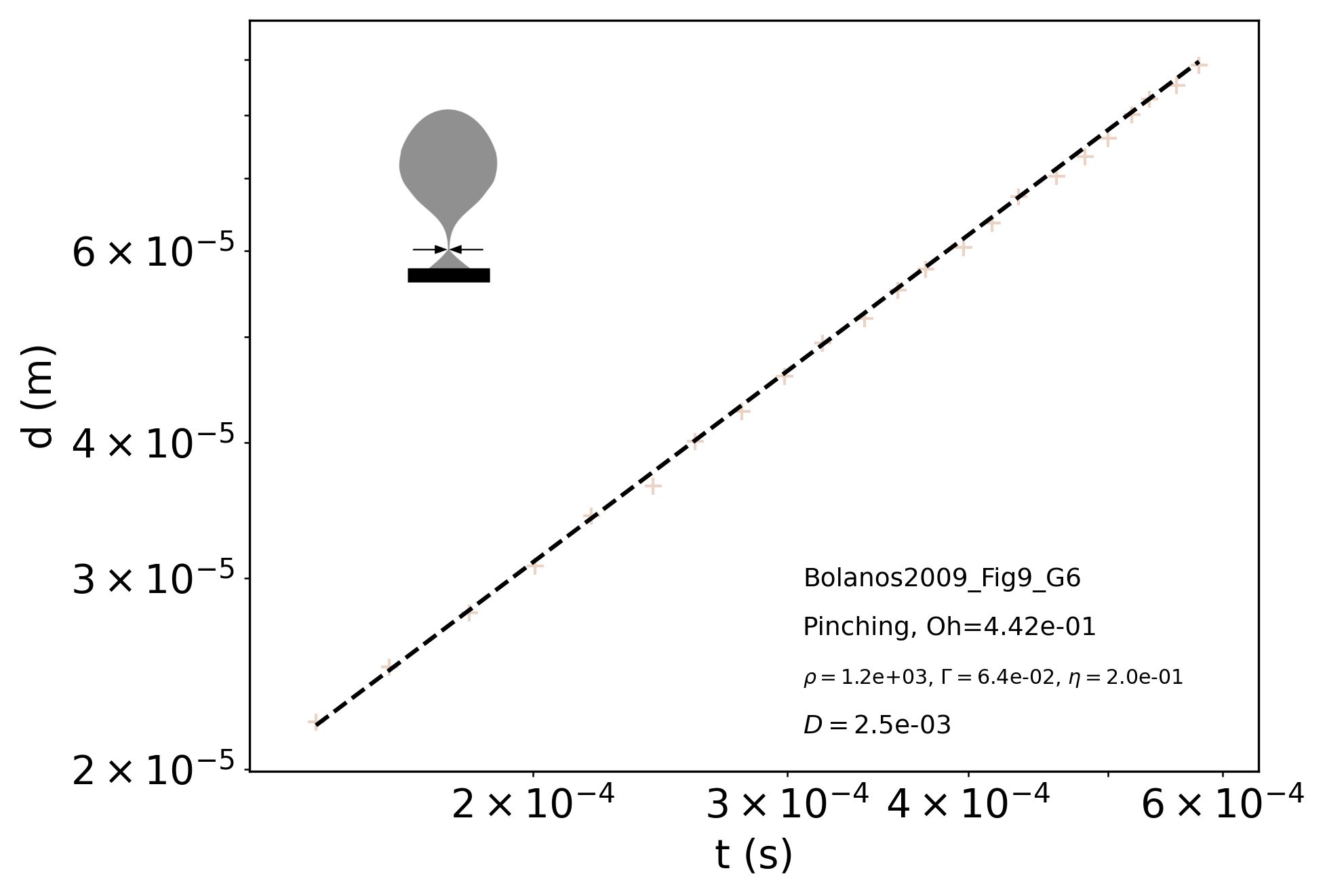} 
 \end{minipage} 
 \\ 
Air bubble in water-glycerol mixture. \newline Density and viscosity are that of the outer fluid. & Air bubble in water-glycerol mixture. \newline Density and viscosity are that of the outer fluid.\\ \hline \hline 
\textbf{Bolanos2009 Fig6 O1} & \textbf{Bolanos2009 Fig6 O3}  \\ 
 \begin{minipage}{.5\textwidth} 
 \includegraphics[width=\linewidth]{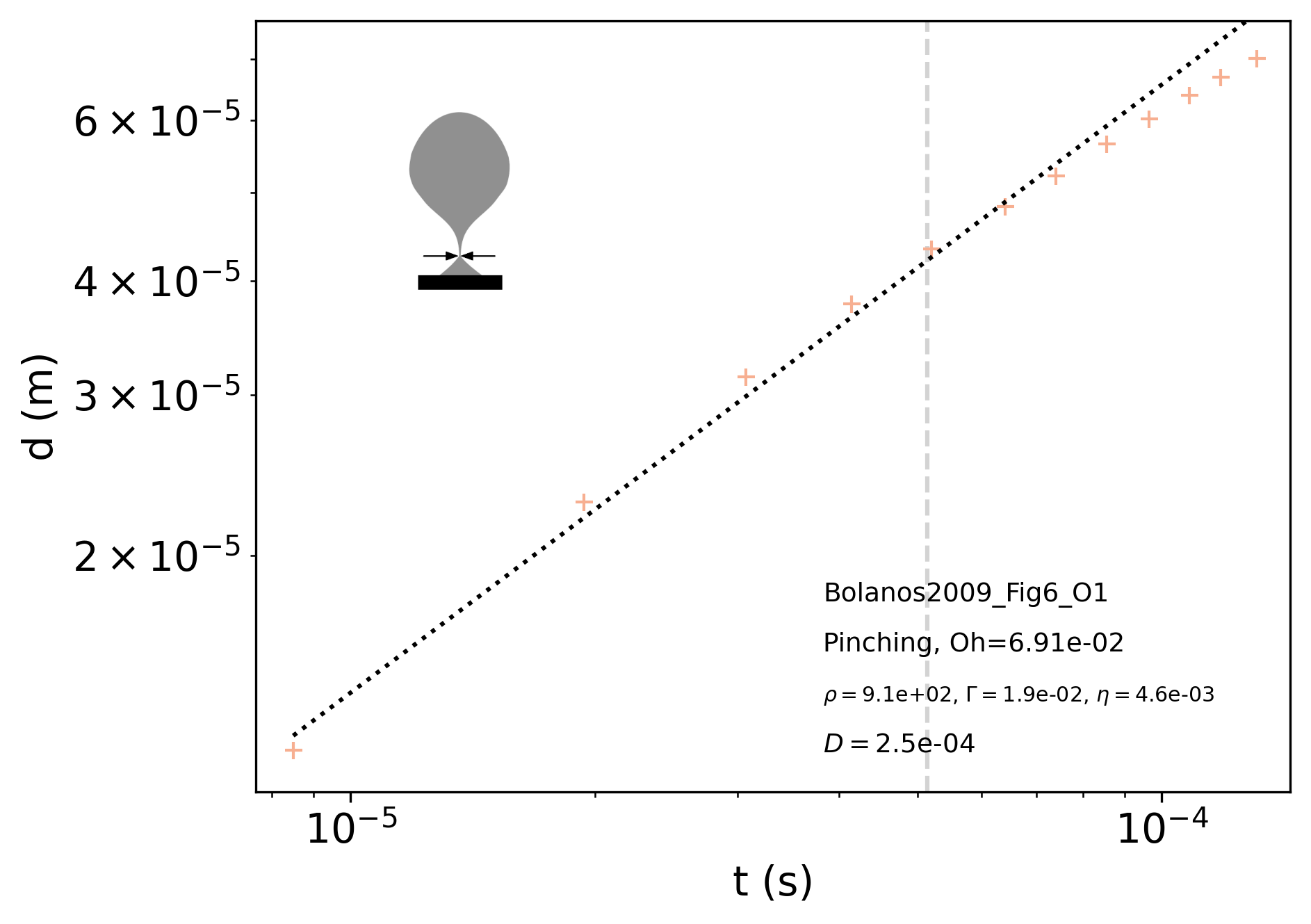} 
 \end{minipage}
 & 
 \begin{minipage}{.5\textwidth} 
 \includegraphics[width=\linewidth]{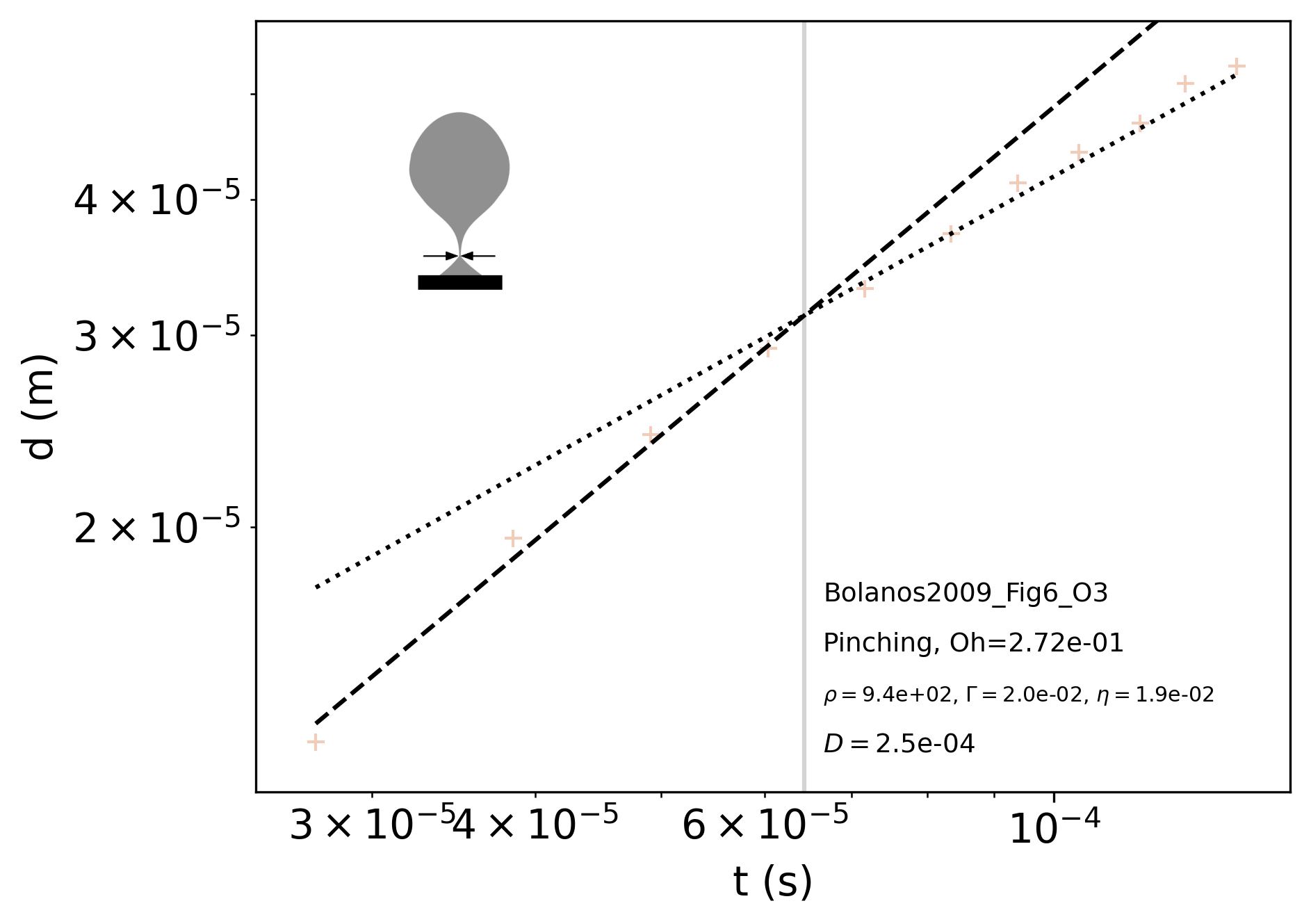} 
 \end{minipage} 
 \\ 
Air bubble in water-glycerol mixture. \newline Density and viscosity are that of the outer fluid. & Air bubble in water-glycerol mixture. \newline Density and viscosity are that of the outer fluid.\\ \hline \hline 
\textbf{Bolanos2009 Fig9 G8} & \textbf{Goldstein2010 Fig5}  \\ 
 \begin{minipage}{.5\textwidth} 
 \includegraphics[width=\linewidth]{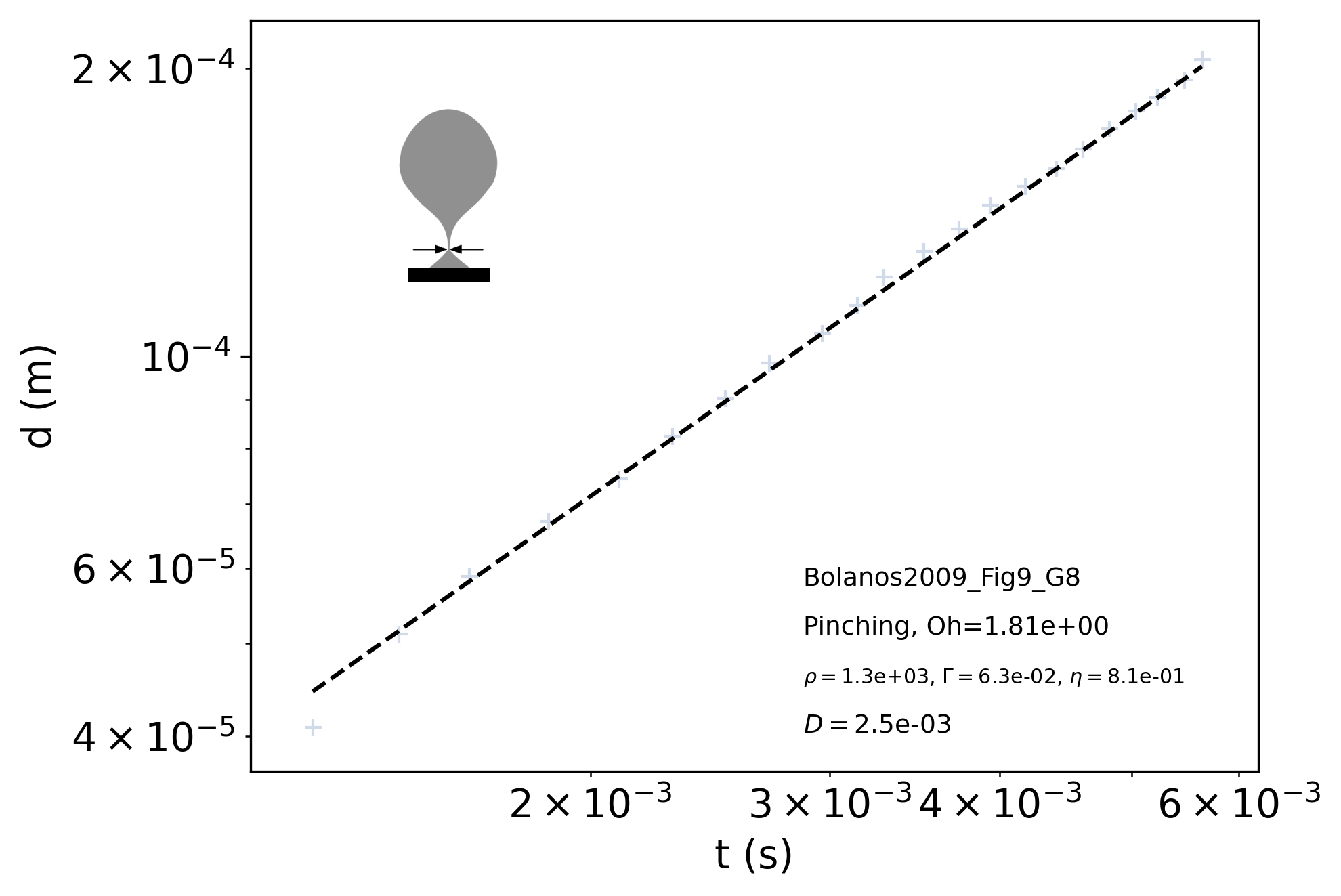} 
 \end{minipage}
 & 
 \begin{minipage}{.5\textwidth} 
 \includegraphics[width=\linewidth]{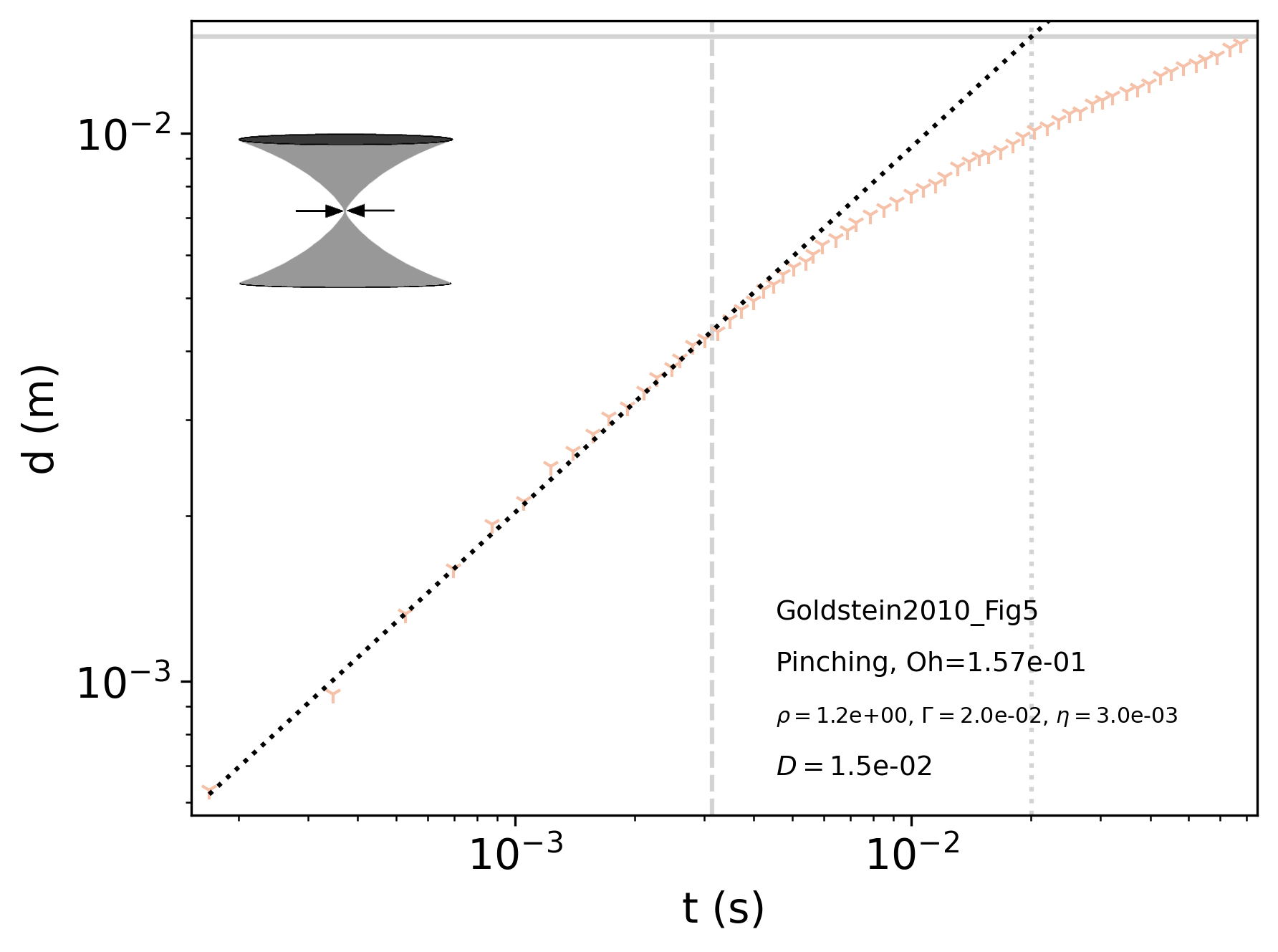} 
 \end{minipage} 
 \\ 
Air bubble in water-glycerol mixture. \newline Density and viscosity are that of the outer fluid. & Soap film on Mobius strip. \newline Values of material parameters estimated since not provided. \newline Density is that of air, viscosity is that of soap film.\\ \hline \hline 
\end{tabular} 
 \end{table}